\documentclass[twoside,12pt]{article}
\usepackage{epsfig}
\usepackage{setspace}
\usepackage{amssymb}
\usepackage{amsmath}
\usepackage{graphicx}
\usepackage{amsfonts}
\usepackage{braket}
\usepackage{xcolor}
\usepackage{bm}
\usepackage[normalem]{ulem}
\usepackage{import}
\usepackage{lipsum}
\usepackage{mathtools, cuted}
\usepackage{tikz}
\usepackage{lipsum}

\newcommand{\be}{\begin{equation}}
\newcommand{\ee}{\end{equation}}
\newcommand{\bea}{\begin{eqnarray}}
\newcommand{\eea}{\end{eqnarray}}

\newcommand{\vecrp}{\vec{r}\mkern2mu\vphantom{r}'}

\newcommand{\vecK}{{\vec K}}

\newcommand{\vecR}{{\vec R}}
\newcommand{\vecRp}{\vec{R}\mkern2mu\vphantom{R}'}

\newcommand{\nuc}[2]{\hbox{$^{#1}$#2}}

\newcommand\copyrighttext{ \footnotesize \textcopyright 2020. Licensed under the Creative Commons CC-BY-NC-ND 4.0 license http://creativecommons.org/licenses/by-nc-nd/4.0/.}
\newcommand\copyrightnotice{
\begin{tikzpicture}[remember picture,overlay]
\node[anchor=south,yshift=10pt] at (current page.south) {\fbox{\parbox{\dimexpr\textwidth-\fboxsep-\fboxrule\relax}{\copyrighttext}}};
\end{tikzpicture}%
}

\begin{document}

\title{\vspace{1cm}Quenching of single-particle strength from direct reactions with stable and rare-isotope beams}
\author{T. Aumann,$^{1,2}$ C. Barbieri,$^{3,4,5}$ D. Bazin,$^{6,7}$ C. A. Bertulani,$^8$ A. Bonaccorso,$^9$ \\
W. H. Dickhoff,$^{10}$ A. Gade,$^{6,7}$ M. G\'omez-Ramos,$^{1,11}$ B. P. Kay,$^{12}$ A. M. Moro,$^{11,13}$\\ 
T. Nakamura,$^{14}$ A. Obertelli,$^1$ K. Ogata,$^{15,16}$ S. Paschalis,$^{17}$ T. Uesaka,$^{18}$\\
\\
\small $^1$Technische Universität Darmstadt, Fachbereich Physik, Institut für Kernphysik, 64289 Darmstadt, Germany\\
\small $^2$ GSI Helmholtzzentrum f\"ur Schwerionenforschung, 64289 Darmstadt, Germany\\
\small $^3$ Department of Physics, University of Surrey, Guildford GU2 7XH, United Kingdom\\
\small $^4$ Dipartimento di Fisica, Università degli Studi di Milano, Via Celoria 16, I-20133 Milano, Italy\\
\small $^5$ INFN, Sez. di Milano, Via Celoria 16, I-20133 Milano, Italy\\
\small $^6$National Superconducting Cyclotron Laboratory, Michigan State University, East Lansing, MI 48824, USA\\ 
\small $^7$Department of Physics \& Astronomy, Michigan State University, East Lansing, MI 48824, USA\\
\small $^8$Department of Physics and Astronomy, Texas A\&M University-Commerce, Commerce, Texas  75429, USA\\
\small $^9$INFN, Sez. di Pisa, Largo B. Pontecorvo 3, 56127 Pisa, Italy\\
\small $^{10}$Department of Physics, Washington University, St. Louis, MO 63130, USA\\
\small $^{11}$Departamento de F\'{\i}sica At\'omica, Molecular y Nuclear, \\
\small Facultad de F\'{\i}sica, Universidad de Sevilla, Apartado 1065, E-41080 Sevilla, Spain\\
\small $^{12}$Physics Division, Argonne   National Laboratory, Argonne, IL 60439, USA\\
\small $^{13}$Instituto Interuniversitario Carlos I de F\'isica Te\'orica y Computacional (iC1), Apdo.~1065, E-41080 Sevilla, Spain\\
\small $^{14}$ Department of Physics, Tokyo Institute of Technology, 2-12-1 O-Okayama, Meguro, Tokyo, 152-8551, Japan\\
\small $^{15}$ Research Center for Nuclear Physics (RCNP), Osaka University, Ibaraki 567-0047, Japan\\
\small $^{16}$ Department of Physics, Osaka City University, Osaka 558-8585, Japan\\
\small $^{17}$Department of Physics, University of York, York,  YO10 5DD, UK\\
\small $^{18}$ RIKEN Nishina Center, 2-1 Hirosawa, Wako, Saitama 351-0198, Japan\\
}

\maketitle

\copyrightnotice

\begin{abstract}
In this review article we discuss the present status of direct nuclear reactions and the nuclear structure aspects one can study with them. We discuss the spectroscopic information we can assess in experiments involving transfer reactions, heavy-ion-induced knockout reactions and quasifree scattering with $(p,2p)$, $(p,pn)$, and $(e,e'p)$ reactions. In particular, we focus on the proton-to-neutron asymmetry of the quenching of the spectroscopic strength.
\end{abstract}

\section{Introduction}
\label{sec1}
The structure of nuclei and its evolution with the number of constituent nucleons, in particular with neutron-to-proton asymmetry, has been a fascinating and multifaceted area in nuclear physics for a long time. It has become even more so recently due to the vast advances in new and planned accelerator facilities and experimental techniques as well as in nuclear theory. While these recent developments permitted already great advances in our understanding \cite{hebeler2015,otsuka2020}, a global picture of nuclear structure and the driving mechanisms responsible for the observed phenomena remain elusive. Major efforts are being made to realize the development of new or upgraded rare-isotope facilities such as the RIBF \cite{sakurai2010}, FRIB \cite{gade2016}, FAIR \cite{aumann2007}, ISOLDE \cite{borge2016}, GANIL-SPIRAL2 \cite{gales2011}, ARIEL \cite{dilling2014}, and SPES \cite{andrighetto2018}, and new experimental techniques that allow for spectroscopic studies of nuclei in unexplored regions of the nuclear landscape. In parallel, new theoretical techniques enabled the calculations of nuclear states from an {\it ab initio} point of view reaching medium-mass nuclei (see , for example, Ref. \cite{hagen2010,soma2014,hergert2016}). Such advances in both the laboratory and theoretical modeling have a direct impact on our understanding of astrophysical nucleosynthesis and compact objects of the universe.

Direct reactions have been one of the key experimental probes to build up our understanding of the nuclear shell structure since the  1950s (see, {\it e.g.}, Ref.~\cite{Burrows50}). At incident energies close to the Fermi momentum, nucleon-transfer reactions can populate particle and hole states via nucleon-adding and nucleon-removing reactions, respectively. At higher energies, nucleons, assumed to be quasifree, can be knocked out in nucleus-nucleus, proton-nucleus, or electron-nucleus collisions (a comprehensive review is published in  \cite{Jacob:1966}). The measured cross sections to individual final states give access to  single-particle ($s.p.$) configuration mixing and identify the main configurations of the wavefunction quantified traditionally through spectroscopic factors and $s.p.$ energies.

\begin{figure}[!ht]
\begin{center}
\includegraphics[trim=0cm 4cm 0cm 0cm,clip,width=0.43\textwidth]{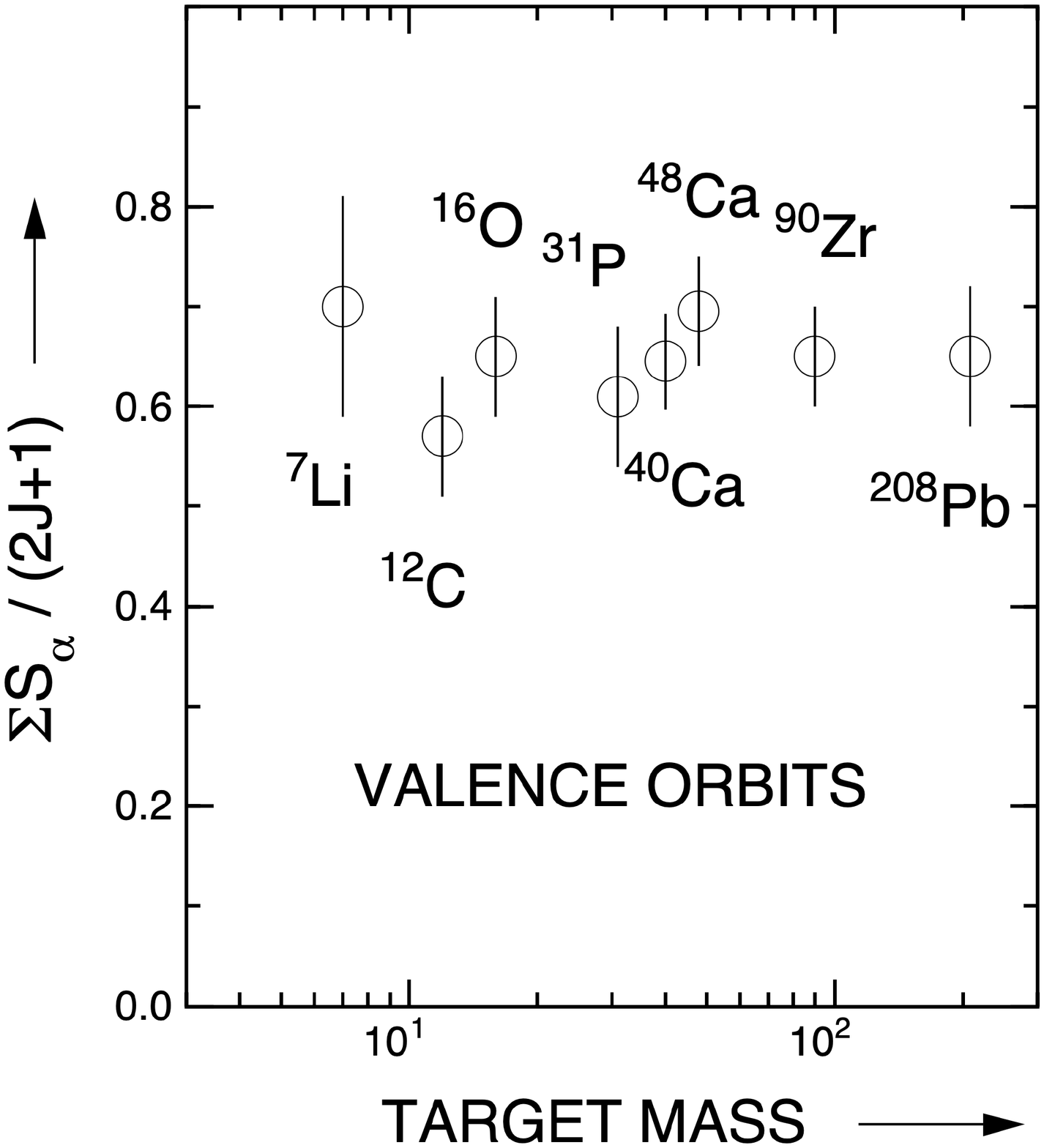}
\caption{\label{fig:louksf_intro} Normalized spectroscopic factors from the $(e,e'p)$ reaction as a function of target mass.
The independent-particle model predicts a normalized value of 1. Reprinted from Ref.~\cite{Lapikas:1993}, with permission from Elsevier.}
\end{center}
\end{figure}

Figure \ref{fig:louksf_intro} shows a summary of spectroscopic factors (SF) extracted from electron-induced ($e,e'p$) knockout reactions on different stable nuclei spanning a wide mass range~\cite{Lapikas:1993}. A reduction of the spectroscopic $s.p.$ strength to around 65\% relative to the independent particle model has been found for all stable nuclei. The reduction is generally understood to originate about equally from long-range correlations (LRC) and short-range correlations (SRC) beyond the pairing correlations around the Fermi surface taken into account in the independent-particle shell model.

The nuclear shell structure and the so-called  $s.p.$ energies, {\it i.e.}, the energy orbital sequence at the mean-field level, can be extracted from the excitation energy of physical (correlated) states and the measurement of nucleon removal and addition cross sections, following the Baranger sum rule~\cite{baranger1970}. Although  $s.p.$ energies are framework dependent and not strictly observables~\cite{furnstahl2002,duguet2012}, they are accepted as meaningful quantities in a specified theoretical framework \cite{furnstahl2010}. 
Removal energies as encountered in high-resolution direct-reaction experiments to discrete final states are observable and are associated with the results of the $(e,e'p)$ experiments shown in Fig.~\ref{fig:louksf_intro}.
Different reactions are sensitive to different parts of the radial overlap functions between the initial and final states, though. However, experiments using different reactions and/or beam energies should lead, after reaction-model-based interpretations, to consistent nuclear-structure information, and this for all nuclei, spanning over a wide range of nuclear binding energies, angular momentum, and neutron-to-proton asymmetries. 

Most of the reaction models are based on the assumption that the structure information necessary to predict cross sections is contained in an overlap function between projectile/target and residue, quantified either by Spectroscopic Factors (SFs), Asymptotic Normalization Coefficients (ANCs), or Electromagnetic Matrix (EM) elements. A reaction is called direct if it proceeds directly from the initial to the final state without the formation of an intermediate compound nucleus, and therefore can be factorized into a nuclear-structure term, quoted $C^2S$ in the following (the notation will be explained later in the review), and a reaction term $\sigma_{\text{sp}}$ corresponding to the cross section in case of  $s.p.$ states. The factorization can be expressed as
\begin{equation}
\label{factorization}
\sigma = C^2S \sigma_{\text{sp}}.
\end{equation}
Recent developments in reaction theory for direct reactions at high bombarding energies ($E_\text{lab} \gtrsim 10$~MeV/nucleon) have been devoted to (a) the treatment of the many-body continuum in coupled-channels calculations using the continuum-discretized coupled-channels (CDCC) method \cite{Kamimura1986,AUSTERN1987,YahiroPTP2012}, (b) the treatment of clusters with near exact solutions of high-energy few-body scattering such as the solution of the Alt-Sandhas-Grassberger (AGS) equations \cite{AGS,Oryu1994,OgataPRC942016}, or (c) the inclusion of short-range correlations~\cite{BERTULANI2006155,Fomin2017ARNPS,hen17}. In practice, in many reaction models, such as in most distorted-wave Born-approximation (DWBA) formulations, the theoretical cross section can be ``satisfactorily'' factorized into a product of two parts, one dependent on the reaction dynamics and another part involving form factors for an operator probing the nuclear structure. In some cases, when the interaction is amenable to an exact treatment such as electron-induced knockout reactions, the theoretical approach can be fully realized without a factorization so that the reaction mechanism and the nuclear-structure input are described on the same footing~\cite{Frois1987,SUDA20171}. Ideally, measured cross sections involving hard probes should be interpreted directly by comparison to theoretical values computed by solving the full many-body problem of the collision, treating consistently structure and reaction in a unified framework and a unique Hamiltonian. This is today numerically impossible, with the exception of few-body systems at low incident energy~ (see, {\it e.g.}, \cite{hupin2015}). An intermediate step provided by the dispersive optical model (DOM) connects structure and reaction information~\cite{Mahzoon:2014,Dickhoff:17,Dickhoff19}, while nucleon-nucleus optical potentials from {\it ab initio} theory are being developed but have not yet reached the necessary sophistication for predictive power~\cite{barbieri2005,PhysRevC.86.021602,PhysRevC.95.024315,idini2019}.

It is important to clarify the notation used in this review. PWBA (DWBA) in electron-induced proton knockout is used for plane (distorted) waves for the electron. On the other hand, the notation PWIA (DWIA) often refers to the plane (distorted) wave for the outgoing proton. The notations DWBA and DWIA are often used as synonyms in hadronic interactions, such $(p,2p)$ reactions. Plane waves are rarely used in theoretical calculations of $(p,2p)$ reactions or heavy-ion induced knockout reactions, although we show an example later for clarification of some physics aspects in the reaction. Additionally, eikonal methods often imply that the distorted waves in DWBA (or DWIA) are taken as eikonal waves. Whereas PWBA and DWBA models are perturbative, mostly first-order perturbation theory, CDCC methods include all orders. Two methods are usually called Transfer to the Continuum. One is the semiclassical model of Bonaccorso and Brink \cite{Bonaccorso:1988}, in the following renamed STC. The other (cf.Sec. \ref{QTC}) is a quantum mechanical model QTC of breakup which considers the nucleon-target full states generated by a real potential. In the STC instead the final nnucleon-target states are generated by an energy dependent, complex, optical potential. Finally, in the spirit of the Glauber lectures in physics \cite{glauber59} and the work of Akhiezer and  Sitenko (see Ref. \cite{akhiezer1957} and references therein),``Glauber method" is often used to describe multi-step collisions using eikonal waves for initial, intermediary, and final waves. In Glauber approaches the eikonal waves are often disguised in the form of profile functions, which include all multiple nucleon-nucleon scattering effects for a nucleus-nucleus collision at impact parameter~$b$. Eikonal, or Glauber methods are also amenable to a proper inclusion of relativistic kinematics (conservation laws) and of relativistic dynamical corrections (interaction modification) \cite{BertulaniPRL94.072701,OgataPTP.123.701}.

\begin{figure}[!ht]
 \begin{center}
        \includegraphics[width=0.6\textwidth]{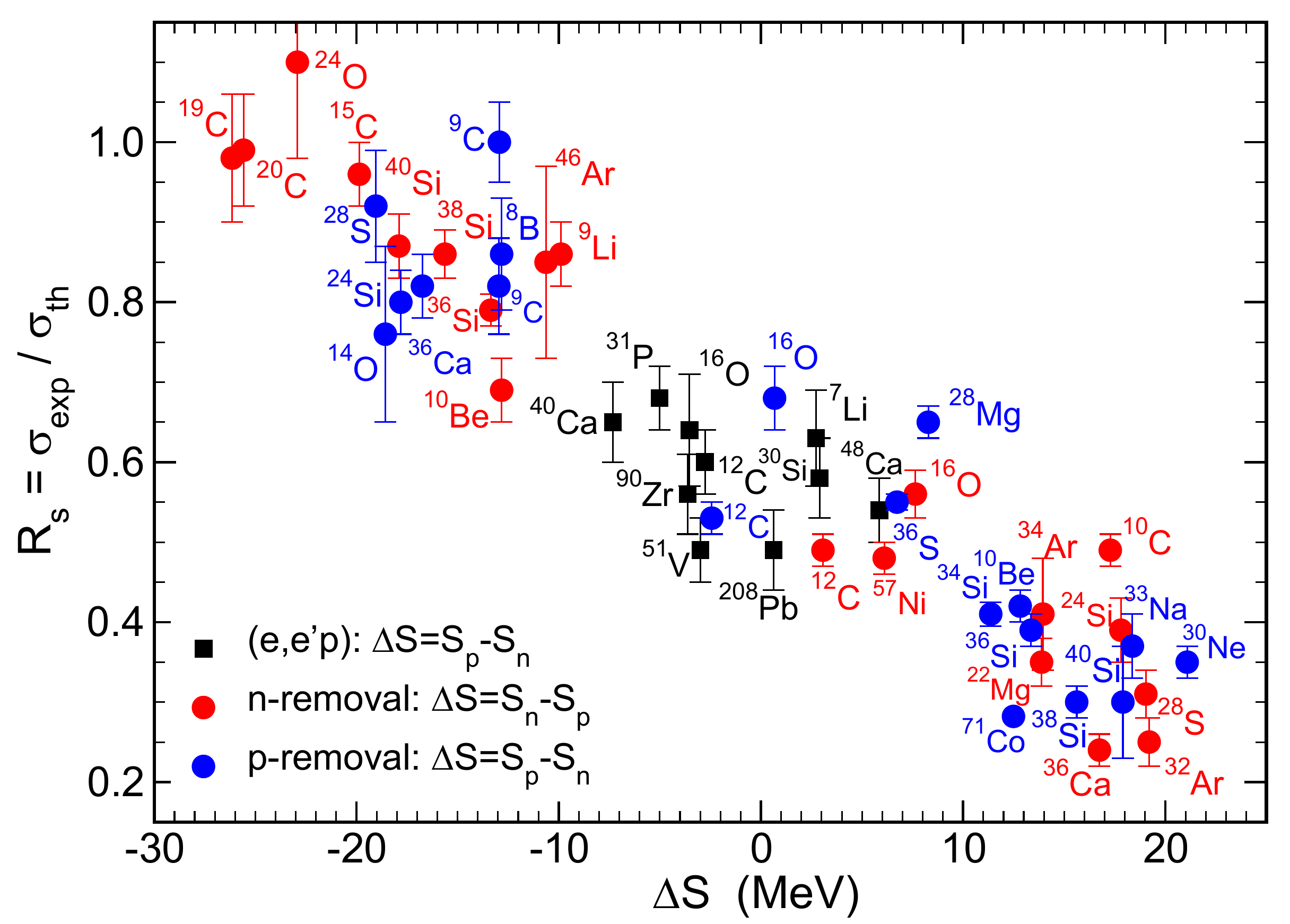}
\caption{\label{Rsplot} (Color online) Compilation of the ratios $R_s$ of the measured and calculated inclusive one-nucleon-removal cross sections for each of the labeled projectile nuclei. $R_s$ is displayed as a function of $\Delta S$, a measure of the asymmetry of the neutron and proton Fermi surfaces. The red (blue) points are for neutron(proton)-removal cases. The solid (black) squares, deduced from electron-induced proton knockout data, are identical to the earlier compilations of~\cite{gade08b,lee06}. The figure is adapted and updated from Ref.~\cite{tostevin14} -- courtesy of J.A. Tostevin (2016); the added data points for \nuc{24}{O}, \nuc{30}{Ne}, \nuc{33}{Na}, \nuc{36}{S} and \nuc{71}{Co} were then preliminary based on the now-published Refs.~\cite{murray19,gade19,divaratne18,lee16,mutschler16}.}
  \end{center}
\end{figure}

The present review emerges from the necessity to improve the interpretation of experimentally determined direct-reaction cross sections. The starting point is a 15-year old puzzle in the comparison of experimental one-nucleon removal cross sections obtained from experiments performed at the NSCL, USA, in inverse kinematics using light ($^9$Be, $^{12}$C) reaction targets and beam energies of about 70-120 MeV/nucleon, together with theoretical calculations based on shell-model spectroscopic factors and a $S$-matrix reaction formalism based on the eikonal approximation~\cite{gade04b,gade08b,tostevin14}. While it is expected that the calculated cross sections are 30-50\% larger than experimental values due to missing correlations in the shell model~\cite{Lapikas:1993} as discussed above (in relation to Fig. \ref{fig:louksf_intro}), a systematic trend has been observed as a function of the asymmetry of the projectile in terms of proton to neutron separation energy ($\Delta {\text S} =  \epsilon \vert {\text S}_{\text n}-{\text S}_{\text p} \vert$, $\epsilon=+1$ for proton removal, $\epsilon=-1$ for neutron removal). This trend is illustrated in Fig.~\ref{Rsplot}. Note that this trend is not observed in the analysis of transfer data from $^{14,16,18}$O oxygen isotopes \cite{flavigny12}, although this conclusion relies on one data set and depends on the inputs of the model \cite{Flavigny18}.  A previous transfer study on $^{34,46}$Ar data did not suggest a large $\Delta {\text S}$ dependency \cite{Lee10,lee2011} either, while another analysis of the same data concluded that it is possible to corroborate the neutron-proton asymmetry dependence reported from knockout measurements \cite{nunes2011}. Recent experimental developments at GSI \cite{atar2018} and RIKEN \cite{Kaw18} allowed quasifree scattering experiments to be performed in inverse kinematics with high-energy radioactive beams. The method has been applied to proton-knockout reactions on oxygen isotopes in a wide range of asymmetry. Both experiments do not confirm such a strong dependence in the $^{\text A}\text{O}(p,2p)$ reduction factors, although analysed by using different reaction models, as will be discussed in Sec. \ref{sec6}. Several recent developments have been made in view of analyzing quasifree scattering data~\cite{aumann2013,ogata2015,cravo2016}.

The above puzzle emerged from the study of isotopes far from stability; it may be seen as a proton-to-neutron asymmetry or binding-energy effect in nuclear structure models or the reaction dynamics, noting that both domains involve uncharted territory. A dependence of SRC to the proton-to-neutron asymmetry, as predicted by {\it ab initio} nuclear-matter theory~\cite{frick05,PhysRevC.79.064308}, has been extracted from large-momentum-transfer $(e,e'pN)$ data \cite{Subedi08,hen17,Duer18,duer2019} but its amplitude does not account for the effect of Fig.~\ref{Rsplot}. Consistently, modern {\it ab initio} structure theories predict a correlation dependence with nucleon separation energy \cite{cipollone2015,jensen11}, which reflects the trend but does not account for the magnitude of the slope and cannot explain the discrepancies among the conclusions obtained in the different experimental probes, suggesting the need for improved reaction descriptions. Part of the community therefore believes that the approximations in the modeling of the reactions should be refined for reaction models involved.

Solving this puzzle in the near future is crucial: with the current state-of-the-art reaction models, structure information extracted from direct reactions at different incident energies or with different targets will differ, clouding our ability to understand the underlying nuclear structure. It is thus essential to understand this discrepancy in view of the goal of obtaining consistent nuclear-structure information for neutron-rich nuclei from reactions performed at the existing and under-construction rare-isotope facilities. The interplay of structure and reaction mechanism in the interpretation of cross sections should also be further treated~\cite{furnstahl2010,more2015} and has been raised as a key issue in nuclear physics to be solved in the coming years in the NUPECC European Long Range Plan (2017) \cite{lrp2017}.

In this review, we propose to give up-to-date answers to the following questions:
\begin{itemize}
\item What is the expected effect of correlations on the  $s.p.$ strength in stable and radioactive nuclei? 
\item What are the main assumptions and model-dependent inputs to calculate direct-reaction cross sections? 
\item What accuracy and precision can be reached from direct-reaction data to obtain structure information?
\item What are the possible explanations for the discrepancy between theory and experiment for light-ion induced deeply-bound-nucleon removal at incident energies of about 100 MeV/nucleon?
\item What is the consensus within the community, and which particular points are still a source of controversy?
\item What new experimental data are called for to solve this problem and to guide theory?
\item What theoretical developments should be put forward towards a consistent treatment of the reaction mechanism and structure?
\end{itemize}
With these objectives in mind, a theoretical background and an overview of past works on short- and long-range correlations in nuclei is given in section \ref{sec2}. In particular, the concepts of  $s.p.$ Green's function, self energy, spectroscopic factors, and spectral distributions are detailed. 
Analysis of the nucleon self-energy and qualitative comparison with results from the $(e,e'p)$ reaction are presented.
In the following sections, the most common direct-reaction mechanisms used for nuclear-structure studies --- namely low-energy nucleon transfer (Sec. \ref{sec3}),  heavy-ion induced nucleon removal (Sec. \ref{sec5}), and proton-induced quasifree scattering (Sec. \ref{sec6}) --- are introduced together with their main approximations. Other direct reactions used to probe weakly-bound nuclei, such as Coulomb breakup \cite{aumannnakamura2013} and resonant elastic scattering \cite{artemov1990}, are not discussed in the following since they are not applicable to a wide range of $\Delta {\text S}$, which is the focus of the review. The consistency of the studied reaction mechanisms in terms of extracting structure information is tested against a unique benchmark (section \ref{sec7}): the proton removal from $^{16}$O.  The comparison of inclusive cross-section data with predictions as a function of the proton-to-neutron asymmetry $\Delta {\text S}$ is addressed for all the above-mentioned reaction mechanisms. Conclusions and still-open questions from past works are emphasized, as well as possible interpretations for the origin of the so-called "reduction factor" at the heart of the present review. Perspectives for both theory and experiment are given in section \ref{sec8}.


\section{Single-particle spectral functions and correlations}
\label{sec2}
\subsection{Introduction}
\label{sec:intro_theory}
In this section, a summary is compiled of some of the theoretical concepts that play a role in extracting information from direct reactions that intend to elucidate the behavior of individual nucleons in nuclei. 
The chosen language here is that of the Green's function method~\cite{Dickhoff:08} as it simultaneously covers the domain of the continuum relevant for nucleon scattering as well as structure information related to nucleon removal. In addition it is a suitable method that can be utilized to perform \textit{ab initio} calculations of relevant quantities~\cite{Dickhoff04}.
In Sec.~\ref{sec:theory} the nucleon Green's function or propagator is introduced which contains at negative energy the nucleon removal amplitudes both to discrete as well as continuum final states. 
It is often referred to as the $s.p.$ spectral function (for removal).
While such a quantity is also available for nucleon addition both to discrete and continuum states, it is more practical in this energy domain to consider the equivalent information encoded in the nucleon self-energy which in earlier practical incarnations has been referred to as the optical potential.

The link between the energy domains related to nuclear structure and reactions is briefly emphasized in Sec.~\ref{sec:DOM} by discussing the dispersion relation connecting them.
This approach forms the basis of the dispersive optical model (DOM) which provides an ideal framework to simultaneously treat reactions and structure and can also perform an intermediate role between experiment and \emph{ab initio} theory.
With this theoretical background (pedagogically illustrated in appendix \ref{sec:self}) it is possible to understand how to employ removal amplitudes to valence hole states as probed in $(e,e'p)$ reactions to determine the extraction of their normalization or so-called spectroscopic factors which are the main topic of this review.
Calculations of spectroscopic factors and the physical ingredients that determine their quantitative value are presented in Sec.~\ref{sec:specdis} with attention to the role of both long-range and short-range correlations.
Recent results from the DOM are discussed as they provide a framework to utilize the Green's function method to let experimental data constrain the nucleon propagator and self-energy and thereby generate implied quantitative results for spectroscopic factors.
This efficacy is illustrated in Sec.~\ref{sec:eepX} which provides an essential confirmation of the analysis of the $(e,e'p)$ by the NIKHEF efforts~\cite{Lapikas:1993} as well as provides insight into the nucleon asymmetry dependence of proton spectroscopic factors in Ca isotopes. We will complete our discussion of the theoretical aspects of correlations by discussing the most recent results from {\it ab initio} theory, in Sec.\ref{sec:abinitio}.

\subsection{Single-particle Green's function and the self-energy}
\label{sec:theory}
We start with a brief summary of relevant results from the Green's function formulation of the many-body problem~\cite{Dickhoff:08}.
The $s.p.$ propagator in a many-particle system is defined as
\begin{equation}
G (\alpha ,\beta ; t- t' ) = -\frac{i}{\hbar}
\bra{\Psi^A_0} {\mathcal{T}} 
[ a_{\alpha_H}(t) a^\dagger_{\beta_H}(t') ] \ket{\Psi^A_0} .
\label{eq:5.1}
\end{equation}
The expectation value with respect to the exact ground state of the system of
$A$ particles, samples an operator that represents both particle as
well as hole propagation. 
The state $\ket{\Psi^A_0}$ is the normalized, nondegenerate
Heisenberg ground state for the $A$-particle
system and $E^A_0$ the corresponding eigenvalue.

The particle addition and removal operators in the definition of the sp
propagator are given in the Heisenberg picture and the labels $\alpha$ or $\beta$ refer to an appropriate complete set of quantum numbers associated with a $s.p.$ basis. The time-ordering operation $\mathcal{T}$ is defined here to include a sign change
when two fermion operators are interchanged.
The propagator depends only on the time difference $t-t'$.
In the following we employ the completeness of the exact eigenstates of $\hat H$ for both
the $A+1$ as well as the $A-1$ system, together with the corresponding eigenvalues of the Hamiltonian.
Using the integral representation of the
step function, the Fourier transform of the propagator can be expressed as:
\begin{eqnarray}
G (\alpha ,\beta ; E )
 =  \sum_m \frac{\bra{\Psi^A_0} a_{\alpha}
\ket{\Psi^{A+1}_m} \bra{\Psi^{A+1}_m} a^\dagger_{\beta} \ket{\Psi^A_0}
}{ E - (E^{A+1}_m - E^A_0 ) +i\eta } 
 +   \sum_n \frac{\bra{\Psi^A_0} a^\dagger_{\beta} \ket{\Psi^{A-1}_n}
\bra{\Psi^{A-1}_n} a_{\alpha} \ket{\Psi^A_0} }{
E - (E^A_0 - E^{A-1}_n) -i\eta} 
. 
\label{eq:5.8}
\end{eqnarray}
The expression is known as the Lehmann representation~\cite{lehm} of the sp
propagator.
Note that any $s.p.$ basis can be used in this
formulation of the propagator.
Continuum solutions in the $A\pm1$ systems are also implied in the completeness relations but are not explicitly included to simplify the notation.
At this point one assumes that a meaningful nuclear Hamitonian $\hat{H}$ exists that contains a two-body component that describes nucleon-nucleon scattering up to a chosen energy, usually the pion production threshold, as well as relevant bound-state data.
There is sufficient experimental evidence and theoretical insight suggesting that at least a three-body component should also be present in the Hamiltonian. 

Maintaining the present general notation with regard to the $s.p.$ quantum numbers, we introduce the spectral functions associated with particle and hole propagation.
At energy $E$, the hole spectral function represents the combined 
probability density for
removing a particle with quantum numbers $\alpha$ from the ground state,
while leaving the remaining $A-1$-system at an energy
$E^{A-1}_n=E^A_0- E$.
The relation to the imaginary part of the diagonal
element of the $s.p.$ propagator is given by
\begin{eqnarray}
S_h (\alpha , E) = \frac{1}{\pi}\
\textrm{Im}\ G(\alpha ,\alpha ; E ) 
= \sum_{n}
\Bigl| \bra{\Psi^{A-1}_n} a_{\alpha} \ket{\Psi^A_0} \Bigr|^2
\delta(E-(E^A_0 - E^{A-1}_n)) ,
\label{eq:5.9}
\end{eqnarray}
for $E \le \varepsilon^-_F$ identifying the Fermi energy for the removal of a particle.
The probability density 
for the addition of a particle with quantum numbers
$\alpha$, leaving the $A+1$-system at energy $E^{A+1}_m=E^A_0+E$
similarly reads
\begin{eqnarray}
S_p (\alpha , E )
= -\frac{1}{\pi}\ \textrm{Im}\
G(\alpha ,\alpha ; E ) 
= \sum_{m} \Bigl| \bra{\Psi^{A+1}_m} a^\dagger_{\alpha} \ket{\Psi^A_0}
\Bigr|^2 \delta(E-(E^{A+1}_m - E^A_0)) 
\label{eq:5.10}
\end{eqnarray}
for $E \ge \varepsilon^+_F$.
Equation~(\ref{eq:5.10})
defines the particle spectral function.
The Fermi energies introduced in Eqs.~(\ref{eq:5.9}) and
(\ref{eq:5.10}) are given by
\begin{equation}
\label{eq:5.11a}
\varepsilon^-_F = E^A_0 - E^{A-1}_0
\end{equation}
and
\begin{equation}
\varepsilon^+_F = E^{A+1}_0 - E^A_0 ,
\label{eq:5.11b}
\end{equation}
respectively.

The occupation number of a $s.p.$ state $\alpha$ can be generated
from the hole part of the spectral function by evaluating
\begin{equation}
n (\alpha) = \bra{\Psi^A_0} a^\dagger_\alpha a_\alpha
\ket{\Psi^A_0} 
= \sum_{n} \Bigl| \bra{\Psi^{A-1}_n} a_\alpha \ket{\Psi^A_0} \Bigr|^2
= \int_{-\infty}^{\varepsilon^-_F} \!\!\! dE\ S_h(\alpha, E) .
\label{eq:5.13}
\end{equation}
The depletion (or emptiness) number is determined by the particle part
of the spectral function
\begin{equation}
d (\alpha) = \bra{\Psi^A_0} a_\alpha a^\dagger_\alpha 
\ket{\Psi^A_0} =  
\sum_{m} \Bigl| \bra{\Psi^{A+1}_m} a^\dagger_\alpha \ket{\Psi^A_0}
\Bigr|^2 
= \int_{\varepsilon^+_F}^{\infty} \!\!\! dE\ S_p(\alpha, E) .  
\label{eq:5.14}
\end{equation}
An important sum rule exists for $n(\alpha)$ and $d(\alpha)$ which can be
deduced by employing the anticommutation relation for $a_\alpha$ and 
$a^\dagger_\alpha$
\begin{equation}
n(\alpha) + d (\alpha) =
\bra{\Psi^A_0} a^\dagger_\alpha a_\alpha \ket{\Psi^A_0} 
+ \bra{\Psi^A_0} a_\alpha a^\dagger_\alpha \ket{\Psi^A_0} 
= \langle \Psi^A_0 | \Psi^A_0 \rangle = 1 .
\label{eq:5.15}
\end{equation}
The distribution between occupation and emptiness of a sp
orbital in the correlated ground state is a sensitive measure of the
strength of correlations, provided a suitable $s.p.$ basis is chosen.

The $s.p.$ propagator generates the expectation value of any
one-body operator in the ground state
\begin{equation}
\bra{\Psi^A_0} \hat O \ket{\Psi^A_0} =
\sum_{\alpha,\beta} \bra{\alpha} O \ket {\beta} \bra{\Psi^A_0}
a^\dagger_\alpha a_\beta \ket{\Psi^A_0} 
= \sum_{\alpha,\beta} \bra{\alpha} O \ket {\beta} n_{\alpha\beta} .
\label{eq:5.16}
\end{equation}
Here, $n_{\alpha\beta}$ is the one-body density matrix element which can
be obtained from the $s.p.$ propagator
using the Lehmann representation
\begin{equation}
n_{\beta\alpha} =
\int \! \frac{dE}{2\pi i}\ e^{iE\eta}\ G(\alpha,\beta;E) .
\label{eq:5.17}  
\end{equation}
The convergence factor ensures that only removal amplitudes contribute.
Knowledge of $G$ in terms of $n_{\beta \alpha}$, therefore
yields the expectation value of any one-body operator in the correlated ground
state according to Eq.~(\ref{eq:5.16}).
An important recent application of this result concerns the nuclear charge density which is measured in detail for stable closed-shell nuclei, providing important constraints on the properties of the $s.p.$ propagator. 
If neutron properties related to scattering are constrained and isospin symmetry is invoked, it is also be possible to make predictions for the neutron distribution and as a consequence the neutron skin.

The perturbation expansion of the $s.p.$ propagator is discussed in various textbooks, \textit{e.g.}, Refs.~\cite{Dickhoff:08,Abrikosov1965}.
It is necessary to order the expansion into the so-called Dyson equation to obtain a meaningful nonperturbative link between the in-medium potential experienced by a nucleon, the so-called irreducible self-energy, and the propagator.
The resulting equation is
\begin{equation}
G(\alpha,\beta;E) = G^{(0)}(\alpha,\beta;E)
+ \sum_{\gamma,\delta} G^{(0)}(\alpha,\gamma;E)
\Sigma(\gamma,\delta;E) G(\delta,\beta;E) .
\label{eq:10.2}
\end{equation}
The noninteracting propagator $G^{(0)}$ can be chosen depending on the particular problem~\cite{Dickhoff04,Dickhoff:08} and most often incorporates global conservation laws especially those associated with rotational symmetry and parity in nuclear applications.
We note that for the purpose of performing approximate calculations of the nucleon self-energy often a noninteracting propagator is chosen that corresponds to localized nucleons~\cite{Dickhoff04} which is accomplished by a corresponding term in the irreducible self-energy so that its effect ultimately cancels out.
Ab initio calculations in infinite systems naturally proceed from the noninteracting Fermi gas starting point.
For applications of the DOM, it is convenient to work in a coordinate or momentum-space $s.p.$ basis suitably accompanied by conserved quantum numbers.
The corresponding quantum numbers are the orbital and total angular momentum of the nucleon, which can therefore also be employed to label the propagator and the irreducible self-energy.


When calculating the nucleon propagator with respect to the $A$-body ground state in the $s.p.$ basis with good radial position (or momentum), orbital angular momentum (parity) and total angular momentum, the numerators of the particle and hole components of the propagator in the Lehmann representation in Eq.~(\ref{eq:5.8}) include the products of overlap functions associated with adding or removing a nucleon from the $A$-body ground state.
For the present discussion the noninteracting propagator is assumed to involve only kinetic energy contributions.
The nucleon self-energy contains all linked diagrammatic contributions that are irreducible with respect to propagation represented by $G^{(0)}$.
All contributions to the propagator are then generated by the Dyson equation itself.
The solution of the Dyson equation generates all discrete poles corresponding to bound $A\pm1$ states explicitly given by Eq.~(\ref{eq:5.8}) that can be reached by adding or removing a particle with quantum numbers $r \ell j$.
The hole spectral function is obtained from
\begin{equation}
S_{\ell j}(r;E) = \frac{1}{\pi}  \textrm{Im}\ G_{\ell j}(r,r;E)  
\label{eq:holes}
\end{equation}
for energies in the $A-1$ continuum.
The total spectral strength at $E$ for a given $\ell j$ combination, 
\begin{equation}
S_{\ell j}(E) = \int_{0}^\infty dr\ r^2\ S_{\ell j}(r;E) ,
\label{eq:specs}
\end{equation}
yields the spectroscopic strength per unit of energy.
For discrete energies as well as all continuum ones, overlap functions for the addition or removal of a particle are generated as well.

For discrete states in the $A-1$ system one can show that the overlap function obeys a Schr{\"o}dinger-like equation~\cite{Dickhoff:08}.
Introducing the notation
\begin{equation}
\psi^n_{\ell j}(r) = \bra{\Psi^{A-1}_n}a_{r \ell j} \ket{\Psi^A_0} ,
\label{eq:overlap}
\end{equation}
for the overlap function for the removal of a nucleon at $r$ with discrete quantum numbers $\ell$ and $j$, one finds
\begin{eqnarray}
\left[ \frac{ p_r^2}{2m} +
 \frac{\hbar^2 \ell (\ell +1)}{2mr^2}\right]   \psi^{n}_{\ell j}(r) 
+   \int \!\! dr'\ r'^2\ 
\Sigma_{\ell j}(r,r';\varepsilon^-_n)  \psi^{n}_{\ell j}(r') =
\varepsilon^-_n \psi^{n}_{\ell j}(r) ,
\label{eq:DSeq}
\end{eqnarray}
where $\varepsilon^-_n=E^A_0 -E^{A-1}_n$
and in coordinate space the radial momentum operator is given by $p_r = -i\hbar({\partial}/{\partial r} + {1}/{r})$.
Discrete solutions to Eq.~(\ref{eq:DSeq}) exist in the domain where the self-energy has no imaginary part and these are normalized by utilizing the inhomogeneous term in the Dyson equation.
For an eigenstate of the Schr{\"o}dinger-like equation [Eq.~(\ref{eq:DSeq})], the so-called quasihole state labeled by $\alpha_{qh}$, the corresponding normalization or spectroscopic factor is given  by~\cite{Dickhoff:08}
\begin{equation}
\mathcal{Z}^n_{\ell j} = \bigg( {1 - 
\frac{\partial \Sigma_{\ell j}(\alpha_{qh},
\alpha_{qh}; E)}{\partial E} \bigg|_{\varepsilon^-_n}} 
\bigg)^{-1} ,
\label{eq:sfac}
\end{equation}
which is the discrete equivalent of Eq.~(\ref{eq:specs}).

The particle spectral function for a finite system can be generated by the calculation of the reducible self-energy $\mathcal{T}$ which also yields direct access to all elastic-scattering observables in the case of neutrons.
For protons a direct solution of the relevant differential equation is usually employed to generate scattering observables.
In some applications relevant for elucidating correlation effects, a momentum-space calculation~\cite{PhysRevC.84.044319} can be employed.
In an angular-momentum basis, iterating the irreducible self-energy $\Sigma$ to all orders, then yields
\begin{eqnarray}
\label{eq:redSigma1}
\mathcal{T}_{\ell j}(k,k^\prime ;E) = \Sigma_{\ell j}(k,k^\prime ;E) 
  +  \!\!       \int \!\! dq\ q^2\ \Sigma_{\ell j}(k,q;E)\ G^{(0)}(q;E )\ \mathcal{T}_{\ell j}(q,k^\prime ;E) ,
\end{eqnarray}
where $G^{(0)}(q; E ) = (E - \hbar^2q^2/2m + i\eta)^{-1}$ is the free propagator.
The propagator is then obtained from an alternative form of the Dyson equation in the following form~\cite{Dickhoff:08}
\begin{eqnarray}
G_{\ell j}(k, k^{\prime}; E) = \frac{\delta( k - k^{\prime})}{k^2}G^{(0)}(k; E) 
 \label{eq:gdys1}  
+ G^{(0)}(k; E)\mathcal{T}_{\ell j}(k, k^{\prime}; E)G^{(0)}(k'; E)  .	
\end{eqnarray}
The on-shell matrix elements of the reducible self-energy in Eq.~(\ref{eq:redSigma1}) are sufficient to describe all aspects of elastic scattering like differential cross sections, reaction cross sections, and total cross sections as well as polarization data~\cite{PhysRevC.84.044319}.
This connection between the nucleon propagator and elastic-scattering data identifies the nucleon elastic-scattering $\mathcal{T}$-matrix with the reducible self-energy obtained by iterating the irreducible one to all orders with $G^{(0)}$~\cite{Dickhoff:08,Bell:59,vill67, blri86}.

The spectral representation of
the particle part of the propagator, referring to the $A+1$ system, appropriate for a treatment of the continuum and possible open channels is given by~\cite{Mahaux:91}
\begin{eqnarray}
G_{\ell j}^{p}(k ,k' ; E)  =  
\sum_n  \frac{ \phi^{n+}_{\ell j}(k) \left[\phi^{n+}_{\ell j}(k')\right]^*
}{ E - E^{*A+1}_n +i\eta }   \label{eq:propp} 
 +  
\sum_c \int_{T_c}^{\infty} dE'\  \frac{\chi^{cE'}_{\ell j}(k) \left[\chi^{cE'}_{\ell j}(k')\right]^* }{
E - E' +i\eta} ,
\end{eqnarray}
generalizing the discrete formulation of Eq.~(\ref{eq:5.8}).
Overlap functions for bound $A+1$ states are given by $ \phi^{n+}_{\ell j}(k)=\bra{\Psi^A_0} a_{k\ell j}
\ket{\Psi^{A+1}_n}$, whereas those in the continuum are given by $ \chi^{cE}_{\ell j}(k)=\bra{\Psi^A_0} a_{k\ell j} \ket{\Psi^{A+1}_{cE}}$, indicating the relevant channel by $c$ and the energy by $E$.
Excitation energies in the $A+1$ system are with respect to the $A$-body ground state $E^{*A+1}_n = E^{A+1}_n -E^A_0$.
Each channel $c$ has an appropriate threshold indicated by $T_c$ which is the experimental threshold with respect to the ground-state energy of the $A$-body system.
The overlap function for the elastic channel can be explicitly calculated by solving the Dyson equation while it is also possible to obtain the complete spectral density for $E>0$ 
\begin{eqnarray}
S_{\ell j}^{p}(k ,k' ; E) 
=
\sum_c \chi^{cE}_{\ell j}(k) \left[ \chi^{cE}_{\ell j}(k') \right]^* .
\label{eq:specp}
\end{eqnarray}
In practice, this requires solving the scattering problem twice at each energy so that one may employ
\begin{eqnarray}
\!\! S_{\ell j}^{p}(k ,k' ; E) 
= \frac{i}{2\pi} \left[ G_{\ell j}^{p}(k ,k' ; E^+) - G_{\ell j}^{p}(k ,k' ; E^-) \right]
\label{eq:specpp}
\end{eqnarray}
with $E^\pm =E\pm i\eta$, and only the elastic-channel contribution to Eq.~(\ref{eq:specp}) is explicitly known.
Equivalent expressions pertain to the hole part of the propagator $G_{\ell j}^{h}$~\cite{Mahaux:91}.

Calculations are performed in momentum space according to Eq.~(\ref{eq:redSigma1}) to generate the off-shell reducible self-energy and thus the spectral density by employing Eqs.~(\ref{eq:gdys1}) and (\ref{eq:specpp}).
Because the momentum-space spectral density contains a delta-function associated with the free propagator, it is convenient to also consider the Fourier transform back to coordinate space 
\begin{eqnarray}
S_{\ell j}^{p}(r ,r' ; E) = \frac{2}{\pi} \label{eq:specpr}  \int \!\! dk k^2 
  \! \int \!\! dk' k'^2 j_\ell(kr) S_{\ell j}^{p}(k ,k' ; E) j_\ell(k'r') ,
\end{eqnarray}
which has the physical interpretation for $r=r'$ as the probability density $S_{\ell j}(r;E)$ for adding a nucleon with energy $E$ at a distance $r$ from the origin for a given $\ell j$ combination.
By employing the asymptotic analysis to the propagator in coordinate space following, \textit{e.g.}, Ref.~\cite{Dickhoff:08}, one may express the elastic-scattering wavefunction that contributes to Eq.~(\ref{eq:specpr}) in terms of the half on-shell reducible self-energy obtained according to
\begin{eqnarray}
\chi^{el E}_{\ell j}(r) = \left[ \frac{2mk_0}{\pi \hbar^2} \right]^{1/2} \left\{ j_\ell(k_0r) 
 +   \int \!\! dk k^2 j_\ell(kr) G^{(0)}(k;E) \mathcal{T}_{\ell j}(k,k_0;E) \right\} ,
 \label{eq:elwf} 
\end{eqnarray}
where $k_0=\sqrt{2 m E}/\hbar$ is related to the scattering energy in the usual way.

The constraints imposed on the nucleon self-energy by elastic scattering data also indirectly determine the presence of strength in the continuum associated with mostly-occupied orbits (or mostly empty but $E<0$  orbits). This strength is obtained by double folding the spectral density in Eq.~(\ref{eq:specpr}) in the following way
\begin{eqnarray}
\!\!\! S_{\ell j}^{n+}(E) 
=  \int \!\! dr r^2 \!\! \int \!\! dr' r'^2 \phi^{n-}_{\ell j}(r) S_{\ell j}^{p}(r ,r' ; E) \phi^{n-}_{\ell j}(r') ,
\label{eq:specfunc}
\end{eqnarray}
using an overlap function 
\begin{equation}
\sqrt{S^n_{\ell j}} \phi^{n-}_{\ell j}(r)=\bra{\Psi^{A-1}_n} a_{r\ell j} \ket{\Psi^{A}_0} , 
\label{eq:overm}
\end{equation}
with $\phi^{n-}_{\ell j}(r)$ normalized to 1 and for valence orbits, $S^n_{\ell j}$ their spectroscopic factor~\cite{PhysRevC.84.044319}.
In practice, any normalized bound-state wavefunction can be utilized.

In the case of an orbit below the Fermi energy, this strength identifies where the depleted strength resides in the continuum.
The occupation number of this orbit is given by an integral over a corresponding folding of the hole spectral density
\begin{eqnarray}
\!\!\!\!\!\!\! S_{\ell j}^{n-}(E) 
= \!\! \int \!\! dr r^2 \!\! \int \!\! dr' r'^2 \phi^{n-}_{\ell j}(r) S_{\ell j}^{h}(r ,r' ; E) \phi^{n-}_{\ell j}(r') ,
\label{eq:spechr}
\end{eqnarray}
where $S_{\ell j}^{h}(r,r';E)$ provides equivalent information below the Fermi energy as $S_{\ell j}^{p}(r,r';E)$ above.
An important sum rule is valid for the sum of the occupation number $n_{n \ell j}$ for the orbit characterized by 
$n \ell j$
\begin{equation}
n_{n \ell j} = \int_{-\infty}^{\varepsilon_F} \!\!\!\! dE\ S_{\ell j}^{n-}(E)
\label{eq:nocc}
\end{equation}
and its corresponding depletion number $d_{n \ell j}$
\begin{equation}
d_{n \ell j} = \int_{\varepsilon_F}^{\infty} \!\!\!\! dE\ S_{\ell j}^{n+}(E).
\label{eq:depl} 
\end{equation}
It is simply given by~\cite{Dickhoff:08}
\begin{eqnarray}
\!\! 1 =  n_{n \ell j} + d_{n \ell j} \!\! =\bra{\Psi^A_0} a^\dagger_{n \ell j} a_{n \ell j} +a_{n \ell j}a^\dagger_{n \ell j}  \ket{\Psi^A_0} ,
\label{eq:sumr} 
\end{eqnarray}
reflecting the properties of the corresponding anticommutator of the operators $a^\dagger_{n \ell j}$ and $a_{n \ell j}$.
It is convenient to employ the average Fermi energy
\begin{equation}
\varepsilon_F \equiv \frac{1}{2} \left[
\varepsilon_F^+  - \varepsilon_F^- \right] =  \frac{1}{2} \left[ (E^{A+1}_0-E^A_0) + (E^A_0 - E^{A-1}_0) \right]
\label{eq:FE}
\end{equation}
in Eqs.~(\ref{eq:nocc}) and (\ref{eq:depl})~\cite{Mahaux:91}.

Strength above $\varepsilon_F$, as expressed by Eq.~(\ref{eq:specfunc}), reflects the presence of the imaginary self-energy at positive energies.
Without it, the only contribution to the spectral function comes from the elastic channel.
The folding in Eq.~(\ref{eq:specfunc}) then involves integrals of orthogonal wavefunctions and yields zero.
Because it is essential to describe elastic scattering with an imaginary potential, 
it automatically ensures that the elastic channel does not exhaust the spectral density and therefore some spectral strength associated with bound orbits in the independent-particle model also occurs in the continuum.

\subsection{Aspects of the dispersive optical model linking nuclear structure and reaction domains}
\label{sec:DOM}
We introduce here some additional results from Green's-function theory as it pertains to the application of the dispersive optical model, promoted successfully by Mahaux~\cite{Mahaux:91} (see also Ref.~\cite{Dickhoff:17}).
The link between the particle and hole domain plays a critical role in this formulation.
As a result, a direct connection between the energy domains pertaining to nuclear reactions and structure is established from the start.
The (irreducible) nucleon self-energy in general obeys a dispersion relation between its real and imaginary parts given by~\cite{Dickhoff:08}
\begin{eqnarray} 
 \mbox{Re}\ \Sigma_{\ell j}(r,r';E)\! = \! \Sigma^s_{\ell j} (r,r')\! \label{eq:disprel} 
- \! {\cal P} \!\!
\int_{\varepsilon_T^+}^{\infty} \!\! \frac{dE'}{\pi} \frac{\mbox{Im}\ \Sigma_{\ell j}(r,r';E')}{E-E'}  
+{\cal P} \!\!
\int_{-\infty}^{\varepsilon_T^-} \!\! \frac{dE'}{\pi} \frac{\mbox{Im}\ \Sigma_{\ell j}(r,r';E')}{E-E'} , 
\end{eqnarray}
where $\mathcal{P}$ represents the principal value.
The static contribution arises from the correlated Hartree-Fock (HF) term involving the exact one-body density matrix of Eq.~(\ref{eq:5.17}). The dynamic parts start and end at corresponding thresholds in the $A\pm1$ systems that have a larger separation than the corresponding difference between the Fermi energies for addition $\varepsilon_F^+$ and removal $\varepsilon_F^-$ of a particle.
The latter feature is particular to a finite system and generates possibly several discrete quasiparticle and hole-like solutions of the Dyson equation in Eq.~(\ref{eq:DSeq}) in the domain where the imaginary part of the self-energy vanishes.\\
The standard definition of the self-energy requires that its imaginary part is negative, at least on the diagonal, in the domain that represents the coupling to excitations in the $A+1$ system, while it is positive for the coupling to $A-1$ excitations.
This translates into an absorptive potential for elastic scattering at positive energy, where the imaginary part is responsible for the loss of flux in the elastic channel.
Subtracting Eq.~(\ref{eq:disprel}) calculated at the average Fermi energy [see Eq.~(\ref{eq:FE})], from Eq.~(\ref{eq:disprel}) generates the so-called subtracted dispersion relation 
\begin{eqnarray} 
\mbox{Re}\ \Sigma_{\ell j}(r,r';E)\! &=& \!  \mbox{Re}\ \Sigma_{\ell j} (r,r';\varepsilon_F)  
- \! {\cal P} \!\!
\int_{\varepsilon_T^+}^{\infty} \!\! \frac{dE'}{\pi} \mbox{Im}\ \Sigma_{\ell j}(r,r';E') \left[ \frac{1}{E-E'}  - \frac{1}{\varepsilon_F -E'} \right]  \nonumber  \\
&+& {\cal P} \!\!
\int_{-\infty}^{\varepsilon_T^-} \!\! \frac{dE'}{\pi} \mbox{Im}\ \Sigma_{\ell j}(r,r';E') \left[ \frac{1}{E-E'}
-\frac{1}{\varepsilon_F -E'} \right]  .
 \label{eq:sdisprel}
\end{eqnarray}
The beauty of this representation was recognized by Mahaux and Sartor~\cite{Mahaux:1986,Mahaux:91} since it allows for a link with empirical information both at the level of the real part of the nonlocal self-energy at the Fermi energy (probed by a multitude of HF calculations) and also through empirical knowledge of the imaginary part of the optical potential (constrained by experimental data) that consequently yields a dynamic contribution to the real part by means of Eq.~(\ref{eq:sdisprel}).
In addition, the subtracted form of the dispersion relation emphasizes contributions to the integrals from the energy domain nearest to the Fermi energy on account of the $E'$-dependence of the integrands of Eq.~(\ref{eq:sdisprel}). 
By choosing standard functionals with relevant parameters it is possible to constrain the nucleon self-energy employing experimental observables.
Recent DOM applications reviewed in \cite{Dickhoff:17} include experimental data up to 200 MeV of scattering energy and are therefore capable of determining the nucleon propagator in a wide energy domain as all negative energies are included as well.
The DOM approach is therefore able to simultaneously describe nuclear structure information related to the ground state as well as elastic scattering data thereby providing both overlap functions as well as distorted wave for the removal and addition of nucleons, respectively.
In Sec.~\ref{sec:eepX} these ingredients are employed to describe the $(e,e'p)$ reaction on ${}^{40}$Ca to the valence hole states in ${}^{39}$K.
The DOM self-energy is also a useful interface between experiment and ab initio as discussed in Sec.~\ref{sec:SLRC}.

\subsection{Analysis of the $(e,e'p)$ reaction}
\label{sec:eep}
Review articles on this subject can be found in Refs.~\cite{FM84,deWitt1,deWitt2,deWitt3,Lapikas:1993,Pandharipande97}.
Most high-resolution $(e,e'p)$ experiments were performed at NIKHEF in Amsterdam  which were interpreted successfully by employing the distorted-wave impulse approximation (DWIA) to describe the reaction.
This description is expected to be particularly good when kinematics is used that emphasizes the longitudinal coupling of the excitation operator, which is dominated by a one-body operator.
The NIKHEF group was able to fulfill this condition by choosing kinematical conditions in which the removed proton carried momentum parallel or antiparallel to the momentum of the virtual photon.
Under these conditions, the transverse contribution involving the spin and possible two-body currents is suppressed. Therefore the process can be interpreted as requiring an accurate description of the transition amplitude connecting the resulting excited state to the ground state by a known one-body operator.
This transition amplitude is contained in the polarization propagator which can be analyzed with a many-body description involving linear response~\cite{Dickhoff:08}.
Such an analysis demonstrates that the polarization propagator contains two contributions.
The first term involves the propagation of a particle and a hole dressed by their interaction with the medium, but not each other.
The other term involves their interaction. The latter term will dominate at low energy when the proton that absorbs the photon participates in collective excitations like surface modes and giant resonances. When the proton receives on the order of 100 MeV it is expected that the excited state that is created can be well approximated by the dressed particle and dressed hole excitation~\cite{Brand90}.
In fact, when strong transitions are considered, like in the present work, two-step processes have only minor influence~\cite{Kramer:1989,Gerard_12C}. 

This interpretation forms the basis of the DWIA applied to exclusive $(e,e'p)$ cross sections obtained by the NIKHEF group.
The ingredients of the DWIA therefore require a proton distorted wave describing the outgoing proton at the appropriate energy and an overlap function with its normalization for the removed proton.
The distorted wave was typically obtained from a standard global optical potential like Ref.~\cite{Schwandt:1982}, which is local.
The overlap function was obtained by adjusting the radius of a local Woods-Saxon potential to the shape of the $(e,e'p)$ cross section while adjusting its depth to the separation energy of the hole.
Its normalization was obtained by adjusting the calculated DWIA cross section to the actual data~\cite{Lapikas:1993}.  
Standard nonlocality corrections were applied to both the outgoing and removed proton wavefunctions~\cite{Perey:63}, in practice making the bound-state wavefunction the solution of a nonlocal potential.
Such corrections are $\ell$-independent and therefore different from when nonlocal potentials are employed as in the recent DOM implementation~\cite{Atkinson:2018}.

In order to describe the $(e,e'p)$ reaction, the incoming electron, the electron-proton interaction, the outgoing electron, and the outgoing proton must therefore be addressed. The cross section is calculated from the hadron tensor, $W^{\mu\nu}$,  which contains matrix elements of the nuclear charge-current density, $J^\mu$~\cite{ElectroResponse}. Using the DWIA, which assumes that the virtual photon exchanged by the electron couples to the same proton that is detected and the final-state interaction can be described using an optical potential~\cite{Giusti:1988,Boffi:1980}, the nuclear current can be written as
   \begin{equation}
      J^\mu(\bm{q}) = \int d\bm{r}e^{i\bm{q}\cdot\bm{r}}\chi^{(-)*}_{E\ell j}(\bm{r})(\hat{J}^\mu_{\text{eff}})_{E\ell j}(\bm{r})\psi^n_{\ell j}(\bm{r})\sqrt{\mathcal{Z}^n_{\ell j}},
      \label{eq:current}
   \end{equation}
   where $\chi^{(-)*}_{E\ell j}(\bm{r})$ is the outgoing proton distorted wave~\cite{ElectroResponse}, 
   $\psi^n_{\ell j}(\bm{r})$ is the overlap function, $\mathcal{Z}^n_{\ell j}$ its normalization, $\bm{q} = \bm{k_f} - \bm{k_i}$ is the electron three-momentum transfer, and 
   $\hat{J}^\mu_{\text{eff}}$ is the effective current operator~\cite{ElectroResponse}. 
   The incoming and outgoing electron waves
   are treated within the Effective Momentum Approximation, where the waves are represented by plane waves with effective momenta to account for distortion from the interaction with the target 
   nucleus~\cite{Giusti:1987}
   \begin{equation}
      k_{i(f)}^{\text{eff}} = k_{i(f)} + \int d\bm{r}V_c(\bm{r})\phi_{\ell j}^2(\bm{r}),
      \label{eq:effective}
   \end{equation}
   where $V_c(\bm{r})$ is the Coulomb interaction.
   This alters Eq.~(\ref{eq:current}) by replacing $\bm{q}$ with $\bm{q}_{\text{eff}}$.

   In the plane-wave impulse approximation (PWIA), in which the outgoing proton wave is approximated by a plane wave, the $(e,e'p)$ cross section can be factorized into an off-shell electron-proton cross section and the sp-eps-converted-to.pdfectral function~\cite{ElectroResponse}, 
   \begin{equation}
      S(E_m,\bm{p}_m) = \frac{1}{k\sigma_{ep}}\frac{d^6\sigma}{dE_{e'}d\Omega_{e'}dE_pd\Omega_p}.
      \label{eq:momdist}
   \end{equation}
   A simple derivation of the PWIA and the relation to the spectral function can also be found in~\cite{Dickhoff:08}.
   The off-shell electron-proton cross section, $\sigma_{ep}$, is approximated from the on-shell one using the $\sigma_{\text{cc1}}$ model as proposed in~\cite{deForest:1983}.
  This separation does not hold true for the DWIA, but the displayed cross sections, both the experimental and theoretical ones, have been divided by the $\sigma_{\text{cc1}}$ cross section.
  Note that Eq.~(\ref{eq:momdist}) is equivalent to the diagonal element of the imaginary part of the propagator below the Fermi energy as in Eq.~(\ref{eq:holes}).
  In principle, corrections due to two-step processes could be considered but they are estimated to make negligible contributions for the transitions to valence hole states.

\begin{figure}[t]
\begin{center}
\includegraphics[clip,trim=3cm 0cm 2.5cm 0cm, width=6.5cm,angle=-90]{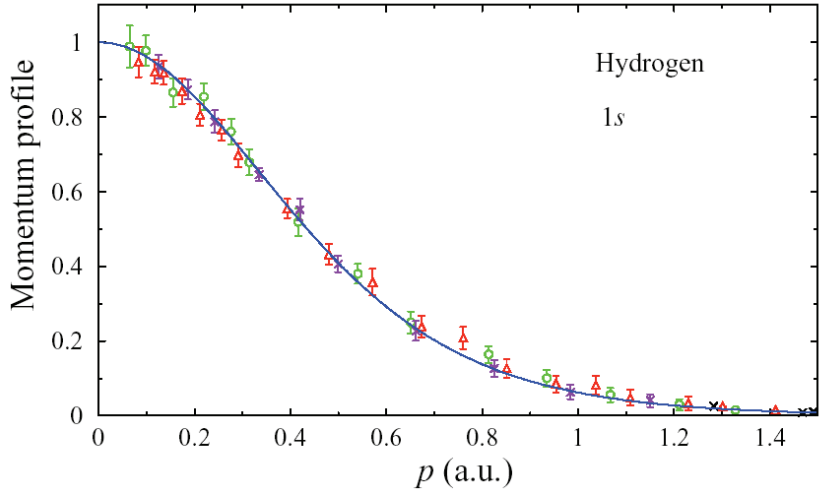}
\caption{\label{fig:hydroe2e}(Color online) Comparison of the normalized
$(e,2e)$ cross section (momentum profile)
from hydrogen with the square of the $1s$
wavefunction in momentum space, adapted 
from~\protect\cite{lohm81}. The momentum $p$ is given in atomic units (a.u.) ($\hbar=m=e=1$).
The solid line represents $(1+p^2)^{-4}$.
The measurements were performed at 1200 eV (crosses), 800 eV (circles),
and 400 eV (triangles). 
Figure adapted from Ref.~\cite{lohm81}.}
\end{center}
\end{figure}

We note that the corresponding reaction on atoms provides similar information. 
This is illustrated with a discussion of $(e,2e)$ data on the hydrogen atom.
The Schr\"{o}dinger equation for hydrogen yields a ground-state wavefunction in momentum space given in atomic units by
\begin{equation}
\phi_{1s}(\bm{p}) = \frac{2^{3/2}}{\pi} \frac{1}{(1+p^2)^2} .
\label{eq:5.34}
\end{equation}
An $(e,2e)$ experiment on hydrogen was reported in~\cite{lohm81}.
The ability of the $(e,2e)$ reaction to extract the square of the ground-state 
wavefunction, is demonstrated in Fig.~\ref{fig:hydroe2e}.
The cross section was obtained at several incident energies, all high enough 
to ensure that the PWIA accurately describes the
reaction. 
When the appropriate electron-electron (Mott) cross section is divided out
at these different energies, the result should become independent of
energy.
The comparison for these different energies with the momentum
profile, given by the square of $(1+p^2)^{-2}$
[see Eq.~(\ref{eq:5.34})], convincingly
demonstrates the correctness of this interpretation.
The $(e,2e)$ experiments on the hydrogen atom therefore come as close
as practically possible to measuring the (square) of the electron
wavefunction. 
Similar experiments on closed-shell atoms reveal that not all the strength of valence orbits can be found in ground state to ion ground-state transition even though typically only a few percent of the strength is missing~\cite{McCW91}.
Furthermore, the missing strength is typically found in small fragments not far removed in energy so that the occupation of the valence strength is still nearly 100\%.

\begin{figure}[t]
\begin{center}
\includegraphics[origin=bl,width=0.6\textwidth,angle=-90]{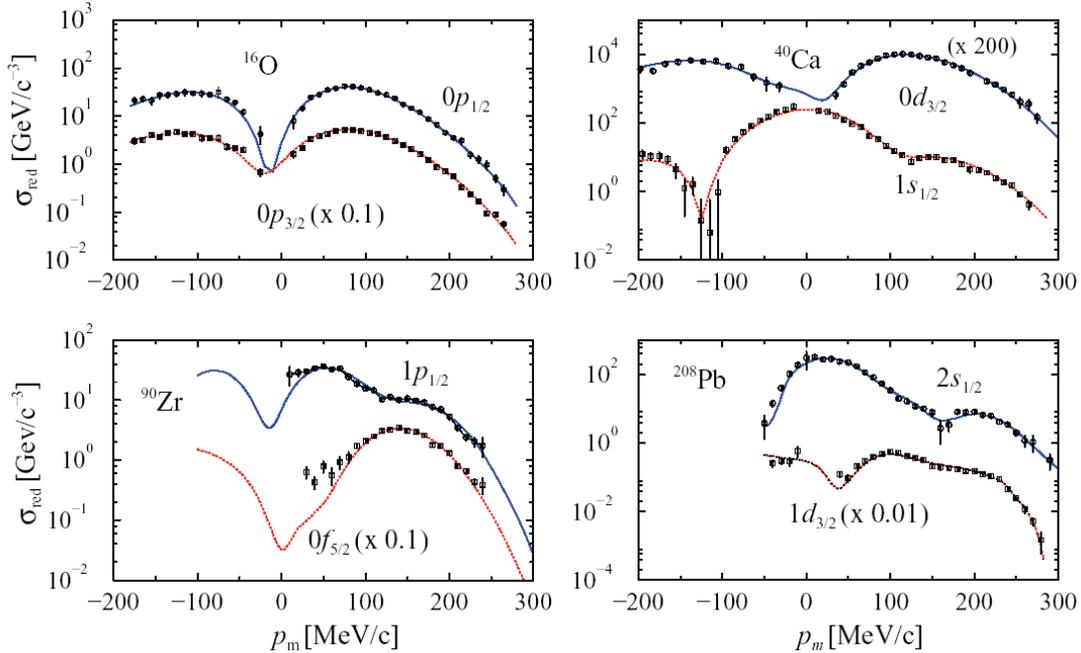}
\caption{\label{fig:loukall}(Color online) Momentum profiles as
a function of missing momentum $p_m$, for valence holes in several
closed-shell nuclei. Appropriate scale factors have been applied to
allow the representation of the data for different orbitals.
The experiments were performed at 
the NIKHEF facility in Amsterdam. Figure adapted from Ref.~\protect\cite{Lapikas:1993}.}
\end{center}
\end{figure}

The situation is different in nuclei as illustrated in the following and already discussed in Sec.~\ref{sec1}.
The momentum dependence of the $(e,e'p)$ cross section for a specific
final state, is dominated
by the $s.p.$ wavefunction associated with the corresponding orbital.
It is always necessary to reduce the theoretical cross section
by a spectroscopic factor to yield the best overall fit to the data.
This spectroscopic factor is found to be substantially less than 1 as illustrated in Fig.~\ref{fig:louksf_intro}.
Examples of this analysis for several closed-shell nuclei,
are shown in Fig.~\ref{fig:loukall}.
So-called reduced cross sections are plotted, which
have been divided by the elementary
electron-proton cross section at the appropriate kinematic conditions, as done in Eq.~\eqref{eq:momdist}.

Figure~\ref{fig:loukall} demonstrates that the shapes of the valence nucleon 
wavefunctions accurately describe the observed cross sections.
Such wavefunctions have been employed for years in nuclear-structure 
calculations, which have relied on the independent-particle model.
The description of the data in Fig.~\ref{fig:loukall}, however, requires
a significant reduction in terms of the appropriate spectroscopic factor.
A compilation for the spectroscopic factor of the last valence orbit for
different nuclei is shown in Fig.~\ref{fig:louksf_intro}~\cite{Lapikas:1993}.
The results shown in Fig.~\ref{fig:louksf_intro} indicate that there is an
essentially global reduction of the $s.p.$ strength of about
35 \% 
for these valence holes in most nuclei.
Such a substantial deviation from the prediction of the independent-particle
model, requires a detailed explanation on the basis of the correlations
that dominate in nuclei.

\begin{figure}[t]
\begin{center}
\includegraphics[width=0.3\textwidth,angle=-0]{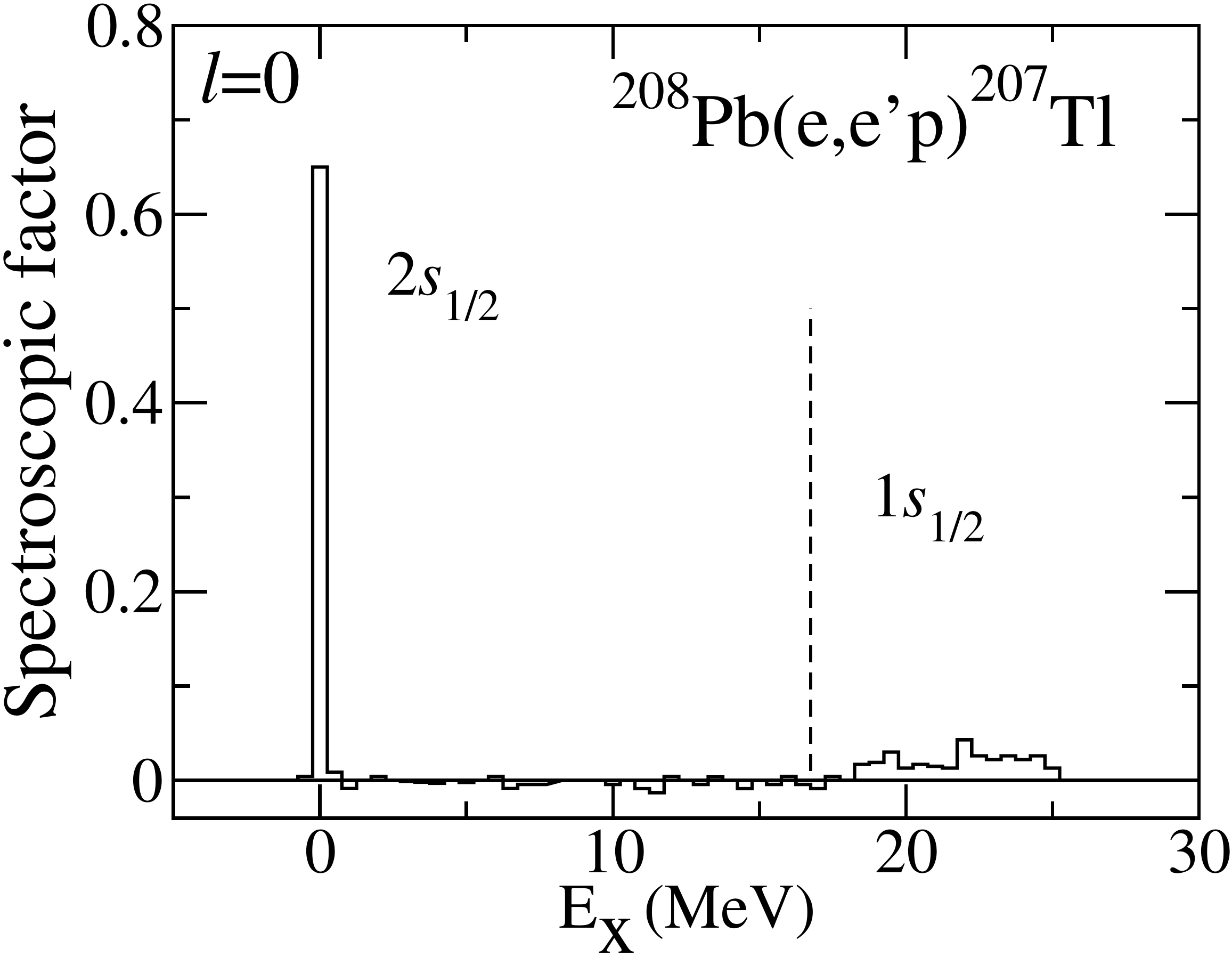}
\includegraphics[width=0.3\textwidth,angle=-0]{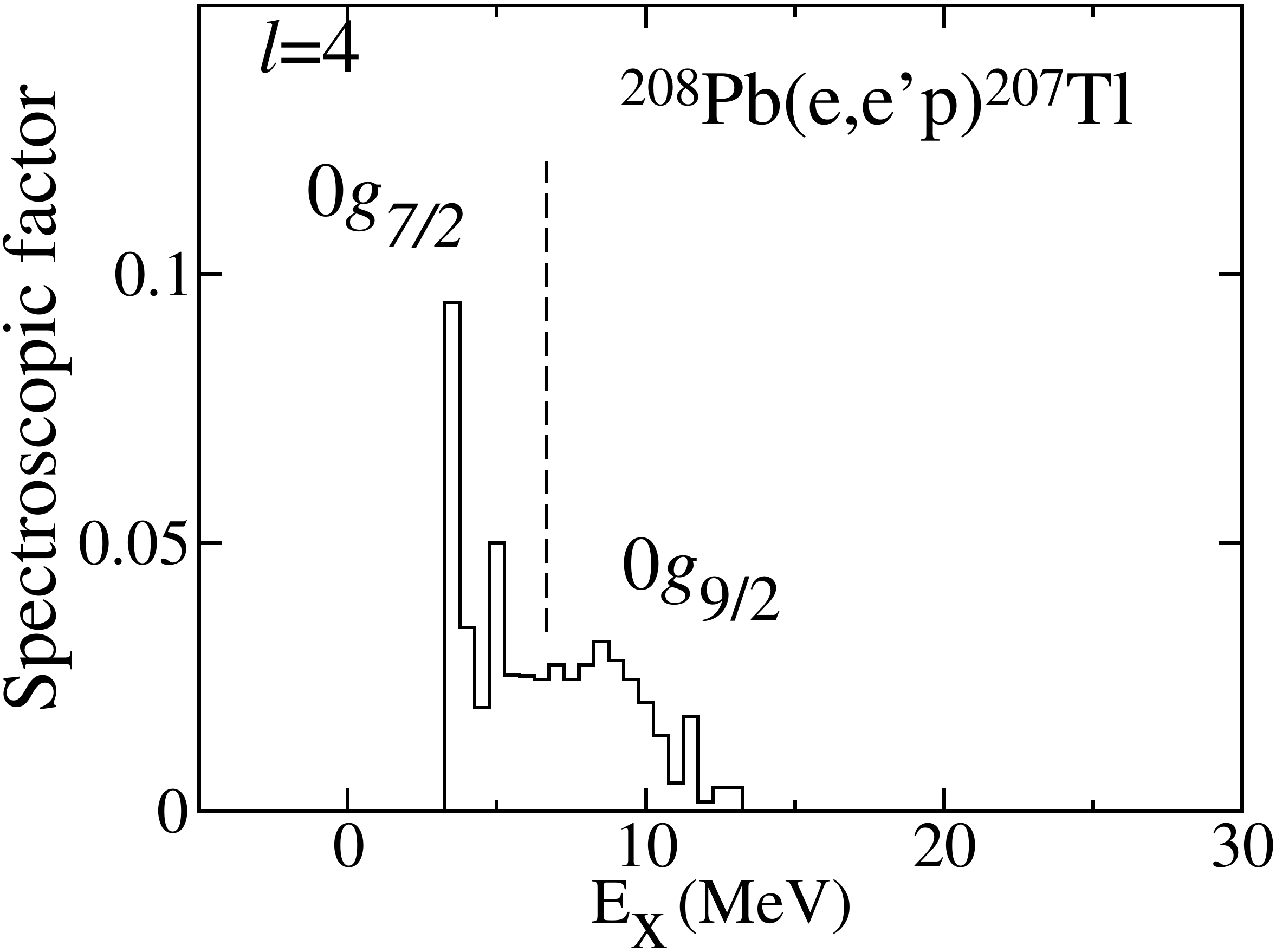}
\includegraphics[width=0.3\textwidth,angle=-0]{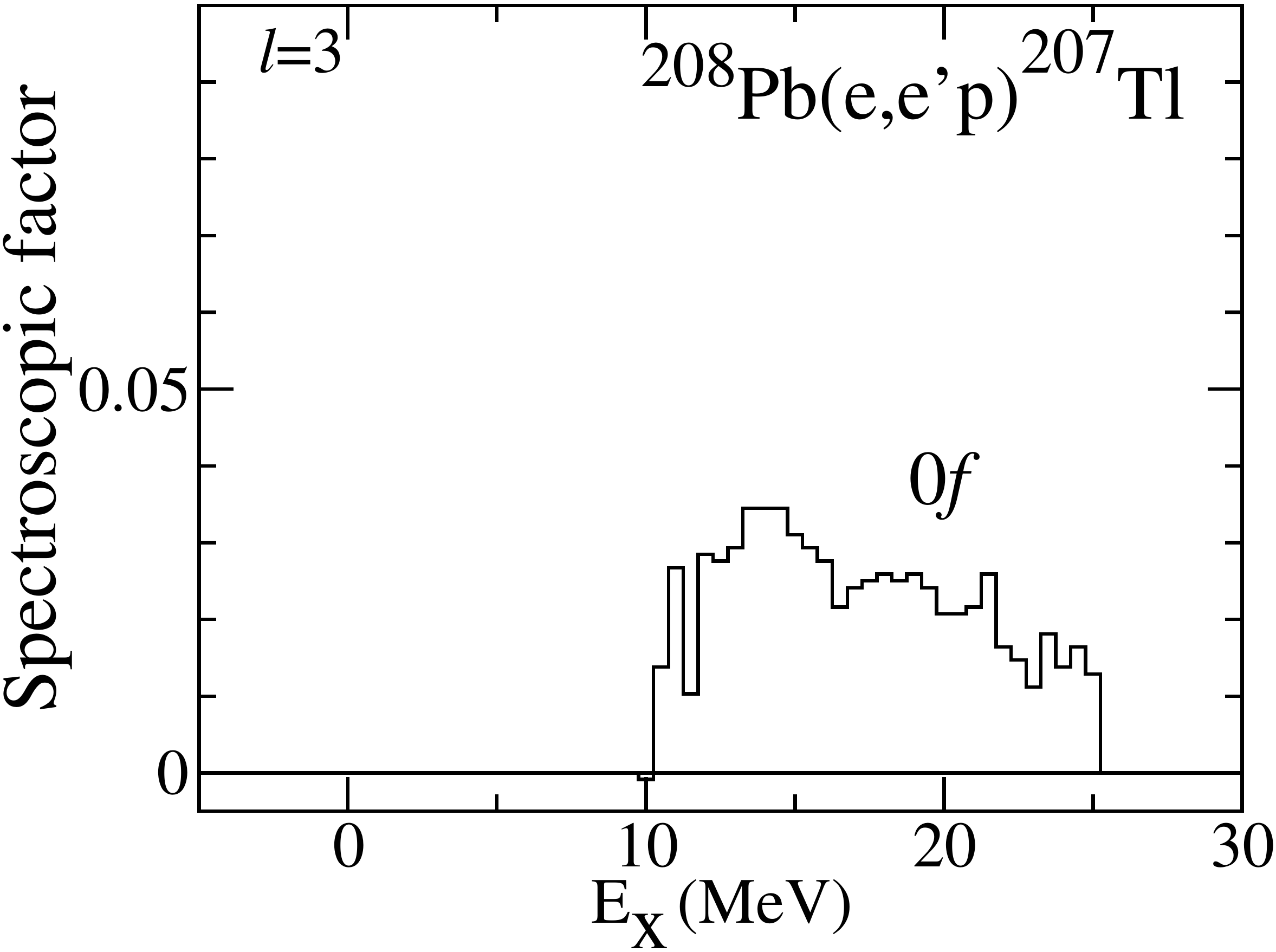}
\caption{\label{fig:quint}
Results for the spectroscopic strength for ${}^{208}$Pb as a function of
excitation energy, $E_X$, in ${}^{207}$Tl for $\ell =0,4$ and 3, obtained from the
$(e,e'p)$ reaction, adapted from Ref.~\protect\cite{qui}.
The spectroscopic factor for the valence $2s_{1/2}$ has been adjusted to
0.65 in accord with the analysis of~\protect\cite{deWitt3}.
No error bars are shown, to emphasize the character of the 
observed fragmentation patterns. 
This vertical dashed line in the central figure identifies the start of the strong fragmentation of the strength associated with the $g_{9/2}$ orbit.
The right figure illustrates the strong fragmentation of the strength associated with the deeply bound $f$ orbits.
}
\end{center}
\end{figure}

An additional feature, obtained in the $(e,e'p)$ reaction, is the fragmentation
pattern of the more deeply bound orbitals in nuclei.
This exhibits single isolated peaks only
in the immediate vicinity
of the Fermi energy, whereas for more deeply bound states a stronger
fragmentation of the strength is observed with increasing distance from
$\varepsilon_F$.
This is beautifully illustrated by the ${}^{208}\mathrm{Pb}(e,e'p)$ data 
from~\cite{qui}, shown in Fig.~\ref{fig:quint}.
Whereas the $2s_{1/2}$ orbit exhibits a single peak shown in the left panel, there is a
substantial fragmentation of the $0f$ strength in the right panel, which corresponds to the
most deeply bound strength considered.
Intermediate results are extracted for orbits in between these two extremes,
as illustrated in the central panel for $\ell =4$ strength.
Additional information about the occupation number of the $2s_{1/2}$ orbit 
is also available.
By analyzing elastic electron scattering cross sections of 
neighboring nuclei~\cite{wag86},
the occupation number for the $2s_{1/2}$ proton orbit of 0.75 is extracted,
which is about 10\% 
larger than the spectroscopic factor~\cite{grab92}.
An occupation number less than 1, requires a different explanation than
for the observed pattern of fragmentation.
The latter pattern can be understood on the basis of the substantial
mixing of the valence hole states with 1p2h states.
In the case of atoms, it is permissible to continue to treat the
ground state as a Slater determinant, even in the presence of electron-electron
interactions.
Such a treatment is not valid for nuclei, since the mutual interaction of
nucleons is much stronger, particularly at short interparticle distances.
Indeed, this repulsive interaction will reduce the wavefunction of
the relative motion of two nucleons substantially.
This reduction requires the admixture of high-momentum components in the
relative wavefunction, corresponding to states at high excitation energy.
The strong short-range repulsion of the interaction is therefore capable to admix
high-lying two-particle--two-hole (2p2h) states into the correlated ground state
$\ket{\Psi^A_0}$ of the nucleus.
Such admixtures lead to a much more complicated ground state which includes
2p2h and additional $n$p$n$h components.
The removal of a valence particle is not possible from these contributions
to the correlated ground state, leading to a reduced occupation number.
The depletion of the Fermi sea must of course be accompanied by
the occupation of states that are empty in the independent-particle 
description, such that the total number of particles is conserved.
The importance of short-range correlations (SRC) suggests that the occupation
of high-momentum states may figure
prominently in accounting for all the particles in the nucleus.

One of the last $(e,e'p)$ experiments performed at the NIKHEF facility
before it was decommissioned, explored
the removal of all the protons in the energy and momentum domain,
corresponding to the independent-particle model.
The experiment was performed on ${}^{208}$Pb~\cite{bat}.
The complete energy and momentum dependence of the cross section was
analyzed in terms of the contribution of all the proton orbits occupied in
the independent-particle model.
For this purpose, energy distributions like those in Fig.~\ref{fig:quint}
were suitably parametrized
and combined with momentum profiles from a standard Woods-Saxon potential
for this nucleus.
As fit parameters to the data, the overall occupation numbers
associated with these orbits, were employed.
The resulting occupation numbers exhibit a depletion of about 20\% for most proton orbits that are full in the independent-particle model while those near the Fermi energy exhibit slightly more 
depletion.
This behavior strongly suggests that SRC play an important role in depleting mean-field orbits.

\subsection{Calculations of spectroscopic factors and spectral distributions}
\label{sec:specdis}
In this section a brief review of the physical ingredients determining the distribution of spectral strength is given and illustrated with examples from the literature.

\subsubsection{Influence of long-range correlations}
\label{sec:LRC}
The importance of low-lying collective states in fragmenting $s.p.$ strength near the Fermi energy has been well understood as documented, {\it e.g.}, in \cite{MAHAUX19851}.
Since these excitations involve surface vibrations, it is natural to introduce the nomenclature to associate their effect with long-range correlations (LRC).
Such a choice of language would also be natural for an infinite system where low-energy excitations are associated with long wavelengths.
The self-energy picture belonging to this type of analysis replaces the noninteracting particle-hole propagators contained in the second-order expression of Eq.~\eqref{eq:13.16} by Tamm-Dancoff (TDA), random phase approximation (RPA), or phenomenological phonons that generate the fragmentation of the strength associated with this coupling.
An example of the neutron $i_{11/2}$ particle strength obtained from a particle-vibration calculation is shown in Fig.~4.37c of Ref.\cite{MAHAUX19851}.
It displays the dependence of the spectroscopic factors of the $i_{11/2}$ neutron orbit upon the excitation energy $E_X$ in ${}^{209}$Pb. The spectroscopic factor (0.85) of the lowest state is equal to the quasiparticle strength. The spectroscopic factors of the higher-lying excitations have been multiplied by 50. The sum of the spectroscopic factors shown in this figure is equal to $1-n(i_{11/2})$ with an occupation corresponding to 0.06 in this case.
The distribution continues to be dominated by a single fragment with the rest of the strength distributed in both the hole and particle domain qualitatively similar to the discussion in appendix \ref{sec:self}.

Low-energy spectra of open-shell nuclei can be well described by large scale shell-model calculations.
These calculations require two-body matrix elements but do not depend on radial wavefunctions explicitly.
Such calculations generate spectroscopic factors associated with the orbits in the chosen configuration space
like, {\it e.g.}, the $sd$-shell.
Transfer reactions are discussed in Sec.~\ref{sec3} and can be employed to compare theoretical spectroscopic factors with those obtained from relevant transfer cross sections.
The standard procedure to obtain spectroscopic factors, is to calculate the distorted waves using optical-model potentials. The overlap functions are typically described by $s.p.$ states in a Wood-Saxon potential with the depth adjusted to give the correct binding energy. Typically the radius and diffuseness parameters are taken to be the  standard values ($r_{0}\sim$1.25~fm and $a\sim$0.65). The dependence on the choice of the optical-model parameters is important, the extracted spectroscopic factor can vary substantially. Schiffer \textit{et al.} argued that one should use a parameter set specifically fitted to elastic-scattering data at the energy of interest with caution \cite{Schiffer:1967}. 
Rather one should employ global parametrizations. 
There are still significant differences with using different sets of global parameters with changes of order 20\% coming from different choices \cite{Liu:2004}. To compare spectroscopic factors, it is best to calculate the cross section in a consistent manner, \textit{i.e.}, with the same global optical model parameters, and the overlap state with the same radius and diffuseness parameters.  Lee, Tsang, and Lynch have recently reanalyzed a large body of  ($d$,$p$) and ($p$,$d$) cross section data to this end and found consistency of the spectroscopic factors from these two reactions~\cite{Lee07}.
Large spectroscopic factors from shell-model calculations tend to correlate quite well with the results from the experimental analysis as illustrated in Ref.~\cite{Tsang09} for the $sd$ and $pf$ shells. A more in depth discussion is provided in Sec.~\ref{sec3}.

\begin{figure}[t]
\begin{center}
\includegraphics[clip,trim=0cm 3cm 0cm 3cm,,width=0.6\textwidth]{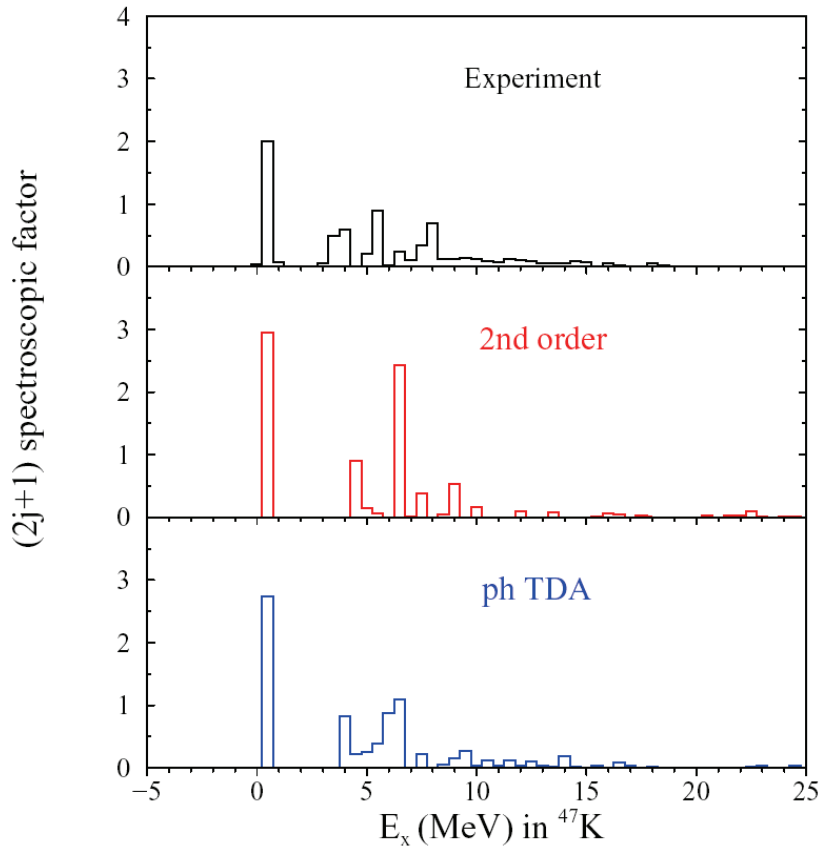}
\caption{\label{fig:VU48} (Color online) Spectroscopic strength for $\ell = 2$ removal compared to extracted experimental strength from the ${}^{48}$Ca$(e,e'p)$ reaction~\cite{Kramer:2001}. Calculated strength in the middle panel is generated from a second-order self-energy calculation~\cite{Brand91} using a local effective interaction that was obtained from a realistic interaction in a nuclear-matter calculation~\cite{Dickhoff83}. 
Correlating the particle-hole propagator in the second-order diagram to include all forward-going diagrams in the TDA results in the bottom panel~\cite{Rijsdijk92}. Figure adapted from Ref.~\cite{Rijsdijk92}.
}
\end{center}
\end{figure}

An example of a self-consistent calculation of the nucleon self-energy employing the second-order self-energy is presented in Ref.~\cite{Dim1991} for ${}^{48}$Ca employing a Skyrme effective interaction.
By using a binning procedure it is indeed possible to achieve self-consistency and generate smooth strength distribution except for valence holes.
The comparison of this type of calculation with experiment confirms that the main qualitative features of the strength are already contained in second-order calculations although the valence spectroscopic factors are substantially larger than extracted from experiment.
This is also the case when a more realistic effective interaction~\cite{Dickhoff83} is employed as shown in Fig.~\ref{fig:VU48}. 
The top panel of Fig.~\ref{fig:VU48} represents the experimental $\ell = 2$ distribution~\cite{Kramer:2001} for the proton removal from the ground state of ${}^{48}$Ca. 
The middle panel exhibits the theoretical strength using the second-order approximation~\cite{Brand91} employing a $\mathcal{G}$-matrix effective interaction~\cite{Dickhoff83} which accommodates SRC at the two-body level (but not in the self-energy).
The bottom panel includes additional low-energy correlations at the TDA level for the particle-hole propagation inside the self-energy~\cite{Rijsdijk92}.
The experimental strength for the lowest fragment corresponds to the removal of the valence $d\scriptstyle \frac{3}{2}$ proton whereas the rest of the experimental strength is presumed to be of $d\scriptstyle \frac{5}{2}$ character. 

The configuration space for the calculation shown in Fig.~\ref{fig:VU48} comprises three major shells above and all shells below the Fermi energy. 
Such calculations therefore encompass the energy domain of both collective low-lying surface excitations as well as giant resonances.
The resulting strength distribution exhibits many similarities to those of quasiparticles in a Fermi liquid as pioneered by Landau~\cite{Land1,Land2,Land3}.
As in a Fermi liquid the $s.p.$ excitations near the Fermi energy behave like particles from the mean field but renormalized by the spectroscopic strength with the remainder distributed to energies both above and below the Fermi energy. 
For $s.p.$ excitations farther from the Fermi energy a substantial fragmentation is observed illustrated, {\it e.g.}, in Fig.~\ref{fig:quint} for proton $\ell = 3$ strength in ${}^{208}$Pb and Fig.~\ref{fig:VU48} for proton $d\scriptstyle \frac{5}{2}$ strength in ${}^{48}$Ca.
Noteworthy conclusions from the calculations are that often additional fragmentation of the strength is required and a further reduction of the spectroscopic factors associated with the valence holes is needed which can be ascribed to the effect of SRC.
In Ref.~\cite{Brand90} this feature was confirmed when excited states were calculated in the same space including the same $s.p.$ fragmentation and therefore explicitly the coupling to 2p2h states.

\subsubsection{Influence of short-range and tensor force correlations}
\label{sec:SRC}

There has been a long history of constructing nucleon-nucleon interactions in the literature.
Ultimately, such an interaction has to be based on Quantum Chromo Dynamics (QCD) and some groups perform lattice gauge calculations to clarify this link as, {\it e.g.}, Ref.~\cite{ishii07}).
An important message from this work is that when more realistic quark masses are employed, and therefore better pion masses implied, a substantial repulsive core emerges that describes the nucleon-nucleon interaction at short distances.
This is in agreement with empirical considerations that recognized long ago that the $S$-wave phase shifts suggest an easy interpretation if it is assumed that a strong repulsive core is responsible for their dependence on scattering energy.
An example of such a phenomenological interaction is the Reid soft-core potential~\cite{Reid68}, later followed by a more accurate Reid93 version~\cite{reid93}, and culminating in the widely used and very accurate Argonne AV18 interaction~\cite{PhysRevC.51.38}.
Simultaneous with these developments, a very successful physical picture emerged in which the exchange of the experimental bosonic excitations of QCD was employed to model the nucleon-nucleon interaction.
This meson-exchange description also resulted in a very accurate so-called CD-Bonn interaction~\cite{PhysRevC.68.041001}.
It is noteworthy that such a physical picture is also relevant to describe in-medium interactions in many-body systems.
The exchange of lattice phonons between electrons is an important example as it is responsible for the phenomenon of superconductivity.

More recently, interactions based on a power counting scheme proposed by Weinberg and incorporating chiral symmetry have become very popular. 
Reviews are available in \cite{MACHLEIDT20111,RevModPhys.81.1773}.
All of these chiral interactions are however quite soft and in the mean time a large number of different versions has proliferated.
This implies that these modern interactions do not do justice to the coupling of the nucleon-nucleon two-body state at low energy to the states with the same overall quantum numbers at higher energy as experimentally documented, {\it e.g.}, in the nonvanishing of total cross sections.
As a result, these interactions do not generate a lot of high-momentum components.
Nevertheless, a local version of such a chiral interaction including three-body forces~\cite{Piarulli15} has been utilized to successfully calculate the spectra of light nuclei~\cite{Piarulli18} with the same quality as has been accomplished with the AV18 supplemented with a phenomenological three-body interaction~\cite{RevModPhys.87.1067}.
While chiral interactions and their further softening with the similarity renormalization group~\cite{Bogner:2010} can provide sensible many-body calculations, it should be noted that the short-distance behavior of the nucleon-nucleon interaction as documented by nonvanishing total nucleon-nucleon cross sections is not incorporated in this approach as the starting point cannot account for such experimental data.
When strong short-range repulsion is included in the interaction, such a phenomenological approach can at least account for this type of physics and therefore adequately treat SRC, as discussed in the following.

The treatment of SRC including the effect of the nucleon-nucleon tensor force for interactions like AV18 can be accomplished for light nuclei most prominently with Monte Carlo techniques~\cite{RevModPhys.87.1067}.
For heavier nuclei other techniques can generate sensible results.
An example is provided by the Green's function method.
Recent work on a direct calculation of the nucleon self-energy for ${}^{40}$Ca with an emphasis on SRC was reported in Ref.~\cite{PhysRevC.84.044319}.
Such a microscopic calculation of the nucleon self-energy proceeds in two steps, as employed in Refs.~\cite{PhysRevC.49.R17,PhysRevC.51.3040,POLLS1995117}.
A diagrammatic treatment of SRC always involves the summation of ladder diagrams.
When only particle-particle (pp) intermediate states are included, the resulting effective interaction is the so-called $\mathcal{G}$-matrix.
Employing such an interaction calculated in nuclear matter, it is possible to iterate the difference between this result and the appropriate finite nucleus quantity thereby generating a sensible approximation to the finite nucleus self-energy that adequately treats SRC.

\begin{figure}[t]
\begin{center}
\includegraphics[origin=bl,width=0.6\textwidth,angle=-90]{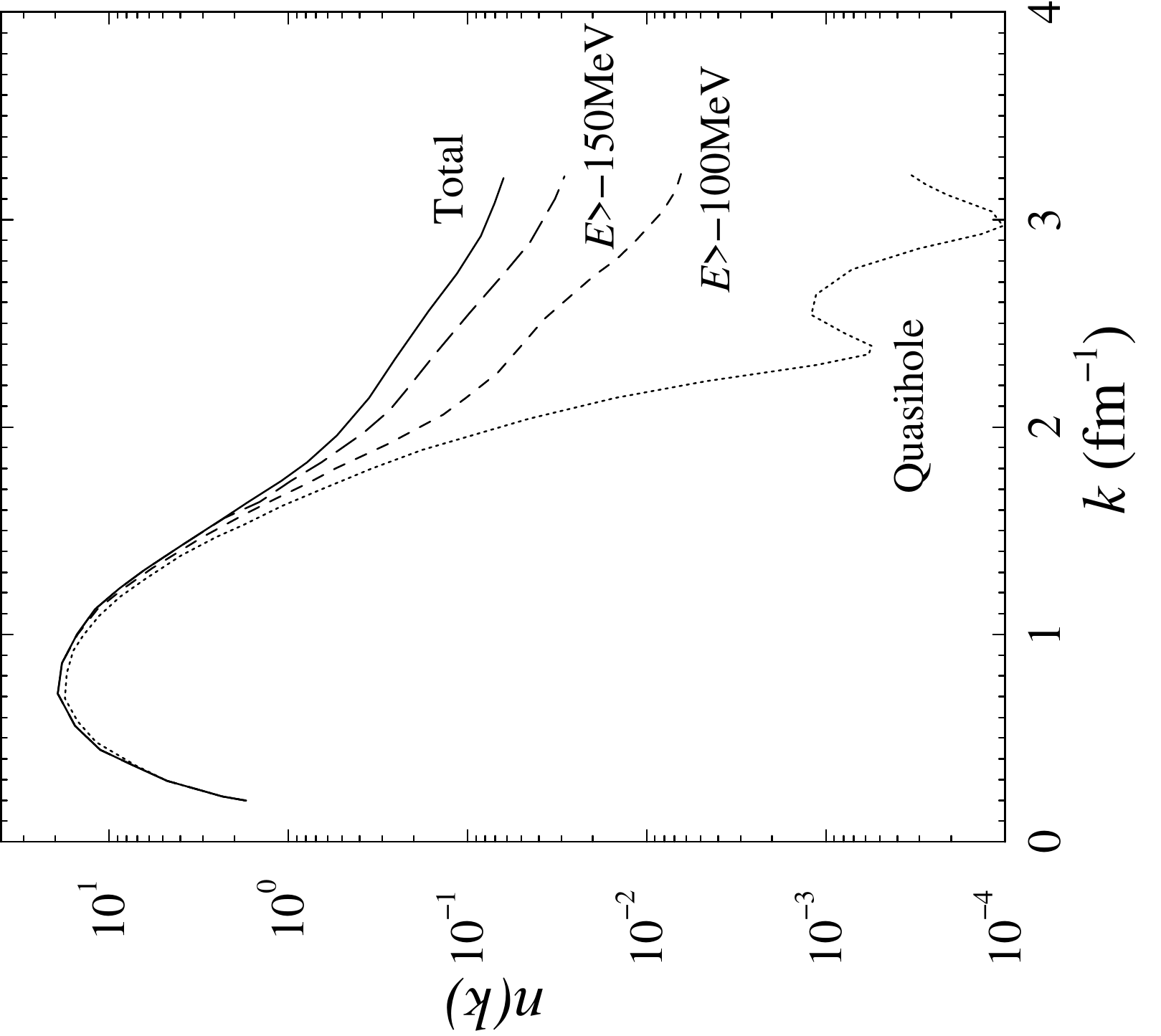}
\caption{\label{fig:16Ohighk} The momentum distribution of ${}^{16}$O adapted from Ref.~\cite{PhysRevC.51.3040}. Shown
are also the quasihole contribution and the results obtained
with various energy cutoffs in the integration of the spectral
functions. The units for $n(k)$ are in fm and therefore contain already the factor $k^2$.}
\end{center}
\end{figure}
The momentum distribution from such a calculation for ${}^{16}$O is shown in Fig.~\ref{fig:16Ohighk}.
The quasihole part corresponds to the energy domain for
$(e,e'p)$ cross sections with small energy transfer, 
\textit{i.e.}, 
leading to the ground state of the final nucleus and excited states
up to $\approx 20$ MeV. The curve denoted by $E>-100$~MeV reflects
the momentum distribution, including all states of the final
nucleus up to around 80 MeV excitation energy, \textit{etc.} 
As a consequence, the high-momentum components of the momentum
distribution due to short-range correlations including the effect of the tensor force (inclusively denoted by SRC in the following) can be
observed mainly in
knockout experiments with an energy transfer of the order of 100 MeV
or more.
Such high-momentum components were indeed not found experimentally near the Fermi energy~\cite{Bobel94}.
The momentum distribution
for $^{16}{\rm O}$ calculated with different approaches and different
interactions, all of which  properly include the effects of SRC, yield quantitatively similar results.
The other approaches include: local density 
approximation (LDA)~\cite{BENHAR1994493}, Fermi hypernetted chain (FHNC)~\cite{CO199473}
and Variational Monte Carlo (VMC)~\cite{PhysRevC.46.1741}.

To understand Fig.~\ref{fig:16Ohighk}, it is important to recall
that the appearance of high-momentum components at a certain
energy in the $A-1$ system is related to the self-energy contribution
containing 1p2h states at that energy.
From energy conservation it is then clear that at low energy
it is less likely to find such states with a high-momentum particle,
than at high energy.  
The same feature is observed in nuclear matter where the peak
of the $s.p.$ spectral function for momenta above $k_F$
increases in energy as $k^2$.
Hence, the hole strength in nuclear matter as a function
of momentum, exhibits the tendency that higher momenta become more dominant
at higher excitation energy. 

The location of high-momentum components was confirmed in \cite{Rohe04} with an $(e,e'p)$ experiment on ${}^{12}$C while also demonstrating that about 10\% of the protons exhibited this high-momentum behavior.
It is noteworthy that this also implies that the reduction of $s.p.$ strength due to SRC associated with removing strength also roughly corresponds to this value.
Experimental NIKHEF data suggest that a substantial additional reduction is required to describe the valence hole distributions of ${}^{16}$O~\cite{Leu94}.
Such an additional reduction must therefore be associated with LRC.

\begin{figure}[t]
\begin{center}
\includegraphics[origin=bl,angle=-90,width=0.5\textwidth]{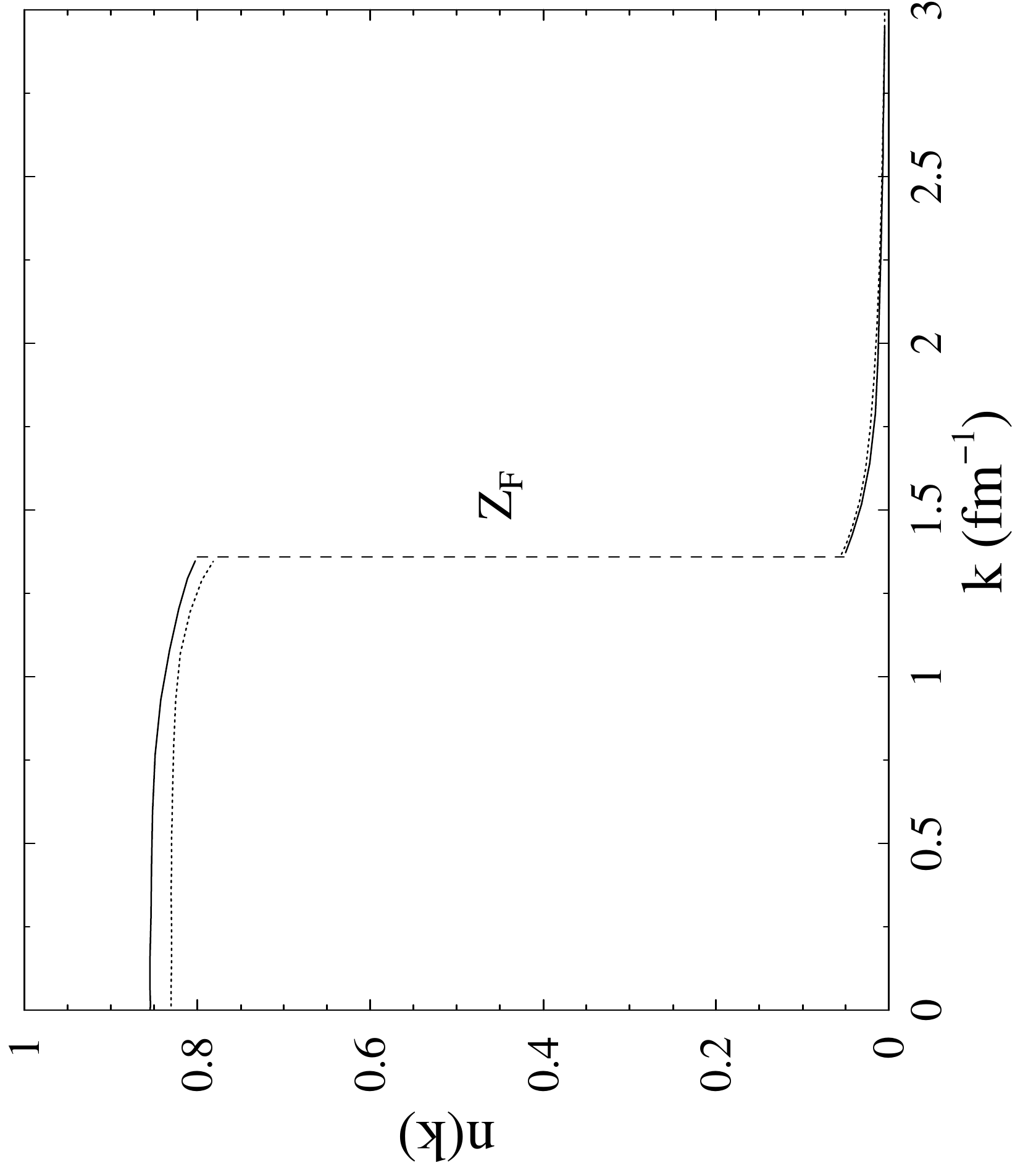}
\caption{\label{fig:nkreidb}
Occupation probability for nuclear matter at 
equilibrium density calculated by integrating hole spectral functions
obtained with mean-field propagators as input (dotted line).
Self-consistent determination of the $s.p.$ propagators yields the solid line~\cite{Libth}.
}
\end{center}
\end{figure}

A complete treatment of SRC can actually be accomplished in nuclear matter calculations.
The self-energy diagrams of Fig.~\ref{fig:SEDB}(a) of the Appendix \ref{sec:self} can be calculated completely when $\Gamma^{4pt}$ is approximated by the sum of ladder diagrams which ensures a proper treatment of SRC.
In turn the ladder diagrams are obtained by including the fully off-shell propagating particles in the medium.
This incorporates the important physics that particles propagate with knowledge of the correlated ground state in which occupation below the Fermi momentum is depleted and compensated by the occupation of high-momentum states.
The self-consistent implementation of such a scheme has been fully realized in recent finite temperature calculations that avoid the technical complications of having to deal with possible pairing instabilities as discussed, {\it e.g.}, in Refs. \cite{PhysRevC.79.064308,PhysRevC.89.044303}).
An approximate result at zero temperature was obtained in \cite{Libth} employing the Reid soft-core interaction~\cite{Reid68}.
The resulting momentum distribution is shown in Fig.~\ref{fig:nkreidb} and compared with an earlier calculation that propagated mean-field particles in the summation of ladder diagrams.
While these two distributions are quite similar, the corresponding spectral function exhibit substantial differences that influence the saturation properties of nuclear matter~\cite{PhysRevLett.90.152501}.
On the basis of such calculations it was predicted that a similar depletion of orbits in finite nuclei would be observed~\cite{DICKHOFF1994119} which was later confirmed in Ref.~\cite{bat}.

Near $k=0$, 
$n(k)$ becomes fairly constant with a value of 0.85.  
Investigating the influence of the tensor force, it may be concluded that roughly $1/3$ of the 15\% 
depletion is due to the effect of tensor correlations in the ladder equation.  
Another $1/3$ is due the to
high-energy tail in the particle spectral function at energies above 500~MeV~\cite{PhysRevC.44.R1265,VONDERFECHT19931}.  Other
many-body methods such as Brueckner theory~\cite{GRANGE1987365} and \cite{PhysRevC.41.1748}
and correlated basis functions
(CBF) theory~\cite{BENHAR1989267}, using other realistic interactions,
produce very similar occupation near $k=0$.  
In the work of~\cite{GRANGE1987365} and \cite{PhysRevC.41.1748}, 0.82 is
reported for the Paris potential.  
All these calculations for different
interactions, using different methods, give similar values for
$n(0)$ which appears to be mostly determined by the strength of the repulsive core and the inclusion of a realistic tensor force.
This is encouraging, since it implies that nonrelativistic many-body
calculations yield stable results in the region, 
where one would like to compare to finite nuclei.  
In contrast to $n(0)$,
the occupation at $k_F$ varies significantly between methods.  
The extra depletion in $n(k)$ as $k\rightarrow k_F$
arises from the enhanced ability of the $s.p.$ state to couple to low lying 2p1h
excitations, as its energy approaches these states and therefore reflects the effect of LRC in nuclear matter. 

\begin{figure}[t]
\begin{center}
\includegraphics[origin=bl,angle=-0,width=0.5\textwidth]{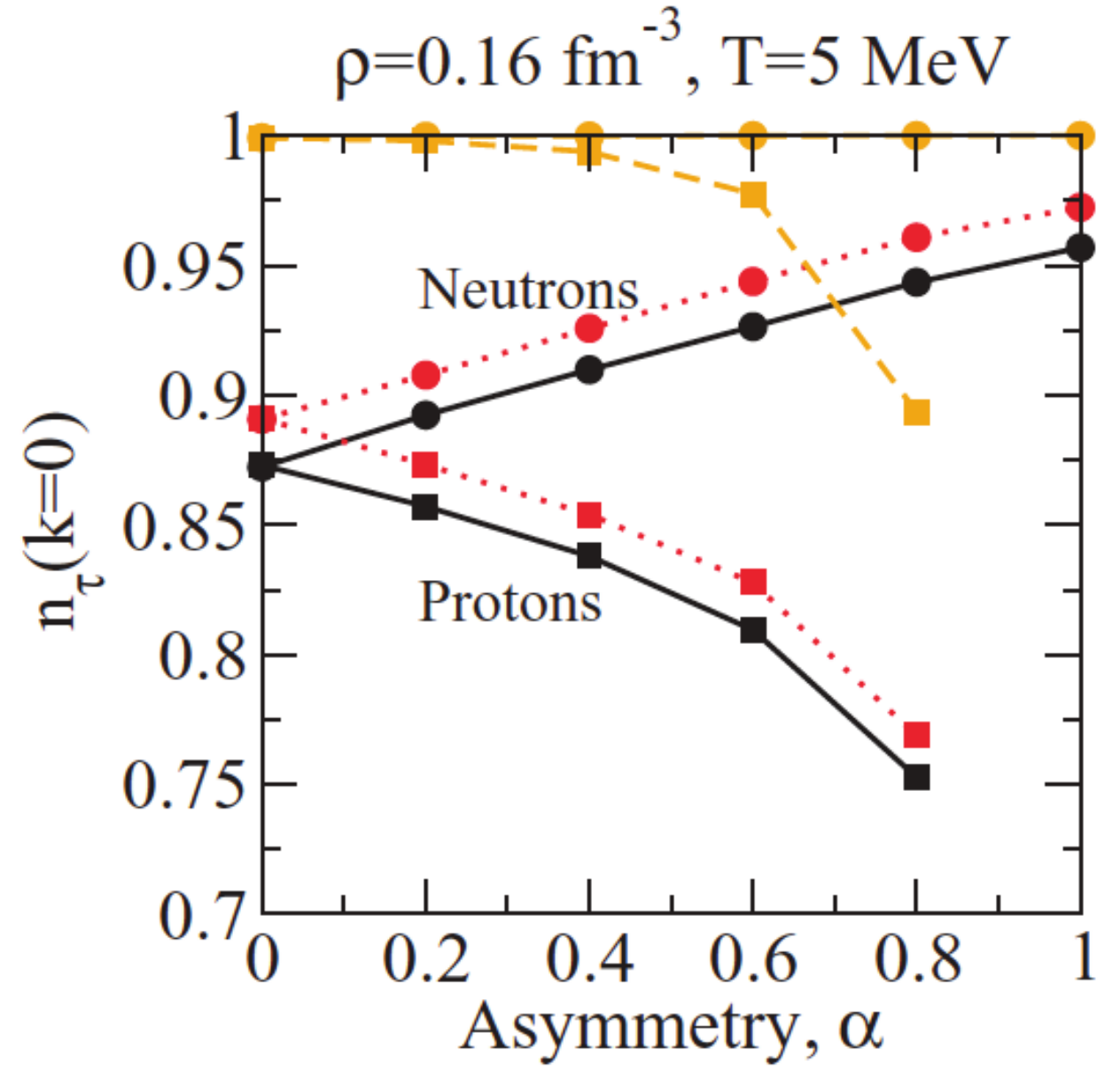}
\caption{\label{fig:rios1}
(Color online) Isospin asymmetry dependence of the neutron (filled circles) and proton (squares) lowest momentum occupation at $T=5$~ MeV and density of 0.16~fm$^{-3}$. Correlated results for the AV18 (solid lines) and CD-Bonn (dotted lines) interactions are compared to the free Fermi gas (dashed lines) predictions. 
Reprinted figure with permission from Ref.~\cite{PhysRevC.79.064308} \textcopyright2009 by The American Physical Society.
}
\end{center}
\end{figure}

These conclusions are essentially maintained when more recently developed interactions that contain a substantial repulsion like AV18~\cite{PhysRevC.51.38} and CD-Bonn~\cite{PhysRevC.68.041001} are employed.
These interactions generate depletions at $k=0$ of 13\% and 11\%, respectively~\cite{PhysRevC.79.064308}.
This information is displayed in Fig.~\ref{fig:rios1}.
The figure displays the occupation of the $k = 0$ momentum state in matter at a density of 0.16 fm$^{-3}$ and temperature $T = 5$ MeV as a function of nucleon asymmetry $\alpha = (N-Z)/(N+Z)$ for protons and neutrons, as well as for the CD-Bonn (dotted lines) and AV18 (solid lines) interactions obtained from a fully self-consistent calculations as outlined above.
The inclusion of the Fermi gas result indicates that the proton results become temperature dependent at asymmetries beyond $\alpha = 0.6$ clarifying the earlier result of Ref.~\cite{frick05}.
For the correlated depletion, the occupation of the zero momentum state is an increasing (decreasing) function of the asymmetry for the more (less) abundant component. The behavior is very similar for both nucleon-nucleon interactions, although the occupations for AV18 are systematically smaller than those for CD-Bonn.

The physical effect of this increasing difference between neutron and proton depletion and the implied occupation of high-momentum states must be sought in the increasing (decreasing) importance of the ${}^3S_1$-${}^3D_1$ channel for protons (neutrons) with increasing nucleon asymmetry.
The effect is therefore associated with the tensor force.
\begin{figure}[t]
\begin{center}
\includegraphics[origin=bl,angle=-0,width=0.5\textwidth]{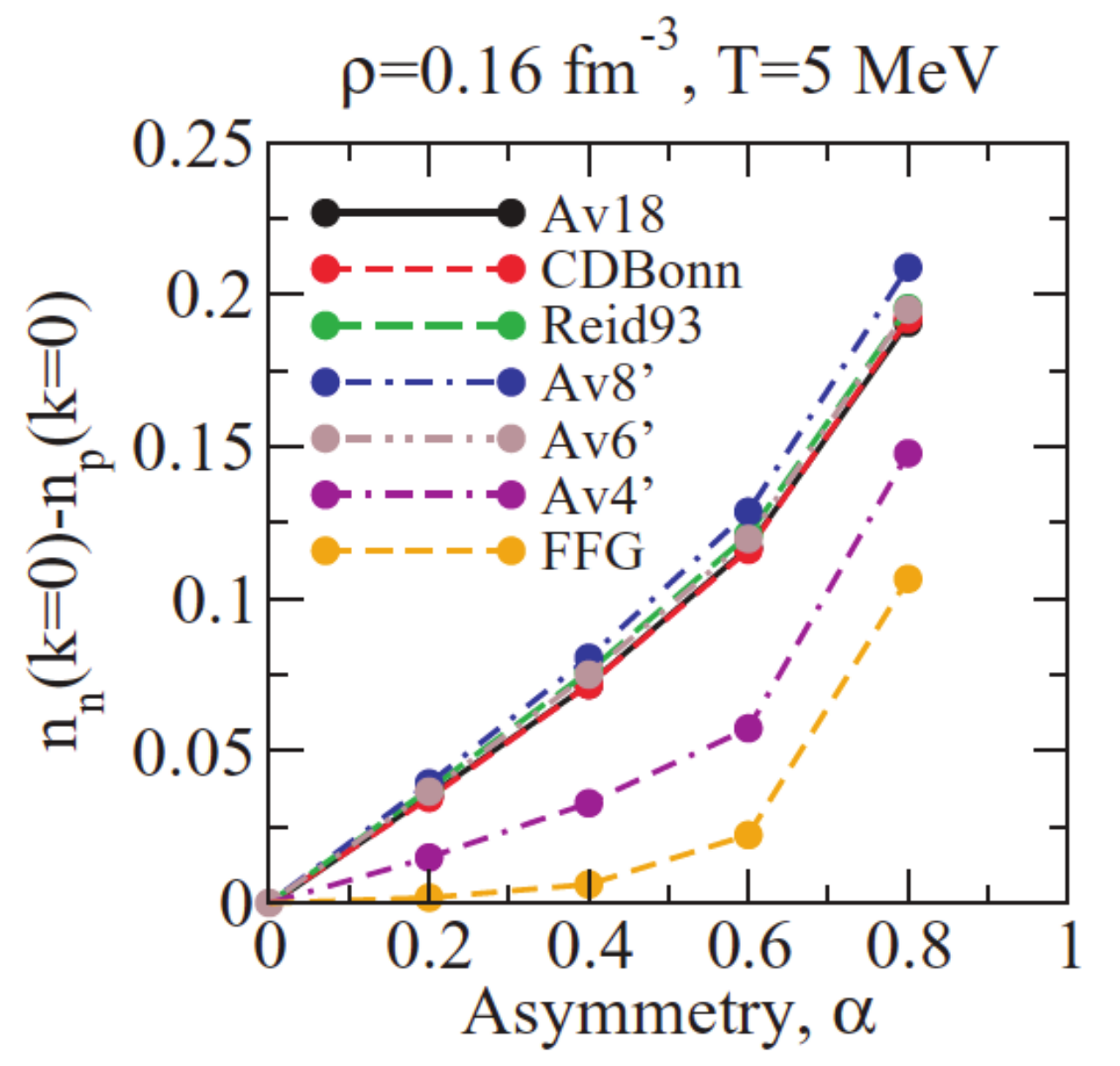}
\caption{\label{fig:rios2}
(Color online) Difference of neutron and proton occupation of the lowest momentum state as a function of isospin asymmetry for different interactions at $T = 5$ MeV for a density of  0.16 fm$^{-3}$.
The curve labeled FFG refers to the free Fermi gas results while the other curves refer to fully self-consistent results for various nucleon-nucleon interactions.
Reprinted figure with permission from Ref.~\cite{PhysRevC.79.064308} \textcopyright2009 by The American Physical Society.
}
\end{center}
\end{figure}
In spite of the differences observed in both symmetric nuclear and pure neutron matter  for the AV18 and CD-Bonn results, the asymmetry dependence of the $k = 0$ occupation is very similar at all densities. This is surprising because both forces have a rather different short-range behavior and tensor structure. Given the almost linear dependence of the depletion with asymmetry, a better insight into these differences can be gained by plotting the difference between the $k =0$ occupation of neutrons and protons. In Fig.~\ref{fig:rios2}, this iso-depletion is shown for different interactions as a function of the asymmetry again at $T = 5$ MeV. 
This difference is the same for a wide variety of modern nucleon-nucleon potentials, independent of their short-range or operatorial structure. This seems to suggests that the iso-depletion is fixed by the phase shifts, most probably via their isospin dependence. This is corroborated at higher densities by comparing the very similar AV18 and CD-Bonn predictions~\cite{PhysRevC.79.064308}.

Two results in Fig.~\ref{fig:rios2} fall below the main iso-depletion line. Immediately below most of the interactions, one finds the results corresponding to the AV4$'$ potential. 
This interaction has an extremely simplified operatorial structure, with only a spin-isospin part and no tensor components, and is fitted to reproduce the binding energy of the deuteron~\cite{PhysRevLett.89.182501}. The fact that it lies significantly below the other results shows the importance of tensor effects for isospin asymmetric systems. It appears that the tensor force tends to increase the difference between neutron and proton momentum distributions as asymmetry increases. Moreover, the comparison with AV6$'$ and AV8$'$ suggests that, once the tensor components are included in a force, the iso-depletion will remain almost the same independently of the extra spin-orbit terms. 
This fact is very surprising, particularly if one considers the fact that the momentum distributions of neutrons and protons can be different for each potential.
These results appear to be relevant for the discussion of the presence of high-momentum components of protons and neutrons when there is a neutron majority as in ${}^{48}$Ca and ${}^{208}$Pb~\cite{Duer18}.
In addition, this increased correlation for protons with increasing nucleon asymmetry must ultimately play a role in the removal probability of valence orbits when large nucleon asymmetries are encountered.
The importance of the nuclear tensor force is also well documented in the dominance of proton-neutron events in experiments where two nucleons are removed~\cite{Subedi08}.

\subsubsection{Combined effects of LRC and SRC}
\label{sec:SLRC}
The role of SRC is important but not sufficient to account for the extracted spectroscopic factors from the $(e,e'p)$ reaction.
The effects of SRC on the spectroscopic factors of  $^{16}$O
have been computed in Refs.~\cite{PhysRevC.50.3010,PhysRevC.63.044319,PhysRevC.49.R17}. 
All these approaches treat essentially only SRC and yield spectroscopic factors of about 0.90 for the knockout of a proton for the $p_{1/2}$ and $p_{3/2}$ shells in the $(e,e'p)$ reaction.
 These results disagree with the extracted experimental value of 
about 0.65 for these orbitals employing the standard NIKHEF analysis~\cite{Leu94}.
In Ref.~\cite{PhysRevC.53.2207} a successful
attempt to combine the treatment of low-energy
long-range and short-range (and tensor) correlations was achieved using the energy dependence of the $\mathcal{G}$-matrix to account for the depleting effect of SRC.
The final result yielded a 
$p_{3/2}$ spectroscopic factor of 76\%. 
This points to the importance of LRC which were included
by taking into account the interaction between the hole and particles
propagating in the system at the level of the TDA and therefore cannot
go beyond the 2h1p level.
Calculations reported in \cite{Rijsdijk92} can account for the
coupling of quasiparticles to RPA collective modes and therefore
go beyond 2h1p contributions,
 although it is limited by the difficulties in including
the coupling to 
collective excitations in \textit{both} the pp(hh) and ph channels.
A procedure to include both ph and pp(hh) excitations at the RPA level employing a Faddeev technique (FRPA) in the construction of the self-energy was proposed and implemented in
Refs.~\cite{PhysRevC.63.034313,PhysRevC.65.064313}.
The results of these calculations only slightly reduce the above mentioned spectroscopic factors but provide an excellent strategy for accurate calculations in finite electronic systems~\cite{PhysRevA.76.052503,PhysRevA.83.042517} and other nuclei (see, {\it e.g.}, Ref.~\cite{barbieri2009}).
An important observation is that the configuration space needed for the incorporation of long-range (surface) correlations, including the coupling to giant resonances, is much larger than the space that can be utilized in large-scale shell-model diagonalizations.
FRPA calculations for ${}^{40}$Ca and ${}^{48}$Ca and ${}^{60}$Ca were performed to shed light on the \textit{ab initio} self-energy properties in medium-mass nuclei including the nucleon asymmetry dependence~\cite{PhysRevC.84.034616}.
The main goal of this work was to clarify whether substantial nonlocal contributions should be expected when optical potentials for elastic scattering are considered.
In particular, one may expect to extract useful information regarding the functional form of the DOM potentials from a study of the self-energy for a sequence of calcium isotopes. 
The resulting analysis was intended to provide a microscopic underpinning of the qualitative features of empirical optical potentials.
Additional information concerning the degree and form of the nonlocality of both the real and imaginary parts of the self-energy were also addressed.

\begin{figure}[t]
\begin{center}
\includegraphics[width=0.5\textwidth]{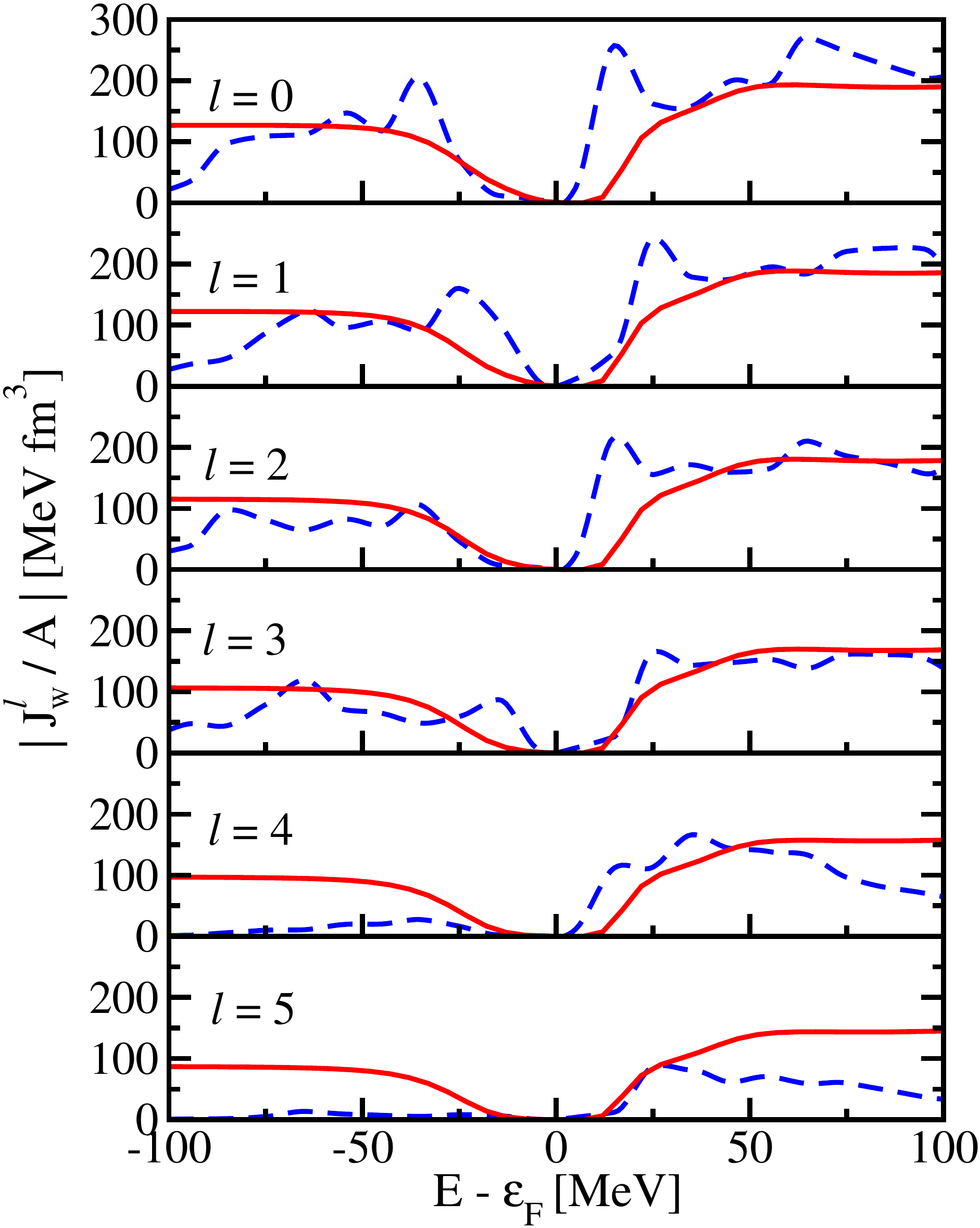}
\caption{(Color online)  Imaginary volume integrals $J^{\ell}_W$ of the ${}^{40}$Ca self-energy for neutrons. The dashed curves represent the FRPA results. The results of the nonlocal DOM fit of Ref.~\cite{Mahzoon:2014} are shown by the solid lines. Figure adapted from Fig.~11 of Ref.~\cite{Dickhoff:17}.} 
\label{fig:Jw_review}
\end{center}
\end{figure}

Early implementations of the DOM employed local potentials and an energy dependence of its imaginary part symmetrically centered around $\varepsilon_F$~\cite{Mahaux:91}. Such features are not obtained in the FRPA
as illustrated in Fig.~\ref{fig:Jw_review}, where for $\ell$-values up to 5, the volume integrals of the FRPA calculation are displayed by the dashed lines.
The domain of the DOM fit extends beyond those shown in Fig.~\ref{fig:Jw_review}, but is limited to those energies for which a meaningful comparison with FRPA results is possible.
We therefore also include the result of the nonlocal fit of Ref.~\cite{Mahzoon:2014} shown by the solid lines in Fig.~\ref{fig:Jw_review} which confirms this assessment.

Since the absorption above the Fermi energy is strongly constrained by elastic-scattering data, it is encouraging that the $\ell$-dependent FRPA result is reasonably close to the DOM fit in the domain where the FRPA is expected to be relevant on account of the size of the chosen configuration space. 
Note that the calculated $J_W$ values decreases quickly at energies $E-\varepsilon_F >$100~MeV due to the truncation the model space. Instead,  it correctly remains sizable even at higher energies in  the DOM.
Also at negative energies, the FRPA results do not adequately describe the admixture of high-momentum components that occur at large missing energies.

An important conclusion is therefore that ab initio calculations most likely will not succeed in generating accurate optical potentials that can be used to analyze nuclear reactions. 
The conclusion also holds when the coupled-cluster approach is employed~\cite{PhysRevC.86.021602,PhysRevC.95.024315}.
It is for this reason that the DOM provides a suitable interface between experimental data and ab initio theory.
Theory informs the functional forms employed in the DOM analysis but cannot describe elastic scattering data in detail. 
The potentials generated by the DOM can however be employed to analyze nuclear reactions as will be clarified in Sec.~\ref{sec:eepX}.
\begin{figure}[t]
\begin{center}
\includegraphics[width=0.6\textwidth]{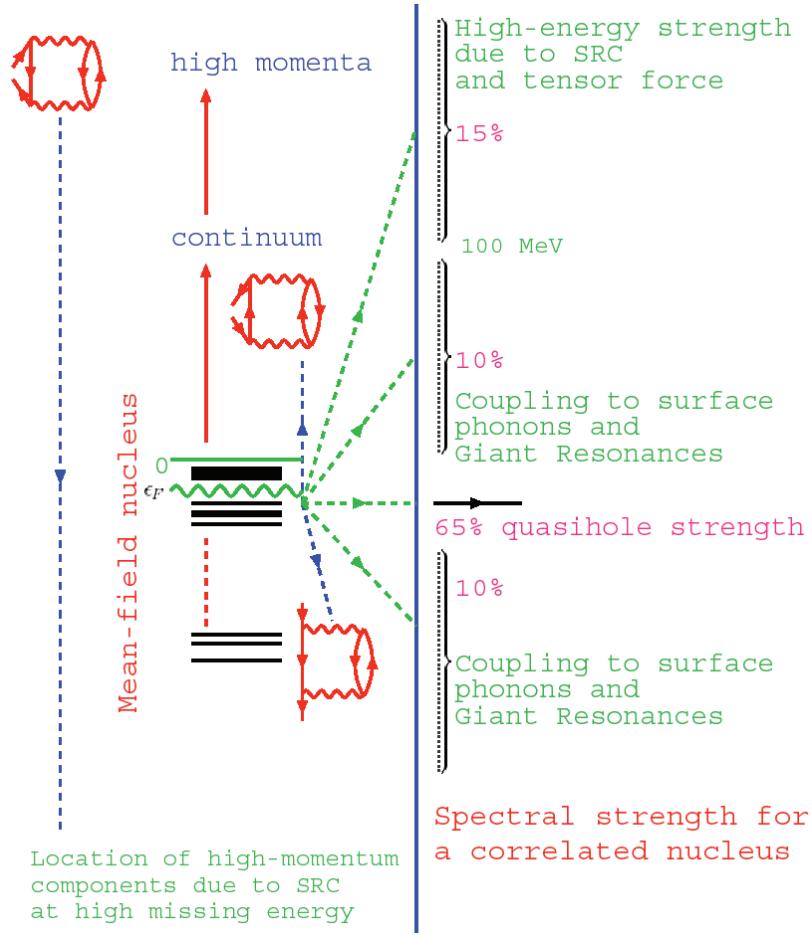}
\caption{(Color online) The distribution of $s.p.$ strength in a nucleus like
${}^{208}$Pb. The present summary is a synthesis of experimental and
theoretical work discussed in this section.
A slight reduction (from 15\% 
to 10\%)
of the depletion effect due to SRC 
must be considered for light nuclei like ${}^{16}$O. 
Reprinted figure with permission from \cite{Dickhoff04} \textcopyright2004 by Elsevier.
\label{fig:strx} }
\end{center}
\end{figure}
As discussed in Ref.~\cite{Dickhoff04}, knowledge obtained from the $(e,e'p)$ reaction and elastic electron scattering combined with some theoretical ingredients shed light on the properties of protons
in the nucleus and may be summarized as follows.
The consequences of SRC in nuclear systems 
appear to be theoretically well understood, while
they have also become available for experimental scrutiny~\cite{Rohe04}.
These results demonstrate that the effect of SRC is two-fold.
First, it involves the depletion
of spectroscopic strength from the mean-field domain.
The data in from Ref.~\cite{bat} indicate that the depletion in heavy nuclei 
corresponds to a little over 15\% for all the deeply bound proton levels.
This was predicted quite some time
ago based on nuclear-matter calculations as discussed above~\cite{DICKHOFF1994119}.
For lighter nuclei all theoretical work suggests that this amount may be
closer to 10\%, also discussed earlier.
Accompanying this information is the realization that valence shells
near the Fermi energy will not contain substantial amounts of high-momentum
components.
This has been experimentally confirmed and clarifies the other role played by 
SRC in nuclei, {\it i.e.}, the admixtures of high-momentum components
at high missing energy that account for the missing protons removed from the
mean-field location.
The location of these high-momentum components~\cite{Rohe04} broadly
conforms with the mechanism
that admixes these correlations with 1p2h states
at large missing energies.

Being able to identify high-momentum components in addition to locating all
the $s.p.$ strength, associated with the mean-field orbits~\cite{bat},
completes the identification of the properties of protons in the
ground state of double closed-shell nuclei like ${}^{16}$O, ${}^{40}$Ca, ${}^{48}$Ca, and ${}^{208}$Pb.
The latter understanding is illustrated in Fig.~\ref{fig:strx}, where 
several generic diagrams are identified
that have unique physical
consequences for the redistribution of the $s.p.$ strength. 
The middle column characterizes the mean-field
picture that is used as a starting point for the theoretical description.
The right column identifies the location of the $s.p.$ strength of the orbits, just below the Fermi energy, when correlations
are included.
One may apply this picture, for example, to the $2s\scriptstyle \frac{1}{2}$ 
proton orbit in ${}^{208}\textrm{Pb}$.
The physical mechanisms responsible for the correlated strength distribution
are also identified.
The strength of this orbit, remaining at the quasihole energy, is
about 65\%.
Long-range correlations are responsible for the loss of 20\% 
of the strength
due to the coupling to nearby 2p1h and 1p2h and more complicated states.
This loss is roughly symmetrically distributed above and below the Fermi energy
and is physically represented by the coupling to low-lying surface modes
and higher-lying giant resonances.
The resulting occupation number of the orbit therefore corresponds to about 75\%.
More deeply bound nucleons have higher occupation numbers corresponding
to about 85\%.
This is true for all the
deep-lying orbits, and is consistent with a global depletion due to 
SRC of 15\%.
The corresponding location of this strength is identified
at very high energy in the particle domain and is due to the SRC induced by a realistic
nucleon-nucleon interaction with a substantial repulsive core.
The left column depicts  
the generic diagram that is responsible for the admixture
of high-momentum components in the ground state.
The energy domain of these high-momentum nucleons is at large missing
energies as discussed for Fig.~\ref{fig:16Ohighk}.

This rather complete picture of the properties of a proton in 
the nucleus is unique to the field of nuclear physics.
Indeed, unlike other fields with strong correlation effects, like particle or
condensed matter physics, it is possible in nuclear physics to 
state that the properties of the constituent protons inside the nucleus
are identified experimentally and understood in global theoretical terms.
While an update of these considerations will be presented in Sec.~\ref{sec:eepX}, the crucial question how these correlations for protons change when neutrons are added or removed, and, conversely, what happens with neutrons when they are in the majority remains the main concern of the present work.
Some glimpses have been provided by the study of SRC as discussed for Figs.~\ref{fig:rios1} and \ref{fig:rios2}.

\subsubsection{Some results from the dispersive optical model}
\label{sec:DOMr}
A recent, more detailed overview of the DOM work described in this section can be found in Ref.~\cite{Dickhoff:17} while a more general review of the optical model is provided in Ref.~\cite{Dickhoff19}.
The Washington University group in St.~Louis first performed local dispersive optical-model fits to understand the $N-Z$ asymmetry dependence of correlations in nuclei~\cite{charity06,Charity:2007,Mueller:2011}. The forms of the potential were largely consistent with those used by Mahaux and Sartor~\cite{Mahaux:91}. 
Fits for neutron and proton elastic-scattering angular distributions, total and absorption cross sections, $s.p.$ energies and spectroscopic factors deduced from ($e,e'p$) reactions were generated. For protons, the long-range correlations associated with the surface imaginary potential showed a strong $N-Z$ asymmetry dependence, with the proton correlations increasing in neutron-rich systems. This was consistent with the assumed asymmetry dependence of  this potential in most global optical-model fits. On the other hand, neutrons seemed to have very little, if any, dependence. The optical model parametrizations were later used to provide distorted waves and overlap functions for the analysis of transfer reactions~\cite{Nguyen:2011}.

The connection of the local DOM optical potential and the self-energy in the Green's-function formalism can be made if the energy-dependent real potential in the local optical model fits is replaced by an energy-independent nonlocal one. Efforts in this regard were made for the case of $^{40}$Ca~\cite{Dickhoff:2010} where the fitted energy-dependent real potential was replaced with such a nonlocal potential. This limited the binding of the lowest $s_{1/2}$ in better agreement with  experiment. However, it was not possible to describe the charge distribution generating too much charge at small radii, as well as particle number.  These problems were cured when a fully nonlocal fit to $n,p$+$^{40}$Ca data was performed with nonlocality in both the real and imaginary potentials~\cite{Mahzoon:2014}. As well as the positive-energy reaction data, this fit also included the experimental $s.p.$ energies, the experimental charge distribution and the neutron and proton particle numbers. 
The introduction of nonlocality in the imaginary potential changes the absorption profile as a function of angular momentum and was essential in reproducing the particle numbers and the experimental charge distribution. This fit allowed for the reproduction of the yield of high-momentum nucleons determined by data from Jefferson Lab~\cite{Mahzoon:2014}.
We note that these high-momentum components involve the single-nucleon spectral function as extracted from measurements reported in Refs.~\cite{Rohe04,Rohe04A}.
The DOM is therefore not yet capable of addressing the relative momentum distribution as probed by knockout experiments involving two removed nucleons~\cite{Subedi08}.

\begin{figure}[t]
\begin{center}
\includegraphics[width=.6\textwidth]{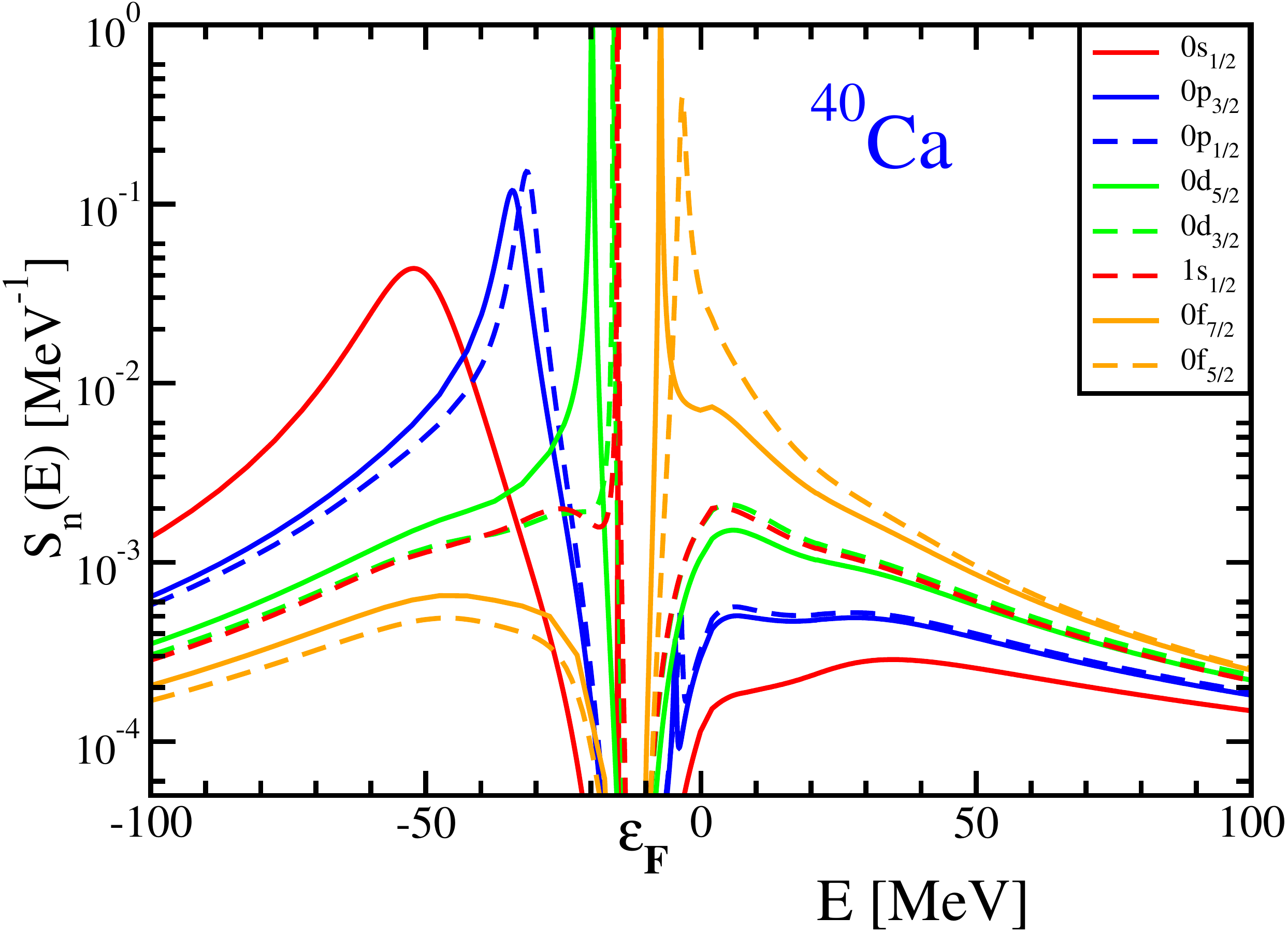}
\caption{(Color online) Spectral function for neutron $s.p.$ orbitals in $^{40}$Ca obtained by the St. Louis group. The sequence of peaks follows the standard order of the independent particle model. Reprinted figure with permission from~\cite{Dussan:2014} \textcopyright2014 by The American Physical Society.}
\label{fig:Dussan2}
\end{center}
\end{figure}
Nonlocal DOM potentials allow for a calculation of the complete off-shell elastic nucleon-scattering matrix in momentum space, as discussed in Sec.~\ref{sec:theory}.
Using Eqs.~(\ref{eq:specfunc}) and the appropriate complementary information below the Fermi energy given by Eq.~(\ref{eq:spechr}), it is possible to generate the complete spectral function for suitably chosen bound $s.p.$ wavefunctions.
By identifying the wavefunctions associated with the main peaks of the spectral distribution for each $\ell j$-combination, including those that correspond to valence hole states near the Fermi energy, it is possible to map out the complete distribution of these states as a function of energy.
From the nonlocal DOM potentials it is thus possible to generate the spectral function for bound orbitals at both negative energies~\cite{Mahzoon:2014} and at positive energies~\cite{Dussan:2014}, as shown in Fig.~\ref{fig:Dussan2} for neutron orbits in ${}^{40}$Ca. 
For plotting purposes the small imaginary part near the Fermi energy was employed giving the peaks a small width.
It is particularly important to note that the strength of bound orbits that resides in the particle continuum is constrained by the fits to experimental data.
While the distributions follow the expected behavior observed in $(e,e'p)$ reactions, the particle strength displays considerable sensitivity to the location of the main peak relative to the continuum.
The deeply bound $0s\scriptstyle \frac{1}{2}$ exhibits a depletion of a few percent (with about 5\% above 200 MeV) implying an occupation of about 90\%.
The valence $1s\scriptstyle \frac{1}{2}$ orbit exhibits an additional 5\% depletion and is therefore 85\% occupied in this analysis.
For mostly empty orbits with main peaks near the continuum threshold considerably more strength is identified in the continuum as shown in Fig.~\ref{fig:Dussan2}.
This behavior suggests that considerable $s.p.$ strength may be located in the immediate vicinity of the main fragment of the distribution when this occurs near the continuum.
This observation may be of paramount importance for properties of the $s.p.$ strength in exotic nuclei. 

The nonlocal DOM treatment has also been extended to the $n,p$+$^{48}$Ca reactions and used to deduce the neutron-matter and weak-charge density distributions. 
The results imply a neutron skin of 0.249$\pm$0.023~fm \cite{Mahzoon:2017} which is larger than the prediction of 0.12-0.15~fm from the coupled-clusters formalism \cite{Hagen:2016}. Future experiments at Jefferson Lab with parity-violating electron scattering should be able to resolve this disagreement \cite{PhysRevLett.108.112502}.   

\subsection{Reaction mechanism}
\label{sec:eepX}
Quite recently, an updated version of the St.-Louis nonlocal DOM parametrization for $^{40}$Ca was used to calculate the cross sections for the $^{40}$Ca($e,e'p$)$^{39}$K reaction employing the DWIA~\cite{Atkinson:2018}. 
In this case the proton distorted waves, the radial-dependence of the overlap functions, their normalization (spectroscopic factor) all were calculated from the DOM potentials and therefore no additional fit parameters were employed.    

\begin{figure}[t]
\begin{center}
\includegraphics[scale=.9]{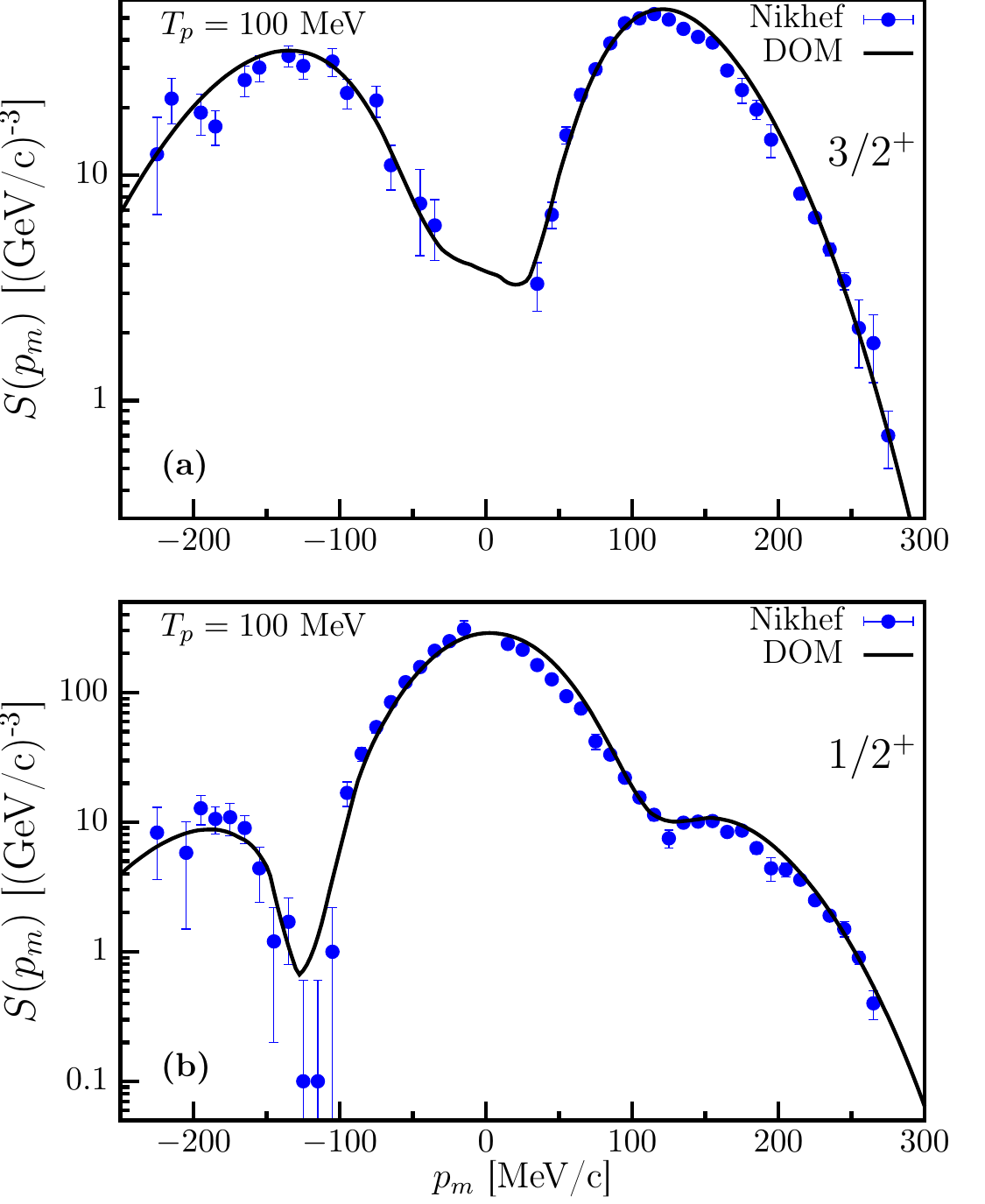}
\caption{(Color online) Comparison of the spectral functions measured at NIKHEF for outgoing proton energies of 100~MeV to DWIA calculations using the proton distorted waves, overlap function and its normalization from a nonlocal DOM parametrization.
Results are shown for the knockout of a $0d\scriptstyle \frac{3}{2}$ and $1s\scriptstyle \frac{1}{2}$ proton in the upper and lower panels respectively.
Reprinted figure with permission from \cite{Atkinson:2018} \textcopyright2018 by the American Physical Society.} 
\label{fig:mack5}
\end{center}
\end{figure}

To accurately calculate the $(e,e'p)$ cross section in DWIA, it is imperative that the DOM self-energy describe not only scattering data but bound-state information as well. 
This is due to the fact that the shape of the cross section is primarily determined by the bound-state overlap function~\cite{Kramer:1989}. Thus, not only should the experimental charge radius be reproduced, but the charge density should match the experimental data.
   The present DOM self-energy leads to the spectral strength distributions with large peaks for the main fragments. 
   The experimental distributions from Ref.~\cite{Kramer90} for $\ell = 0$ clearly show that the 
   strength is already strongly fragmented at low energies. 
   The DOM does not yet include the details of the low-energy fragmentation of the valence hole states which requires the introduction of pole structure in the self-energy~\cite{Dickhoff04}.
   The spectroscopic factor of Eq.~(\ref{eq:sfac}) corresponds to the main peak of each distribution. 
   It is calculated directly from the ${}^{40}$Ca DOM self-energy resulting in values of 0.71 and 0.74 for the $0\textrm{d}\frac{3}{2}$ and $1\textrm{s}\frac{1}{2}$ peaks, respectively.

Agreement with data obtained for the electron-induced knockout of the $0\textrm{d}\frac{3}{2}$ and $1\textrm{s}\frac{1}{2}$ orbitals with outgoing proton energy to 100 MeV shown in Fig.~\ref{fig:mack5}, is as good as, or better than, previous descriptions employing the Schwand \textit{et al.}'s local global OM potential~\cite{Schwandt:1982} and overlap functions from Wood-Saxon potentials where the radius and  normalization factor are adjusted to fit the data. 
The net effect of nonlocality, is in the opposite direction to the transfer study of Ross \textit{et al.}~\cite{Ross:2015}, and allows for slightly larger spectroscopic factors.  
Presumably this is related to the role of knockout from the interior of the nucleus in the ($e,e'p$) case.
Figure~\ref{fig:mack5} shows the agreement of these DWIA calculation with the distorted spectral functions obtained from the ($e,e'p$) data.
A slight increase of the spectroscopic factors obtained from the DOM description compared to the values of the standard NIKHEF analysis~\cite{Lapikas:1993} is obtained otherwise validating the procedure of the NIKHEF group~\cite{Atkinson:2018}.  
The agreement with the $\ell = 0$ cross section is achieved by taking into account the experimental fragmentation at low energy which generates a spectroscopic factor of 0.60 for the main fragment.

 In order to estimate the uncertainty for the DOM spectroscopic factors, the bootstrap method from Ref.~\cite{Varner91} was followed, which was also employed in Ref.~\cite{Mahzoon:2017} to assess the uncertainty for the neutron skin in ${}^{48}$Ca.
New modified data sets were created from the original data by randomly renormalizing each angular distribution or excitation function within the experimental error to incorporate fluctuations from the systematic errors. Twenty such modified data sets were generated and refit. The resulting uncertainties are listed in Table~\ref{table}.
  \begin{table}[bt]
  \begin{center}
   \caption{Comparison of spectroscopic factors deduced from the previous analysis~\cite{Kramer:1989} using the Schwandt optical potential~\cite{Schwandt:1982} to the normalization of the corresponding overlap functions obtained in the present analysis from the DOM including an error estimate as described in the text.
   Reprinted table with permission from \cite{Atkinson:2018} \textcopyright2018 by the American Physical Society.}
   \vspace{0.5cm}
         \begin{tabular}{ c c c } 
            \hline
            $\mathcal{Z}$ & $0\textrm{d}\frac{3}{2}$ & $1\textrm{s}\frac{1}{2}$\\
            \hline
            \hline
            Ref.~\cite{Kramer:1989} & $0.65 \pm 0.06$ & $0.51 \pm 0.05$\\
            \hline
            DOM & $0.71 \pm 0.04$ & $0.60 \pm 0.03$ \\
            \hline
         \end{tabular}
   \label{table} 
   \end{center}
\end{table}

  \begin{figure}[t]
      \begin{minipage}{\linewidth}
         \makebox[\linewidth]{
            \includegraphics[scale=0.9]{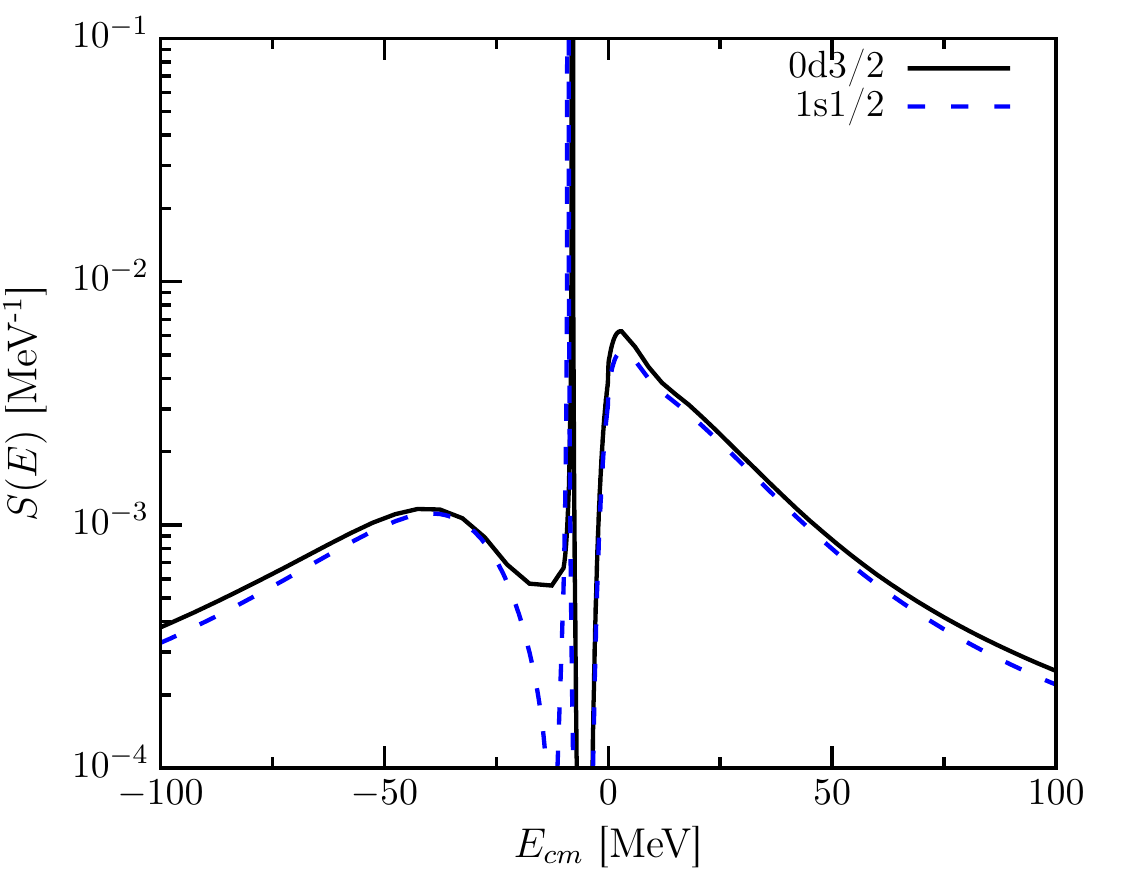}
         }
      \end{minipage}
      \caption{(Color online) Spectral distribution of the $0\textrm{d}\frac{3}{2}$ and $1\textrm{s}\frac{1}{2}$ orbits in ${}^{40}$Ca as a function of energy. Additional strength outside this domain is not shown.
      Reprinted figure with permission from \cite{Atkinson:2018} \textcopyright2018 by the American Physical Society.}
   \label{fig:spectralsd}
\end{figure} 
The DOM results also generate the complete spectral distribution for the $0\textrm{d}\frac{3}{2}$ and $1\textrm{s}\frac{1}{2}$ orbits according to Eqs.~(\ref{eq:specfunc}) and (\ref{eq:spechr}).
These distributions are displayed in Fig.~\ref{fig:spectralsd} from -100 to 100 MeV.
The energy axis refers to the $A-1$ system below the Fermi energy and the $A+1$ system above.
As in Fig.~\ref{fig:Dussan2}, a small imaginary part near the Fermi energy was employed to give the main peaks a small width.
The occupation probabilities are obtained from Eq.~(\ref{eq:nocc}) and correspond to 0.80 and 0.82 for the $0\textrm{d}\frac{3}{2}$ and $1\textrm{s}\frac{1}{2}$ orbits, respectively.
The strength at negative energy not residing in the DOM peak therefore corresponds to 9 and 7\%, respectively.
This information is constrained by the proton particle number and the charge density.
The strength above the Fermi energy is constrained by the elastic-scattering data and generates 0.17 and 0.15 for the $0\textrm{d}\frac{3}{2}$ and $1\textrm{s}\frac{1}{2}$ orbits, respectively, when Eq.~(\ref{eq:depl}) is employed up to 200 MeV.
The sum rule given by Eq.~(\ref{eq:sumr}), associated with the anticommutation relation of the fermion operators, therefore suggests that an additional 3\% of the strength resides above 200 MeV, similar to what was found in Ref.~\cite{Dussan:2014}.
Strength above the energy where surface physics dominates can be ascribed to the effects of short-range and tensor correlations.
The main characterizations of the strength distribution shown in Fig.~\ref{fig:strx}~\cite{Dickhoff04} are therefore confirmed for ${}^{40}$Ca.

Application of this strategy to ${}^{48}$Ca has also generated excellent results for the corresponding $(e,e'p)$ cross sections to valence hole states.
The importance of proton reaction cross sections is confirmed as they were essential for the success of Ref.~\cite{Atkinson:2018}.
The ${}^{48}$Ca results were published in Ref.~\cite{Atkinson:2019} and demonstrate that a non negligible nucleon asymmetry dependence is obtained for the spectroscopic strength of valence hole states when 8 neutrons are added to ${}^{40}$Ca.
  The smaller spectroscopic factors in $^{48}$Ca shown in Table~\ref{table:sf_comp} are consistent with the
experimental cross sections of the $^{48}$Ca$(e,e'p)^{47}$K reaction. The comparison of $\mathcal{Z}_{48}$ and $\mathcal{Z}_{40}$ in Table~\ref{table:sf_comp} reveals that both orbitals experience a
reduction. This indicates that strength from the spectroscopic factors is pulled to the continuum in $S(E)$ when eight neutrons are added to $^{40}$Ca. 
The stronger coupling to surface excitations in $^{48}$Ca, documented by the larger proton reaction cross section when compared to $^{40}$Ca~\cite{Atkinson:2019}, contributes to the quenching of the proton
spectroscopic factor and is also consistent with the $np$-dominance observed in the nucleon pair momentum distribution~\cite{Subedi08,hen17} which is associated with the nucleon-nucleon tensor force as discussed in Sec.~\ref{sec:SRC}. 

\begin{table}[h]
   \caption[Comparison of DOM spectroscopic factors in $^{48}$Ca and $^{40}$Ca]{Comparison of DOM spectroscopic factors in $^{48}$Ca and $^{40}$Ca. These factors have been renormalized to
   account for low-energy fragmentation as observed experimentally~\cite{Atkinson:2019}.}
   \vspace{0.5cm}
   \begin{minipage}{\linewidth}
      \makebox[\linewidth]{
         \begin{tabular}{ c c c } 
            \hline
            $\mathcal{Z}$ & $0\textrm{d}\frac{3}{2}$ & $1\textrm{s}\frac{1}{2}$\\
            \hline
            \hline
            $^{40}$Ca & $0.71 \pm 0.04$ & $0.60 \pm 0.03$ \\
            \hline
            $^{48}$Ca & $0.58 \pm 0.03$ & $0.55 \pm 0.03$ \\
            \hline
            \hline
         \end{tabular}
      }
   \end{minipage}
   \label{table:sf_comp} 
\end{table}

The results presented in Table~\ref{table:sf_comp} document a nucleon asymmetry dependence that is weaker than suggested by Fig.~\ref{Rsplot} but stronger than the ($p,2p$) results on Oxygen isotopes~\cite{atar2018,Kaw18} and transfer reactions.
The present results thus suggest that it is possible to generate a consistent picture of the strength distributions of valence orbits in Ca isotopes employing all the available experimental constraints.
Recent DOM results for ${}^{208}$Pb confirm this observation~\cite{Atkinson:2020}.
We therefore conclude that it is indeed quite meaningful to employ concepts like spectroscopic factors and occupation probabilities when discussing correlations in nuclei.

\subsection{\emph{Ab initio} computations of correlations}
\label{sec:abinitio}

Several aspects of nuclear correlations, and the quenching of $s.p.$ strength in particular, have been addressed through first principles computations in recent years. In practical applications one seeks for a nearly exact (or the most accurate possible) solution of the many-nucleon Schr\"odinger equation, while exploiting a realistic nuclear Hamiltonian which should be derived as closely as possible from the underlying QCD. The quality of such predictions is in constant evolution as new theory and computational advances are introduced. However, it is generally expected for an \emph{ab initio} approach that the uncertainties due to both the many-body computations and the Hamiltonian can be estimated, so that a direct comparison between theory and experiment (each with their own uncertainties) is meaningful.

To better grasp the discussion below, it should be clarified that almost all modern \emph{ab initio}  applications exploit nuclear Hamiltonians obtained through some effective field theory (EFT). The latter can be seen as a low-energy and low-momentum expansions of QCD, hence limiting phenomenological inputs. The implicit pre-diagonalization of short-distance degrees of freedom leads to very soft nuclear Hamiltonians, which have made it relatively easy for a class of post-HF many-body approaches to reach {\it ab initio} predictions with mass numbers up to $A\sim140$~\cite{Bind2014,Hagen:2016,Arth2020prl}. This approach doesn't affect the low-energy structure, which is the topic of this review, because of a clear separation of scales between long- and short-range physics.
From the discussion of Sec.~\ref{sec:specdis} we know that direct signatures of SRC should be seen through experiments at large momentum transfer, that is outside the realm of applicability of EFT. Due to the separation of scales, the SRC effects for low-energy states are limited to an overall \emph{partial} reduction of occupations and SFs that turns out to be about 10-15\%---one should also keep in mind that SRC \emph{do not} account for all of the quenching.
Since EFTs wash out short-range physics by construction, this reduction effect disappears from typical \emph{ab initio} computations. However, EFT Hamiltonians are all expected to reproduce the same, experimentally observed, cross sections. Thus, any difference in the SFs must be compensated through changes in the final state interactions and reaction mechanism.

\subsubsection{Role of shell-model and particle-vibration coupling in LRC}
\label{sec:abinitio_SM}

\begin{figure}[t]
\centering
 \includegraphics[width=1.0\linewidth]{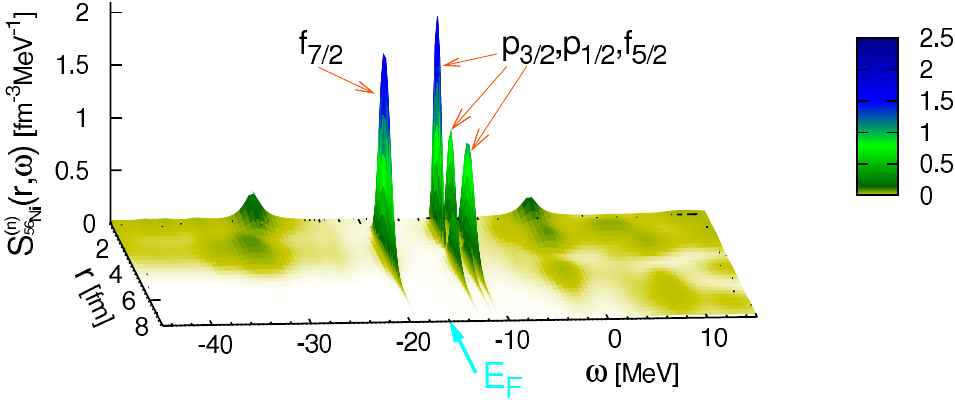}%
  \caption{\label{fig:Sf_Ni56_3d} (Color online) Single particle spectral function, $S(r,\omega)$, for the removal of addition of a neutron from/to $^{56}$Ni at a distance $r$ from its center, from SCGF-FRPA computations. Quasiparticle states of the $f_{7/2}$ hole and the $p_{3/2}$, $p_{1/2}$ and $f_{5/2}$ particle are clearly seen. Adapted form Ref~\cite{barbieri2009}.}
\end{figure}

First principle computations of the nuclear spectral function are possible through the self-consistent Green's function (SCGF) approach, as already introduced at the beginning of this section and in appendix \ref{sec:self}. One normally truncates the many-body expansion using the most complete Faddeev random phase approximation (FRPA) introduced in Refs.~\cite{PhysRevC.63.034313,PhysRevA.76.052503,PhysRevA.83.042517} or the thrid-order algebraic diagrammatic construction [ADC(3)]~\cite{Shir1982,BarbCarb2017,Raim2018}.
The two approaches are closely related and both compute the nuclear self-energy including particle-vibration couplings systematically in all possible channels. While model spaces include several oscillator shells  and are large enough to account for coupling to all giant resonances (GRs), the configuration mixing effects are limited to 2p1h and 2h1p states and cover only part of a standard shell-model computation. Fig.~\ref{fig:Sf_Ni56_3d} displays the computed particle and hole spectral functions, of Eqs.~\eqref{eq:5.9} and \eqref{eq:5.10} together, for a neutron on $^{56}$Ni and for the case were the radial coordinate $r$ is chosen instead of a general basis $\{\alpha\}$. This plot is interpreted as the  probability distributions of adding or removing a neutron at distance $r$ from the center while transferring an energy $\pm\omega$ to the system. It could be compared to the schematic representation of Fig.~\ref{fig:strx}. The spikes near the Fermi surface are the (square of the) quasiparticle overlap functions for the $pf$-shell orbits, the heavy fragmentation of $sd$ and $sdg$ shells is visible further away in both directions, while for energies $\omega>0$ the $n$+$^{56}$Ni system is in the continuum and $S(r,\omega)$ extends to infinity as expected for an elastic-scattering wavefunction.

\begin{table}         
\begin{center}
\begin{tabular}{rlcccccccc}
\hline                 
 &                 &  \multicolumn{3}{c}{10 h.o. shells}   & Exp.~\cite{Yur.06}  &~&  \multicolumn{3}{c}{$1p0f$ space}      \\
 &                 &   FRPA     &   full   &      FRPA         &            &~& FRPA   &   SM        & $\Delta Z_\alpha$ \\
 &                 &   (SRC)    &   FRPA   &+$\Delta Z_\alpha$ &            & &        &             &                   \\
\hline                 
\hline                 
$^{57}$Ni: &
   $\nu 1p_{1/2}$    &    0.96    &   0.63   &  0.61             &            & & 0.79  &   0.77     &  -0.02    \\
& $\nu 0f_{5/2}$    &    0.95    &   0.59   &  0.55             &            & & 0.79  &   0.75     &  -0.04    \\
& $\nu 1p_{3/2}$    &    0.95    &   0.65   &  0.62             & 0.58(11)   & & 0.82  &   0.79     &  -0.03    \\
$^{55}$Ni: &
   $\nu 0f_{7/2}$    &    0.95    &   0.72   &  0.69             &            & & 0.89  &   0.86     &  -0.03    \\
$^{57}$Cu: &
   $\pi 1p_{1/2}$    &    0.96    &   0.66   &  0.62             &            & & 0.80  &   0.76     &  -0.04    \\
& $\pi 0f_{5/2}$    &    0.96    &   0.60   &  0.58             &            & & 0.80  &   0.78     &  -0.02    \\
& $\pi 1p_{3/2}$    &    0.96    &   0.67   &  0.65             &            & & 0.81  &   0.79     &  -0.02    \\
$^{55}$Co: &
   $\pi 0f_{7/2}$    &    0.95    &   0.73   &  0.71             &            & & 0.89  &   0.87     &  -0.02    \\
\hline                 
\end{tabular}
\end{center}
 \caption[]{Spectroscopic factors (given as a fraction of the IPM)
    for valence orbits around $^{56}$Ni.  For the SC-FRPA
    calculation in the large harmonic oscillator space, the values shown
    are obtained by including only SRC, SRC and LRC from particle-vibration
    couplings (full FRPA), and by SRC, particle-vibration couplings and extra
    correlations due to configuration mixing (FRPA+$\Delta Z_\alpha$).
     The last three columns give the results of SC-FRPA and SM in the restricted
    $1p0f$ model space. The $\Delta Z_\alpha$ are the differences between the last
    two results and are taken as corrections for the SM correlations
    that are not already included in the FRPA formalism.    
    }
\label{tab:Ni56SFs}
\end{table}

Ref.~\cite{barbieri2009} used the computations from Fig.~\ref{fig:Sf_Ni56_3d} to disentangle the different effects of nuclear correlations. The outcome of this analysis for $^{56}$Ni is shown in Table~\ref{tab:Ni56SFs}. The computation was performed in FRPA, covering 10 major nuclear shells and using the chiral N3LO interaction of Ref.~\cite{PhysRevC.68.041001}. This early EFT Hamiltonian is only partially soft and contains mild short-range repulsion effects. Moreover, it was augmented by a phenomenological correction, set to constrain the particle-hole (p-h) gap, that corrected for the then missing three-nucleon forces (see Ref.~\cite{Barb2009Ni} for details). 
As shown in the table, we first estimated the SRC reduction of SFs by dynamically diagonalising two-body ladders, as demonstrated in Refs.~\cite{PhysRevC.53.2207,PhysRevC.65.064313}. This amounts to a 5\% effect, which is in accordance with the partially-soft nature of the nuclear force used and is the same across all valence orbits.

The full spectroscopic factors predicted by  FRPA are given in the fourth column and display a reduction of approximately 30-40\% that can vary for different isotopes and final states. Then, the additional 25-35\% reduction should be interpreted as LRC effects due to the coupling of $s.p.$ orbits to vibrations in the giant resonance (GR) region. Shell-model calculations can handle more complex configuration mixings, beyond 2p1h and 2h1p, but are typically restricted to a single valence shell were GRs cannot be accounted for.  To estimate the missing correlations both FRPA and shell model calculations were performed restricted to the $pf$  shell. The same $s.p.$ energies and effective interactions where used in both cases, and were obtained projecting the \emph{ab initio} quasiparticle states from Fig.~\ref{fig:Sf_Ni56_3d} into the $pf$ model space (see also Ref.~\cite{PhysRevC.100.024317} for more details).
The comparison is shown in the last columns of Tab.~\ref{tab:Ni56SFs}.  The shell-model, including up to 8p8h configurations, yields basically the same SFs of the 2p1h or 2h1p FRPA. However, the reduction is only 20\% because of the limited model space.  Clearly this difference is minimal for this particular case because $^{56}$Ni was already predicted to be a rather good closed shell form full microscopic FRPA. Hence, additional configuration mixings are unimportant. For other open shell isotopes, the shell-model would actually add additional correlations. 
Completely similar results are seen for other isotopes, such as $^{48}$Ca~\cite{barbieri2009}.

The above comparison leads to two  considerations. First, correlation effects that can be ascribed to the configuration mixing in the shell model ({\it e.g.}, pairing effects and deformation) are not completely orthogonal to other low-energy mechanisms that can affects SFs. All of these can be thought of being included in the class of LRC, as opposed to very localised short-range interactions that are by nature decoupled to high momenta. The second consideration is that the quenching of SFs that can be computed from the shell-model usually cannot account for all correlation effects. This should raise a warning in cases were the experiment is compared to shell-model based predictions to extract $R_s$ factors. While this is sometimes necessary, for example if there is no other way to estimate relative weights in inclusive measurements, additional layer of uncertainty is introduced that depends on the choice of shell model effective interaction, model space, level of p-h truncation, and so on.

\subsubsection{Isospin asymmetry dependence of spectroscopic factors}

\begin{figure}
\centering
\includegraphics[width=0.45\linewidth]{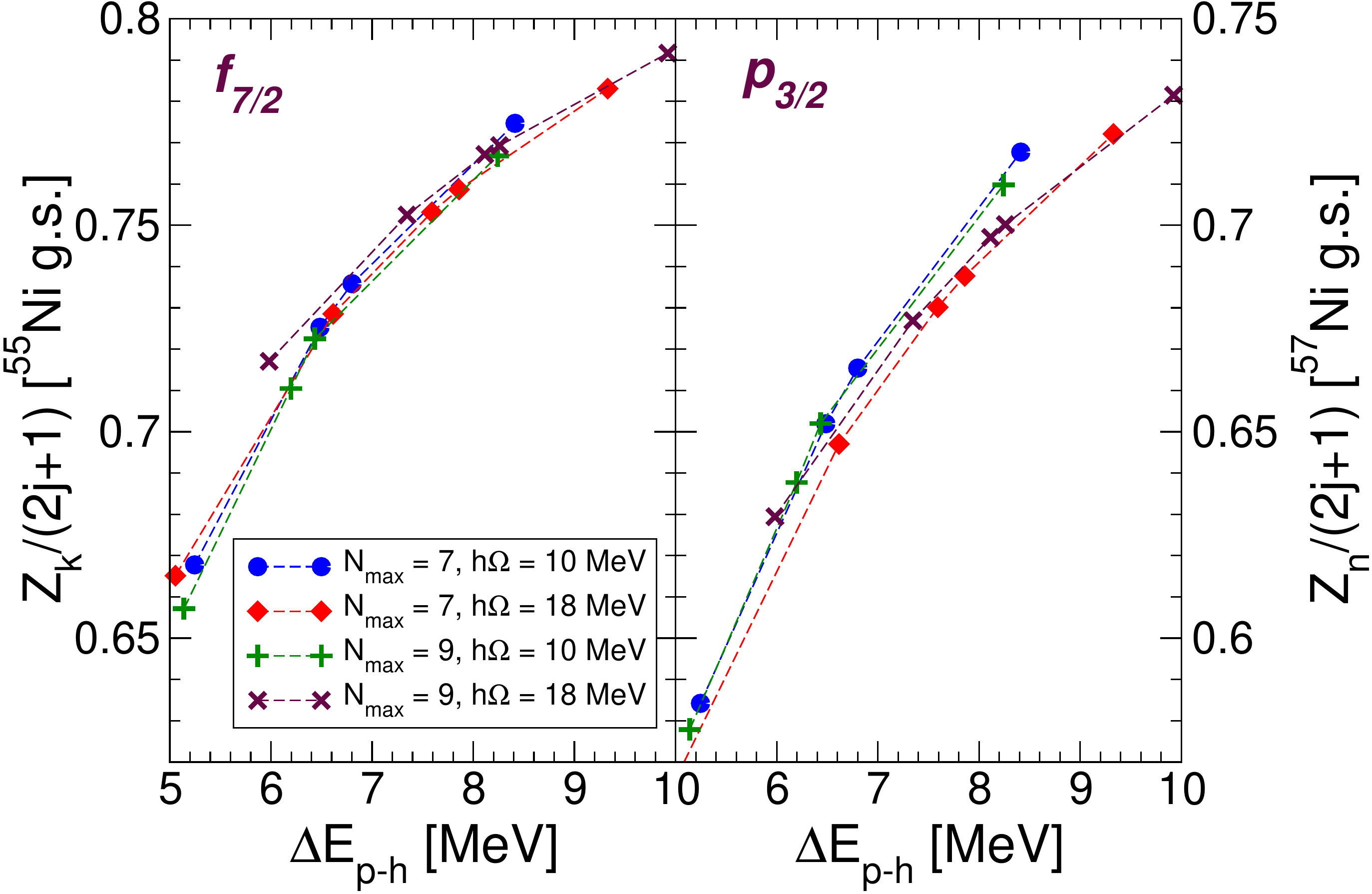} \hspace {0.7 cm}
  \includegraphics[width=0.40\linewidth]{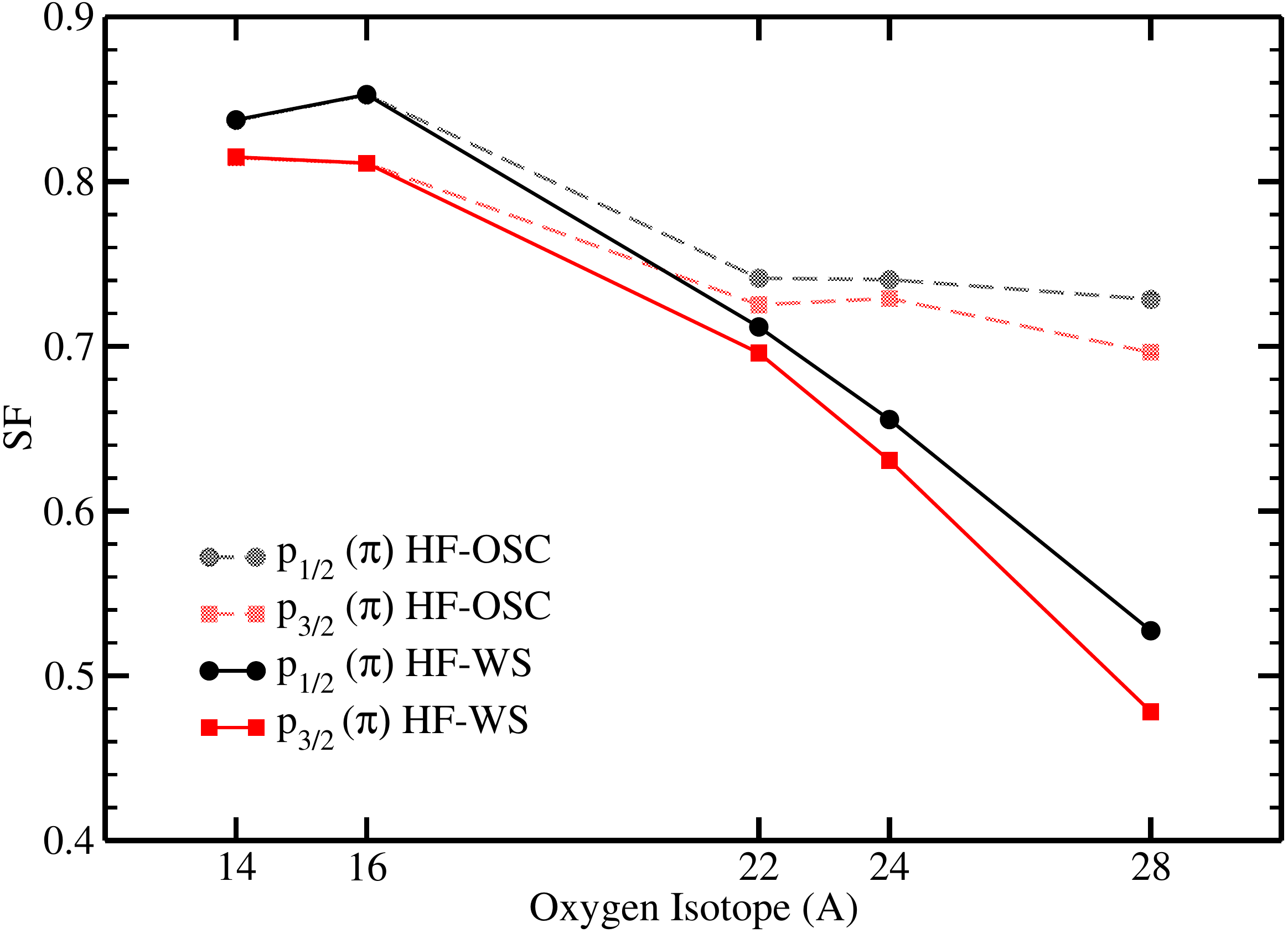}%
  \caption{\label{fig:Theo_SF_vs_gap} (Color online)  (Left) dependence of spectroscopic factors on the  p-h energy gap of $^{56}$Ni, for the g.s.- to g.s. transition in removing a neutron from $^{56}$Ni and $^{57}$Ni. Reprinted figure with permission from Ref.~\cite{Barb2009Ni}\textcopyright2009 by the American Physical Society. (Right) Effect of the coupling to the continuum for SFs of the dominant  $^{\rm A}$O quasiholes. Reprinted figure with permission from Ref.~\cite{jensen11}\textcopyright2011 by the American Physical Society.}
\end{figure}

An important feature of LRC is a direct connection to the p-h energy gap. This is demonstrated by the left panel of Fig.~\ref{fig:Theo_SF_vs_gap} where the SFs of dominant quasiparticle and quasiholes states is plotted against the gap, for the same $^{56}$Ni computation discussed above. When the calculated gap is tuned by the \emph{ad hoc} (3NF) correction, all SFs fall on a correlation line that is independent of model space parameters. Hence, the SFs near the Fermi surface might be heavily constrained by the (observable) energy gap at the Fermi surface. This effect can be simply understood observing that smaller excitation energies allow for stronger configuration mixing. However, the mechanism is likely to be more important for opposite isospins due to the nuclear force being particularly strong in the proton-neutron tensor channel, as seen for the SFs for proton removal from oxygen isotopes shown on the right panel. This coupled cluster computation used  the same N3LO interaction from Ref.~\cite{PhysRevC.68.041001} (but without 3NF corrections) and  found slightly stronger quenching of protons near the neutron dripline, where excitations of neutrons cost little energy. By properly including the continuum states (solid lines) the p-h neutron gap is effectively reduced, leading to stronger quenching~\cite{jensen11}. 

This above discussion suggests that separation energies may not be the optimal parameter to gauge the evolution of correlations effects proposed in Fig.~\ref{Rsplot}. Rather, isospin asymmetry acts indirectly by forcing small (large) excitations gaps near to (far from) the driplines.

\begin{figure}
\centering  \includegraphics[width=0.8\linewidth]{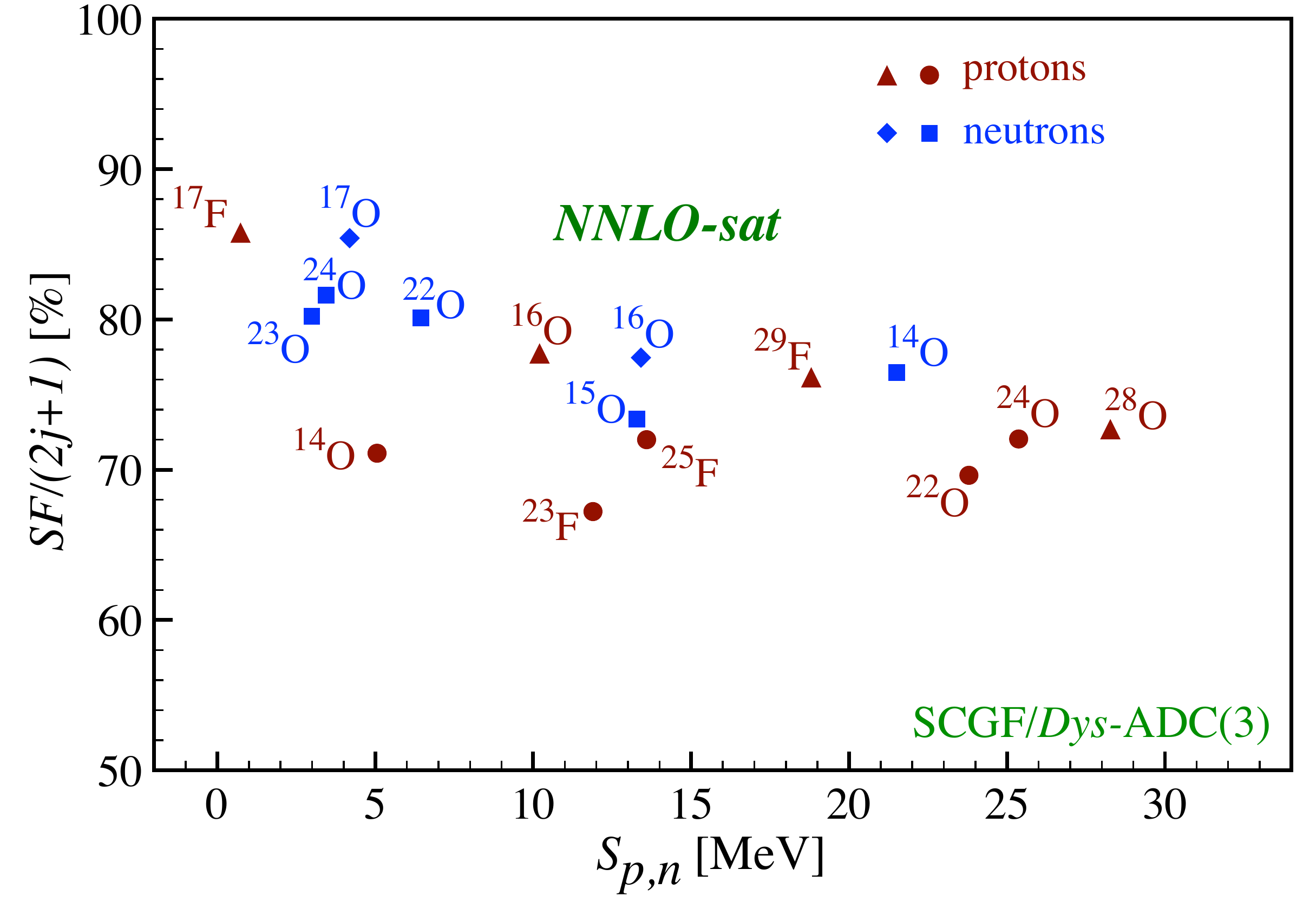}
  \caption{\label{fig:Oxy_SFs_NNLOsat} (Color online)  Spectroscopic factors for removal of a nucleon around oxygen isotopes, as a function of the separation energy.  Each point refers to the g.s. to g.s. transition for the removal od a nucleon from the isotope indicated nearby. This plot is for NNLOsat with the SCGF-ADC(3) and does not include coupling to the continuum effects.}
\end{figure}

Following these considerations, the SFs for g.s. to g.s. transitions along the oxygen chain have been computed using SCGF-ADC(3) theory and different Hamiltonians in Refs.~\cite{Barb2009IMP,cipollone2015} and Fig.~\ref{fig:Oxy_SFs_NNLOsat}. The dependence on separation energies is mild and the heavier quenching for proton removal form $^{14}$O to $^{23}$F can be related to small p-h gaps obtained in the same computations. 
It should be noted that early computations form Ref.~\cite{cipollone2015} where based on a softer chiral Hamiltonian, which underestimated radii and was unable to predict the correct energy separation across nuclear shells. These yielded larger SFs, in analogy with the correlation lines shown in Fig.~\ref{fig:Theo_SF_vs_gap}.  For this reason the computation was repeated using a
NNLO$_{\rm sat}$ interaction from Ref.~\cite{Ekst2015} that is known to predict correct quasiparticle energies to within $\approx$1~MeV.  The SFs shown in Fig.~\ref{fig:Oxy_SFs_NNLOsat} were found to agree well with the ($p$,$2p$) QFS  data discussed in Sec.~\ref{sec6}.

\subsubsection{Observability of spectroscopic factors}

For a given a nuclear Hamiltonian, SFs are exactly determined though the solution of the many-body problem and the overlap integral of Eq.~\eqref{eq:specpp} and~\eqref{eq:norm_overlap}. However, their \emph{non observability} has been stressed by a number of authors on the basis of renormalization group arguments~\cite{furnstahl2002,Duguet15}: As discussed for example in Sec.~\ref{sec:DWBA}, measured data always comes in the form of cross sections that are convolutions of an overlap function (with its SF), the reaction mechanism, and final state (scattering) interactions. Unitary transformation of the Hamiltonian can alter the wavefunction and values of the SFs without changing the observed cross sections.   On the other hand, there is a body of experimental and theoretical evidence which suggests that SFs are useful `measures' of the effect of correlations in atomic nuclei, as discussed in previous sections. Their correlation with p-h gaps displayed in Fig.~\ref{fig:Theo_SF_vs_gap} is an example. 

The unobservability feature is likely to contribute to the wide range of results, sometimes contradictory, extracted from experimental observations over the years. However, this can \emph{only be one part} of the problem since the lack of knowledge in the reaction mechanism and the missing consistency between structure and reactions have been plaguing the interpretation of data for decades. In fact, the question of how uncertain the value of a SF can be, given a realistic description of the nucleus, is still widely unanswered and it might not be settled until complete theories (that is, handling both structure and reactions aspects) become available.

In view of these considerations, a firm point can come by realising that once a truly first principle Hamiltonian is chosen then it must be able to describe all the low-energy nuclear phenomena, irrespective of the kinematics or reaction mechanisms. Therefore, while the nuclear force and the associated SFs might not be observable, it is reasonable to seek for at least one combination of an overlap function and an optical model that are capable to describe the several types of reaction measurements, such as those discussed in this review. Such a global analysis is largely missing in the current literature, however, it is important to realise that it offers a valuable opportunity to put our understanding of nuclear correlations on a firmer ground.
 
\subsubsection{Optical potentials}

\begin{figure}
  \includegraphics[width=0.45\linewidth]{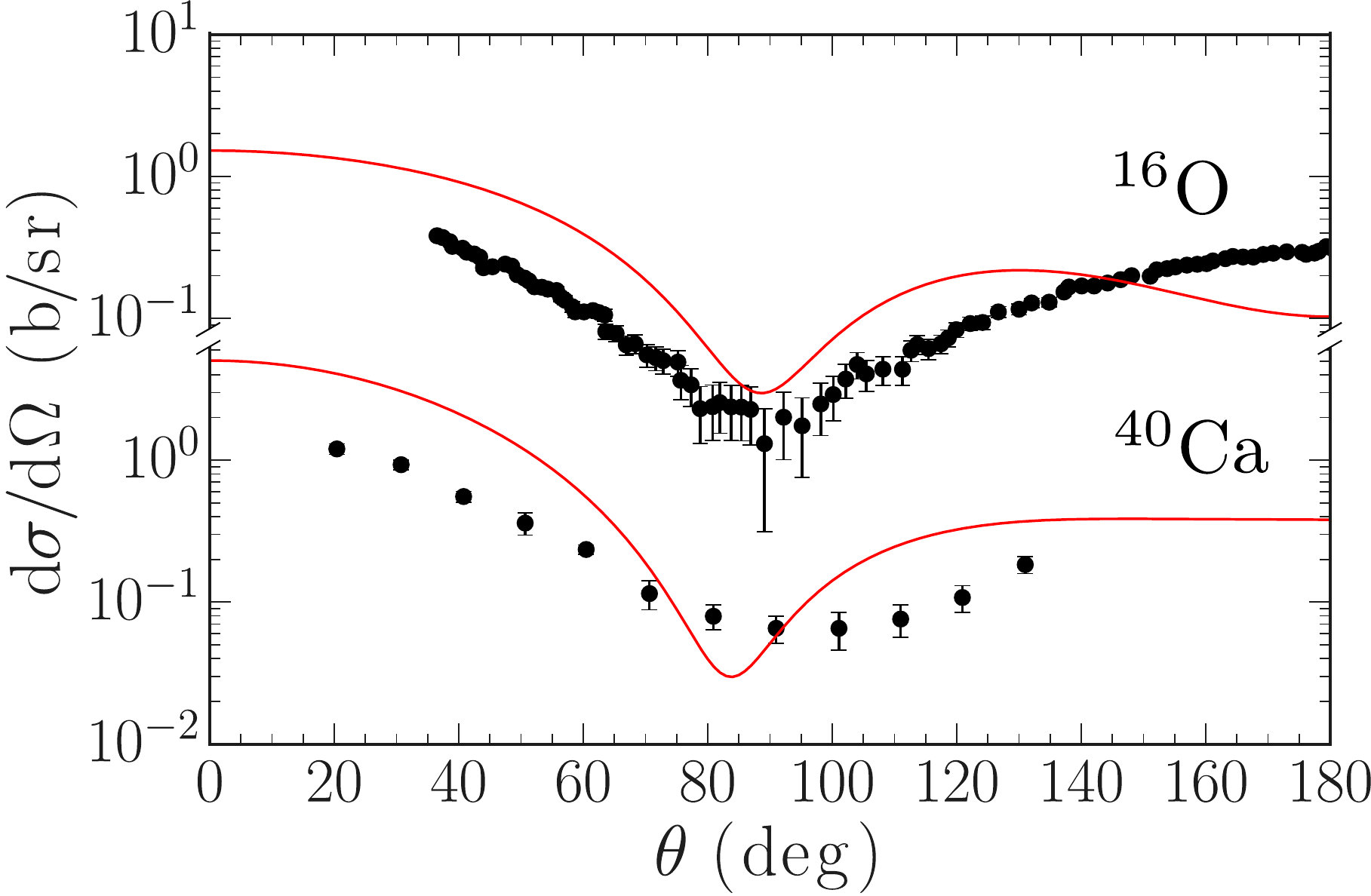}  \hspace {0.7 cm}
  \includegraphics[width=0.47\linewidth]{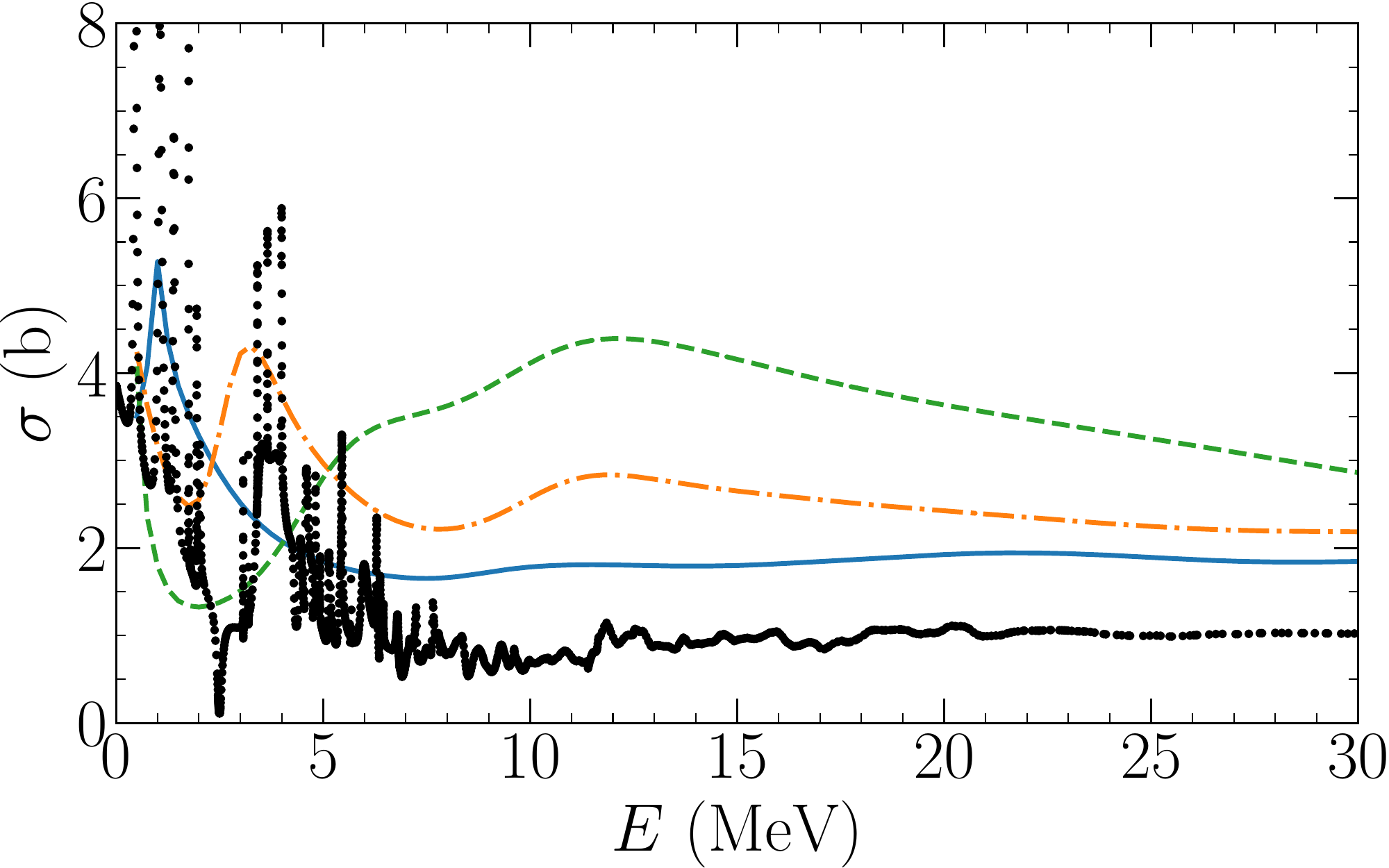}%
  \caption{\label{fig:ADC3_OptPot} (Color online)  Computed elastic differential cross section for a neutron off $^{16}$O  ($^{40}$Ca)  at 3.286 (3.2) MeV scattering energy (left) and total elastic cross section for $n$-$^{16}$O (right). In the latter the panel, dashed, dot-dashed and full lines display respectively the result obtained from a pure mean field potential (all 2p1h configurations frozen), by retaining half of the 2p1h states, and the full ADC(3) computation. All results are from SCGF-ADC(3) optical potentials. Figures reprinted with permission from Ref.~\cite{idini2019} \textcopyright2019 by The American Physical Society. }
\end{figure}

The consistency between the computation of overlaps (structure) and reaction is crucial to the above quest for understanding the degree of `observability' of SFs. So far the DOM allowed the most important steps forward. This  potential is still in part based on phenomenological assumptions but it is fitted `globally' around specific areas of the nuclear chart, hence constrained by a large body of data.   Having an \emph{ab initio} counterpart will further reduce phenomenology and allow bias-free interpretations, including the investigations of cutoff dependence from EFT Hamiltonians.
Optical potentials derived from \emph{ab initio} are still at an early of development stage but hold a promise to fill this gap. There are two main paths, both based on Green's function theory.  The coupled cluster method has been used to compute the one-nucleon Green's function and then to extract the optical potential (that is, a scattering wavefunction is first obtained and then used to invert the scattering equations)~\cite{Rotu2020}. Conversely,  SCGF computes the self-energy to directly obtain the optical potential and the corresponding overlaps, avoiding ambiguities in the inversion process~\cite{barbieri2005,PhysRevC.84.034616}.

Figure~\ref{fig:ADC3_OptPot} shows the quality of currently attainable predictions. Ref.~\cite{idini2019} successfully benchmarked to no-core sell model calculations, that are possible for light nuclei, and extended the scope of applications to medium masses finding qualitatively good differential low-energy elastic cross sections. However, lack of absorption is found at larger scattering energies. This is demonstrated by the total neutron-$^{16}$O cross section in  the left panel, where missing configurations beyond 2p1h can be identified as being responsible for the discrepancy with the experiment.  This  is a difficult challenge but one that could be resolved by introducing novel advances to SCGF theory in the mid-term. \\

Gathering all the above considerations, we can summarise our theoretical understanding of correlations from \emph{ab initio} as follows:
\begin{itemize}
\item The pairing and configuration mixing effects (described by the shell model), as well as particle-vibration coupling are the main mechanisms that induce correlations at low energy. Since their resolution scales overlap, they cannot be cleanly disentangled and should be considered as parts of a general class of \emph{Long-Range Correlations}.
\item $R_s$ reduction factors, defined through a comparison with a structure calculations, such as the shell model, should always be accompanied by a detailed description of the model calculations. Since this introduces an additional layer of model dependence in interpreting data, it is advisable to avoid it whenever possible and to use unquenched overlaps instead. In any case, it is generally incorrect to interpret $R_s$ as a measure of SRC effects.
\item The quenching of spectroscopic factors for quasiparticles near the Fermi surface is strongly correlated to the particle-hole energy gap.
\item \emph{Ab initio} theory does not support the strong dependence of the spectroscopic factors on isospin asymmetry, as it could be suggested from the eikonal analysis of heavy-ion-induced knockout data at energies around 100~MeV/nucleon, as illustrated in Fig.~\ref{fig:Rs}. However, approaching the driplines, the vicinity of the continuum affects the quenching through reduced excitation gaps and a strong proton-neutron tensor channel, consistently illustrated in Fig.~\ref{fig:rios1}, \ref{fig:rios2} and \ref{fig:Theo_SF_vs_gap}. The presently available theoretical predictions suggest a weaker effect.
\item After short-range degrees of freedom are integrated out, the quenching of spectroscopic factors is still sizeable and it is determined by low-energy LRC effects.  It is reasonable to imagine that the \emph{intrinsic uncertainty} of SFs, as due to their non-observable character, is rather small. However, the answer to this question is still largely unknown. This is uncharted territory that will require better consistency between structure and reactions and data from different energies and reaction mechanisms to be explored.
\end{itemize}


\section{Nucleon transfer reactions}
\label{sec3}
\subsection{Introduction}
In the early 1950s it was recognized that the angular distributions of proton yields following the $^{16}$O(d,p)$^{17}$O nucleon transfer reaction at 8~MeV showed characteristic forward-focused shapes, which reflected the orbital angular momentum of the transferred nucleon~\cite{Burrows50,Butler50}. This was followed by a remarkable amount of activity, both experimentally and theoretically. The proliferation of higher energy tandem Van de Graaff accelerators~\cite{Bromley74} and cyclotrons \cite{craddock2008}, coupled with the development of high resolution magnetic spectrographs, led to a wealth of experimental data on transfer-reaction cross sections.

Theoretically, major developments in nuclear structure and nuclear reactions were also occurring. The nuclear shell model suggested that the motion of nucleons in the nuclear medium is relatively unimpeded~\cite{Mayer49,Haxel49}, and the connection between this and direct reactions to $s.p.$ states became apparent, which built on a series of major works too numerous to discuss here. Those include connections to the early studies of resonances~\cite{Breit36,Wigner46}, which set the conceptual framework to develop a model that describes the overlap between nuclear states ---providing a connection to reaction observables such as cross sections. The many experimental and theoretical developments quickly led to the concepts of spectroscopic amplitudes and SFs. The experimental data on nucleon transfer reactions were highly instructive, and arguably formed the skeleton of our understanding of $s.p.$ nuclear structure as we know it today---providing overlaps, angular momentum assignments and $s.p.$ energies. It was of course limited to stable systems. It is also true that only a fraction of the studies resulted in published cross sections. Often, only the model-dependent spectroscopic factors were reported in much of the earlier works. The model dependencies of spectroscopic factors have been the subject of scrutiny for the last 50 years or so.

There is considerable debate about uncertainties on spectroscopic factors and nuances in the models, and the degree to which spectroscopic factors are observable. They are now accepted to be not observable ~\cite{furnstahl2002}. However, arguments about how big is the intrinsic uncertainty due to this feature ~\cite{Jennings11,Duguet15} are still debated. For example, there are indications of strong correlation between the values of SFs and observable quantities such as the three-point formula (mass differences) of particle-hole gaps at the Fermi energy~\cite{barbieri2009}. It is acknowledged by many that cross sections from transfer (and other) reactions, when analyzed in a given framework, provide reliable information on the $s.p.$ structure of nuclei, albeit with some (often well understood) limitations and uncertainties. In this section, a brief overview of the model framework, reaction mechanism, and spectroscopic factors is given and how it relates to our current understanding of the quenching of $s.p.$ cross sections, with emphasis on stable beam works and some of the few recent works with radioactive-ion beams.

The body of work on nucleon-transfer reactions using radioactive-ion beams is quite small compared to the work on intermediate-energy knockout reactions with radioactive-ion beams. This is in part due to the fact that there are only a few accelerator facilities capable of delivering radioactive-ion beams at the ideal energies of a few MeV/u above the Coulomb barrier at sufficient intensities, though the number is growing: there are now several facilities around the world that have, or are moving towards having, radioactive-ion beams available at these energies (around 10~MeV/u), such as ATLAS, ISOLDE at CERN, GANIL-SPIRAL2 and several others.

It is clear that the next decade or so will lead to a wealth of transfer-reaction data as new facilities come online and spectrometer techniques mature. Not only is this led by a desire to understand nuclear structure at the extremes of isospin and stability, but also essential to advancing nuclear astrophysics~\cite{Bardayan16}. There are a handful of recent review articles on nucleon transfer reactions with radioactive-ion beams~\cite{hansen03,Wimmer18}. There are many papers and books on the subject of nucleon transfer, many well established over the decades.

This review concerns the degree to which $s.p.$ motion in nuclei is quenched due to correlations. This is probed via the observation of lower reaction cross sections to $s.p.$ states than expected were $s.p.$ motion unimpeded. Where does the data from transfer reactions fit into the picture raised by the nucleon-removal data and analysis from the NSCL~\cite{tostevin14}? Given the vast amount of literature defining the theories of transfer reactions, in particular the most commonly used approach of the distorted wave Born approximation (DWBA) and related models, this section of the review focuses on the interplay between the experimental choices, the key ingredients to the parametrization of the theory, and the method by which the data are analyzed.

\subsection{Choice of experimental conditions}
Whether planning an experiment or scrutinizing existing data, the decisions made in carrying out the experiment, from the choice of reactions such as ($d$,$p$) versus ($\alpha$,$^3$He), both of which result in adding a neutron to the initial nucleus, the incident beam energy, the angles measured, and the method used to determine the cross sections all play an essential role in how credible the resulting analysis in terms of extracting spectroscopic factors will be. Further, it might dictate what analysis methods should be used.

\begin{figure*}
\centering
\includegraphics[scale=0.9]{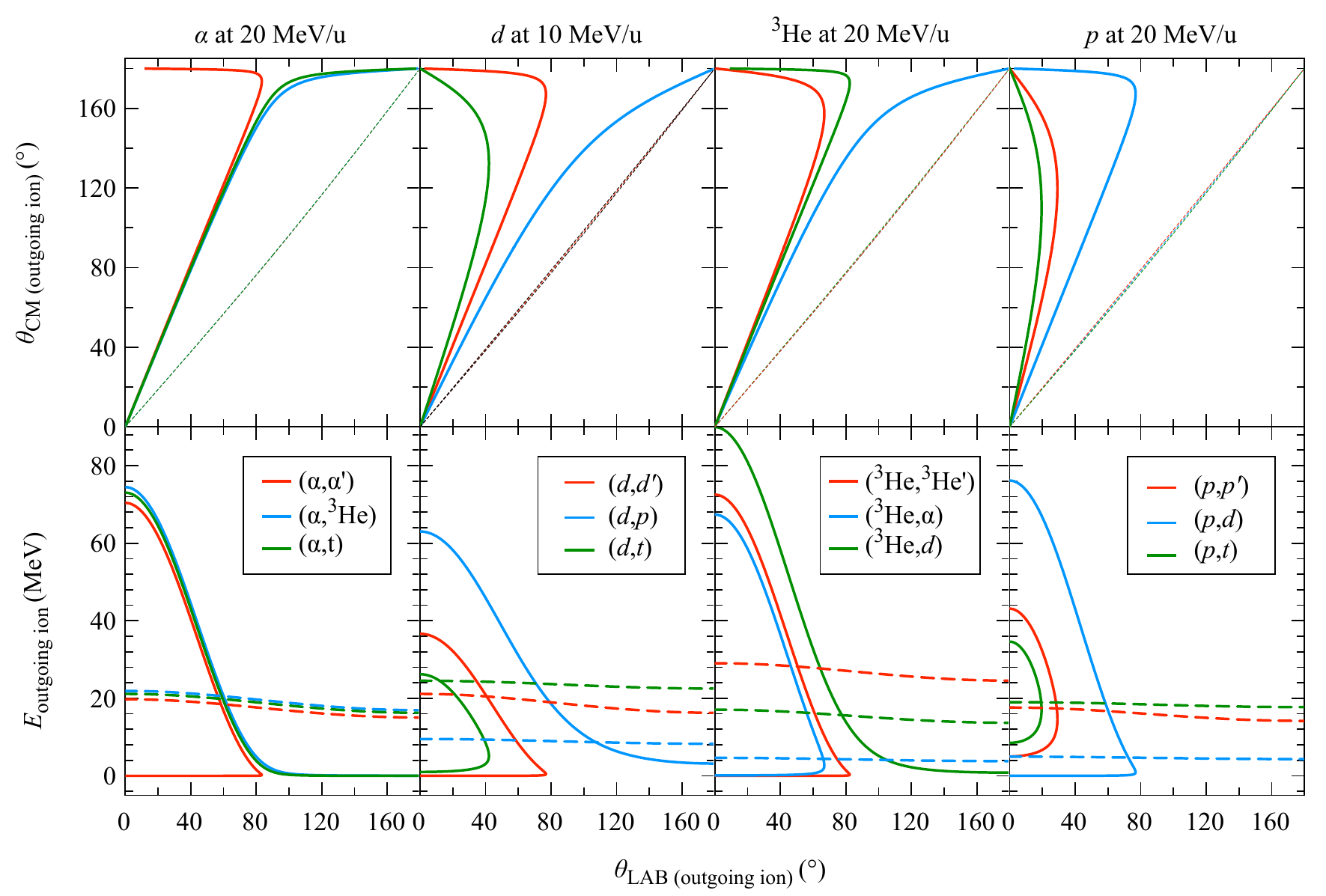}
\caption{\label{fig_kine} (Color online) The kinematics of various reactions in so-called normal (dashed lines) and inverse kinematics (solid lines). In each case the target (normal) or beam (inverse) is assumed to be $^{60}$Ni. For each reaction, the final state is a fictitious 1-MeV state in final nucleus. The calculations assumed a target of negligible thickness and a beam of infinitesimal emittance and size.}
\end{figure*}

 In general, it can be challenging to determine absolute cross sections from reaction studies in inverse kinematics at incident energies around the Coulomb barrier. To determine the luminosity, both the target thickness and the number of beam particles need to be known. With low-intensity beams, nominally less than $\sim$10$^5$~ions per second, the beam can be counted directly with the appropriate detector. With more intense beams Faraday cups can be used. Further, the thin plastic targets used can degrade, particularly with heavy-ion beams~\cite{Rehm98,Kay11}. This makes measurements with heavier beams challenging, as the rate of degradation of the target has to be measured simultaneously. Measuring scattering reactions in the Rutherford regime, where cross sections are known, tends to involve measurements near $\theta_{\rm lab}=90^{\circ}$, where the outgoing ions have small energies. An example of the kinematics of various reactions in `normal' and inverse kinematics can be seen in Fig.~\ref{fig_kine}. Further, at lower beam energies, where the scattering angles are more favorable, resonances can dramatically skew the yields as is well known from stable-beam studies (e.g., Ref~\cite{Evans63}).

\subsubsection{Beam energy and momentum matching}
\label{benergy}

A simple approach to choosing the energy regime for a transfer-reaction study is to choose energies where calculations with the reaction model indicates that the sensitivity of the cross sections to the incident energy, to reaction $Q$ value, or to the target ($A$,$Z$), is small. The corollary of this is that if the cross section varies sharply with energy, then it is also likely to be sensitive to the choice of parameters. Extracting spectroscopic factors from cross sections determined in this regime is likely to result in significant uncertainties---much beyond what is estimated in ideal situations (the nominal 20\% is discussed in Section~\ref{models}).

\begin{figure}[t]
\centering
\includegraphics[trim=0cm 0cm 0cm 0cm,clip,scale=0.9]{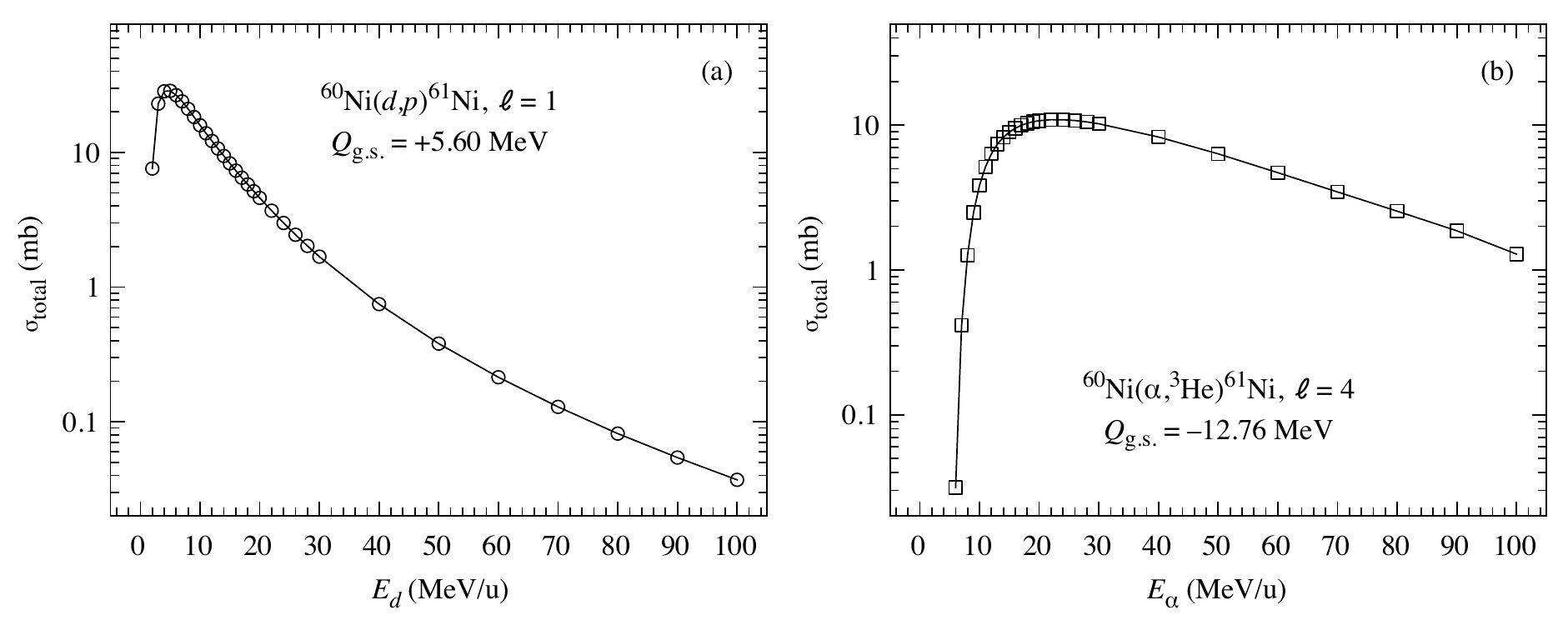}
\caption{\label{fig_energy} Calculations of the total cross sections for the (a) $^{60}$Ni($d$,$p$)$^{61}$Ni$_{\rm g.s., \ell=1}$ and (b) $^{60}$Ni($\alpha$,$^3$He)$^{61}$Ni$_{\rm g.s., \ell=4}$ reactions as a function of incident beam energy carried out in a DWBA framework using modern global optical-model parameterizations and form factors.}
\end{figure}

Figure~\ref{fig_energy}(a) indicates that the energy dependence of (total) cross sections is relatively flat for $(d,p)$ reactions at deuteron energies between 5--15 MeV per nucleon ($E_{d}=10$--30~MeV) on light-mid mass nuclei, arbitrarily chosen for illustrative purposes. One notes that in this regime, both the incoming deuteron and outgoing proton are  above the Coulomb barrier by over an MeV/u, or so. At both lower and higher energies than this the transfer cross sections change rapidly. By 50 MeV/u the peak cross sections are decreased by an order of magnitude. Figure~\ref{fig_energy}(b) shows another type of neutron-adding reaction, ($\alpha$,$^3$He). Here, the reaction has a large, negative $Q$ value. A similar behavior is seen, where the cross section is relatively flat but in this case at around 10--50 MeV/u ($E_{\alpha}=40$--200~MeV). While the energy is higher, the same criteria holds that both the incoming and outgoing ions are a few MeV/u above the Coulomb barrier.

\begin{figure*}[t]
\centering
\includegraphics[scale=0.9]{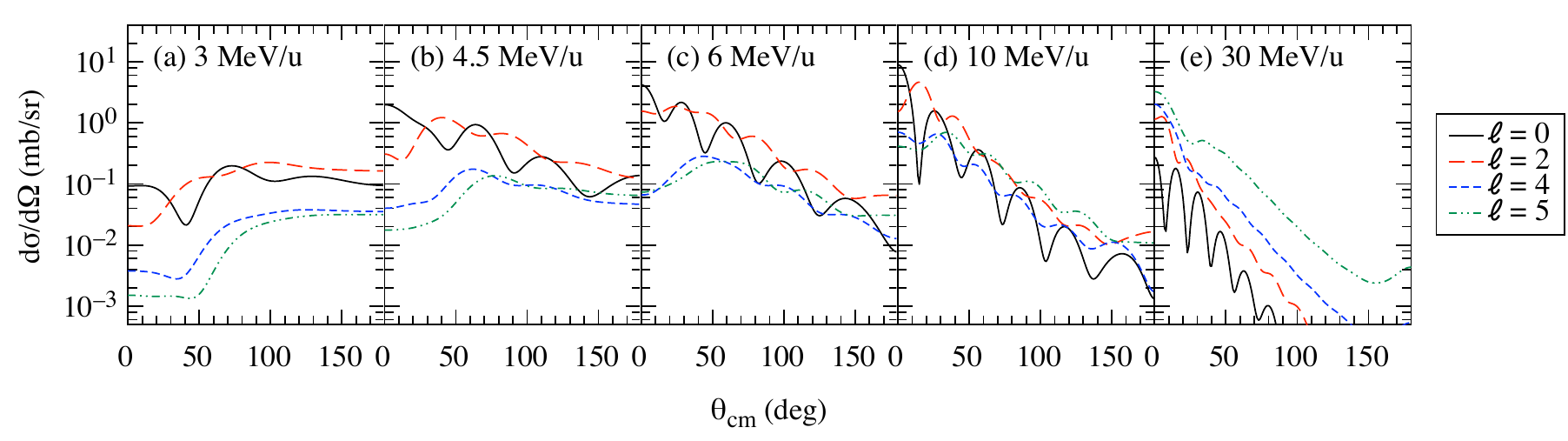}
\caption{\label{fig_sn_energy} (Color online) Calculations of $\ell=0$, 2, 4, and 5 transfer in the $^{120}$Sn($d$,$p$)$^{121}$Sn reaction at energy below (a), around (b), just above (c), a few MeV/u above (d) and well above (e) the Coulomb barrier in both the entrance and exit channel. The calculations were carried out in a DWBA framework using modern global optical-model parameterizations and form factors.}
\end{figure*}

Figure~\ref{fig_sn_energy} illustrates the different angular momenta transfer for the $^{120}$Sn($d$,$p$)$^{121}$Sn reaction, arbitrarily chosen, carried out at energies below the barrier (3~MeV/u), close to the barrier (4.5~MeV/u), and above the barrier at energies of 6, 10, and 30~MeV/u. 

At low incident energies below the Coulomb barrier, transfer reactions have little to no sensitivity to $\ell$ transfer (see, e.g., Ref. ~\cite{Jeans69}, and the similarity of the shapes in Fig.~\ref{fig_sn_energy}(a)), but are highly sensitive to other parameters such as radii and the tail of the wavefunction. This makes cross sections from such measurements less relevant to the discussion of overlaps and quenching. While sometimes secondary to a measurement, being able to do reliable spectroscopy (such as the assignment of $\ell$ values) can be essential to the measurement, especially when exploring new systems as is common with RIB measurements.  At high energy, either poor matching or multistep processes can be problematic. Some models attempt to compensate for this, as will be briefly mentioned in Section~\ref{models}.

From Fig.~\ref{fig_sn_energy}, it can be seen that around 4.5--10~MeV/u is likely optimal for such measurements. At these energies, the low-$\ell$ transfer is forward peaked and distinctive, suggesting these are reactions happening dominantly at the nuclear surface, and the cross sections a few orders of magnitude larger than at sub-barrier energies. The individual shapes for different $\ell$ transfer can be understood in terms of momentum matching, discussed below.

The bulk of the data on stable nuclei had been gathered at energies between $\sim$5-20 MeV, because these were the energies available from the tandem Van de Graaff accelerators and cyclotrons that were built largely for the exploration of nuclear structure and were suitable for high-resolution measurements with magnetic spectrographs.

Similar figures to those shown in Fig.~\ref{fig_sn_energy} have been used to motivate the development of radioactive-ion beam facilities, demonstrating the ideal energy for transfer reactions, and especially for stressing the importance of beam energy for new facilities (e.g., REX-ISOLDE (3~MeV/u) upgrading to HIE-ISOLDE (10~MeV/u), and ReA3 (nominally 3~MeV/u) upgrading to ReA6 and potentially beyond).
 
There have been several recent pioneering transfer-reaction measurements that were carried out at (too) low (or too high) energies that provided essential glimpses into the systems in question---in many cases, $\ell$ identification and other nuclear structure properties can be gleaned, but perhaps not with precise (cross sections) spectroscopic factors, nor have many of those measurements been on systems with large $\Delta S = \epsilon \vert {\text S}_{\text n}-{\text S}_{\text p} \vert$, $\epsilon$=+1 for proton removal, $\epsilon$=-1 for neutron removal.

\begin{figure}[t]
\centering
\includegraphics[trim=0cm 0cm 0cm 0cm,clip,width=18cm]{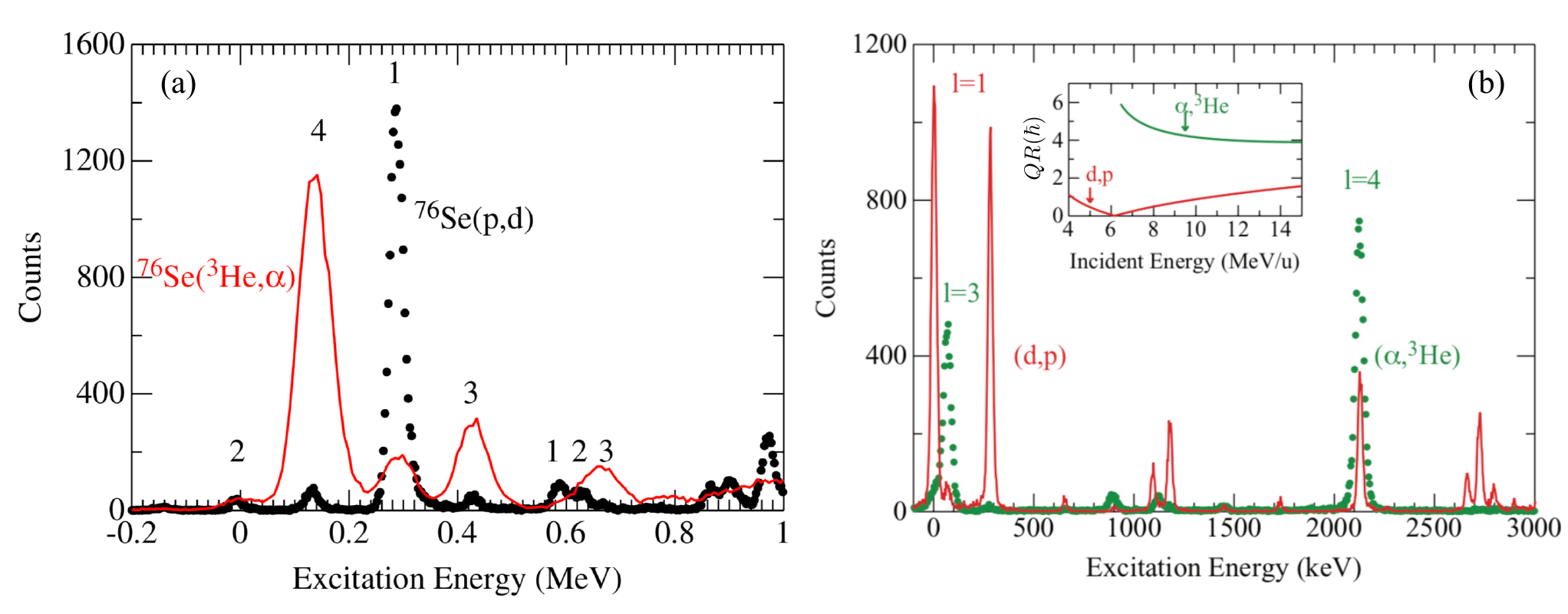}
\caption{\label{fig_matching} (Color online) Examples of (a) neutron-removing ($p$,$d$) and ($^3$He,$\alpha$) reactions~\cite{Schiffer08} at 23 and 26~MeV, respectively and (b) neutron-adding ($d$,$p$) and ($\alpha$,$^3$He) reactions~\cite{Schiffer13} at 10 and 38~MeV, respectively, demonstrating the role of momentum matching in nucleon transfer reactions. Reprinted figures with permission from Refs. ~\cite{Schiffer08} and ~\cite{Schiffer13}\textcopyright2013 by the American Physical Society.}
\end{figure}

As mentioned above, if the experiment can be carried out under appropriate conditions in terms of beam energy, the transfer mechanism can be considered a one-step process happening in the proximity of the nuclear surface, populating $s.p.$ states in the target nucleus. Under such circumstances, the approximations made in the reaction model are likely to be far more reliable. A key consideration, beyond beam energy, is momentum matching. A simple, semi-classical take on this relates the change ($Q$) in incoming ($p_{\rm in}$) and outgoing ($p_{\rm out}$) momentum at the nuclear surface ($R$) to the transferred orbital angular momentum such that
\begin{equation}
\label{eq_lmatch}
|{\bf Q} \times {\bf R}| =  |({\bf p}_{\rm in} - {\bf p}_{\rm out})\times {\bf R} |   \simeq \ell \hbar
\end{equation}
(with the height of the Coulomb barrier subtracted, since the reaction takes place near the surface, and close to the barrier height), mimicking ${\bf \ell}={\bf r} \times {\bf p}$.

Figure~\ref{fig_matching} shows two examples of momentum matching, (a) contrasting the neutron-removing ($p$,$d$) and ($^3$He,$\alpha$) reactions on $^{76}$Se~\cite{Schiffer08} and (b) the neutron adding ($d$,$p$) and ($\alpha$,$^3$He) reactions on $^{60}$Ni~\cite{Schiffer13}. For the latter example, the beam energies were 5 MeV/u for the ($d$,$p$) reaction and 9.5~MeV/u for the ($\alpha$,$^3$He) reaction, which are both near the peak cross sections as shown in Fig.~\ref{fig_energy}.

From Fig.~\ref{fig_matching}(b), the spectra from the two reactions, ($d$,$p$) and ($\alpha$,$^3$He), appear very different. The momentum matching, shown in the inset, indicates $Q\times R\approx1$ for the $(d,p)$ reaction and $\approx4$ for the $(\alpha,^3$He) reaction, and the ratios of cross sections between the two reactions for $\ell$=1 or 4 differ by about two orders of magnitude depending on their $\ell$ value. This is an astonishing demonstration of momentum matching.

The cross sections from the respective reactions are low (high) because of poor (good) matching. When the momentum matching is poor, the contributions from more complicated, indirect (multi-step) pathways may contribute more significantly, and the interpretation of the cross section in terms of a simple one-step process becomes more questionable---and as a result the spectroscopic factors will  not be reliable. 

\subsubsection{Detection angles} 
Strongly coupled to the choice of beam energy and the role of momentum matching (Section~\ref{benergy}), is the angular range or singular angle at which the cross section was measured, and thus the spectroscopic factors are determined. An example of the kinematics of various reactions in `normal' and inverse kinematics can be seen in Fig.~\ref{fig_kine}. The preceding discussion points to the reliable cross sections being determined from forward peaked (in the c.m. frame), surface-dominated reactions. In this regime, reliable spectroscopic factors are determined from the peak cross sections, or as close to the peak as is practical. While a general reproducibility of the angular distributions gives some faith in the distorting potential used, the details of the shape are sensitive to the model. Attempting to extract spectroscopic factors from the second maxima or angles near minima is likely to result in misleading results. It is noted that the angle of the peak is related to momentum transfer ($QR\simeq k_{cm}\sqrt{2(1-\cos\theta)}$), which naturally leads to the peaks being more forward-focused at higher energies, as evidenced by plots in Figure~\ref{fig_sn_energy}.

For studies in inverse kinematics, particularly for neutron-rich $sd$-shell nuclei, where many of the systems with large $\Delta S$ lie, the challenge is to determine cross sections for $\ell=0$ transfers. Often, detector systems are highly capable of detecting outgoing ions over 10--40$^{\circ}$ and to larger c.m. angles. To measure close to 0$^{\circ}$ is something that can be challenging even in normal kinematics, where detector systems can become rate limited at angles less than $\sim$5$^{\circ}$ in the lab frame (similar to the c.m.\ frame in stable beam measurements at these modest energies). It is possible in some circumstances. For example, the outstanding measurement by Margerin {\it et al.}~\cite{Margerin15}, who measured the $^{26}$Al($d$,$p$)$^{27}$Al reaction with the luxury of a very intense beam, allowing for the placement of annular Si detectors at large distances from the target. This resulted in an angular coverage of around $2^\circ<\theta_{\rm c.m.}<15^\circ$, essentially capturing the peak $\ell=0$ cross section. This was a system with small $\Delta S$. Other examples of measurements carried out with radioactive-ion beams that captured relatively far-foward c.m.\ angles are Flavigny {\it et al.}~\cite{Flavigny13} for $\ell=1$ transfer and Schmitt {\it et al.}~\cite{Schmitt12} for $\ell=0$.

\subsubsection{Comment on heavy-ion transfer} 

The availability of heavy ion beams opened up the possibility to explore heavy ion transfer. It was soon realized that in comparing different heavy-ion transfer reactions, a notable $j$ dependence, or selectivity, was observed at energies both above and below the Coulomb barrier. It was Brink~\cite{Brink72} who looked at the kinematics of heavy-ion transfer and commented on both the $Q$-value dependence and the $j$ dependence. Changes to both the reaction choice and thus $Q$-value significantly impact the yields to given final states.  A beautiful demonstration is shown via heavy-ion transfer on $^{54}$Fe carried out at $\sim$5~MeV/u~\cite{Pougheon73} to final states in $^{55}$Co. The ($^{16}$O,$^{15}$N) and ($^{14}$N,$^{13}$C) reactions both transfer a $p_{1/2}$ proton which has $j_i=\ell_i-s$ and therefore one would expect strong population of $j_f=\ell_f+s$ states (Ref.~\cite{Brink72}) relative to the $j_f = \ell_f - s$ (if seen at all). These are contrasted with the ($^{12}$C,$^{11}$B) reaction, where a $p_{3/2}$ proton is transferred. For the ($^{16}$O,$^{15}$N) reaction, the $f_{7/2}$ ground state ($j=\ell+s$) is strongly populated, and relatively little yield is seen the $f_{5/2}$ state ($j=\ell-s$). It is, however, less pronounced an effect for the ($^{14}$ N,$^{13}$C) reaction, which has smaller $Q$ value. 

With careful choices of pairs of reactions, robust spin assignments can be made with heavy-ion transfer reactions. This has particular value for high-$j$ states, where the matching conditions are more favorable in heavy-ion transfer. In this vain, the Oak Ridge group have demonstrated the advantages of sub-barrier heavy-ion transfer in inverse kinematics. They used complementary ($^{13}$C,$^{12}$C) and ($^{9}$Be,$^{8}$Be) reactions on isotopes of $^{136}$Xe, $^{134}$Te~\cite{Allmond12,Radford05a,Radford05b}, and $^{132}$Sn~\cite{Allmond14} to help determine the excitation of high-$j$ states, in particular the ($j_>$) 13/2$^+$ and ($j_<$) 9/2$^-$ states, and extract useful spectroscopic information. In most instances, such measurements are coupled with coincidence gamma-ray measurements, which can support the assignments. While similar techniques are likely to be a powerful technique both at current and next generation facilities, there remains unresolved discrepancies in describing the absolute cross sections for heavy-ion transfer using DWBA. Pieper {\it et al.}~\cite{Pieper78} and Olmer {\it et al.}~\cite{Olmer78} explored this in some detail via the ($^{16}$O,$^{15}$N) and ($^{16}$O,$^{17}$O) reactions on $^{208}$Pb, noting that while relative cross sections (and spectroscopic factors) at different energies, both above and below barrier, appear in quite good agreement with light-ion transfer, the absolute magnitude of the cross sections does not scale well with energy. The discrepancies are as much as a factor of a 2-3 when compared to light-ion transfer. Both works~\cite{Pieper78,Olmer78} noted that the discrepancies could not be corrected by simple changes to optical-model or bound-state parameters, which is a finding corroborated by Winfield {\it et al.}~\cite{Winfield85} in a study of the ($^9$Be,$^{10}$B) reaction on isotopes of O, Mg, Fe, Ca, and Cu at energies around 5~MeV/u. Further studies by Winfield {\it et al.}~\cite{Winfield89}, at higher energies, found similar discrepancies with respect to light-ion transfer overlaps. The issue appears not to have been resolved.

\subsection{Reaction theory for transfer reactions}\label{models}
\subsubsection{The distorted-wave Born approximation (DWBA)}
\label{sec:DWBA}
Given the nearly sixty years of history behind the DWBA formalism, and its extensive use during this period, only a very brief description of this model is given here. References~\cite{Satchler83,Glendenning83} contain detailed descriptions of the model.
Significant effort has been spent on the model uncertainties and parameterizations, much of it leading to better understandings (a recent example, relevant for this work is Ref.~\cite{Flavigny18}), but the importance of carrying out the measurement in a regime that is suitable for the model and the parameters remains essential, as discussed in the previous section. There is a strong connection between the experimental data uncertainties and the analyses uncertainties.

Using the description in Satchler's book~\cite{Satchler83}, the scattering amplitude for a reaction $A(a,b)B$ [in the center-of-mass frame, where normal and inverse kinematics can be treated in the same way], can be written as
\begin{equation}
\label{integration}
f_{\rm DWBA}(\theta,\phi)= -\dfrac{\mu_\beta}{2\pi\hbar^2}\int\chi^{(-)}_{\beta}({\rm\bf k_{\beta}},{\rm\bf r_{\beta}})^{\ast} \langle {\rm b},{\rm B}\left|V'\right|{\rm a},{\rm A}\rangle\chi^{(+)}_{\alpha}({\rm\bf k_{\alpha}},{\rm\bf r_{\alpha}}){\rm d\bf r_{\beta}},
\end{equation}
where the functions $\chi_{\alpha}$ and
$\chi_{\beta}$ are distorted waves describing the elastic scattering of the particles in  
the entrance ($\alpha={\rm a}+{\rm A}$) and exit ($\beta={\rm b}+{\rm B}$) channels, with momentum and  relative coordinates  ${\rm\bf k_{\alpha,\beta}}$ and ${\rm\bf r_{\alpha,\beta}}$
respectively. $V'$ is the interaction inducing the transition, specific to the reaction, and $\mu_\beta$ is the reduced mass in the exit channel.

It follows that the cross section for a $s.p.$ state, with a given angular momentum and orbital angular-momentum transfer is
\begin{equation}
\frac{d\sigma(\theta)}{d\Omega}=\frac{v_{\beta}}{v_{\alpha}}
\left|f_{\rm DWBA}(\theta)\right|^2,
\end{equation}
where $v_{\alpha}$ and $v_{\beta}$ are the c.m.\ velocities in the entrance and outgoing exit channels.

In Eq.~\ref{integration}, the  matrix element $\langle {\rm b},{\rm B}\left|V'\right|{\rm a},{\rm A}\rangle$ involves the integral over the internal coordinates of the many-body wavefunctions of the incident and outgoing particles. It is usually assumed that $V'$ does not depend on these internal coordinates.  For example, if $B$ results from the addition of one neutron to the target $A$, one needs to perform the overlap integral
\begin{equation}
\label{eq:ABoverlap}
 \int d\xi \; \Psi^{*}_B (\xi, {\bf r}) \Psi_A(\xi)    \equiv \psi^{\ell,j}_{AB}({\bf r}) ,
\end{equation}
where $\xi$ stands for the internal coordinates of $A$ and ${\bf r}$ that of the additional neutron.This overlap integral can be identified with that in  Eq.~(\ref{eq:overlap}) of Sec.~\ref{sec:theory}, and  is proportional to the probability amplitude of finding the state $A$ when a nucleon is removed from $B$. In general $\psi^{\ell,j}_{AB}({\bf r})$ is not normalized to one. In fact, its normalization gives the so-called {\it spectroscopic factor},
\begin{equation}
\label{eq:norm_overlap}
\int d {\bf r} |\psi^{\ell,j}_{AB}({\bf r})|^2 = S^{\ell,j}_{AB} 
\end{equation}

Very often, in practical calculations of the DWBA method the overlap function is approximated by the a $s.p.$ wavefunction, obtained as a solution of a Schr\"odinger equation with some mean-field potential (typically of Woods-Saxon type), with the appropriate quantum numbers  $\ell,j$ and separation energy.  Since the $s.p.$ wavefunction is unit normalized, one then writes
\begin{equation}
\label{eq:sp_overlap}
\psi^{\ell,j}_{AB}({\bf r}) \approx \sqrt{S^{\ell,j}_{AB}} \psi^{\ell,j}_{sp}({\bf r})
\end{equation}
where $\psi^{\ell,j}_{sp}({\bf r})$ is the $s.p.$ wavefunction.

When angular momentum is explicitly introduced, additional Clebsh-Gordan coefficients appear. Furthermore, if  the isospin formalism is used to express the states $A$ and $B$,  and additional isosopin coefficient ($C$) appears as well which is sometimes  singled out from the definition of the spectroscopic factor and hence written explicitly ($C^2 S$). The use of Clebsch-Gordan coefficients in the context of transfer reactions in discussed in Ref.~\cite{Schiffer69} and an explicit example is given in Ref.~\cite{Szwec16}. $C^2$ is often 1 and not discussed at length, or sometimes intentionally or unintentionally ignored.\\

If the $s.p.$ overlap Eq.~(\ref{eq:sp_overlap}) is used in the scattering amplitude Eq.~(\ref{integration}), one may express the differential cross section as  a $s.p.$ cross section multiplied by the corresponding spectroscopic factor
\begin{equation}\label{abss}
\frac{d\sigma(\theta)}{d\Omega} = g C^2 S_i \left . \frac{d\sigma(\theta)}{d\Omega} \right |_{\rm sp}
\end{equation}
where $S_i$  is the spectroscopic factor for a specific state $i$.  Note that, if both the projectile and target overlaps are expressed in terms of $s.p.$ overlaps, a product of the corresponding spectroscopic factors will appear in Eq.~(\ref{abss}). A statistical factor $g$ is needed, which is $(2j+1)$ for adding an $1$ for removing.

Many codes have been written to perform the calculations (the full integration of Eq.~(\ref{integration})) and are generally available. Examples include Ptolemy~\cite{Ptolemy78}, FRESCO~\cite{Thompson88}, TWOFNR \cite{TWOFNR}, etc. Finite-range calculations do not approximate the integration and if the bound state overlaps are known, the resulting cross section should reflect the experimental cross section without normalization. A number of the available codes allow for modifications to simple DWBA calculations, such as the performance of coupled-channels Born approximation (CCBA) and coupled-reaction-channels (CRC) calculations, or the use of the adiabatic distorted wave approximation (ADWA)~\cite{Johnson74}, which is able to include the breakup of the weakly-bound deuteron in $(p,d)$ and $(d,p)$ reactions in a very computationally-efficient way. The ADWA approximation has been compared to the more sophisticated continuum-discretized coupled-channel (CDCC) \cite{chazono17} and Faddeev/AGS \cite{nunes2011,nunes11b,upadhyay12}, finding differences of less than $\sim 20\%$ at energies of $\sim $10~MeV/nucleon, for which the cross sections of $(p,d)$ and $(d,p)$ reactions are highest, although the differences increase with increasing energy and decreasing cross section \cite{upadhyay12}.

As stressed above, the current discussion concerns analyses using the DWBA framework, although for $(d,p)$ and $(p,d)$ reactions the similar ADWA framework is also commonly used. In the 1960s there was considerable work done with stable systems (modest $\Delta S$) in validating the DWBA framework. Perhaps one of the earliest and most cited being the study of the $^{40}$Ca($d$,$p$) reaction by Lee {\it et al.}~\cite{Lee64}. As is so often concluded, that work summarized that so-called `absolute' spectroscopic factors could be extracted with an accuracy of around 20\% with numerous caveats, such as the choice of bound-state potentials and optical-model parameterizations. It is these choices that are discussed below.
 
It is only recently that compelling efforts have been made towards an {\it ab initio} approach to nuclear reactions and nuclear structure, both aspects of which are considered in the same framework \cite{Raimondi16}. This approach is likely to be the crowning achievement of reaction theory in the coming decade or so, and in particular its application to very weakly or tightly bound systems.

\subsubsection{Optical potential} 
Optical potentials\footnote{The name optical model arose from the analogy of light scattering from a cloudy crystal ball.} are key ingredients of any reaction calculation and transfer reactions are not an exception.  In a DWBA calculation,  optical potentials are meant to describe the elastic scattering  of the $a+A$ and $b+B$ systems. Flux is removed from the elastic-scattering channel as the reaction occurs, and this is modeled using imaginary components in the optical-model potential.  As described in Sec.~\ref{sec2}, optical potentials are also intrinsically nonlocal, as well as energy and angular-momentum dependent but, most commonly, they are approximated by more simple local representations, $U(\mathbf{r})$.  Moreover, they are assumed to be described by a central potential, $U(r)$, which is  frequently parameterized as follows \cite{Satchler90}:
\begin{equation}
\label{eq:omp_pot}
U(r)=V_{\rm C}(r) - V f(r,R_0,a_0)-i W f(r,R_w,a_w) - i W_D g(r,R_d,a_d ) 
\end{equation}
where $V_C(r)$ is the Coulomb potential, $f(r,R,a)$ is the Woods-Saxon  form factor, given by  $f(r,R,a)=1/[1+{\rm exp}((r-R)/a)]$, and $g(r,R,a)=4 \, a\, (d/dr)f(r,R,a)$. The constants  $V$, $W$ and $W_D$ define the strengths of the  real volume, imaginary volume, and imaginary surface potentials, respectively, whereas  $R_x$ and $a_x$ are the radius and diffuseness parameters defining their geometry.

When either the projectile or target have nonzero spin, a spin-orbit component is usually added. For that, the Thomas form is commonly adopted:
\begin{equation}
\left ( \frac{\hbar}{m_\pi c} \right )^2
\frac{1}{r}\frac{df(r,R_{so},a_{so})}{dr} 2 \bf{ L} \cdot {\bf S}
\end{equation}
where  $\left ( \hbar  / m_\pi c \right )^2  \approx 2$~fm$^2$ is introduced for convenience. 

Optical-model parameterizations can be specific or global. The concept of global parameterizations was established very early in the progression of the field as a natural extension of understanding the systematics of scattering reactions. It was clear that for nuclei near stability and particularly those greater than somewhere between around mass say $A\sim16$, the trends in terms of mass, volume, and $N-Z$ show a smooth variation, with a classic analysis performed in Ref.~\cite{Perey63}. Over the last fifty or so years, a wealth of elastic scattering data has led to very well evolved global optical-model parameterizations, particularly for protons~\cite{KD03,Varner91} (and neutrons) and deuterons~\cite{An06}. The situation is improving for $^3$He (e.g.,~\cite{Pang09}) and $^4$He, were fewer data have been collected. 
\begin{figure*}[t]
\centering
\includegraphics[scale=1.2]{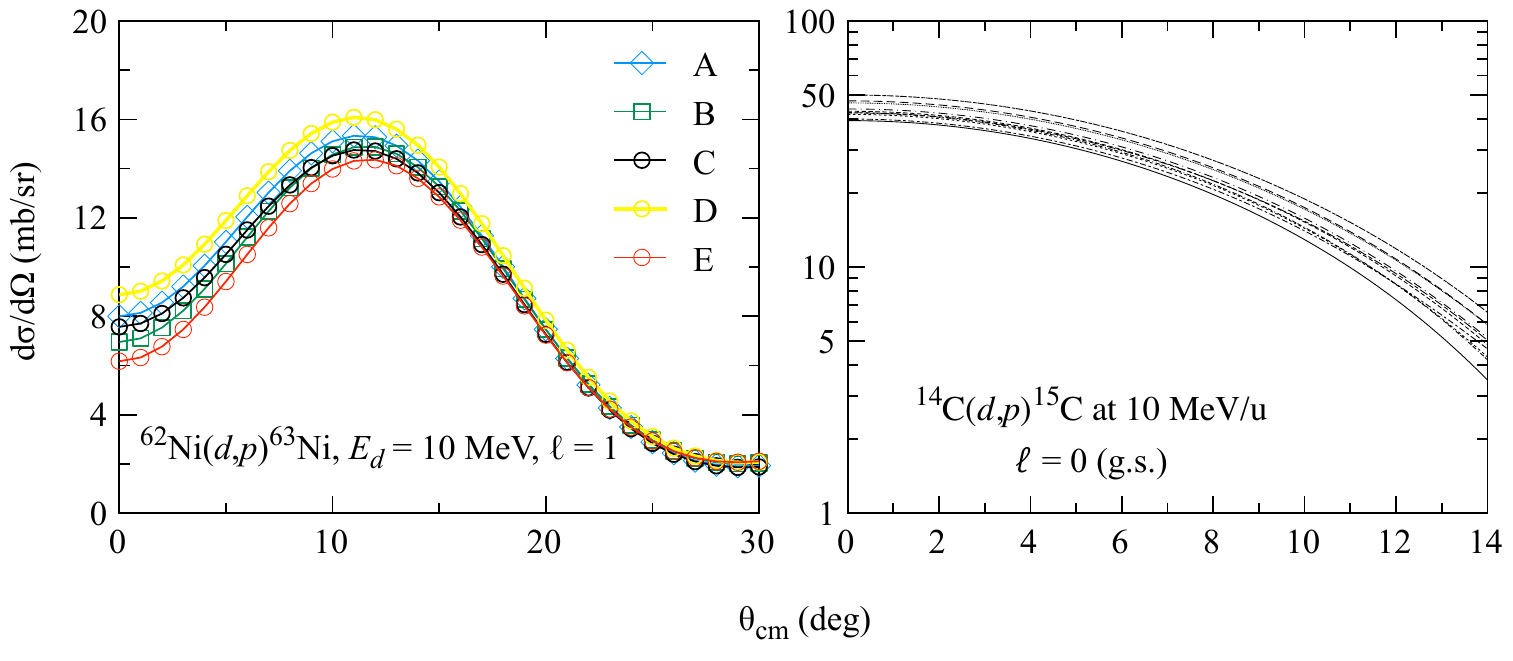}
\caption{\label{fig_omp} (Color online) Calculated angular distributions (left) for the five different pairing of optical-model potential parameterizations for protons and deuterons in the $\ell=1$ ($d$,$p$) reaction at 5~MeV/u on $^{62}$Ni and (right) for ten different pairings of parameters for the $\ell=0$ ($d$,$p$) reaction at 10~MeV/u on $^{14}$C. }
\end{figure*}

For light nuclei, the situation is poorer. Many global potentials are defined for heavier nuclei, typically above C or O. Further, for exotic systems, such as halo nuclei or other very weakly bound systems, there is still work to be done with regards to optical-model potentials (or parameterizations). This is likely to go hand in hand with new and numerous transfer-reaction studies at future radioactive-ion beam facilities.
For heavier systems, global optical-model potentials are indispensable. They are particularly useful when exploring a range of nearby systems or a chain or isotopes or isotones. A rigorous discussion of this was presented in Ref.~\cite{Schiffer13} and is presented in more detail below.

The optical-model parameters can modify the magnitude and particularly the shape of the angular distribution. The parameters used for the bound-state form factors can have a large impact on the magnitude of the calculated cross section. However, these can typically be well constrained. Figure~\ref{fig_omp}(a) shows calculated angular distributions for the ($d$,$p$) reaction on $^{62}$Ni to a hypothetical $\ell=1$ state at 1~MeV using five different combinations of optical-model parameters (for both proton and deuteron) that are well defined in this mass and energy region. The rms spread in the magnitude of the cross section at the peak cross section is $<11$\% . Figure~\ref{fig_omp}(b) shows calculated angular distributions for ten different combinations (again, for both proton and deuteron) of optical-model parameters for a system that has a relatively large $\Delta S$, and one for which the optical-model parameters are not necessarily well defined, the $^{14}$C($d$,$p$)$^{15}$C$_{\rm g.s.}$ reaction. Here the rms spread at the peak is $<8$\%. In each case, the bound-state potentials were kept the same. This gives some sense of the systematic uncertainties in the optical model parameterizations. There are, however, detailed studies that explore the propagation of uncertainties from the optical potentials obtained from elastic scattering to transfer cross sections, covering all angles, that suggest larger uncertainties. See, for example, Refs.~\cite{lovell15,lovell17,king18}.

In the study of ($d$,$p$) and ($p$,$d$) reactions, the deuteron wavefunction has typically used the Reid potential~\cite{Reid68}, or more recently the Argonne $\nu_{18}$ potential~\cite{PhysRevC.51.38}. For the more complex projectiles such as $t$, $^3$He, and $^4$He, there was the recent development of Green's Function Monte Carlo derived parameterizations~\cite{Brida11}. In general, transfer reactions at low energies depend mostly on volume properties of the overlaps $\langle p \vert d \rangle$, $\langle d \vert ^3He \rangle$, so they are rather insensitive to the models used to describe them, as long as they are sensible.

\subsubsection{Overlap functions}
 For the overlap function of the target and residual nuclei, as discussed before, this is usually approximated by a $s.p.$ wavefunction, computed in some mean-field potential, whose depth is varied to match the experimental separation energy of the transferred nucleon for the given final state. The wavefunction is often generated in a Woods-Saxon potential, which can be defined by radial parameters that are consistent with, for example, the $(e,e'p)$ data discussed at length in Sec.~\ref{sec:intro_theory}. From Ref.~\cite{Kramer01}, for most nuclei above mass 20, the radius parameter is $r_0\sim1.28$~fm and the diffuseness $a=0.65$~fm. A spin-orbit term is essential, with the depth often around 6~MeV and the radius term around 6/7 of the volume term, or $r_{\rm so0}\sim1.1$~fm. In analyses of both $(e,e'p)$ and transfer reaction data involving stable nuclei, parameterizations similar to this have proven consistent~\cite{Kramer01,Kay13a}. However, it is likely not the case far from stability and for removal of nucleons from deeply-bound orbitals. The sensitivities of these various parameters have been explored in detail by several groups (for recent examples related to the $\Delta S$ dependence, see Ref.~\cite{nunes2011,Flavigny18}). In many cases, variations based on `sensible' input parameters are of the order of 20\%. In low-energy transfer reactions, the absolute value of the cross section is mostly proportional to the asymptotic normalization coefficient (ANC) so the whole dependence on the bound state can be assigned to the ANC. The following sections approach the analyses of these data in the context of the parameters and analysis types.

\subsection{Analyses of transfer-reaction data}\label{analysis}

\subsubsection{Relative and absolute spectroscopic factors from transfer reactions}

\begin{figure*}[t]
\centering
\includegraphics[scale=0.75]{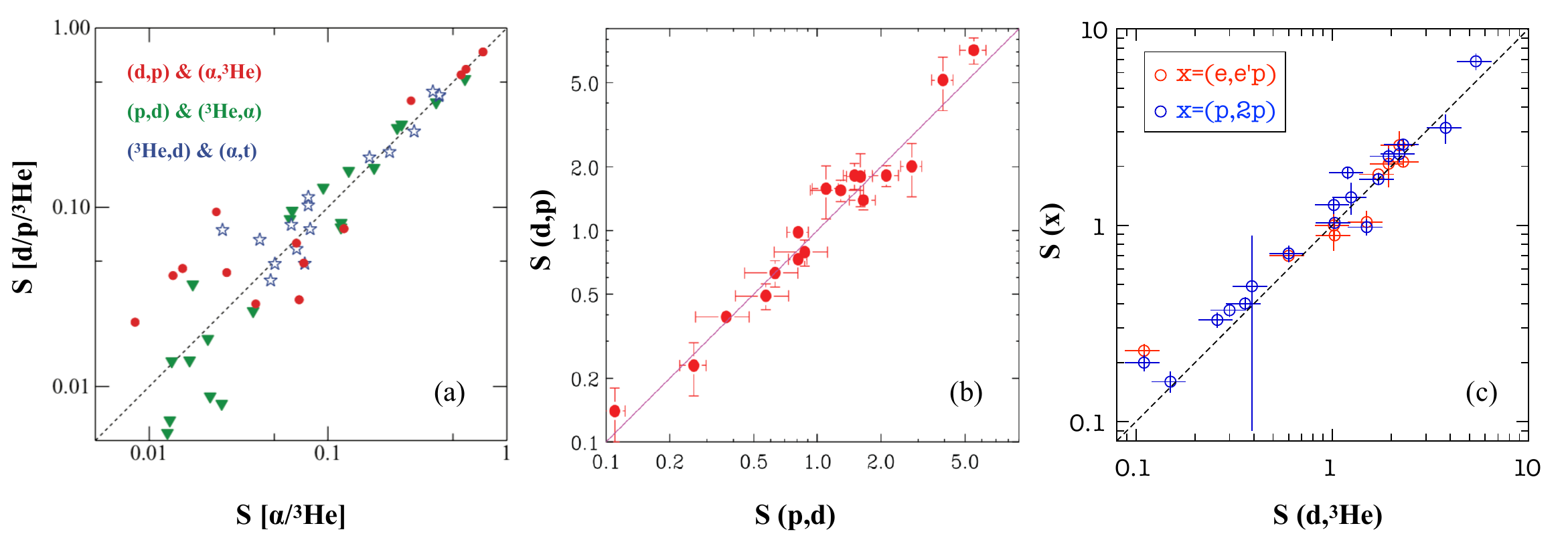}
\caption{\label{fig_various_s}(Color online)  Various comparisons of spectroscopic factors for (a) different reactions to the same final states for $\ell=3$ transfer on the stable Ni isotopes modified with permission from Ref.~\cite{Schiffer13}\textcopyright2013 by the American Physical Society, (b) the ($d$,$p$) reaction from, and ($p$,$d$) to, the same final state for many nuclei studied in  Ref.~\cite{Lee07}, adapted with permission from  Ref.~\cite{Lee07}\textcopyright2007 by the American Physical Society, and (c) different proton knockout and transfer reactions to the same final states (see e.g., data summarized in~\cite{Kramer01,Wakasa17}) for nuclei between Li and Pb.}
\end{figure*}

Relative spectroscopic factors are not important to the topic of this review. In general, for a given set of parameters for the form factors and the optical-model potentials, the relative strength of states with a given $j^{\pi}$ are reliable to better than 5\% or so~\cite{Schiffer12}. This is in a large part due to the fact that the DWBA is handling the kinematics and distortions, resulting in only a modest dependence on reaction $Q$ value. Determining the centroid (effective $s.p.$ energy) of an orbital that is fragmented into several states is thus likely to be quite insensitive to the choice of DWBA parameters (in fact, it can be shown that there is often only a small difference between the spectroscopic-factor weighted centroid and the cross-section weighted centroid).

The focus here is on absolute spectroscopic factors. That is, assuming absolute cross sections have been determined in a suitable experiment in terms of beam energy, momentum matching (choice of reaction), and at a suitable angle, and suitable parameterizations have been used in a DWBA (or alternative) calculation, then Eq.~(\ref{abss}) gives the absolute spectroscopic factor for a given state with respect to a pure independent $s.p.$ configuration ($S=1$).

It is common for individual states to be scrutinized, typically certain low-lying excitations, often the ground state. This is particularly the case for knockout reactions that are central to the topic of this review. As such, it can be instructive to perform a shell-model calculation, or similar, to determine the extent to which the state of interest is a pure ($S=1$) configuration, or some fraction thereof. Computing the ratio between the shell-model spectroscopic factor and the absolute spectroscopic factor from the DWBA calculation can define the extent to which the experimental cross section relates to that of a simple ``shell-model'' state---if we understand the motion of nucleons in the mean-field, this should be unity. This is not the case, and central to the topic of this paper, discussed shortly.

The accuracy of spectroscopic factors was previously mentioned to be around 20\%. Figure~\ref{fig_various_s} shows various ways in which one might deduce this in a somewhat empirical manner. Figure~\ref{fig_various_s}(a) from Ref.~\cite{Schiffer13} shows the spectroscopic factors deduced from the population of identical $0f_{5/2}$ final states outside of the stable, even Ni isotopes with different reactions. The figure suggests that for reasonably strong states, greater than about 10\% of the total $s.p.$ strength, the correlations between different reaction mechanisms (and thus different bound-state and optical-model parameterizations) is robust. Below this fraction, the scatter is large and thus so are the uncertainties. Figure~\ref{fig_various_s}(b) from Ref.~\cite{Lee07} shows a robust agreement between complementary adding and removing reactions probing the same final state, subtlety different from Fig.~\ref{fig_various_s}(a). The scale of the axes is slightly different, but highlighting that for a state with $S>10$\%, the agreements are good, certainly of the order of 20\% or better. Finally, Fig.~\ref{fig_various_s}(c) shows the comparison of the proton-removing ($d$,$^3$He) transfer reaction to the same final states as populated in the $(e,e'p)$ and $(p,2p)$ proton knockout reactions (with data summarized in Refs.~\cite{Kramer01,Wakasa17}). Again, here there is an excellent agreement.

Essentially all of the examples shown here are for stable nuclei, with modest $\Delta S$. This is intentional to set the scene. As mentioned in the preamble to this section, it reflects the dearth of transfer reaction data from radioactive-ion beams. In many of the cases described throughout this review, spectroscopic factors are derived from single states. Unique to transfer reactions is the ability to carry out both adding and removing reactions and use the nucleon-transfer sum rules. 

\subsubsection{Nucleon-transfer sum rules}\label{sumrule}

A natural extension to the concept of spectroscopic factors, describing the extent to which a given state represents a $s.p.$ excitation, is that the sum of all $s.p.$ excitations with the same quantum configuration should represent the total degeneracy of that orbital. These are the nucleon-transfer sum rules, introduced by Macfarlane and French~\cite{Macfarlane60}. There have been many works in the past that have invoked the sum rules in the analyses of nucleon-transfer cross sections (e.g.,~\cite{Grabmayr92}). Many of the recent results have come from the group of John Schiffer, triggered by detailed studies of nuclei involved in neutrinoless double beta decay~\cite{Ejiri19}. These include works on Ge and Se isotopes~\cite{Schiffer08}, which spawned a detailed work on the interpretation of nucleon-transfer cross sections and the sum rules~\cite{Schiffer12,Schiffer13}, which also explored quenching in some detail.

\begin{figure}[t]
\centering
\includegraphics[trim=0cm 0cm 36cm 0cm,clip,width=8cm]{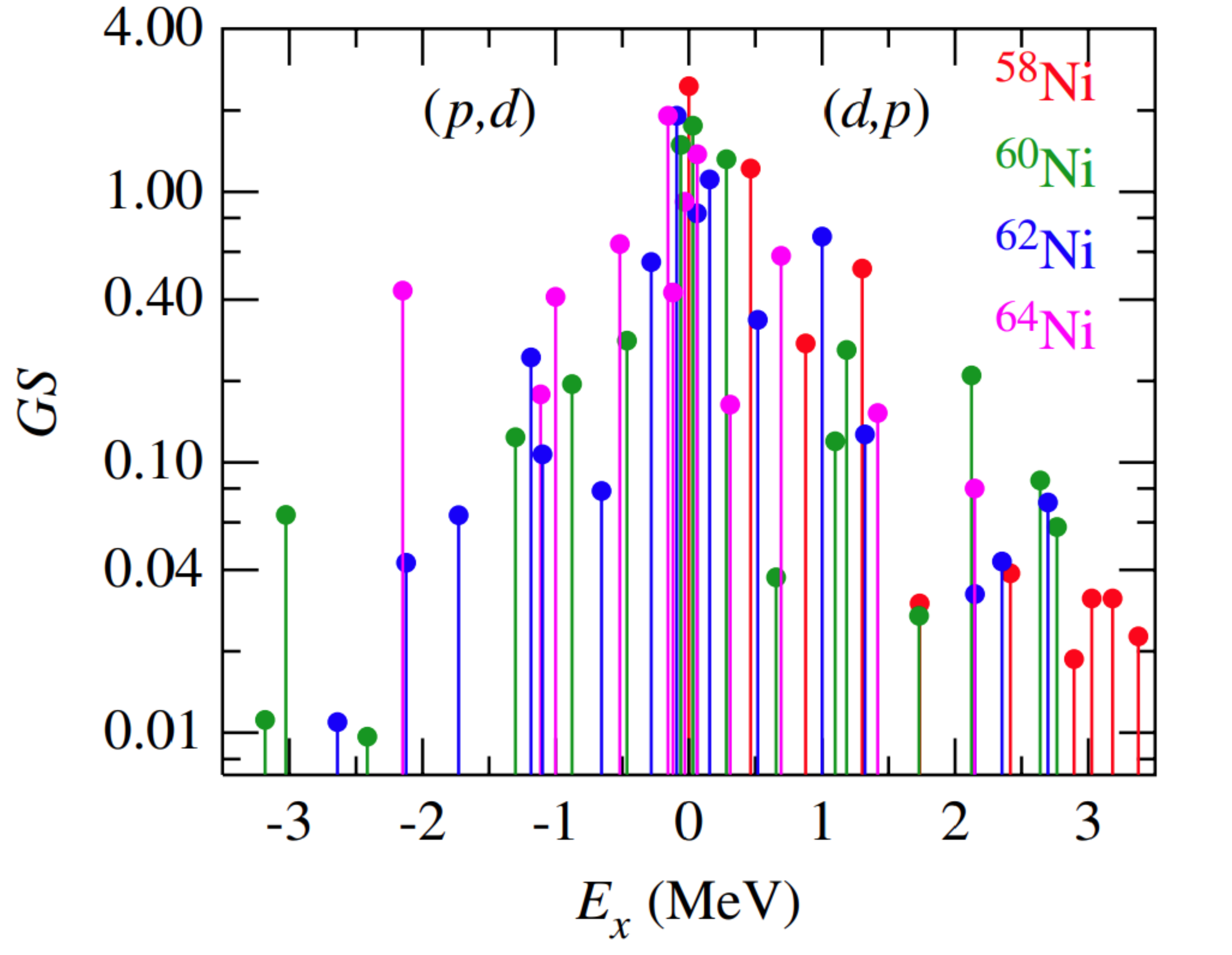}
\includegraphics[trim=36cm 0cm 0cm 0cm,clip,width=8cm]{figures_transfer/Fig1_flipped.pdf}
\caption{\label{fig_missing}(Color online)  (left) Showing the strength of low-lying $\ell=1$ ($p_{3/2}$ and $p_{1/2}$) states populated in the neutron-removing ($p$,$d$) reaction (negative energy) and the neutron-adding ($d$,$p$) reaction (positive energy) and (right) a histogram of sums of the adding and removing strengths in 0.5-MeV bins, which is fitted with a Lorentzian shape. Reprinted figures with permission from Ref. ~\cite{Schiffer12}\textcopyright2012 by the American Physical Society. }
\end{figure}

In the simplest sense, Equation~\ref{abss} can be written explicitly in terms of the spectroscopic factors for adding ($S^+$) and removing ($S^-$) reactions for transfer to final states of a given $j^{\pi}$, and naively the sum should equal $(2j+1)$, or
\begin{equation} \label{sumr}
(2j+1)N_j=\sum(2j+1)C^2S^+_j+\sum C^2S^-_j,
\end{equation}
where $N_j$ is a seemingly {\it ad hoc} normalization factor in this equation. Were our understanding of $s.p.$ motion in nuclei described by the independent $s.p.$ model, then $N_j\equiv1.0$. It should be noted that Eq.~\ref{sumr} can be seen as a simple reexpresion of Eq.~\ref{eq:sumr}.\\

\begin{figure}[t]
\centering
\includegraphics[scale=1]{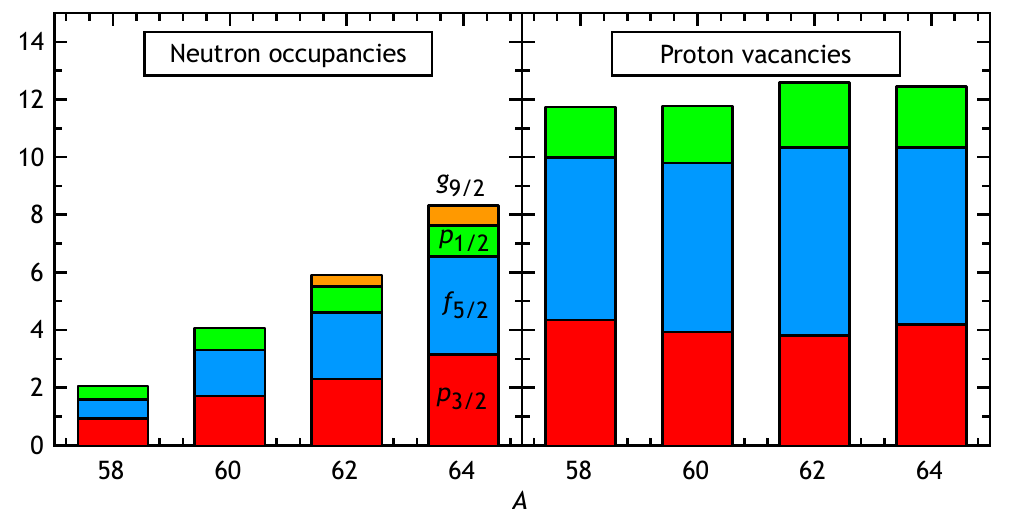}
\caption{\label{fig_nisum}(Color online)   Neutron occupancies and proton vacancies outside of the stable, even Ni isotopes extracted using transfer-reaction data. The data are from Ref.~\cite{Schiffer12}.}
\end{figure}

Below is a simple example taken from Ref.~\cite{Schiffer08}, where the neutron-removing ($p$,$d$) and neutron-adding ($d$,$p$) reactions were carried out on several isotopes of Ge and Se. Taking the $\ell=1$ (to both $p_{3/2}$ and $p_{1/2}$ states, where the $j$ values were previously known) spectroscopic strength for each reaction and adding them up and dividing by the total degeneracy of the $p_{3/2}$ and $p_{1/2}$ orbitals, (4+2), one gets a value of $N_j=0.53$. That is, the normalization factor derived from the sum rules reveals quenching.

\subsection{Quenching of transfer reaction cross sections}\label{quench}

\begin{figure}[h!]
\centering
\includegraphics[scale=1.0]{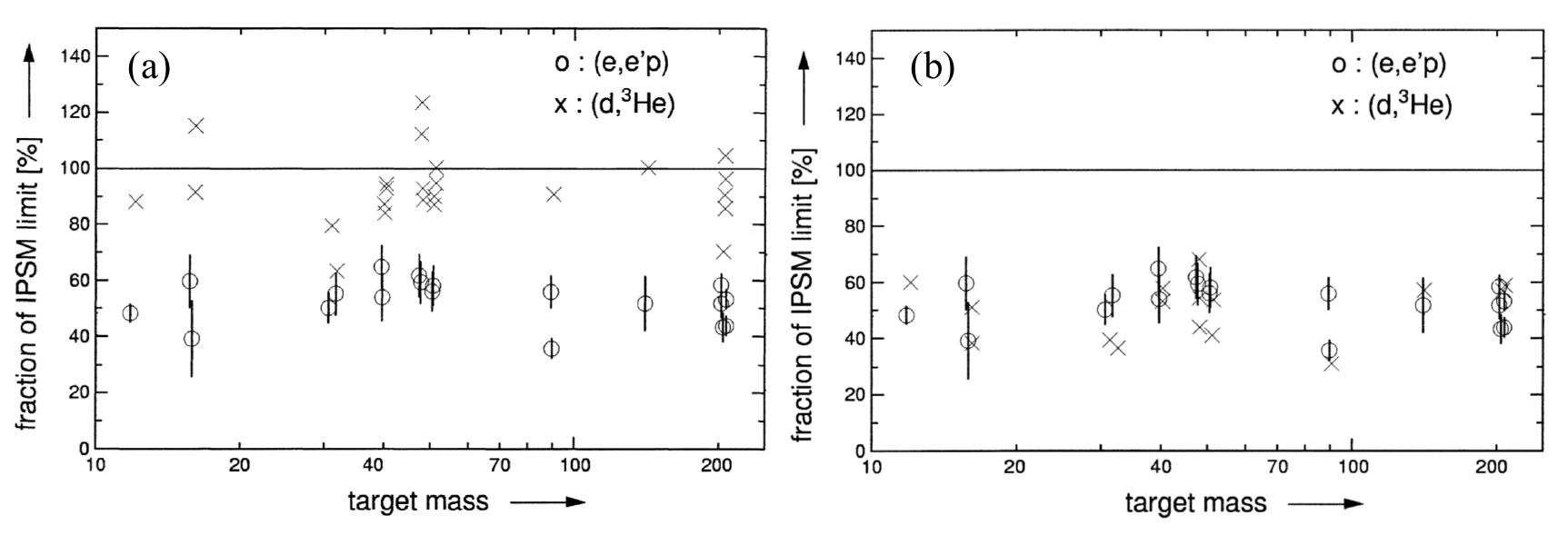}
\caption{\label{fig_kramer} (a) For nuclei between $12<A<208$, the degree to which the proton knockout strength derived from the $(e,e'p)$ reaction totals that of the independent-particle shell-model (IPSM) limit, compared to the same properties determined from the proton-removal ($d$,$^3$He) reaction, analyzed independently, and (b) the same data with careful attention paid to the radial parameters used in the analysis of the transfer-reaction data. Reprinted from Ref.~\cite{Kramer01}, with permission from Elsevier.}
\end{figure}

As highlighted throughout the paper, it was the $(e,e'p)$ reaction studies of NIKHEF group led by Lapik\'as that shone light on this matter. That led to a reanalysis transfer reaction data, presenting in the famous paper of Kramer, Blok and Lapik\'as~\cite{Kramer01}. A summary of their findings and of much of the $(e,e'p)$ work is given there and summarized in Fig.~\ref{fig_kramer}, showing the results before the reanalysis, and afterwards. This work came at a time when transfer reactions as a tool for exploring nuclear structure were not used as frequently as in the 60s and 70s, though it did coincide with the time when people were starting to consider transfer reactions in inverse kinematics in anticipation of future radioactive-ion beam facilities~\cite{Kraus91}---the era we are now in.

It was in the early 2000s that significant interest in quenching came about, prompted by the intermediate energy knockout at NSCL, discussed in Section~\ref{sec5}. While there was much activity on this front, and by definition on unstable systems, other groups started to re-explore transfer reactions on stable systems in the context of (a) detailed determinations of centroids across isotopic and isotonic chains to scrutinize the evolution of $s.p.$ states and the role of the tensor force, and (b) to constrain the calculations of neutrinoless double beta decay ($0\nu2\beta$) nuclear matrix elements. 

\begin{figure}[t]
\centering
\includegraphics[scale=1.1]{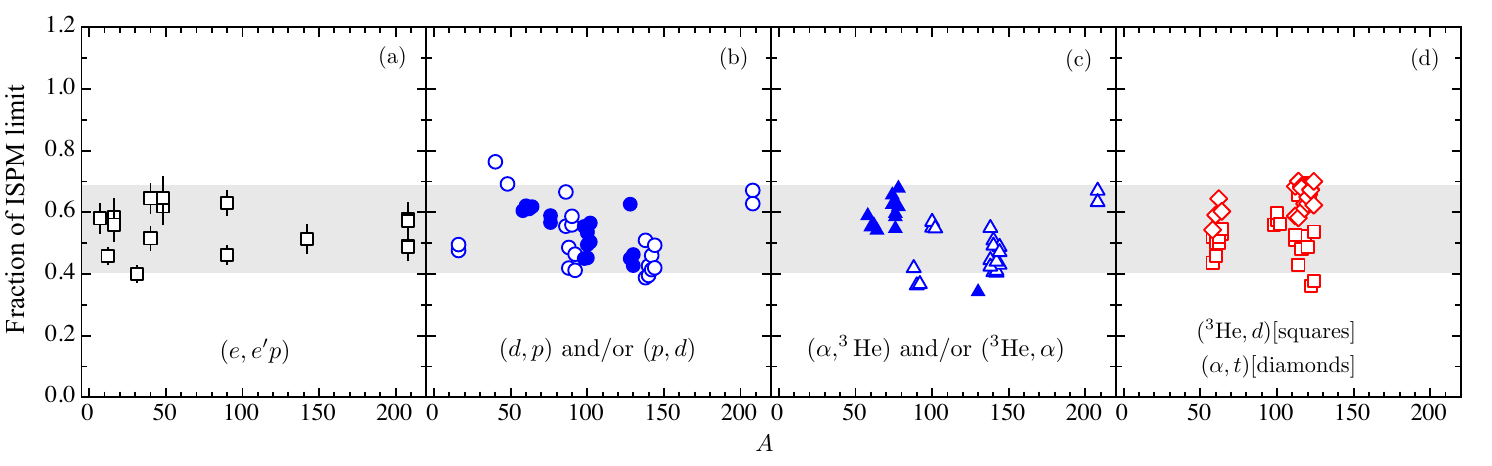}
\caption{\label{fig_transfer} (Color online) Observed $s.p.$ strength compared to that of the independent-particle shell-model (IPSM) limit as a function of mass $A$. The $(e,e'p)$ data in Panel (a) are from Refs.~\cite{Kramer01}. The grey band represents the mean $\pm$2$\sigma$ of the $(e,e'p)$ data to guide the eye. The data in Panels (b), (c), and (d), are from the analysis presented in Ref.~\cite{Kay13a}. Solid symbols are from adding and removing reactions while the empty ones are from just adding or just removing. Reprinted figure with permission from Ref.~\cite{Kay13a} \textcopyright2013 by The American Physical Society. }
\end{figure}

In Ref.~\cite{Kay13}, much of the data from Schiffer's group and also some past works where reliable cross sections had been published, was analyzed consistently using modern global optical-model parameterizations and treatments of the bound-state wavefunctions. The data fell into two broad categories. The first considered studies that had used both the adding and removing reactions, such that the normalization, or quenching factor, could be derived in the way that is described by Eq.~(\ref{sumr}). In each of these cases, the data were derived from studies of several nearby nuclei and common normalizations applied to all. These include the  ($d$,$p$), ($p$,$d$), ($\alpha$,$^3$He), and ($^3$He,$\alpha$) reactions on $^{58,60,62,64}$Ni~\cite{Schiffer12,Schiffer13}, $^{74,76}$Ge,  $^{76,78}$Se~\cite{Schiffer08}, and $^{130}$Te~\cite{Kay13}.

The second category was for studies where only adding or only removing data were available for a given nucleus---typically when looking at trends in particle or hole states outside of closed shell nuclei. This analysis is essentially the same as what was done for the $(e,e'p)$ data~\cite{Kramer01}, requiring that the total strength adds up to the number of vacancies in the closed shell, or the number of particles outside it, such that a simplified version of Eq.~(\ref{sumr}) can be expressed  as 
\begin{equation}\label{eqn2}
N_j\equiv\frac{1}{(2j+1)}\sum(2j+1)gC^2S_j,
\end{equation}
where $N_j$ is used to denote the normalization, which is the quenching factor (similar to the reduction factor) and $g$ hides the statistical factor. For these analyses, a mix of old and new data were considered. These include: $^{16}$O($d$,$p$)~\cite{Alty67}, $^{40}$Ca($d$,$p$)~\cite{Lee64}, $^{48}$Ca($d$,$p$)~\cite{Metz75}, the ($d$,$p$) and ($\alpha$,$^3$He) reactions on $^{88}$Sr, $^{90}$Zr, and $^{92}$Mo~\cite{Sharp13}, the ($d$,$p$), ($p$,$d$), ($^3$He,$\alpha$), and ($^3$He,$d$) reactions on $^{98,100}$Mo and $^{100,102}$Ru~\cite{Freeman17}, the $^{112-124}$Sn($^3$He,$d$) and ($\alpha$,$t$) reactions~\cite{Mitchell}, the ($p$,$d$), ($^3$He,$\alpha$), and ($\alpha$,$^3$He) reactions on the stable even $N=82$ isotones~\cite{Howard12,Kay08}, and the ($d$,$p$)~\cite{Jeans69} and ($\alpha$,$^3$He)~\cite{Perry81} reactions on $^{208}$Pb (this list and further details are available in Ref.~\cite{Kay13a}).

Figure~\ref{fig_transfer} shows a summary of these results, compared to those from the $(e,e'p)$ work. The transfer data are 124 unique data points from $16<A<208$ using proton and neutron adding and removing reactions to final states with $0\leq\ell\leq7$ and a large range of binding energies.\\

Perhaps the most impressive of these results are those of Ref.~\cite{Freeman17}, which was acquired using the high-resolution Munich $Q3D$ spectrograph. Given the outstanding statistics and resolving power, hundreds of states contributed to the sums with quenching factors of 0.618(20), 0.614(48), 0.564(18), and 0.647(25) from different adding and removing reactions on four isotopes, being $^{100,102}$Ru and $^{98,100}$Mo. The uncertainties are considerably smaller than other studies. It is, however, clear from Fig.~\ref{fig_transfer} that in general the spread is modest, around $\pm$10\% rms. This is true for transfer studies and comparable to the $(e,e'p)$ work. It is likely dominated by unknown systematic uncertainties in the experimental cross sections.

\begin{figure}[t]
\centering
\includegraphics[trim=0cm 0cm 0cm 0cm,clip,width=18cm]{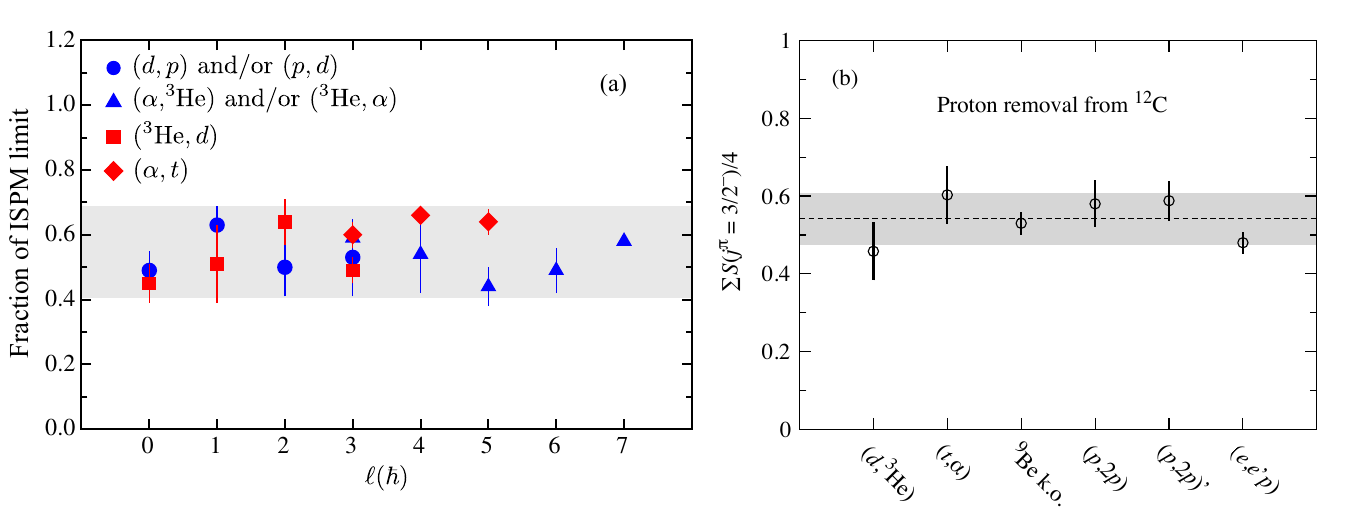}
\caption{\label{fig_ldepc12}(Color online)  (a) Average of the quenching factor for different $\ell$ transfer. The error bars shown represent the rms spread in values. Figure reprinted with permission from Ref.~\cite{Kay13a}\textcopyright2013 by the American Physical Society. (b) The sum of the 3/2$^-$ strength below $^{12}$C as probed via the ($d$,$^3$He) and ($t$,$\alpha$) proton-removal reactions, $^9$Be-induced knockout, proton-induced ($p$,$2p$) knockout in both `normal' and inverse kinematics, and lepton-induced ($e$,$e'p$) proton knockout, as a fraction of the simple shell-model sum rule limit of 4---that is to say, the quenching factor. The average, 0.53(10) (excluding the $^9$Be-induced knockout as the excited state is missing), is indicated by the horizontal dashed line. The data were taken from the Evaluated Nuclear Structure Data File~\cite{ENSDF}. }
\end{figure}

Figure~\ref{fig_ldepc12} emphasizes several additional aspects of the data. The first is clear from Fig.~\ref{fig_transfer}, and that is that the quenching factor seems to be ubiquitous not only in $A$ and independent adding or removing reactions involving protons or neutrons, but also independent of the $\ell$ value of the final state. Further, for a light nucleus where many different reactions have been carried out, $^{12}$C, there is excellent agreement in the strength of the low-lying $p_{3/2}$ excitations between the ($d$,$^3$He), ($t$,$\alpha$), $^9$Be-induced knockout, $(p,2p)$, and $(e,e'p)$ reactions---all of which remove a proton from $^{12}$C. 

What is clear from these stable-beam studies is that while the quenching appears to ubiquitous, these systems typically have $-12\lesssim\Delta S\lesssim12$~MeV. The final part of this section discusses quenching of transfer reactions in the context of $\Delta S$ and other probes. 

\subsection{Transfer reactions and $\Delta \boldmath{S}$}\label{dels}

Figure~\ref{fig_all} attempts to contrast the available transfer reaction data, this time including transfer reactions on systems that have a large $\Delta S$, with data from other probes.\\

\begin{figure}[t]
\centering
\includegraphics[scale=1.4]{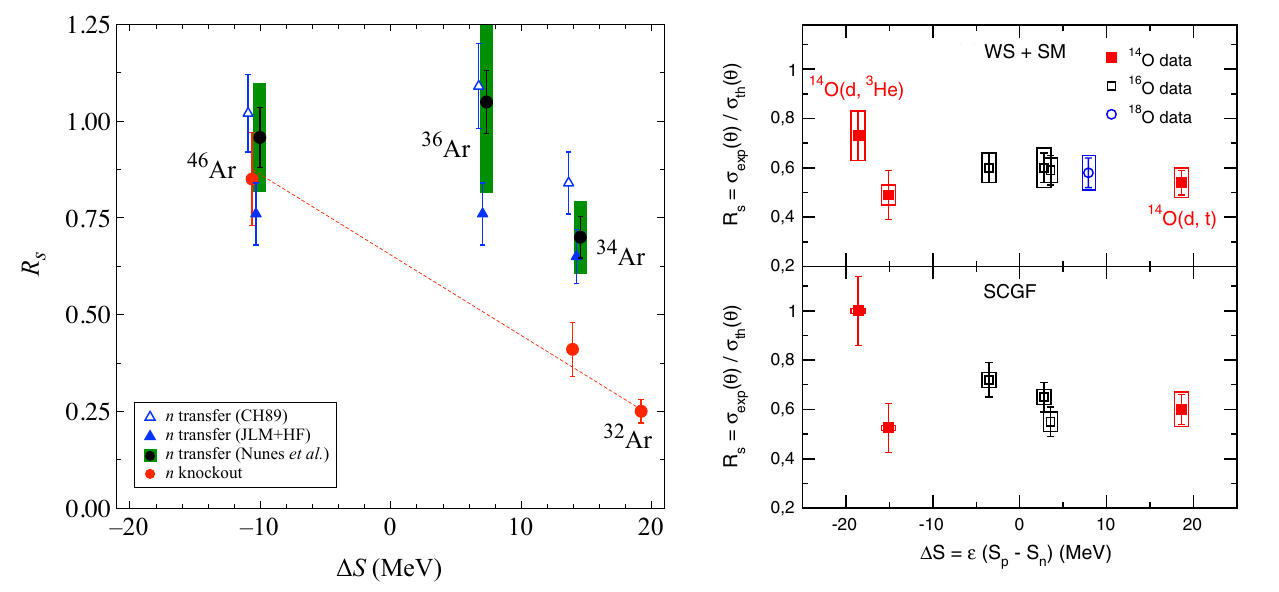}
\caption{\label{fig_rib} Transfer-reaction studies with radioactive-ion beams with large $\Delta S$. (Left) A study of the ($p$,$d$) reaction on isotopes of Ar~\cite{Lee10,nunes2011} compared to results from intermediate-energy knockout reactions and (right) a simultaneous measurement of the ($d$,$^3$He) and ($d$,$t$) reaction on neutron-deficient $^{14}$O~\cite{Flavigny13}. The top panel is obtained with overlap functions computed from a Woods-Saxon and shell-model spectroscopic factors (WS+SM), while the bottom panel considered overlap functions computed within the self-consistent Green's Function (SCGF) theory. Reprinted figure with permission from Ref.~\cite{Flavigny13} \textcopyright2013 by The American Physical Society. }
\end{figure}

There are only a small number of quantitative experimental studies of transfer reactions on nuclei with large $\Delta S$. These include a study of the $^{34,36,46}$Ar($p$,$d$)$^{33,35,45}$Ar reactions~\cite{Lee10}, directly mirroring those studied with intermediate-energy knockout, and the simultaneous measurements of the $^{14}$O($d$,$t$/$^3$He)$^{13}$O/$^{13}$N reactions~\cite{Flavigny13}. Both of these were milestone measurements in response to the observed trends seen in Ref.~\cite{gade08b}. As can be seen from Fig.~\ref{fig_all} (and the original published results are shown in Fig.~\ref{fig_rib}), these results neither firmly corroborate nor contradict the knockout data.

The work of Ref.~\cite{Lee10} studied the ($p$,$d$) reaction at an ideal beam energy of 33~MeV/u on neutron-deficient, stable ($N=Z$), and neutron-rich Ar isotopes spanning $-10\lesssim\Delta S\lesssim12$~MeV, which on the negative (weakly bound) side of the $\Delta S$ plot (Fig.~\ref{fig_all}) is similar to the stable beam data but pushes further on the deeper bound (positive) side. In a simple analysis using the ADWA model, it was not obvious that a trend was seen (Fig.~\ref{fig_rib}) to support the knockout data. Further analyses using more refined models~\cite{nunes2011} led to a similar systematics of reduction factor for the analysed three Ar isotopes (see Fig. \ref{fig_rib}) but concluded that from this data set, a slope similar to that derived from heavy-ion-induced knockout data cannot be excluded. The work of Ref.~\cite{Flavigny13} used the same incident beam, neutron-deficient $^{14}$O at 18.1 MeV/u, to determine the neutron and proton removal cross sections via the ($d$,$t$) and ($d$,$^3$He) reactions, respectively. The measurement probes a large range in $\Delta S$, comparable to that of the majority of the intermediate-energy knockout data. The simultaneous measurement allowed for an absolute comparison of two reaction channels. Using different analyses, both a traditional approach and a more detailed microscopic analysis, as shown in Fig.~\ref{fig_rib}, suggested that there was a weak or negligible dependence of reduction factor on $\Delta S$~\cite{Flavigny13}. This was corroborated by later studies, as discussed earlier~\cite{Flavigny18}.

\subsection{Conclusion}
All probes of $s.p.$ structure quantitatively agree that there is a quenching of $s.p.$ strength for nuclei with $-12\leq\Delta S\leq12$~MeV, and that the value is $\sim$40-70\% of the independent $s.p.$ model. A large body of nucleon-transfer reaction data corroborate this, and it appears to be independent of target mass, choice of transfer reaction, and angular momentum transfer. Data from transfer reactions on systems with large $\Delta S$ are very few at present. With a consistent analysis of the available data for oxygen isotopes $^{14,16,18}$O, no strong $\Delta S$ dependence of the quenching factor was found so far. The published analyses for $^{34,36,46}$Ar did not reproduce the reduction obtained from heavy-ion-induced knockout but could not conclude on the presence of absence of a significant slope with $\Delta S$ either. 

Optical potentials for light ions ($p$, $d$, $^e$He, $t$ mostly) with rare isotopes remain the main source of uncertainties in the cross section analysis. This challenge can be partly mitigated by the measurement of precision elastic scattering, accessible at least for the entrance channel, and transfer angular distribution covering small angles in the center of mass. 

Today, only few high-quality transfer data exist at large ($\geq$15 MeV) $\Delta S$. More data are called for to further assess the slope of $R_s$ with $\Delta S$ for transfer reactions. 
\section{Heavy-ion induced knockout}
\label{sec5}
\subsection{Introduction}
Direct nuclear reactions have long been used as important tools for the spectroscopy -- and, to a lesser degree the quantification -- of the proton and neutron $s.p.$ degrees of freedom in the nuclear wavefunction. On the quest to study the structural evolution of rare isotopes near the nucleon driplines, often only available for experiments as low-intensity, fast beams of ions with energies exceeding 50~MeV/nucleon, \nuc{9}{Be}- or \nuc{12}{C}-induced one-nucleon removal reactions have been introduced and developed by Gregers Hansen and the NSCL team for spectroscopy as a fast-beam, inverse-kinematics alternative to light-ion-induced transfer reactions that remove a nucleon from the nucleus of interest~\cite{navin98,aumann00,tostevin01,hansen03}.  

\begin{figure}[ht]
 \begin{center}
        \includegraphics[width=0.7\textwidth]{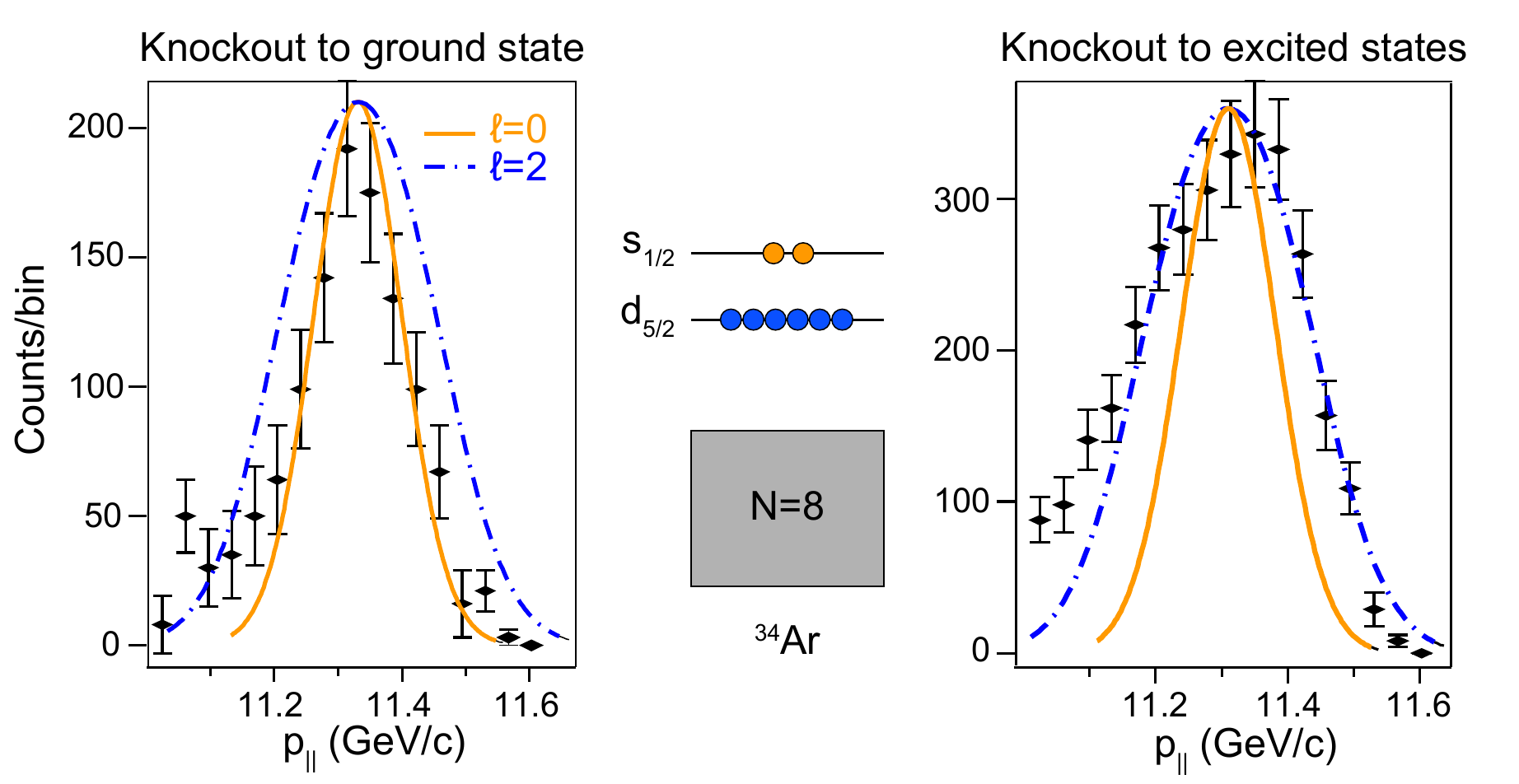}
\caption{\label{fig:knock_example}  The sensitivity to the $s.p.$ degree of freedom shown for the example of the one-neutron knockout reaction  \nuc{9}{Be}(\nuc{34}{Ar},\nuc{33}{Ar}$+\gamma$)X measured  at NSCL~\cite{gade04b}. (left) Momentum distribution for events leading to the  knockout residue \nuc{33}{Ar} in the ground state. (right) Momentum  distribution in coincidence with de-excitation $\gamma$ rays. The calculated  momentum distributions for knockout from $d$ and the $s$ orbitals  confirm the simple picture for \nuc{34}{Ar} (middle), with neutron  removal from the $s_{1/2}$ orbital leading to the ground state and the  knockout from $d$ orbitals populating excited states in \nuc{33}{Ar}. } 
  \end{center}
\end{figure}

For these one-nucleon removal reactions, the channel of interest is one where, in a single step, one proton or neutron is removed from the projectile and the projectile-like residue with $A-1$ nucleons survives. This reaction channel is  dominated by fast, surface grazing collisions of the projectile (with $A$ nucleons) and the target nucleus. From a large body of experiments performed at energies of 50-1600~MeV/nucleon at (rare-isotope) facilities around the world, it has been established that, with large cross sections, the dominant single-hole states relative to the projectile ground state are populated in the projectile-like reaction residue, demonstrating the sensitivity to the $s.p.$ degree of freedom. The residue longitudinal momentum distributions measure the momentum distribution on the nuclear surface \cite{hufner1981}, and encode in their shape and width the information of the orbital angular momentum $\ell$ and separation energy of the removed nucleon~\cite{bertulani92,bertulani04,bertulani06,hansen03}, the result of the predominant single-step direct reaction mechanism. The selectively populated single-hole state cross sections scale with the respective spectroscopic factor or wave-function overlap (see Fig.~\ref{fig:knock_example} for an example). Statistical descriptions of such a reaction will not capture these features. These characteristic features of the reaction observables have been used extensively to explore halo nuclei (e.g., see~\cite{anne90,orr92,bazin95,navin98,simon99,aumann00,cortina02,hwang17}) and details of the shell evolution towards the driplines (e.g., see~\cite{maddalena01,cortina04,terry06a,terry06b,stroberg14,stroberg15,gade16,murray19}).   

Comparisons of such one-nucleon removal data with structure model calculations have been enabled by a direct reaction model description~\cite{tostevin99,tostevin01,hansen03} that uses the sudden (short interaction time) and the spectator-core approximation to many-body eikonal (forward scattering) theory. The $s.p.$ nuclear structure information then enters the calculations through spectroscopic factors, or wave-function overlaps, that scale the calculated cross sections for the removal of one nucleon from the corresponding orbital, as in light-ion-induced transfer reactions. In 2004, the measured cross section to the $5/2^+$ ground state of \nuc{31}{Ar}, the only bound state, extracted from the \nuc{9}{Be}(\nuc{32}{Ar},\nuc{31}{Ar})X reaction performed at NSCL, was found to be only 24(3)\% of that predicted using the many-body shell-model spectroscopy~\cite{gade04a}. This case was the first example of the removal of a strongly-bound neutron ($S_n=22$~MeV) leading to a reaction product that is proton-bound by less than 400~keV, and prompted systematic studies of a large number of nuclei covering essentially all scenarios of nuclear binding realized over the nuclear chart. With remarkable consistency, a strong correlation was observed for the cross-section ratio $R_s=\sigma_{exp}/\sigma_{th}$ with $\Delta S$, where  $\sigma_{exp}$ is the measured, final-state inclusive cross section for the nucleon removal to all bound states in the residue and $\sigma_{th}$ is the calculated cross section, combining eikonal reaction dynamics and shell-model spectroscopic factors, summed over all predicted shell-model bound states~\cite{gade08b}. The quantity $\Delta S$, for one-neutron and one-proton removal is the cross-section weighted separation energy difference $S_n - S_p$ and $S_p - S_n$ for the projectile, respectively, a measure of asymmetry of the proton-neutron Fermi surfaces. $\Delta S$ is large and positive for the removal of a strongly-bound proton(neutron) from a neutron-(proton-)rich nucleus and is negative for the removal of a weakly-bound nucleon. With only four cases beyond $\Delta S \geq 14$~MeV in 2008~\cite{gade08b}, the systematics has been extended significantly~\cite{tostevin14} since, as shown in Fig.~\ref{Rsplot} in the introduction. These accumulated systematics provide a basis for a discussion of the reaction dynamics description and/or the nuclear structure implications of these observations.

The reaction model, outlined above, has been developed so as to treat all systems consistently -- but in the face of very limited experimental information on the colliding systems of interest. It thus leans heavily on available, generic theoretical inputs and, apart from the experimental beam energy and the ground-state to ground-state separation energy, uses shell-model and Hartree-Fock structure input plus the generic forward-scattering description of the nucleon-nucleon amplitudes.  Assuming the validity of the reaction dynamics description across the given range of nucleon binding, the plot suggests that the shell model captures the spectroscopic strength well for the weakly-bound valence orbitals of an excess nucleon species, overestimates the spectroscopic factors in the given energy window by $\sim 40$\% for stable nuclei that have $\Delta S \approx 0$ (in agreement with results from $(e,e'p)$ reactions), and further overestimates the spectroscopic strength for orbitals at the well-bound Fermi surface of a minority nucleon species, as can be probed in isospin-asymmetric nuclei. Near stability, based on the accumulated data from electron-induced proton knockout reactions, the low-lying shell-model configurations are thought to be depleted due to their mixing with higher-lying shells as well as through short-range and long-range correlations not captured by configuration-space-truncated effective-interaction shell-model theory. Taking this idea of missing correlations and truncated model spaces to the extended range of $\Delta S$ afforded by the one-nucleon removal reactions with rare isotopes, Fig.~\ref{Rsplot} suggests that the description of the valence orbitals of the majority nucleon species, e.g. a neutron in a neutron-rich nucleus, is subject to less correlations and the description of the minority nucleon species, e.g. a neutron in a proton-rich nucleus, is more impacted by missing correlations and limited model spaces.

Several theoretical works in the field of low-energy nuclear physics support such a trend, but not the magnitude of the observed correlation. For example, Timofeyuk argues that truncated model spaces and soft interactions can lead to an isospin asymmetry dependence of spectroscopic factors~\cite{timofeyuk09,timofeyuk11}. Early studies using the dispersive optical model potential are consistent with the idea that the valence protons are more correlated in neutron-rich  \nuc{48}{Ca} than in $N=Z$ \nuc{40}{Ca}~\cite{charity06}. Continuum effects were shown to potentially induce an isospin-asymmetry dependence for proton spectroscopic factors along the oxygen isotopic chain~\cite{jensen11}. For nuclear matter, it was shown that the proton states are more strongly depleted when the $N-Z$ asymmetry increases whereas the occupation of the neutron states is enhanced as compared to the symmetric case~\cite{frick05}. In the field of medium-energy physics, nucleon-nucleon correlations have been probed and quantified through extensive and elaborate correlation studies in electron-scattering/knockout data~\cite{hen17} and recent work shows that the fraction of high-momentum, short-range-correlated protons increases markedly with the neutron excess in the nucleus~\cite{Duer18}.

The first case \nuc{32}{Ar}~\cite{gade04a} and the subsequent extensive systematics culminating in the correlation~\cite{gade08b,tostevin14} displayed in Fig.~\ref{Rsplot} have prompted ongoing programs to benchmark the reaction model predictions using further experimental data. These efforts are the topic of the subsequent sections within this chapter.    

\subsection{Ingredients of the $R_s$ plot and their sensitivities}

In the following, we briefly explain the key ingredients to Fig.~\ref{Rsplot}: measured inclusive cross sections, $s.p.$ cross sections from reaction theory, and spectroscopic factors. 

The reader is reminded that all points included in Fig.~\ref{Rsplot} were obtained by applying the same procedure to constrain the reaction model, and to be consistent with chosen nuclear structure input (see section \ref{sec:model}).  $R_s$ is not simply the ratio of ``experimental'' and calculated spectroscopic factors. Rather, the denominator combines calculated $s.p.$ cross sections at a given beam energy from reaction theory, constrained by Skyrme Hartree-Fock calculations, with final-state energies and shell-model spectroscopic factors and thus possesses a complex model dependence. The shell-model calculations are also those appropriate for each nucleus and thus each involve different effective interactions and assumed model spaces. The broad spread of $R_s$ may be viewed as reflecting, to some extent, the combined uncertainties of the reaction dynamics and shell-model description.          

Even in the absence of a detailed understanding of these systematics, the observed correlation in Figure~\ref{Rsplot} can and has been used to extract spectroscopic factors from one-nucleon removal data~\cite{mutschler16,mutschler17,crawford17}. After normalization, the relative spectroscopic factors were found to be in agreement with transfer analyses when data are available, see for example the work presented in~\cite{mutschler16}, using a ($d$,\nuc{3}{He}) transfer reaction.    

\subsubsection{Measured inclusive cross sections}
For one-nucleon removal reactions, the secondary projectile beams are produced in-flight via fragmentation (or fission) and are magnetically separated in fragment separators. The mass $A$ projectiles are then identified and tracked event by event through a variety of detector systems. Low-$Z$ targets, such as Be and C, are preferred to induce nucleon removal reactions at the experimental end station to ensure that effects of the Coulomb interaction are minimized and the projectile-target interaction is mediated by the strong interaction. The projectile-like reaction residue with $A-1$ nucleons is typically identified on an event-by-event basis and characterized using magnetic spectrometers or particle detection systems. A critical quantity to measure for the analysis of nucleon removal reactions is the longitudinal (or parallel) momentum distribution of the residues. Transverse (or perpendicular) momentum distributions can also be useful but have greater sensitivity to Coulomb and diffractive scattering processes. The collection efficiency of the projectile-like residues of interest is often close to 100\%, owing to the forward-focused kinematics at the high beam energies involved. For reactions of light nuclei, several momentum settings might have to be run in a limited-acceptance spectrometer setup, see for example~\cite{grinyer11}.

\begin{figure*}[ht]
 \begin{center}
        \includegraphics[width=0.9\textwidth]{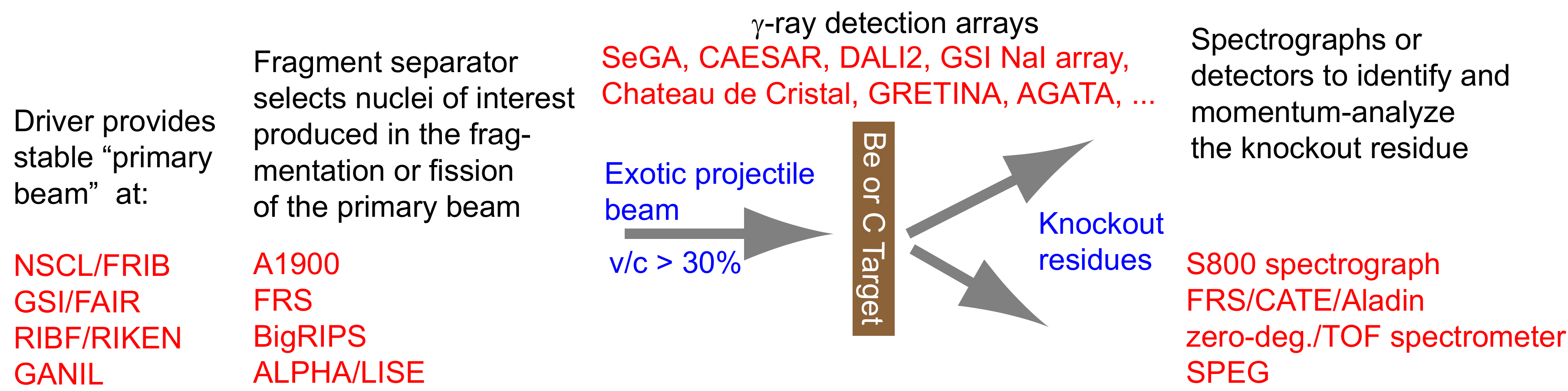}
\caption{\label{fig:scheme} Experimental scheme for Be- and C-induced one-nucleon removal reactions at rare-isotope facilities around the world.}
  \end{center}
\end{figure*}

To tag the bound final states populated in the reaction residue, $\gamma$-ray detection is typically employed, in coincidence with the unambiguously identified one-nucleon removal residues. This allows for the use of thick targets that restore luminosity in measurements near the driplines with beam intensities of only a few particles per second. Since in this scheme the $\gamma$-ray emission occurs in-flight, Doppler reconstruction is required and mandates detection arrays with position sensitivity, for example, through granularity. In this scheme of $\gamma$-ray tagging, unobserved feeding transitions can hamper the extraction of final-state exclusive partial cross sections, in particular for low-lying excited states and the ground state. Figure~\ref{fig:scheme} sketches the experimental scheme as pursued at rare-isotope facilities around the world.

\subsubsection{Theoretical cross sections combining eikonal reaction theory and shell model}
\label{sec:model}

Nucleon removal as described here requires that the projectile-like residue -- the core -- survives the reaction. This requirement of core survival together with the highly absorptive character of the core-target interaction localizes the one-nucleon removal events at the nuclear surface~\cite{hansen03}. In $p$-induced knockout, on the other hand, the greater mean free path and higher energies require a greater understanding of the $NN$ interaction within the nuclear medium, as discussed in section \ref{sec6}.  

The eikonal model description of the Be- and C-induced one-nucleon removal reaction dynamics uses the sudden (short interaction time) and eikonal (forward scattering) approximations. Additionally, the model assumes that in the fast, surface-dominated interaction of the mass $A$ projectile with the target, the (mass $A-1$) core state component $I^{\pi}$ in the projectile wavefunction is a spectator and is unchanged during the collision.

\paragraph{Brief summary of the reaction model and main characteristics} The direct reaction model that leads to the systematics shown in Fig.~\ref{Rsplot} is described in detail elsewhere~\cite{tostevin99,tostevin01,hansen03} and in Annex \ref{eikonal}. The main characteristics are sketched here to provide context to our subsequent summaries of sensitivity studies and experimental benchmarks of some of the assumptions. 

The theoretical cross section in one-nucleon removal from a $s.p.$ orbit with quantum numbers $n,\ell,j$ populating a final state $I^{\pi}$ in the residue, $\sigma(I^{\pi})$, factorizes into the $s.p.$ cross section $\sigma_{sp}$, containing the reaction dynamics for the nucleon removal from that orbit, and the spectroscopic factor, $C^2S$:  

\begin{equation}
\label{eq:knock_xsec}
\sigma_{th}(I^{\pi})=\sum_j C^2S(j,I^{\pi})\sigma_{sp}(j,S_N+E_x(I^{\pi})),
\end{equation} 
with summation over the allowed angular-momentum transfers $j$. The sum $S_N+E_x(I^{\pi})$ represents the effective separation energy of the removed nucleon, where $S_N$ is the nucleon's separation energy from the ground state of the projectile, and $E_x(I^{\pi})$ denotes the excitation energy of the final state of the residue. We note that the same factorization is assumed for transfer reactions reactions (see section \ref{sec3}).

For the interaction with Be and C targets discussed here, inelastic and elastic processes -- stripping and diffraction -- contribute to the $s.p.$ cross section $\sigma_{sp}$. Stripping, or inelastic breakup, refers to reactions where the removed nucleon interacts inelastically with the target (involving the absorptive part of the nucleon-target optical interaction). Diffraction, or elastic breakup, describes the dissociation of the projectile into the nucleon and core through their two-body interactions with the target nucleus. In this latter process, each residue is at most elastically scattered and the target nucleus remains in its ground state. If only the residue final state is measured, then the $s.p.$ cross section is the sum of the inelastic {(\it inel)} and elastic {\it (el)} contributions: 

\begin{equation}
\sigma_{sp}=\sigma_{sp}^{inel}+\sigma_{sp}^{el}. 
\end{equation}

The cross sections are calculated in a spectator-core approximation (see, e.g., Refs. \cite{Hussein:1985,hencken96,tostevin01}). The corrections to the ground-state term,
due to interactions with the target that excite bound excited states of
the projectile rather than leading to breakup, have been estimated in the
case of the weakly-bound \nuc{12}{Be}(-n) reaction~\cite{tostevin01} (and also the well-bound \nuc{28}{Mg}(-p) reaction~\cite{wimmer14}) for which the diffraction dissociation terms are a more significant component of $\sigma_{sp}$. The corrections to $\sigma_{sp}$ for \nuc{12}{Be}(-n) leading to the \nuc{11}{Be}($1/2^+$) and \nuc{11}{Be}($1/2^-$) halo-like states were 5 and 3\%, respectively. These small, excited bound-state contributions are usually neglected. In the inelastic breakup or stripping contribution, the spectator-core approximation entails that, while the removed nucleon is absorbed by the target and excites it, the core scatters elastically.     

The stripping and elastic breakup contributions are calculated independently from the  target-core and target-nucleon $S$ matrices which are presented as a function of the corresponding impact parameter. The $S$ matrices are computed in the optical limit of Glauber's eikonal multiple-scattering theory~\cite{glauber59}. This treatment of the nucleon, core, and target three-body system incorporates the effects of the breakup of the projectile to all orders. A study of non-eikonal corrections to the eikonal $S$ matrices at 50~MeV/nucleon, the lower limit of removal reaction analyses, are presented in~\cite{brooke99}.

The inelastic (stripping) and elastic (diffraction) contributions are given as~\cite{hansen03}:
 \begin{equation}
     \sigma^{inel}_{sp} =\frac{1}{2j+1}\int d\vec{b}\sum_{m} \langle \psi_{jm}|(1-|S_n|^2)|S_c|^2|\psi_{jm}\rangle,
      \label{strip1}
     \end{equation}
     and
     \begin{equation}
      \label{dis1}
\sigma^{el}_{sp} = \frac{1}{2j+1}\int d\vec{b}\sum_{m,m'} [\langle \psi_{jm}||1-S|^2|\psi_{jm}\rangle \delta_{m,m'}-|\langle\psi_{jm'}|(1-S)|\psi_{jm} \rangle|^2],
 \end{equation}
 with $S=S_cS_n$ as the product of the elastic $S$ matrices for the residue–target ($S_c$) and removed-nucleon–target ($S_n$) systems. The formulation of $\sigma_{sp}^{inel}$ is rather descriptively given by the combined probability that a nucleon is removed, $1-|S_n|^2$, and that the core (residue) survives, $|S_c|^2$, while $\sigma_{sp}^{el}$ is less intuitive.   
 
\paragraph{Consistent treatment of the reaction formalism and associated sensitivities} 

A crucial aspect of the accumulated cross-section comparison plot (Fig,~\ref{Rsplot}) is the consistent use of inputs to the reaction model in all cases. The details, reasoning, and main emerging sensitivities are based on work presented in Refs.~\cite{tostevin99,tostevin01,hansen03,gade04a, gade04b,gade08b} and are summarized below.  

Essential parameters used to compute the $S$ matrices are the effective nucleon-nucleon interaction, the assumed core and target matter distributions, and their root-mean-squared ($rms$) radii. The effective interactions are built from empirical free nucleon-nucleon interactions~\cite{charagi90}. The real-to-imaginary ratios of the forward-scattering amplitudes are interpolated from tabulations in Ray~\cite{ray79}. For energies below 300~MeV/nucleon, the interaction is assumed to be Gaussian with a 0.5-fm range~\cite{tostevin99}. This approach reproduces measured reaction cross sections for \nuc{12}{C}+\nuc{27}{Al} at 83~MeV/nucleon~\cite{hansen03,kox87} and for $p$+\nuc{9}{Be} at 60~MeV/nucleon~\cite{hansen03,renberg71}, for example. The use of other empirical  $NN$ interactions and parameterizations was tested over the course of the work presented in~\cite{bertulani06} and found to have minimal impact. 

For completeness, the $S$ matrices are computed in eikonal approximation via 
\begin{eqnarray*}
S(b) &=& \exp{[i\chi(b)]},~\mathrm{with}\\
\chi(b) &=& \frac{1}{k_{NN}}\int_0^{\infty}dq q\rho_p(q)\rho_t(q)f_{NN}(q)J_0(qb)~\mathrm{and}  \\
f_{NN}(q)&=&\frac{k_{NN}}{4\pi}\sigma_{NN}(i+\alpha_{NN})\exp{(-\beta_{NN}q^2)},
\end{eqnarray*}
where $f_{NN}(q)$ is the eikonal nucleon-nucleon scattering amplitude in Ray parameterization~\cite{ray79} and $\rho_{p,t}(q)$ are the Fourier transforms of the nuclear densities of the projectile and the target.

To account for weak-binding effects such as nucleon skins, the $S$ matrix of the core is calculated in the double-folding optical limit where the proton and neutron matter distributions are taken from Skyrme Hartree-Fock (HF) calculations using the SkX parameterization~\cite{brown98}. The choice of this particular Skyrme force is reasoned below. It is worth noting that, as the majority of data sets used in constructing the systematics in Fig.~\ref{Rsplot} involve a Be target and a somewhat limited range of (NSCL) beam energies, the absorptive nucleon-target $S$-matrix, that has a major role in determining the dominant stripping component of the removal cross section, is essentially fixed for the majority of the data points shown, across all values of $\Delta S$.  

\begin{figure}[ht]
 \begin{center}
\includegraphics[width=0.95\textwidth]{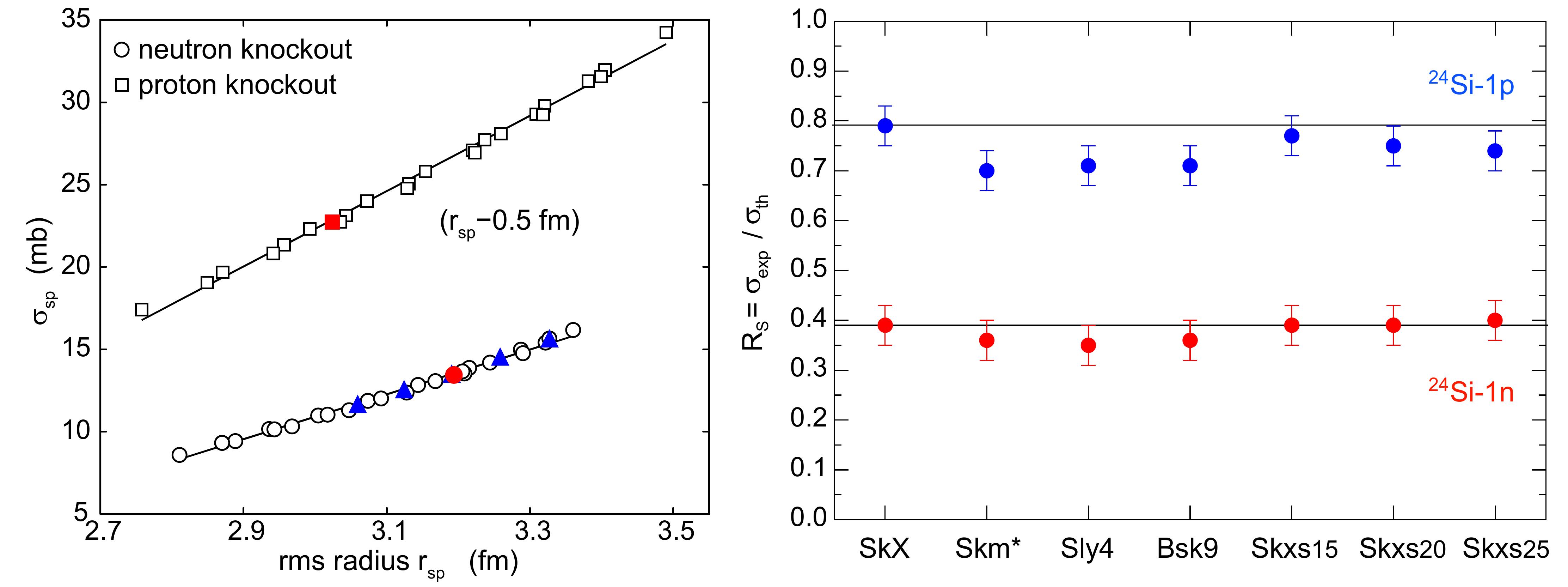}
\caption{\label{fig:sensitive} (left) Single-particle cross sections to the \nuc{23}{Si} and \nuc{23}{Al} ground states, assuming $d_{5/2}$ neutron- (circles) and $d_{5/2}$ proton-removal (squares) from \nuc{24}{Si} at 85.3 MeV/nucleon. Each point assumes a different WS radius $r_0$ and diffuseness $a_0$ parameters for the potential used to calculate the nucleon bound-state wavefunction. The $r_{sp}$ values used in the physical calculations are shown by the filled (red) symbols. Cross sections indicated by blue triangles also include a non-locality $\beta \neq 0$. The proton orbital radii have been offset to smaller values ($r_{sp} - 0.5$~fm) for display purposes. (right) Deduced values of $R_s$ for the reactions \nuc{9}{Be}(\nuc{24}{Si},\nuc{23}{Al})X and \nuc{9}{Be}(\nuc{24}{Si},\nuc{23}{Si})X as obtained using different Skyrme parametrizations as input to the HF calculations used for the reaction methodology. Adapted figure with permission from Ref.~\cite{gade08b}\textcopyright2008 by the American Physical Society.} 
\end{center}
\end{figure}

The nucleon-core relative wavefunctions are derived as eigenstates of an effective two-body Hamiltonian based on a local potential with its depth adjusted to reproduce the effective separation energy of the nucleon from the projectile. Calculations are carried out using a Woods-Saxon geometry. Here again, SkX Skyrme HF calculations are employed to obtain the $rms$ separation, $r_{sp}$, of the nucleon and the core nucleus. The Woods-Saxon (WS) radius parameter, $r_0$, at a fixed diffuseness of $a_0=0.7$~fm is then adjusted to reproduce the $r_{sp}$ predicted from the SkX Skyrme HF calculations. This procedure provides a consistent approach to constrain the geometry of the potential~\cite{gade04a,gade04b}. 

Indeed, in an extensive sensitivity study~\cite{gade08b}, one of the $s.p.$ cross section's main sensitivities was identified to be the $rms$ separation $r_{sp}$ of the nucleon and the core. This is reproduced in Fig.~\ref{fig:sensitive} (left), where each point assumes a different WS radius parameter $r_0$ and diffuseness $a_0$ for the potential used to calculate the nucleon bound-state wavefunction (the case of non-locality is examined as well). This scaling demonstrates that the shape of the potential and non-localities do not impact the calculation of the $s.p.$ cross section, only the resulting value of $r_{sp}$ does. This 2008 analysis underlined the need for a highly consistent treatment of input to the reaction calculations related to nuclear sizes that is not presently available from experiment.     
The SkX Skyrme parameterization~\cite{brown98} was chosen as it was obtained  from a fit to binding energies, $rms$ charge radii, and $s.p.$ energies of both `normal' and exotic spherical nuclei. The SkX Skyrme was shown to reproduce extremely well (i) measured interaction cross sections for a large number isotopes from He to Mg, including exotic ones (see Fig.~3 in Ref.~\cite{brown01}) and (ii) measured charge densities of stable nuclei from Si to Pb (see Fig.~4 of Ref.~\cite{richter03}). The choice of the Skyrme functional was investigated in Ref.~\cite{gade08b}, reproduced in Fig.~\ref{fig:sensitive} (right). As is obvious from the figure, the choice of the particular Skyrme parametrization has little impact on the cross section ratio $R_s$. We remind that the SkX Skyrme HF calculations enter in two places of the model calculations, through the proton and neutron densities used for the double-folding in the construction of the core $S$ matrix and via the constraint of the WS potential geometry to $r_{sp}$ from the HF calculation for the bound-state wavefunction. The consistent choice of these two sizes, the radial extent of the absorption in the optical potentials, that locates the nuclear surface and the localization of the reaction, and the radial extent of the occupied valence orbitals of the removed nucleon is a vital ingredient of the model calculations.

It is noted that deformation effects and Pauli blocking are not explicitly included in the formalism. Both effects appear to not play a major role in the explanation of the $R_s$ systematics as the effect of Pauli blocking is predicted to impact the calculated cross sections by 10-15\% at 50-100 MeV/u incident energy~\cite{flavigny09,bertulani10} and the $R_s$ systematics already includes spherical and deformed nuclei without discernible difference within the spread of $R_s$. We remind that methods have been developed to account for the deformation of projectiles in the description of knockout reactions~\cite{batham05,sakharuk99}.  

\paragraph{Configuration-interaction shell-model spectroscopic factors}
The spectroscopic factors for individual final states are taken from the best available, large-scale shell-model calculations, whenever possible.

Given the spectator-core approximation, the nuclear structure enters the computation of the partial cross section to a given final state via the spectroscopic factor $C^2S$. The spectroscopic factor expresses the parentage of the initial (ground) state in  the projectile with $A$ nucleons with respect to a specific final state in the mass $A-1$ residue coupled to a nucleon with quantum numbers ($\ell,j$)~\cite{brown02b}. The notation $C^2S$ indicates that the isospin coupling coefficient has been taken into account. In a sum-rule limit, the sum of the spectroscopic factors to all final states reached by the removal from a given $s.p.$ orbital is the average occupancy of this orbital. When the spectroscopic factor is taken from many-body shell-model theory in a harmonic oscillator basis, a multiplicative, $A$-dependent  center-of-mass correction of magnitude  $[A/(A-1)]^n$, with the major oscillator quantum number $n$, has to be applied~\cite{dieperink74} (e.g., $n=2$ and $n=3$ for the $sd$ and $fp$ shell, respectively). The inclusive theoretical cross section that enters $R_s$ is thus: 
\begin{equation}
\label{eq:theo_xsec}
\sigma_{th}^{inc}=\sum_{E_x(I^{\pi}) \leq min(S_p,S_n)} \sum_j \left(\frac{A}{A-1}\right)^nC^2S(I^{\pi},j)\sigma_{sp}(j,S_n+E_x(I^{\pi}))=\sum_{E_x(I^{\pi}) \leq min(S_p,S_n)} \sigma_{th}(I^{\pi}).
\end{equation}

We note that the use of inclusive cross sections in the definition of $R_s$ (i) avoids the experimental complication of unobserved feeding and overestimated partial cross sections inherent to measurements that rely on $\gamma$-ray tagging and (ii) probes the shell-model description of the summed spectroscopic strength within the given energy window rather than demanding that the spectroscopic factor of an individual final state is correct, in light of the fragmentation of spectroscopic strength. 

\subsection{Probing the reaction model}
\subsubsection{Stripping vs. diffraction}
\label{sec:mechanisms}
\begin{figure}[ht]
   \centering
    \includegraphics[width=10cm]{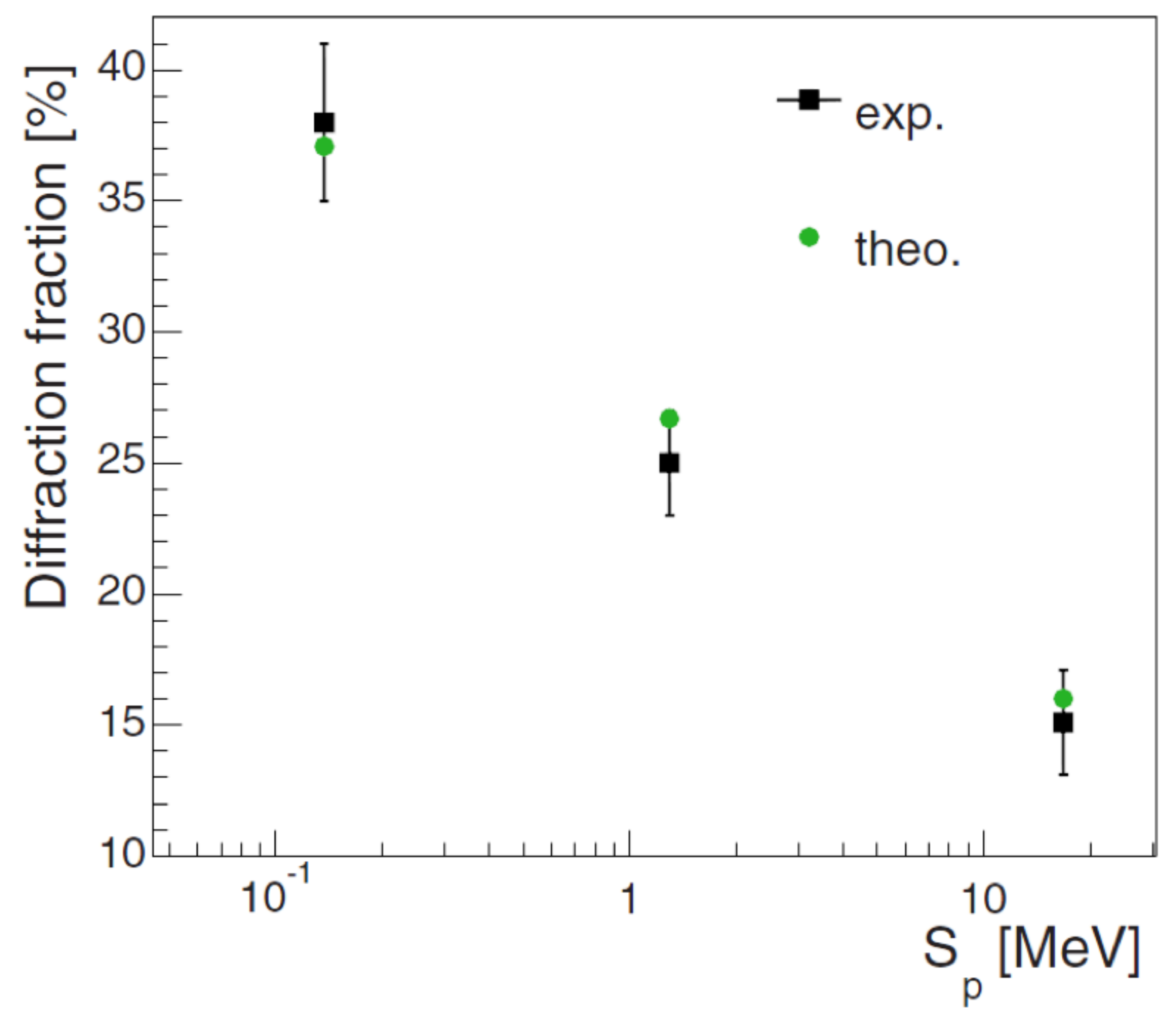}
   \caption{Fraction of elastic removal reaction mechanism in proton knockout reactions on nuclei with various proton binding energies. The approximately logarithmic dependence of the elastic relative cross section is very well reproduced by the eikonal model.Reprinted figure with permission from Ref.~\cite{wimmer14}\textcopyright2014 by The American Physical Society.}
    \label{fig:diffraction}
\end{figure}

The eikonal model presented in section \ref{sec:model} differentiates the reaction mechanism between two distinct processes: (i) elastic removal where core and nucleon interactions with the target dissociate the projectile, the core and target nuclei remaining in initial states, and (ii) inelastic removal where the nucleon is removed as a result of a strong, inelastic collision with the target. Exclusive experiments have been carried out to test and quantify this hypothesis \cite{bazin09}. They rely on the simultaneous detection of both the projectile residue and removed nucleon. The proportion of elastic removal cross section varies significantly with the binding energy of the removed nucleon, therefore, cases with weakly- and strongly-bound nucleons were examined. Figure \ref{fig:diffraction} shows the measured and calculated fractions of elastic removal for proton knockout from $^8$B, $^9$C and $^{28}$Mg with separation energies ranging from 137 keV to 17 MeV \cite{wimmer14}.

Remarkably, the elastic removal signature is still very clear for the deeply-bound valence proton in \nuc{28}{Mg}, and the assumption of the two distinct reaction mechanisms is confirmed experimentally, both qualitatively and quantitatively over two orders of magnitude of separation energy. A similar study performed at lower incident energy (36 MeV/nucleon instead of 80 MeV/nucleon) showed the same result, for the weakly-bound valence proton of \nuc{8}{B} \cite{jin15}. At even lower incident energy, the description of the nucleon removal by the sudden and eikonal approximations might fail because of other reaction mechanisms, such as the towing mode described in \cite{lacroix1999}. Section \ref{sec:kinematic} further explores the limitations of the use of the eikonal approximation for low incident beam energies.

\subsubsection{Core survival and core excitations}
\label{sec:core}
\begin{figure}[ht]
 \begin{center}
        \includegraphics[width=0.7\textwidth]{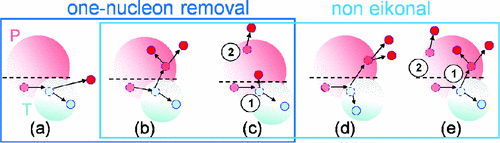}
\caption{\label{fig:INCL} Schematic depiction of events that may contribute to the one-nucleon removal cross section from the projectile [(a) - (c)] and processes that destroy the core and are beyond what is included in the three-body reaction calculations of the core survival. Reprinted figure with permission from Ref.~\cite{louchart11}\textcopyright2011 by The American Physical Society.}  
  \end{center}
\end{figure}

One speculation as to the origin of the small $R_s$ reported for the removal of strongly-bound nucleons is that the effective three-body model reaction treatment overestimates the core survival probability due to the omission of statistical, cascade-model-like processes such as indicated in Fig.~\ref{fig:INCL} ~\cite{louchart11}. Such considerations are naturally most important for weakly-bound cores. Higher-order processes as depicted in (d) and (e), in which the struck nucleon then interacts with and subsequently destroys the $A-1$ core, are not included in a three-body model. Experimental data on such cross sections are not available yet. A first attempt to provide a measure of the core breakup in the one-neutron removal from \nuc{14}{O} was performed~\cite{sun16}. Although only an upper limit could be extracted due to low statistics, the agreement of measured cross sections with intranuclear cascade estimates indicates that experimental work towards quantifying the core breakup is worth pursuing in the future. The higher-order processes illustrated in (b) and (c), whereby the initially struck nucleon then interacts with and excites the core, resulting in the decay of a different nucleon from the $A$ body system, and with a loss of information on the struck nucleon orbital, are also outside of the a three-body model description. Their presence, with significant cross sections, would be visible in the large body of one-nucleon removal data through (i) the observation of core excitations, manifested, for example, in the population of complex-structure final states that can only be reconciled with the knockout from an excited state, for example, and, similarly, (ii) the population of unexpected final states in the reaction residue that do not reflect the $s.p.$ degree-of-freedom selectivity of a direct one-nucleon removal reaction. Evidence for core excitations on the few-\% level relative to the inclusive cross section has indeed been found and will be discussed below. However, for the vast majority of reported measurements, a dominant fraction of the longitudinal momentum distribution, the center and high-momentum part excluding the often-observed low-momentum tail, is described extremely well by the eikonal model calculations, indicating that statistical contributions as depicted in (b) and (c) of Fig.~\ref{fig:INCL} are small. This is supported by the fact that, except for the occasional observation of weakly-populated complex-structure states, the Be- and C-induced one-nucleon removal reactions populate, with large cross sections, single-hole states relative to the projectile initial state.          

An often-seen deviation from the eikonal model description is a low-momentum tail in intermediate-energy one-nucleon removal data (see~\cite{gade05} for one of the most extreme cases). This observation has been attributed to the energy-sharing and dissipative processes occurring in the interaction of the projectile and target. As argued in Ref.~\cite{gade05}, the eikonal-model integrated partial cross sections from the (dominant) stripping term express the joint probability that the nucleon excites the target while the residue, at most, scatters elastically. Through completeness, the expression is inclusive with respect to nucleon-residue relative-motion final states. In contrast, the differential cross section with respect to momentum~\cite{hencken96} is obtained by projection onto a complete set of assumed plane-wave states and involves a Fourier-like transform of the overlap function, modified by the spatial localization produced by the absorptive interactions. The model calculations of the momentum distribution treat only approximately the kinematical and excitation energy sharing in the three-body final states of the residue, target, and nucleon~\cite{gade05}. 

\begin{figure*}[ht]
 \begin{center}
 \includegraphics[width=1\textwidth]{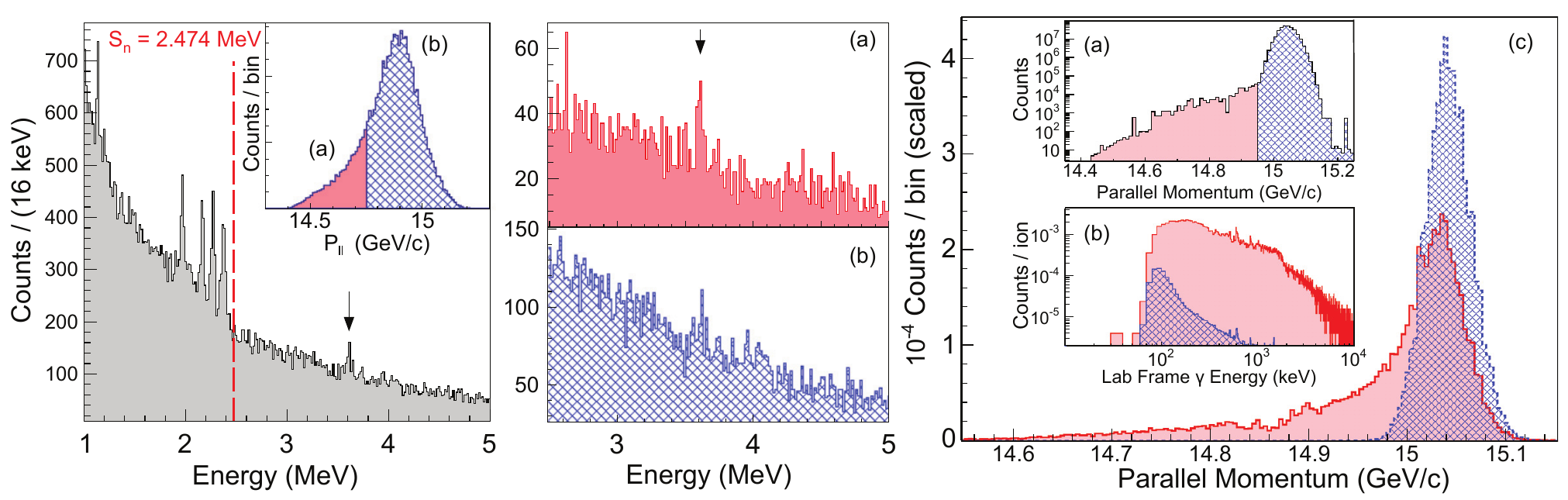}
\caption{\label{fig:core_excitation} (left and middle) The left panel shows the 3611-keV $\gamma$-ray peak detected in \nuc{35}{Si} in one-neutron knockout, with the inset indicating the gates used on the outgoing parallel momentum distribution. The middle panel shows the $\gamma$-ray spectrum gated on the main peak in the momentum distribution (blue hatches) and the low-momentum tail (solid red). The 3611-keV peak seems clearly associated with the tail of the distribution. (right) Exploration of the momentum distribution of (in)elastically scattered \nuc{36}{Si} projectiles. Violent events that may destroy the target nucleus with the debris causing a prompt, high-energy $\gamma$-ray background, is clearly associated with the low-momentum tail of the \nuc{36}{Si} momentum distribution. Figure adapted with permission from Ref.~\cite{stroberg14}\textcopyright2014 by the American Physical Society.}  
  \end{center}
\end{figure*}

Increasing experimental sensitivity has continued to allow measurements of smaller and smaller partial cross sections. The population of higher-spin core-coupled complex-structure states in removal residues was reported for \nuc{35}{Si}, \nuc{35}{P} and the $A=55$ Co and Ni mirrors in Refs.~\cite{stroberg14,mutschler16,spieker19}. While concrete experimental evidence as to the origin of the population of these states is lacking, all of the observed states could, in principle, be reconciled with a picture where a nucleon is removed from the projectile inelastically excited to its first $2^+$ state, for example. Down-shifted momentum distributions in coincidence with transitions from the higher-spin states were reported~\cite{flavigny12,mutschler16,spieker19}, indicating that they occur predominantly in dissipative collisions between projectile and target. Such events may indicate two-step processes involving inelastic excitation. The case of the population of a presumed higher-spin neutron-unbound state in \nuc{35}{Si} observed in the  one-proton removal from \nuc{36}{Si} was examined in more detail, as shown in Fig.~\ref{fig:core_excitation} (left and middle). For the momentum distribution of the inelastically scattered projectiles (see Fig.~\ref{fig:core_excitation} (right)) an observed tail (a) was found to be associated with a large $\gamma$-ray background (b), indicative of the target breaking apart and causing prompt $\gamma$-ray background in the interaction of the debris with surrounding material. Conversely, high-energy $\gamma$-ray background (above 1~MeV), was established to be predominantly associated with the low-momentum tail of the (in)elastic-scattering momentum distribution (c). Taken together, a picture emerges in which two-step processes involving knockout from inelastically excited states may explain the few-\%  population of higher-spin complex-structure states that have been associated with momentum distributions shifted to lower momenta.                    

For \nuc{35}{Si}~\cite{stroberg14} and \nuc{35}{P}~\cite{mutschler16}, the cross section to complex-structure states that may be indicative of two-step or higher-order processes is on the 1\% and 5\% level, respectively. Spieker {\it et al.},~\cite{spieker19} report a higher fraction but the data suffer from the possibility of significant unobserved feeding to the rather low-lying states in question. Complex-structure states were also reported in coincidence with the low-momentum tail for the one-proton removal from \nuc{16}{C} at lower beam energy~\cite{flavigny12}. Although the effect of higher-order contributions appears rather minor in the cases studied, core excitations ending up in the particle continuum and leading to the speculated decrease in core survival should be investigated in future experiments.    

\subsubsection{Kinematic limit for the removal of deeply-bound nucleons}
\label{sec:kinematic}
\begin{figure}[ht]
 \begin{center}
 \includegraphics[width=0.6\textwidth]{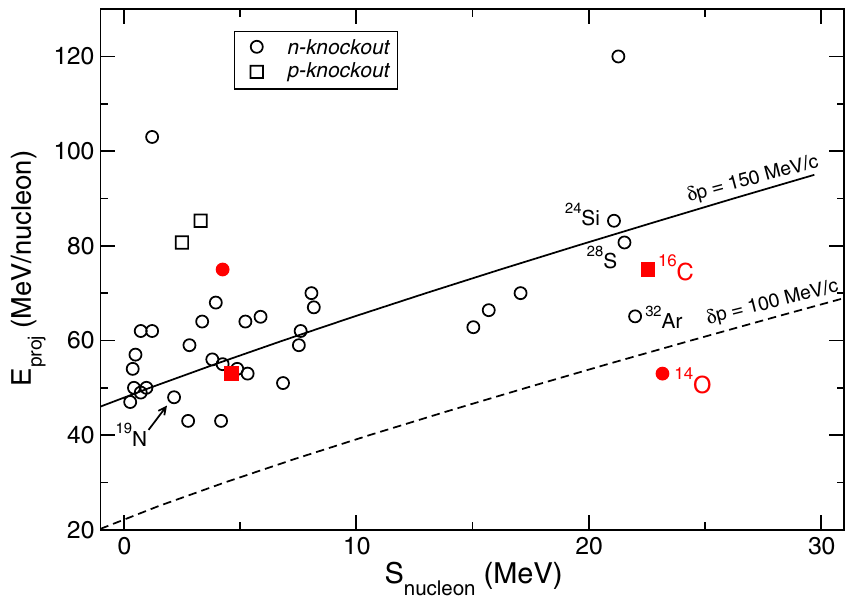}
\caption{\label{fig:kinematic} Kinematic cutoff at 100 MeV/c (dashed) and 150 MeV/c (full) plotted as a function of incident energy and binding energy of the removed nucleon. Depending of the width of the residue's momentum distribution, reactions performed below the cutoff lines are subject to a kinematical cut that lowers the measured cross section. Several reactions from performed experiments are indicated (red filled symbols), including those from the figure's original publication~\cite{flavigny12}. Reprinted with permission from Ref.~\cite{flavigny12}\textcopyright2012 by the American Physical Society.}  
\end{center}
\end{figure}

The limits at which the knockout picture of the removal reaction breaks down were explored experimentally by removing deeply-bound nucleons from projectiles at lower energy than the usual 100 MeV/nucleon or so customarily used. The removal of a proton from \nuc{16}{C} and a neutron from \nuc{14}{O} at energies of 75 MeV/nucleon and 53 MeV/nucleon, respectively, clearly exhibits a cutoff in the upper parts of their respective residues momentum distributions \cite{flavigny12}. This cutoff is easily explained from the kinematical limit for removing the valence proton and neutron in these projectiles, that are bound by 22.5 and 23.2 MeV, respectively, as reproduced in Fig.~\ref{fig:kinematic}. It can be noted in Fig.~\ref{fig:kinematic} that the cutoff depends on both the binding energy of the removed nucleon and the incident beam energy. 

The effect of the cutoff is to lower the measured cross section, thus  a reaction model not-taking into account this kinematical cutoff would get larger cross sections and smaller reduction factors than a model containing the kinematical cutoff. However, this kinematical limit is unable to explain the trend of low reduction factors observed in knockout reaction, as reactions performed well above the limit fall well within the systematics. Two examples were recently measured at energies above 200 MeV/nucleon: proton knockout from $^{30}$Ne at 230 MeV/nucleon \cite{lee16} and from $^{33}$Na at 221 MeV/nucleon \cite{murray19}. The measured cross sections from these two experiments were compared to calculated ones following the same procedure outlined in Sec.~\ref{sec:model} and have been added to the recent update of the quenching plot (Fig.~\ref{Rsplot}). While these nuclei present challenges to the nuclear structure description since they are located in the $N=20$ island of inversion, a very recent measurement of one-proton knockout from \nuc{16}{C} at 240~MeV/u reports a value for $R_s$ consistent with the reduction observed in the same reaction at 75~MeV/u~\cite{zhao19}. \nuc{16}{C} was shown to exhibit no exotic behavior that challenges shell model~\cite{wuosmaa10}. The fact that so far all of these measurements follow the systematics obtained from lower incident energies indicates that the kinematical limit is likely not a predominant cause for the large quenching observed for removal of deeply-bound nucleons. Systematic studies of the energy dependence for a set of nucleon removal reactions from projectiles whose structure is well described by shell model would be useful to solidify the above, initial conclusions. 

Moreover, reduction factors deduced from neutron knockout from various carbon isotopes on a \nuc{9}{Be} target at energies well above 1 GeV/nucleon also fall well into the systematics established at lower energies \cite{volkov11}. These observations assert the robustness of the approximations used in the eikonal reaction model, especially when taking care of avoiding the kinematical limit set by the separation energy of the removed valence nucleon.


\subsection{Putting structure and reactions onto a more similar footing}

A long standing issue in using nuclear reactions to probe nuclear structure properties is the disconnect between reaction and structure models. Reaction models tend to use global description of nuclei such as Hartree-Fock densities, with forces fitted to a large sample of (mostly) stable nuclei, whereas wavefunctions used to describe the static properties of the reacting nuclei are often based on locally adjusted interactions. This unfortunate situation is starting to change, with the introduction of new methods such as ab-initio Monte-Carlo, no-core shell model, or dispersive optical model (see section \ref{sec2}). In the case of the knockout reactions discussed here, an ongoing effort is being pursued to investigate the origin of the observed quenching with experimental data on $p$-shell nuclei for which microscopic wavefunctions from Variational Monte-Carlo (VMC) as well as No-Core Shell Model (NCSM) calculations are now available.

\subsubsection{Incorporating ab-initio densities and overlaps into reaction model}
\label{sec:ab-initio}
These new ab-initio microscopic description of nuclei provide more accurate models of the densities and overlaps that are used in the reaction model framework. They can be incorporated on a case-by-case basis, by fitting them with a Wood-Saxon line-shape for instance, and use the resulting shapes in the eikonal calculations. 
\begin{figure}[ht!]
    \centering
    \includegraphics[width=0.55\textwidth]{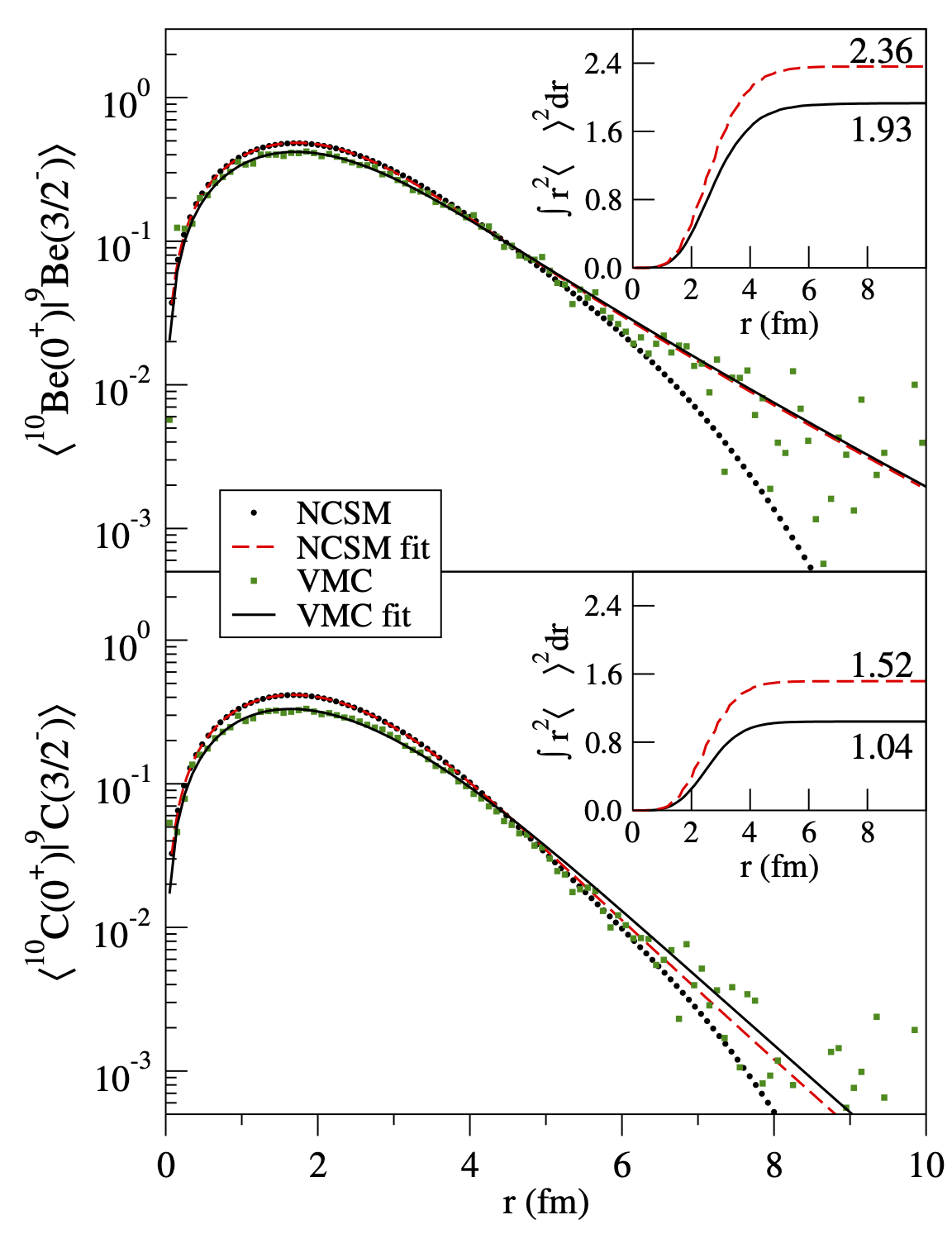}
    \caption{Example of fitted Wood-Saxon line-shapes on VMC and NCSM overlaps for the neutron removal from \nuc{10}{Be} and \nuc{10}{C}. The insets show the integrations of the overlaps that converge to the spectroscopic factors. Reprinted figure with permission from Ref.~\cite{grinyer11}\textcopyright2011 by The American Physical Society.}
    \label{fig:overlaps}
\end{figure}

An example of such a procedure is shown in Fig. \ref{fig:overlaps} for the cases of the one-neutron knockout from \nuc{10}{Be} and \nuc{10}{C} \cite{grinyer11}. The core densities calculated from these ab-initio methods are also directly used in the eikonal calculations to compute the $S$ matrices relevant for the reaction. This self-consistent method is a lot more robust and can actually be used to probe the effects of the reaction model on the observed strength reduction in knockout reactions.

\subsubsection{Ab-initio $p$-shell calculations at face value: \nuc{10}{Be} and \nuc{10}{C}}
The neutron knockout reactions from \nuc{10}{Be} and \nuc{10}{C} were selected as a first attempt to benchmark the eikonal reaction model, because (i) their ab-initio wavefunctions as well as that of their residues \nuc{9}{Be} and \nuc{9}{C} are available, (ii) they have only one final state since both \nuc{9}{Be} and \nuc{9}{C} are only bound in their ground states, therefore only the inclusive cross section needed to be measured, and (iii) the binding energy of their valence neutrons differs greatly, from 6.8 MeV to 21.3 MeV for \nuc{10}{Be} and \nuc{10}{C}, respectively. The corresponding values of the asymmetry energy $\Delta S$ are -12.8 MeV and 17.3 MeV, respectively.

Incorporating the ab-initio densities and overlaps as described in section \ref{sec:ab-initio}, and computing the theoretical cross sections using the ab-initio spectroscopic factors shows that the VMC-deduced cross sections agree the most with the experimental values \cite{grinyer11}. The reduction factors relative to the VMC spectroscopic factors are 1.01(6) and 0.75(5) for \nuc{10}{Be} and \nuc{10}{C}, respectively. Taking the VMC calculations at face value reveals the following: i) no quenching due to the reaction model is observed in the removal of the (relatively) weakly bound neutron of \nuc{10}{Be}, and ii) a 25\% quenching is observed for the removal of the deeply bound neutron of \nuc{10}{C}. This effect may be attributed to the core survival approximation, where in this case it is assumed that the \nuc{9}{C} core, which is bound by only 1.3 MeV, survives after the removal of a 21.3 MeV bound neutron in \nuc{10}{C}. This first attempt at testing the limitations of the reaction model should be expanded to other cases, but can also guide the improvements of the reaction theory.

\subsubsection{The power of mirror reactions}
\label{sec:mirror}
The isospin symmetry of the nuclear force provides an opportunity to disentangle the effects of the reaction model on the observed quenching from the quenching stemming from the limitations of the independent particle model. In so-called mirror reactions, where the initial and final states are identical except for the projection of the isospin $T_Z$, the spectroscopic factors describing the overlap between the initial and final wavefunctions are identical. However, because the number of protons and neutrons in the initial and final states differ within a given mirror reaction pair, the separation energy asymmetry $\Delta S$ and the separation energies of the cores also vary. Examples of such reactions in the $p$-shell region are shown in Table \ref{tab:mirror}.
\begin{table}[ht]
    \centering
    \begin{tabular}{c|c|c|c}
    Reaction     & Initial State & Final State & $\Delta$S (MeV) \\ 
    \hline
    \hline
    (\nuc{7}{Li},\nuc{6}{He}) & \nuc{7}{Li} (3/2$^-$) & \nuc{6}{He} (0$^+$) & 2.72 \\
    (\nuc{7}{Li},\nuc{6}{Li}) & \nuc{7}{Li} (3/2$^-$) & \nuc{6}{Li} (0$^+$) & -2.72 \\
    \hline
    (\nuc{9}{Li},\nuc{8}{Li}) & \nuc{9}{Li} (3/2$^-$) & \nuc{8}{Li} (2$^+$) & -9.88 \\
    (\nuc{9}{C},\nuc{8}{B}) & \nuc{9}{C} (3/2$^-$) & \nuc{8}{B} (2$^+$) & -12.95 \\
    \hline
    (\nuc{10}{Be},\nuc{9}{Li}) & \nuc{10}{Be} (0$^+$) & \nuc{9}{Li} (3/2$^-$) & 12.83 \\
    (\nuc{10}{C},\nuc{9}{C}) & \nuc{10}{C} (0$^+$) & \nuc{9}{C} (3/2$^-$) & 17.28 \\
    \hline
    (\nuc{11}{B},\nuc{10}{Be}) & \nuc{11}{B} (3/2$^-$) & \nuc{10}{Be} (0$^+$) & -0.22 \\
    (\nuc{11}{C},\nuc{10}{C}) & \nuc{11}{C} (3/2$^-$) & \nuc{10}{C} (0$^+$) & 4.43 \\
    \hline
    (\nuc{12}{C},\nuc{11}{B}) & \nuc{12}{C} (0$^+$) & \nuc{11}{B} (3/2$^-$) & -2.76 \\
    (\nuc{12}{C},\nuc{11}{C}) & \nuc{12}{C} (0$^+$) & \nuc{11}{C} (3/2$^-$) & 2.76 \\
    \end{tabular}
    \caption{Examples of mirror reactions listed as pairs. The initial and final states are identical except for the projection of the isospin. The separation energy differences $\Delta S$ vary within each pair because of the different binding of the initial and final nuclei.}
    \label{tab:mirror}
\end{table}
Measuring the knockout cross sections between the initial and final states of these reactions requires identification of the final state of the residues since in some cases they are either excited states, or the residual nucleus has more than one bound state that can be populated by the reaction.
Although an experimental program to study these reactions at various incident energies is under way for Be- and C-induced knockout, they should also be probed with other methods currently used to extract spectroscopic factors, such as transfer reactions or quasifree scattering off proton targets.

\subsection{Other theoretical methods}
\label{sec:other_theory}
The above discussion is based on the analysis with the eikonal reaction theory \cite{YABANA1992295,hencken96,bertulani04,tostevin01}. However, there are alternative methods that can be used. For example, the CDCC \cite{PhysRevC.66.024607} method can be used to calculate elastic breakup (diffraction) or the Ichimura-Autern-Vincent  (IAV) method, recently implemented numerically by \cite{PhysRevC.92.061602}, can be used for both stripping and diffraction. Both these methods are fully quantum mechanical and are based on the core-spectator picture plus a description of the breakup particle final state interaction with the initial core and the target. 

These methods contain energy conservation and the IAV present implementation allows for the use of energy-dependent nucleon-target and core-target optical potentials. The numerical implementation of the CDCC and IAV method are heavy and the calculations are lengthy but feasible on a super-computer. An intermediate method, the semiclassical  transfer-to-the continuum (STC) exists \cite{Bonaccorso:1988,Bonaccorso:1991}. It provides analytical formulae which are straightforward to implement numerically, thanks to the use of analytical Hankel functions for initial and final states. This method is based on a semiclassical treatment of the relative motion within a time dependent "semiclassical transition" amplitude. Thus energy conservation is preserved.
One of the most important characteristics of this method is that it uses an optical-model final-state wavefunction for the breakup particle interacting with the target. Then, from the properties of the energy averaged S-matrix, the total probability is obtained naturally as a sum of a diffraction (elastic breakup) and stripping (inelastic breakup) terms. It allows to use energy-dependent optical potentials and it can include spin couplings. In the high-energy limit, the method provides an eikonal formalism consistent with the \cite{YABANA1992295,hencken96} formalism for the stripping part. The diffraction part will be further discussed in the Appendix.
  
IAV and STC model have been benchmarked and found consistent in the situations in which both can be applied \cite{Jme}. They can calculate the same observables but the P$_\perp$ or core angular distributions which are possible only in the IAV model.
STC and eikonal have been benchmarked \cite{Bonaccorso:2001} and it has been shown that phase space effects which arise from energy conservation can be introduced in the eikonal, thus improving, in principle, its accuracy, which is particularly relevant at low incident energies. In general, it is found that the STC gives somewhat smaller cross sections than the eikonal predictions, because it calculates the n-target S-matrices by the optical model which is more accurate for the low partial waves. Formally, the core-target treatment is the same in the two methods, apart from the choice of the optical potential. The eikonal uses the double folding via a t${\rho\rho}$ treatment. The accuracy of  the STC has been enhanced for the $^9$Be target by the careful fitting of two optical potentials on a large range of energies for n-$^9$Be scattering data \cite{Bonaccorso:2014}. One such potential is derived with the DOM method \cite{Mahaux:1986,Mahaux:1989a}. Furthermore, it has been shown that core-target potentials can be obtained by single-folding appropriate core densities with the n-$^9$Be potential \cite {Bonaccorso:2016,Bonaccorso:2016a,imane}. Such potentials have been used in the benchmark of IAV and STC and shown to provide a satisfying description of the momentum distribution and cross sections obtained in \cite{flavigny12}. More comparisons to data are necessary to benchmark the predictive power of both the IAV and STC models. In order to discuss the quenching and the dependence with the asymmetry, a systematic comparison still has to be done and is the objective of an ongoing project \cite{Jinme2}. 

\subsection{Conclusions}
\label{sec:conclusions}
Single-nucleon removal reactions induced by light nuclear targets have now been used successfully for approximately two decades to probe the $s.p.$ structure of nuclei far from stability. The trend of the experimental to theoretical cross sections ratio $R_s$ observed with cross sections deduced from configuration-interaction models such as the shell model (see Fig.~ \ref{Rsplot}), is now well documented, comprising numerous data points spanning a wide range of separation-energy difference as well as incident beam energy of the projectile.

The theoretical framework used to describe these reactions quantitatively has been investigated experimentally in various ways, and continues to be a focus of ongoing research projects. From these studies, to date, the eikonal reaction model appears robust.

 The least-explored issue at this time is the extent to which the core survival is accurately described within the effective three-body reaction model in cases where the removal is of a strongly-bound nucleon and where the core is weakly bound. Higher order core excitations into the continuum that lead to core breakup may make a small contribution to the quenching in these cases, but beyond-shell-model structure effects are also hinted at by the particular case of a 25\% quenching in neutron removal from \nuc{10}{C} when analyzed in a self-consistent manner using VMC calculations as input to the eikonal model \cite{grinyer11} (this assumes that the VMC model provides the correct description of \nuc{10}{C}). In contrast, there has been little discussion of the lack of quenching for the removal of a weakly-bound nucleon, as compared to the 30-40\% or so observed on nuclei located in the valley of stability, in agreement with observations reported from $(e,e'p)$ reactions. Together with new data from Jefferson Laboratory~\cite{Duer18}, one may speculate that the cross sections ratio trend observed from the removal reactions discussed here could be an indication of a dependence of Short Range Correlations (SRC) on binding energy. In the extreme cases of halo nuclei for instance, the loosely-bound valence nucleon is largely spatially removed from the bulk of the nucleus, and, therefore, may indeed experience less SRC (and other correlation) effects than those occupying orbitals within the nuclear volume.

In this context, a promising prospect of disentangling the effects of the reaction model approximations on the observed cross sections is offered by the use of mirror reactions, as discussed above.
\section{Quasifree $\boldsymbol{(p,2p)$} and $\boldsymbol{(p,pn)$} scattering}
\label{sec6}

\subsection{Introduction}

In quasifree $(p,pN)$ scattering  a proton with energy of several hundred MeV knocks out a bound nucleon $N$. The energy and momentum of the outgoing nucleons provide information on the quantum numbers of the removed nucleon in its initial state in the nucleus. The shape of the angular distributions of the outgoing nucleons, the nuclear recoil momentum, and the momentum distribution are all important observables, connected with the initial state of the knocked-out nucleon.  In the past five decades quasifree scattering (QFS) has been  a great experimental probe of the nucleus and much of this work has been reviewed in seminal papers by Maris and Jacob \cite{Jacob:1966,Jacob:1973}, who fled Europe and settled at the Federal University of Rio Grande do Sul, Brazil. Maris participated in some of the first experiments on quasifree scattering \cite{Tyrenn1957}, for the reaction $^{12}$C$(p,2p)^{11}$B, and is also the creator of a method, the ``Maris effect'', to obtain additional information in nuclear structure using a small detail of QFS scattering with polarized protons \cite{MARIS1958577,JACOB1976517,MARIS1979461,Kitching1985} (described below). The first experiments on $(p,2p)$ reactions were performed at the Berkeley laboratory using a 350 MeV proton beam \cite{ChamberlainSegre:52,Cladis:52} incident on d, Li, C and O targets. 

\begin{figure}[th!]
\centering
\includegraphics[scale=0.5]{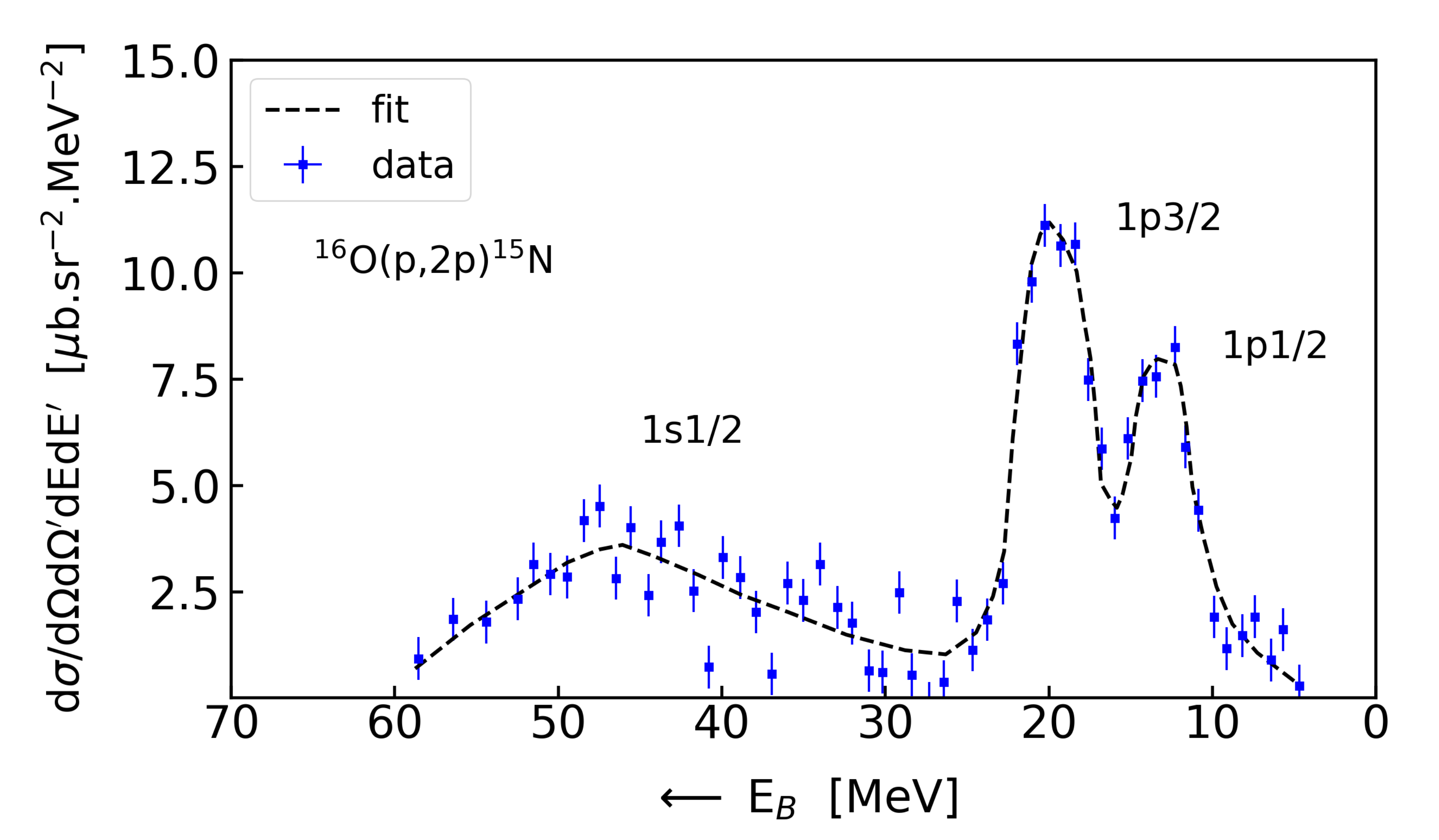}
\caption{(Color online) Energy spectrum for the reaction $^{16}$O$(p,2p)^{15}$N with the binding energy of the removed nucleon $E_B$ increasing from right to left. The dashed curve is a fit to the experimental data \cite{MARIS1958577}.}
\label{oldp2pexp} 
\end{figure}

In Figure \ref{oldp2pexp} we show a typical spectrum of one of the first $(p,2p)$ experiments   for the reaction $^{16}$O$(p,2p)^{16}$N \cite{MARIS1958577}. It is clear that the reaction is an excellent probe of nuclear structure with the identification of the expected energy locations of the nuclear levels. Early on it was also noticed that ``absorption'' of the incoming and outgoing particles, due to multiple collisions, results in an energy loss.  However, the energy losses are smeared out over the whole particle energies and angles, yielding a smooth background in the reconstructed missing-energy spectrum of the residual nucleus, starting at its ground state and rising with excitation energy. Peaks are superimposed to this spectrum caused by clean quasifree collisions. Therefore,  the energy spectrum contains the signature of quasi-elastic collisions even on the presence of a background due to absorption or multiple collisions.  It is thus clear that in order to extract meaningful results for spectroscopic amplitudes, a treatment of multiple collisions needs to be incorporated in the theory. This fact has been known for a long time, e.g., in Refs.~\cite{MARIS1958577} a reduction of $(p,2p)$ cross sections due to absorption is shown to be as large as 1/10 for $^{40}$Ca nuclei, increasing drastically with the atomic number A. 

Following a picture based on the shell model, a hole created by the knockout is filled by the frequency of collision between the outgoing proton and nucleons within upper shells. As an example, for  $1p$ shell nuclei (Li, Be, B, C, N, and O isotopes) one expects the width of the 1s hole state to be directly proportional to the number of proton-nucleon pairs in the $1p$ shell and inversely proportional to the nuclear volume. As summarized in Refs.~\cite{Jacob:1966,Jacob:1973}, this is actually observed in experiments.

\subsection{Distorted wave impulse approximation}
As with other probes in direct reactions,  the distorted wave impulse approximation (DWIA) is the most widely used theoretical method in the study of $(p,pN)$ reactions \cite{Cha77,ChantRooss83}. The DWIA assumes that  the knockout reaction occurs by a single interaction of the incident proton with the  struck nucleon. Multiple scattering with the other nucleons is only accounted for by means of an optical potential, which can also include absorption due to the nuclear excitations. In medium modifications of $pN$ interaction are sometimes accounted for \cite{KERMAN1959551,BERTSCH1977399,PhysRevC.16.80,PhysRevC.24.1073,PTP70459,CHEON1985227,PhysRevC.31.488,PhysRevC.33.2059,PhysRevC.38.51,bertulani10}. Ambiguous treatments of in-medium interactions and in the reaction mechanism still exist,  and multi-step processes are not negligible \cite{MARIS1958577,JACOB1976517,PhysRevC.51.2646,PhysRevC.73.064603,PhysRevC.80.011602}.

Assuming the validity of the DWIA, the quasifree cross section can be shown to factorize as \cite{Jacob:1973,Kitching1985}
\begin{equation}
\frac{d^3\sigma}{dT_N d\Omega^\prime_p d\Omega_N} =K^\prime \frac{d\sigma_{pN}}{d\Omega} |F({\bf Q})|^2,
\label{int}
\end{equation}
where the momentum distribution of the knocked-out nucleon $N$ within the nucleus is given by $|F({\bf Q})|^2$, $d\sigma_{pN}/d\Omega$ denotes the free $pN$ cross section and $K^\prime$ is a kinematic factor. An off-shell $pN$ t-matrix is needed in  Eq.~\eqref{int} and the off-shellness is only treated as a kinematical correction. Therefore, it is only valid if off-shell effects are not relevant, which seems to be the case for high-energy collisions  \cite{Jacob:1973,Kitching1985}.  For the bulk of quasi-elastic scattering, since one is almost on-shell, it does not make too much difference which cross section one uses. On the other hand, for extracting the smaller, more interesting, parts of the spectral function, the high-momentum and large-missing-energy components, the off-shell ambiguity can be quite large.

To clarify how several approximations can be used, we start from the basic DWIA formulation where the transition amplitude for
the A$(p, pN)$B reaction is \cite{Jacob:1973}
\begin{equation}
T_{p,pN}=\sqrt{S(lj)}\left< \chi_{{\bf k}_p^\prime}^{(-)} \chi_{{\bf k}_N}^{(-)}          
\left| \tau_{pN}\right| \chi_{{\bf k}_p}^{(+)}\psi_{jlm}\right> .\label{Tmat}
\end{equation}
Here, $\chi_{{\bf k}_p^\prime}^{(-)}$ ($\chi_{{\bf k}_N}^{(-)}$) denotes the distorted wave for the outgoing proton (knocked-out nucleon) interacting with  the residual nucleus B, whereas $\chi_{{\bf k}_p}^{(+)}$ denotes the incoming proton distorted wave interacting with the target nucleus A. $ \psi_{jlm}$ represents the bound-state wavefunction of the struck nucleon, and
$ \sqrt{S(lj)}$ is the spectroscopic amplitude of the ejected nucleon, with angular momentum quantum numbers ($lj$). 
The energy $E$ entering the the two-body $pN$ scattering matrix $\tau_{pN}$ is described below. 

\begin{figure}[t]
\begin{center}
\includegraphics[width=5in]{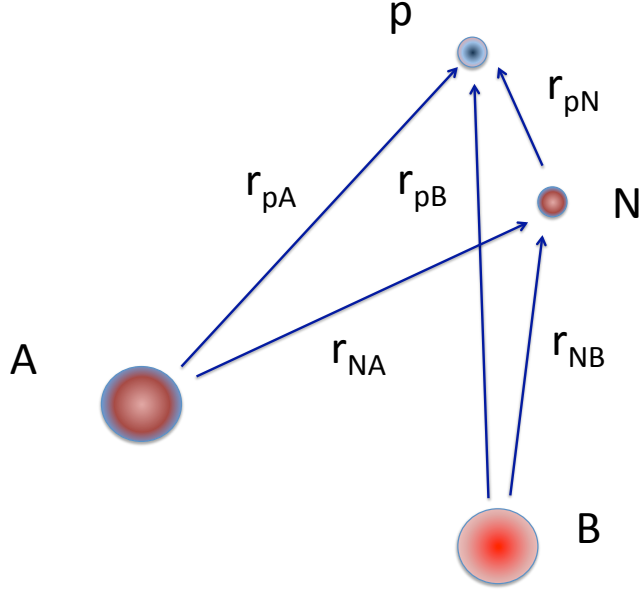}
\caption{(Color online). The coordinates used in the text are shown.}
\label{coord}
\end{center}
\end{figure}

Equation \eqref{Tmat} means
\begin{eqnarray}
T_{p,pN}&=&\sqrt{S(lj)} \int d^3{\bf r}^\prime_{pB} d^3{\bf r}^\prime_{NB}
d^3{\bf r}_{pA} d^3{\bf r}_{NB}   \tau  ({\bf r}_{pB},{\bf r}_{NB};{\bf r}^\prime_{pA}, {\bf r}_{NB}^\prime )
\nonumber \\
&\times&\chi_{{\bf k}_p^\prime}^{(-) *} ({\bf r}^\prime_{pB}) \chi_{{\bf k}_N}^{(-) *}({\bf r}^\prime_{NB}) \chi_{{\bf k}_p}^{(+)} ({\bf r}_{pA})\psi_{jlm}({\bf r}_{NB}),
 \nonumber \\
\label{tmat2}
\end{eqnarray}
with the normalization in the continuum as  
$
\int d^3{\bf r} \chi^*_{\bf k}({\bf r}) \chi_{\bf k^\prime}({\bf r}) =  \delta({\bf k}-{\bf k^\prime})
$ 
and the normalization of the bound-state wavefunction $\psi_{nlj}$ as 
$
\int d^3{\bf r} |\psi_{nlj}({\bf r})|^2=1.
$
The coordinates used above are shown in Fig. \ref{coord}. 

Further approximations have been used in the literature and proved to be numerically accurate. They have been used to reduce the number of integrations in  Eq. \eqref{tmat2}. Notice that the proton and the struck nucleon $N$  coordinates are related by ${\bf r}_{pA}={\bf r}_{pN}+ {\bf r}_{NA}$ and, since the $pN$ interaction has a smaller range than the nuclear size, Eq. \eqref{tmat2} will only include small values of $r_{pN}$ for which $r_{pN} \ll r_{NA}$,  Eq. \eqref{tmat2} for the $T$-matrix becomes an integral over  ${\bf r}_{NB}$  \cite{Jacob:1973,Kitching1985},
\begin{eqnarray}
T_{p,pN}=\sqrt{S(lj)}\tau  ({\bf k}^\prime_{pN},{\bf k}_{pN};E) \int d^3{\bf r}_{NB} \chi_{{\bf k}_p^\prime}^{(-) *} ({\bf r}_{NB}) \chi_{{\bf k}_N}^{(-) *}({\bf r}_{NB}) \chi_{{\bf k}_p}^{(+)} (\alpha {\bf r}_{NB})\psi_{jlm}({\bf r}_{NB}),\nonumber \\
\label{tmat5}
\end{eqnarray}
with $\alpha = (A-1)/A$ and $\tau({\bf k}^\prime_{pN},{\bf k}_{pN};E)$ denoting the Fourier transform of the $pN$ t-matrix appearing in Eq. \eqref{tmat2}.
This equation  has been  used as the standard for numerical calculations of quasifree reactions \cite{Cha77,ChantRooss83}.  Several corrections have been introduced in the literature to study the validity of approximations used (see, e.g. \cite{Jacob:1973,Kitching1985,Jackson1982}).

\subsection{Cross section}
Using the $T$-matrix of Eq. (\ref{tmat5}), the differential cross section becomes \cite{Jacob:1973,Kitching1985,Jackson1982}
\begin{equation}
\frac{d^3\sigma}{dE_p^\prime d\Omega_p^\prime d\Omega_N} =\frac{K}{(2s_p+1)(2J_A+1)}
\sum_\gamma \left| T_{p,pN}(\gamma)\right|^2
\label{cross1}
\end{equation}
where the kinematical factor $K$ is   (with $k_i=p_i/\hbar$) \cite{MARIS1958577}
\begin{eqnarray}
K=\frac{m_p^2m_N c^6}{(\hbar c)^6 (2\pi )^5} \frac{k_p^\prime k_N}{k_p}
\left| 1 + \frac{m_N}{M_B} \left[1- \frac{k_p}{k_N}\cos \theta_N +\frac{k_p^\prime}{k_N}
\cos(\theta_p^{\prime} +\theta_N)\right]\right|^{-1} 
\label{cross2}
\end{eqnarray}
and  the prime indices denote the proton variables in the final channel.

The summation of the transition matrix $T_{p,pN}(\gamma)$ in Eq.~\eqref{cross1} over
\begin{equation}
\gamma = (\mu_p, \mu_p^\prime, \mu_N, M_A, M_B),
\end{equation} 
involves the spin quantum numbers of the incident proton, $s_p$  and $\mu_p$, the target nucleus, $J_A$ and $M_A$, the final proton, $s_p$ and $\mu_p^\prime$, the struck nucleon, $s_N$ and $\mu_N$, and the residual nucleus, $J_B$ and $M_B$.  For single particle states, a simpler notation $jlm$ is often used for the angular momentum quantum numbers of the ejected nucleon.

The missing momentum, ${\bf p}_m$,  missing energy, $\epsilon_m$,  and missing mass, $m_m$,  are defined by
 \begin{eqnarray}
\epsilon_m&=&E_N+ E_{p}^\prime- E_p-m_Nc^{2}\nonumber\\
 {\bf p}_m &=&\hbar({\bf k}_N+{\bf k}_{p}^\prime - \alpha {\bf k}_p) \nonumber \\
 m_m^2c^{4}&=&\epsilon_m^2-p_m^2c^{2}, 
\label{cross2b}
\end{eqnarray}
where $\alpha = (A-1)/A$ accounts for the c.m. correction. They are the momentum and energy taken over by the residual nucleus in the final state.

\subsection{Plane wave impulse approximation}
The $pN$ scattering amplitude (in the $pN$ c.m.) relates to the $pN$ t-matrix by means of \cite{HUSSEIN1991279}
\begin{equation}
f_{pN}(\theta ; E)=-\frac{2\pi^2 m_{N}}{\hbar^2}\tau({\bf k}^\prime_{pN},{\bf k}_{pN};E), \label{ftheta}
\end{equation}
and the elastic scattering cross section  is  given by
\begin{equation}
\frac{d\sigma_{pN}}{d\Omega} = \left|f_{pN}(\theta; E)\right|^2 .
\end{equation}
Using plane waves in Eq. \eqref{tmat5} yields
\begin{eqnarray}
T_{p,pN}^{(PWIA)}=\sqrt{S(lj)}\tau({\bf k}^\prime_{pN},{\bf k}_{pN};E) \int d^3{\bf r} \ e^{-i{\bf Q.r}}\psi_{jlm}({\bf r}),\nonumber \\
\label{cross5}
\end{eqnarray}
where ${\bf Q}=  {\bf p}_m $ is the missing momentum defined  above.
Therefore,  Eq.~\eqref{int} is reproduced where 
\begin{equation}
F({\bf Q})= \int d^3{\bf r} \ e^{-i{\bf Q.r}}\psi_{jlm}({\bf r}) \label{ftpsi}
\end{equation} 
and the kinematic factor in Eq.~\eqref{int} is related to the  kinematic factor of Eq. \eqref{cross1} through the relation
\begin{equation}
K^\prime=\left(\frac{\hbar^2}{2\pi^2 m_{N}}\right)^2 \ K .
\end{equation}  
The distortions due to  absorption and elastic scattering are not accounted for in the PWIA, but can easily be accommodated in the DWIA. The PWIA is still useful for physical interpretation of experimental data. 

\subsection{DWIA vs. PWIA}
\begin{figure}[t]
\begin{center}
\includegraphics[width=5.3in]{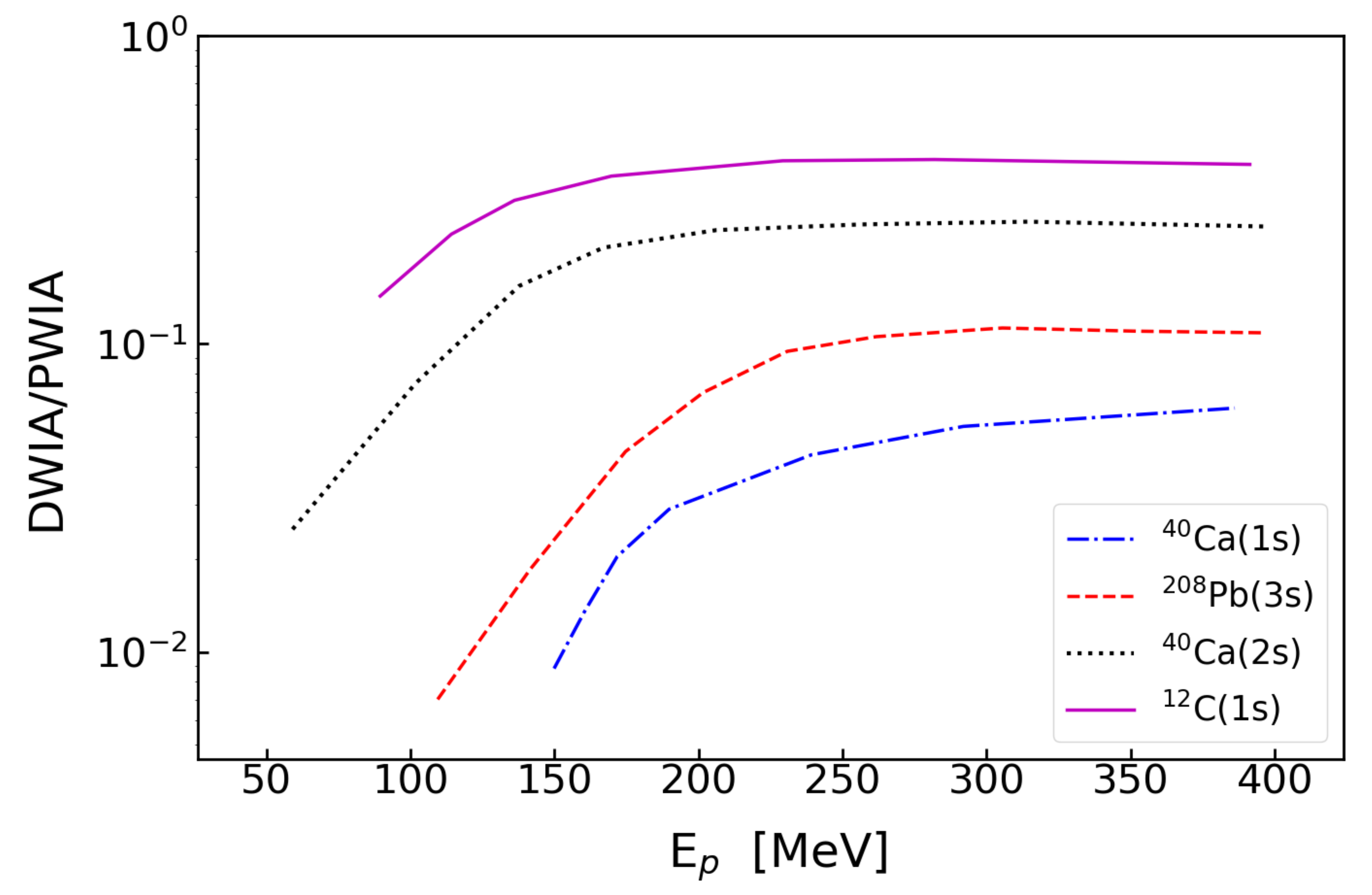}
\caption{(Color online) Ratio of DWIA  and PWIA  results for the cross section of $(p,2p)$ reactions on several nuclei.}%
\label{pwia}%
\end{center}
\end{figure}

In Fig. \ref{pwia} we show the ratio of PWIA  and DWIA  calculated  cross section of $(p,2p)$ reactions on several nuclei. It highlights the importance of distortion and absorption effects.

In Ref. \cite{PhysRevC.80.011602}, eikonal wavefunctions were used to describe the distorted waves. The distorted waves were separated in initial and final channels allowing for a treatment of collisions in which the outgoing nucleons can scatter into large angles, by assuming that the outgoing proton and ejected nucleon suffer minor deflections on their way out of the residual nucleus. In terms of the momentum transfer ${\bf Q}$, the momentum distributions are given by
\begin{eqnarray}
\left( \frac{d\sigma}{d ^3 Q}\right)_{DWIA} =  \frac{1}{(2\pi)^3} \frac{S(lj)}{2j+1} \sum_m \left< \frac{d\sigma_{pN}}{d\Omega} \right>_Q  \left| \int d^3 {\bf r} \ e^{-i{\bf Q.r}}\left<{\cal S}(b)\right>_Q
\psi_{jlm}({\bf r}) \right|^2 , \label{momdis}
\label{strxsec}
\end{eqnarray}
where $\left<{\cal S}(b)\right>_Q$ is the product of the incoming proton and outgoing nucleons $S$-matrices averaged  over all scattering angles leading to the same momentum transfer $Q$. The absorption is included in terms of the in-medium nucleon-nucleon cross sections \cite{bertulani10}. Further integration of this equation allows one to obtain the longitudinal, ${d\sigma/ d Q_{z}}$,  and transverse, ${d\sigma/ d Q_{t}}$ momentum distributions \cite{PhysRevC.80.011602}. 

\begin{figure}[t]
\begin{center}
\includegraphics[width=5.3in]{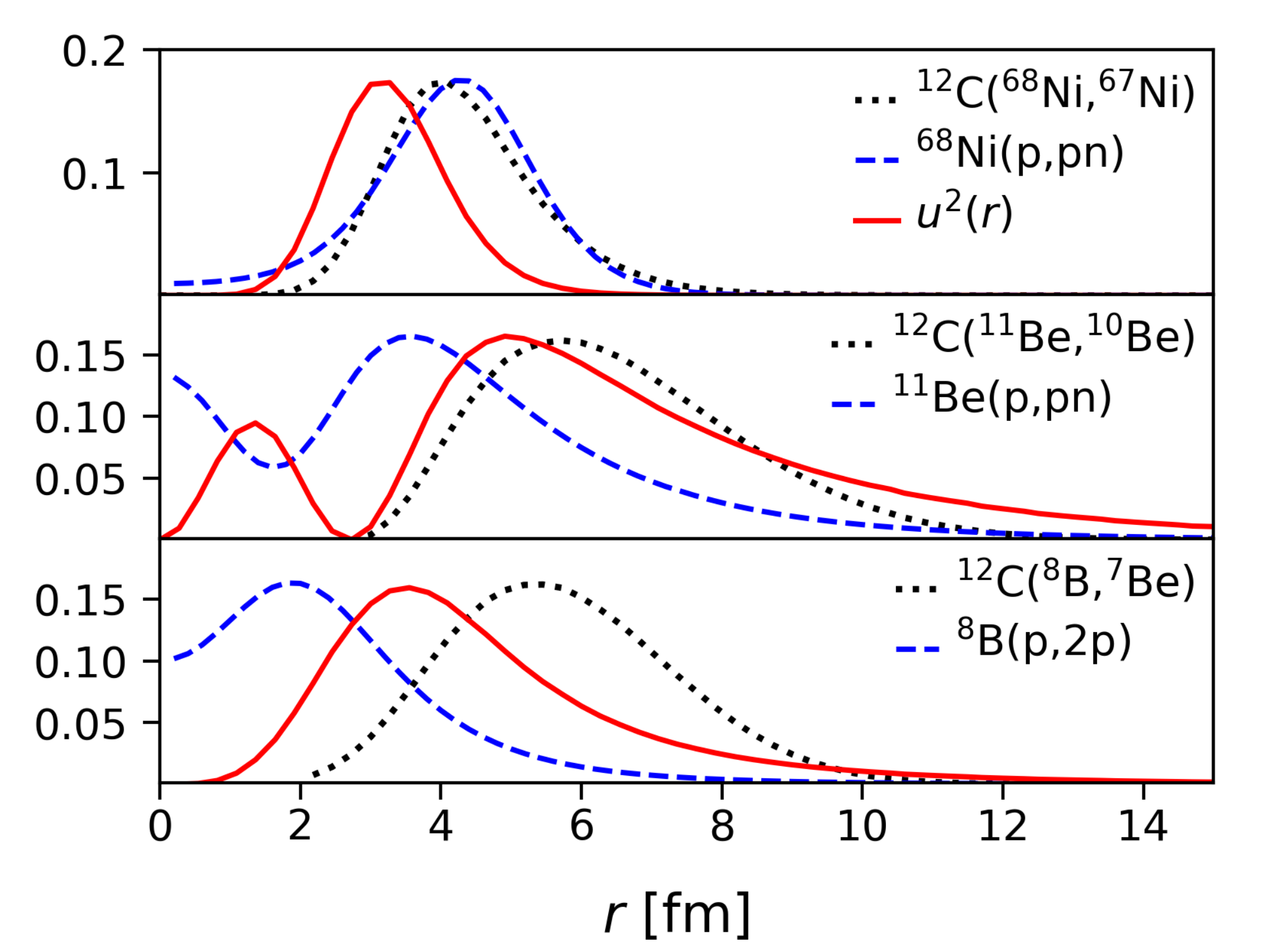}
\caption{(Color online). {\it Upper panel:} The dashed curve shows the probability for removal of a neutron in the reaction $^{12}$C($^{68}$Ni,$^{67}$Ni) at 500~MeV/nucleon as a function of the distance to the c.m. of $^{68}$Ni. The dotted curve represents the removal probability in a $^{68}$Ni$(p,pn)$ reaction at the same energy. For comparison the square of the radial wavefunction $u^2(r)$ is also shown (solid curve). {\it Middle panel:}  Same as the upper panel, but for the reactions $^{12}$C($^{11}$Be,$^{10}$Be)  and $^{11}$Be$(p,pn)$ at 500~MeV/nucleon. {\it Lower Panel:}  Same as upper panel, but for the reactions $^{12}$C($^{8}$B,$^{7}$Be)  and $^{8}$B$(p,2p)$ at 500~MeV/nucleon. Figure adapted with permission from \cite{aumann2013}\textcopyright2013 by the American Physical Society.}
\label{probabilities}
\end{center}
\end{figure}

\begin{figure}[t]
\begin{center}
\includegraphics[width=5.3in]{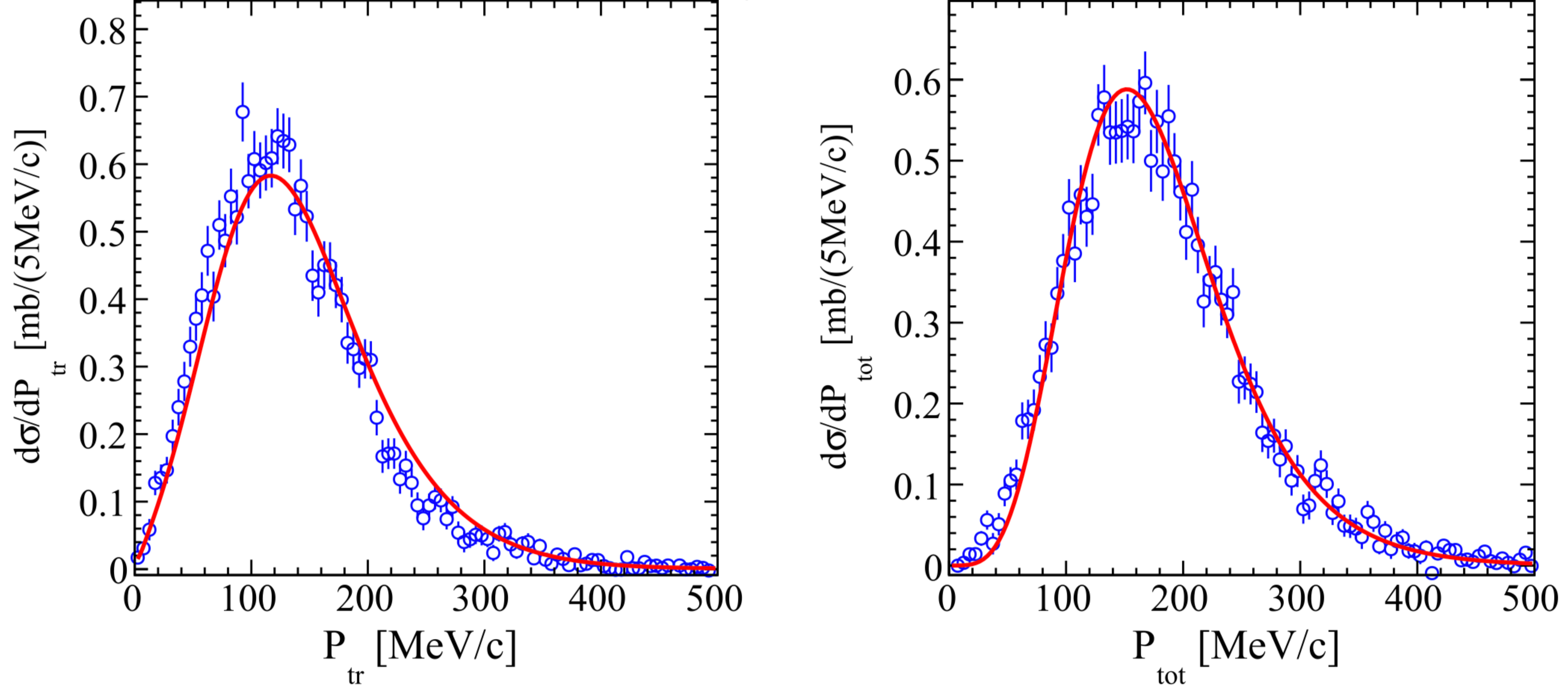}
\caption{ (Color online) Measurements of $^{12}$C$(p,2p)^{11}$B reactions with transverse (left) and total (right) recoil momentum of $^{11}$B in the rest frame of the incident $^{12}$C. The experimental momentum distributions (circles) are compared with theoretical calculations (curves) for $0p$-shell QFS knockout which take into account nuclear absorption effects. The theoretical curves are normalized to the experimental data with the scaling factor of 0.64. Figure adapted with permission from \cite{Pan16}. Copyright by Elsevier.}
\label{panin}%
\end{center}
\end{figure}

One of the advantages of using eikonal distorted waves is the interpretation of the classical impact parameters sampled by heavy-ion knockout reactions and in $(p,2p)$ reactions. For heavy-ion knockout reactions, one obtains from Eq.~(\eqref{dis1}) for the probability to remove a nucleon in orbital ($lj$) located at distance ${\bf b}$ (perpendicular to the collision axis) from the center of the projectile nucleus as
\begin{eqnarray}
 P_{ko}(b) = \frac{S(lj)}{2j+1} \left| S_c(b)\right|^2 \sum_m  \int d^3 {\bf r} \left| \psi_{jlm}({\bf r})\right|^2   \left[ 1- \left|S_n \left( \sqrt{r^2 \sin^2 \theta + b^2 -2rb \sin \theta \cos \phi}\right)\right|^2 \right] . \nonumber \\
 \label{pjl}
 \end{eqnarray}
 Here $S_n$ ($S_c$) is the eikonal matrix amplitude for the scattering  of the nucleon $N$ (core c) on the target, and ${\bf r} \equiv (r, \theta, \phi)$. In the case of $(p,2p)$ reactions, the nucleon-knockout probability is given by integrating Eq. \eqref{strxsec} over the momentum transfer. The integrand of the resulting cross-section formula yields the quasifree scattering probability \cite{PhysRevC.80.011602}
 \begin{eqnarray}
P_{{qfs}}(b)=   \frac{S(lj)}{2j+1} \sum_m \left< \frac{d\sigma_{pN}}{d\Omega} \right>_{o.s.} |{\cal C}_{lm}|^2 
 \left| \left<{\cal S}(b)\right>_{o.s.} \right|^2  \int_{-\infty}^\infty dz  \left| \frac{u_{lj}(r)}{r}P_{lm}(b,z)  \right|^2 , \nonumber \\
 \label{totxprob}
\end{eqnarray}

In Figure \ref{probabilities}, upper panel, the dotted curve shows the probability for removal of a neutron in the reaction $^{12}$C($^{68}$Ni,$^{67}$Ni) at 500~MeV/nucleon as a function of the distance to the c.m. of $^{68}$Ni. The dashed curve represents the removal probability in a $^{68}$Ni$(p,pn)$ reaction at the same energy. For comparison the square of the radial wavefunction $u^2(r)$ is also shown (solid curve). We assume a neutron in the $0f_{7/2}$ orbital in $^{68}$Ni, bound by 15.68 MeV. The figures in the middle panel are for the reactions $^{12}$C($^{11}$Be,$^{10}$Be)  and $^{11}$Be$(p,pn)$ at 500~MeV/nucleon.   We assume a neutron in the $1s_{1/2}$ orbital in $^{11}$Be, bound by 0.54 MeV. The figures in the lower panel are for the reactions $^{12}$C($^{8}$B,$^{7}$Be)  and $^{8}$B$(p,2p)$ at 500~MeV/nucleon. We assume a proton in the $0p_{3/2}$ orbital in $^{8}$B, bound by 0.14 MeV. One observes that the removal cross sections for both knockout and $(p,pn)$ reactions probe the surface part of the wavefunction. This is due to the fact that in  both cases, the absorption is very strong for small impact parameters. For proton or neutron removal from even deeper bound states, with a concentration of the wavefunction closer to the origin,  both reaction mechanisms will probe an even smaller part of the wavefunction tail.

\begin{figure}[t]
\begin{center}
\includegraphics[
width=5.3in
]%
{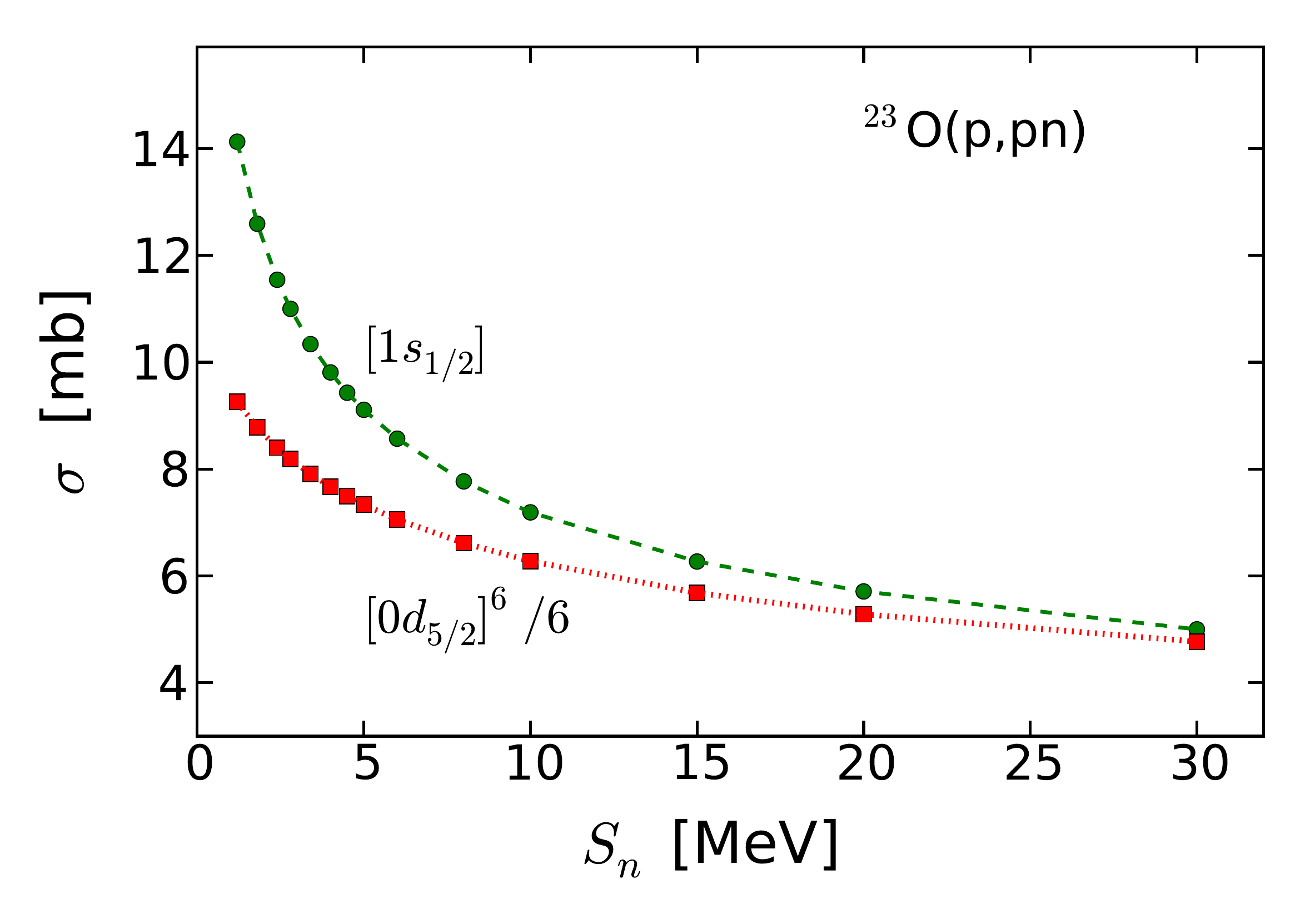}%
\caption{(Color online). Cross sections for neutron removal in $(p,pn)$ reactions on $^{23}$O from $[0d_{5/2}]^6$ and $[1s_{1/2}]$ orbitals as a function of the separation energy. The cross sections for neutron removal from the $[0d_{5/2}]^6$ orbital is divided by the number of the neutrons in the orbital (6). The separation energies are varied artificially. The dashed and dotted curves are guide to the eyes. Figure reprinted with permission from \cite{aumann2013}\textcopyright2013 by the American Physical Society.}%
\label{o23XS}%
\end{center}
\end{figure}

In the middle panel of Fig. \ref{probabilities} one sees that  the part of the wavefunction probed in the $^{12}$C($^{11}$Be,$^{10}$Be) is again limited to the surface of the nucleus, beyond the orbital maximum density. On the other hand, the $(p,pn)$ reaction has a much larger probability of  accessing information on the inner part of the wavefunction, as seen by the dashed curve. 
These results are in agreement with the conclusions drawn in Ref. \cite{PhysRevC.83.054601} for stable nuclei where it has been shown that for light nuclei the average density probed in $(e,e'p$ is comparable to the one probed in $(p,2p)$. There is a strong A dependence, though, and for medium-heavy and heavy nuclei one is rather probing the surface region in $(p,pN)$ reactions.
It is thus clear that knockout reactions with heavy ions and $(p,pN)$ reactions yield complementary nuclear spectroscopic information. For deep bound states the first reaction is only accessible to the tail of the nuclear wavefunction, whereas the $(p,pN)$ reaction process probes the largest part of the  wavefunction for loosely bound nuclei.  

\begin{figure}[t]
\begin{center}
\includegraphics[
width=5.3in
]%
{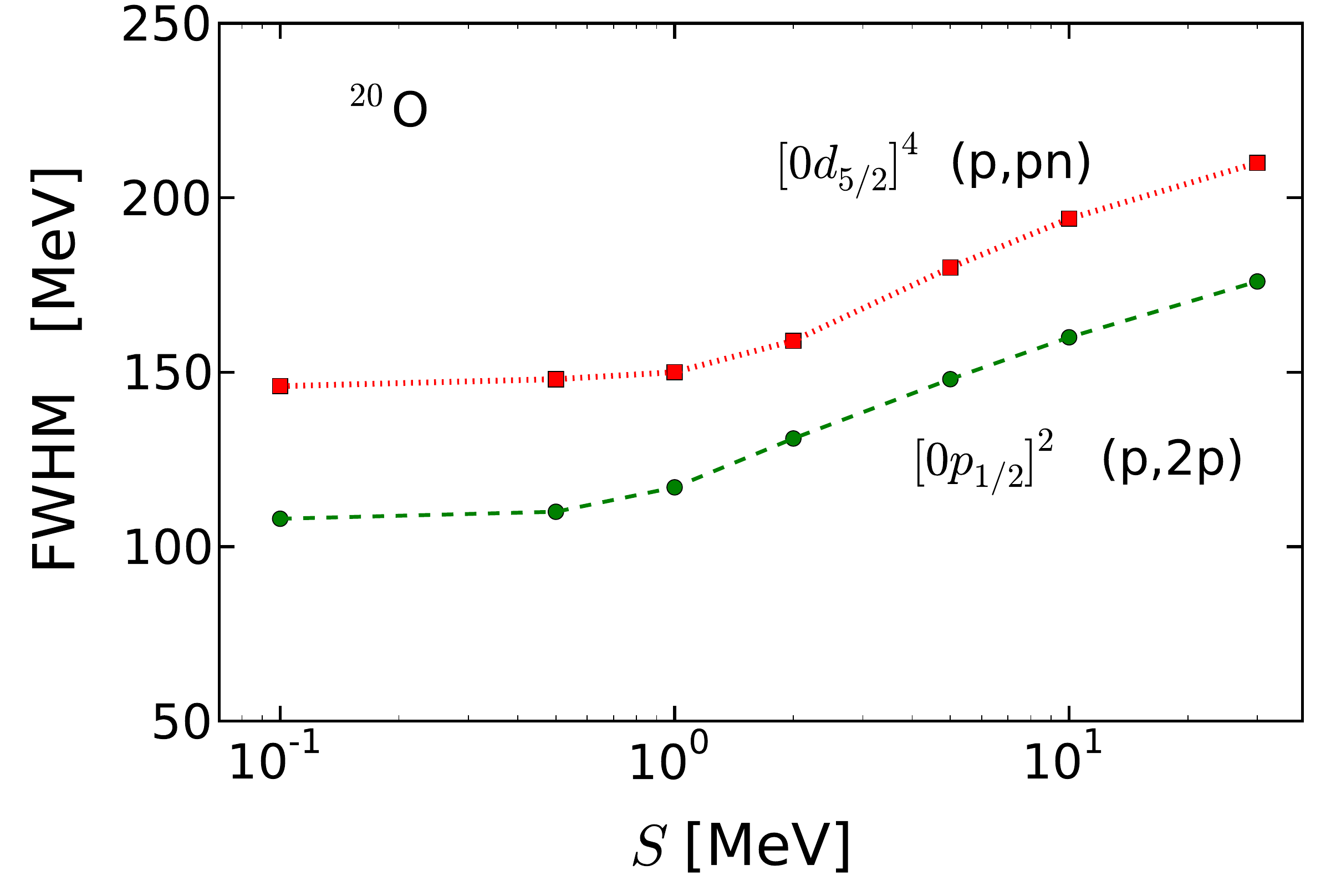}%
\caption{(Color online). Full width at half maximum (FWHM) of the transverse momentum distributions in $(p,2p)$ and $(p,pn)$ reactions of $^{20}$O at 500~MeV/nucleon as a function of the proton and neutron separation energies. The upper curve is for $(p,pn)$ and lower curve for $(p,2p)$ reactions. The separation energies are varied artificially. The dashed and dotted curves are guide to the eyes. Figure reprinted with permission from \cite{aumann2013}\textcopyright2013 by the American Physical Society.}%
\label{width}%
\end{center}
\end{figure}

In recent years, with the advent of radioactive nuclear beams, it it necessary to perform $(p,2p)$ reaction in inverse kinematics. $(p,2p)$ experiments using this technique were measured for the first time with $^{9--16}$C beams at 250~MeV/nucleon at HIMAC~\cite{kobayashi2007}. The missing mass spectra obtained in these measurements show distinct peak structures for $p$-shell and $s$-shell proton knockout. The first exclusive and kinematically complete $(p,2p)$ experiment in inverse kinematics was performed using a $^{12}$C beam at an energy of 400~MeV/nucleon as a benchmark \cite{Pan16}. This new technique has been developed to study the $s.p.$ structure of exotic nuclei in experiments with radioactive-ion beams. The outgoing pair of protons and the fragments were measured simultaneously, enabling an unambiguous identification of the reaction channels and a redundant measurement of the kinematic observables. Both valence and deeply-bound nucleon orbits are probed, including those leading to unbound states of the daughter nucleus. Figure \ref{panin} shows measurements of this $^{12}$C$(p,2p)^{11}$B reaction with transverse (left) and total (right) recoil momentum of $^{11}$B in the rest frame of the incident $^{12}$C \cite{Pan16}. The experimental momentum distributions (circles) are compared with theoretical calculations (curves) for $0p$-shell QFS knockout which take into account nuclear absorption effects. The theoretical curves are normalized to the experimental data with the scaling factor of 0.64. The agreement between this $(p,2p)$ experiment  and the spectroscopic factors deduced from electron-induced knockout reactions demonstrates the validity of the proposed method, which has become a powerful tool for studies with radioactive beams to investigate the $s.p.$ structure of exotic nuclei.

In figure \ref{o23XS} we plot the cross sections for neutron removal in $(p,pn)$ reactions on $^{23}$O from $[0d_{5/2}]^6$ and $[1s_{1/2}]^1$ orbitals as a function of the separation energy. The cross sections for neutron removal from the $[0d_{5/2}]^6$ orbital is divided by the number of the neutrons in the orbital (6). As expected, the cross sections are strongly energy dependent close to the threshold and steadily decrease with increasing separation energy. Close to threshold ($S_n=0$) a large fraction of the wavefunction lies in a region where absorption, or multiple scattering, is smaller, thus increasing the removal probability at larger impact parameters, consequently increasing the knockout cross section. As the separation energy increases it becomes less likely to knockout a neutron without rescattering effects.

In Figure \ref{width} we show the full width at half maximum (FWHM) of the transverse momentum distributions in $(p,2p)$ and $(p,pn)$ reactions of $^{20}$O at 500~MeV/nucleon as a function of the proton and neutron separation energies. The upper curve is for $(p,pn)$ and lower curve for $(p,2p)$ reactions. As expected, the widths for $d$-states are larger than those for $p$-states. 

\subsection{The quantum transfer-to-the-continuum method}
\label{QTC}
Standard analyses of $(p,pN)$ reactions have been performed  within the DWIA approximation. However, in recent years, some alternative methods, which provide additional insights as well as an independent extraction of spectroscopic factors and momentum distributions, have been proposed. One of these  methods is the so-called {\it transfer to the continuum} formalism. This is a fully quantum-mechanical approach based on the direct application of the prior-form transition amplitude for the $(p,pN)$ reaction. For a process of the form $A(p,pN)C$, this transition amplitude can be written as \cite{Mor15}
\begin{equation}
{\cal T}^{3b}_{if} =   \langle \Psi^{3b(-)}_{f} |  V_{pN} + U_{pC} - U_{pA} |\phi_{CA} \chi^{(+)}_{pA} \rangle,  
\label{T3b_uaux}
\end{equation}
where $ V_{pN}$ is the nucleon-nucleon interaction, $U_{pC}$ and  $U_{pA}$ are  effective nucleon-nucleus interactions, $\chi^{(+)}_{pA}$ is the distorted wave generated by the entrance channel auxiliary potential $U_{pA}(\vecR)$ and  $\phi_{CA}$ is the overlap function between the initial and residual nuclei $A$ and $C$. This function could be in principle directly obtained from  {\it ab-initio} calculations of the $A$ and $C$ systems, as discussed in 
Sec.~\ref{sec:ab-initio} in the context of heavy-ion knockout. Since in many cases these are not available, it is a common practice to approximate this overlap by the product of a $s.p.$ wavefunction $\varphi^{\ell,j}_{CA}$ for a $s.p.$ configuration $\ell,j$ (obtained with some mean-field potential) and the square root of the spectroscopic factor $S_{CA,\ell,j}$, {\it i.e.}, $\phi_{CA} \approx 
\sqrt{S_{CA,\ell,j}}  \varphi^{\ell,j}_{CA}$. The spectroscopic factor of the removed nucleon is usually obtained from a small-scale shell model calculation. The  wavefunction $\Psi^{3b(-)}_{f}$  appearing in the final state of (\ref{T3b_uaux})  is the  solution of the effective three-body equation
%
\begin{align}
\label{eqPsi}
[E - K_{r'}- K_{R'}  - V_{pN}-U^\dagger_{pC} - U^\dagger_{NC} ] \Psi^{3b(-)}_{f} (\vecrp,\vecRp)&=0 ,
\end{align}
where the interactions $U_{pC}$ and  $U_{NC}$ are in practice represented by optical model potentials at  half the incident energy per nucleon.

Provided that the exact three-body wavefunction $\Psi^{3b}_{f}$ is used, the  amplitude (\ref{T3b_uaux}) is strictly independent of the choice of $U_{pA}$. In practical calculations, in which $\Psi^{3b}_{f}$ must be approximated somehow, the result will however depend on this potential. The usual choice is to use for $U_{pA}$ an optical potential describing the elastic scattering of the $p+A$ system. With this choice, one expects that the difference $U_{pC} - U_{pA}$ (the so-called {\it remnant term}) will contribute little to the integral and hence the matrix element will be mostly determined by the $V_{pN}$ interaction.

In the applications performed so far with this method, the three-body wavefunction is approximated by an expansion in $p+N$ eigenstates, which is expected to describe accurately the exact three-body wavefunction in the range of the $V_{pn}$ interaction, the main contributor of the $(p,pN)$ reaction. Since these states form a continuum, a procedure of discretization is employed, similar to that used in the Continuum-Discretized Coupled-Channels method (CDCC) \cite{AUSTERN1987},  
\begin{align}
\label{PhiCDCC}
\Psi^{3b(-)}_{f} \approx \Psi^\mathrm{CDCC}_{f}=\sum_{n,J,\pi}  \tilde{\phi}^{J\pi}_{n}(\vecrp) \chi_{n,J,\pi}(\vecK_{n},\vecRp) ,
\end{align}
where $\tilde{\phi}^{J\pi}_{n}(\vecrp)$ represent the discretized $p-n$ states, corresponding to a momentum interval $[k_{n-1},k_n]$ and angular momentum-parity $\{J,\pi \}$ obtained by a linear superposition of scattering states  $\phi_{pN_1}^{J,\pi}(k,\vecrp)$ as
\begin{equation}
   \tilde{\phi}_{n}^{J,\pi}(\vecrp)=
   \sqrt{\frac{2}{\pi N}}\int_{k_{n-1}}^{k_n}\phi_{pN}^{J,\pi}(k,\vecrp)dk. 
\end{equation}
The functions $\chi_{n,J,\pi}(\vecK_{n},\vecRp)$, which describe the relative motion between the outgoing $p-N$ pair, are obtained by solving a set of coupled-differential equations, which result upon insertion of the model wavefunction ($\ref{PhiCDCC}$) into the equation (\ref{eqPsi}). \\

The angular differential cross section of the residual core,  for a given final discretized bin $f = \{ n,J, \pi \}$, is 
\begin{align}
\label{eq:dsdw_bin}
 \frac{d\sigma_{n,J,\pi}}{d\Omega_c}   =   \frac{1}{(2s_p+1)(2J_A+1)} \frac{\mu_i \mu_f}{(2 \pi \hbar^2)^2} \frac{K_n}{K_i} \sum_{\sigma}|{\cal T}^{3b}_{i,f} |^2 ,
\end{align}
with $\mu_{i}$ ($\mu_{j}$) the reduced mass of the initial (final) mass partition and ${\cal T}^{3b}_{i,f}$ the transition amplitude obtained by replacing the CDCC expansion (\ref{PhiCDCC}) in the transition amplitude (\ref{T3b_uaux}). The sum in $\sigma$ includes the spin projections of the outgoing $p+N$ pair and of the residual nucleus $C$. The angle specified by $\Omega_c$ is the scattering angle of the core in the c.m. frame.

The double differential cross section, with respect to the scattering angle of the core  and the internal energy of the $p$+$N$ system, can be obtained at the discretized energies $e_{pN}= e^n_{pN}$ as
\begin{align}
\label{eq:dsdw}
\left . \frac{d^2\sigma_{J,\pi}}{d e_{pN} d\Omega_c} \right |_{e_{pN}=
 e^n_{pN} } \simeq  \frac{1}{\Delta_n} \frac{d\sigma_{n,J,\pi}}{d\Omega_c} ,
\end{align}
where $\Delta_n$ is the width of the bin to which the energy $e_{pN}$ belongs. This can be readily transformed into a double differential cross section with respect to the energy of the outgoing core in the overall c.m. frame ($E_c$), using the usual non-relativistic relation for binary collisions
\begin{align}
\label{eq:ec}
E_c = \frac{m^{*}_{pN}}{m^{*}_{pN} + m_c} E_\mathrm{cm} ,
\end{align}
where $m^{*}_{pN}= m_{p}+m_{N} +e_{pN}$. Thus, 
\begin{align}
\label{eq:dsdwc}
\frac{d^2\sigma_{J,\pi}}{d E_c d\Omega_c} =
\frac{m^{*}_{pN}}{m_c + m^{*}_{pN}} \frac{d^2\sigma_{J,\pi}}{d e_{pN}  d\Omega_c} .
\end{align}

The inclusive cross section will be then obtained summing the contributions from all final $J^\pi$ configurations
\begin{align}
\label{eq:dsdwc-inc}
 \frac{d^2\sigma}{dE_c d\Omega_c}  =  \sum_{J,\pi} \frac{d^2\sigma_{J,\pi}}{d E_c d\Omega_c} .
\end{align}

Note that Eq.~(\ref{T3b_uaux}) resembles the transition amplitude for a transfer process, analogous to that appearing in the standard CCBA method for binary collisions \cite{Satchler83}. Taking advantage of this formal analogy, one can evaluate this transition amplitude using standard coupled-channels codes. Applications performed so far with this method have been done with a modified version of the code {\sc fresco} \cite{Thompson88}. 

We finally note that, although the non-relativistic expressions have been used here, applications of the method to high-energy reactions (hundreds of MeV/nucleon) require the introduction of relativistic kinematics. Further details are  given in Ref.~\cite{Mor15}. 

\subsection{Comparison with the DWIA and Faddeev methods}
\begin{figure}[th]
\centering
\includegraphics[width=6in]{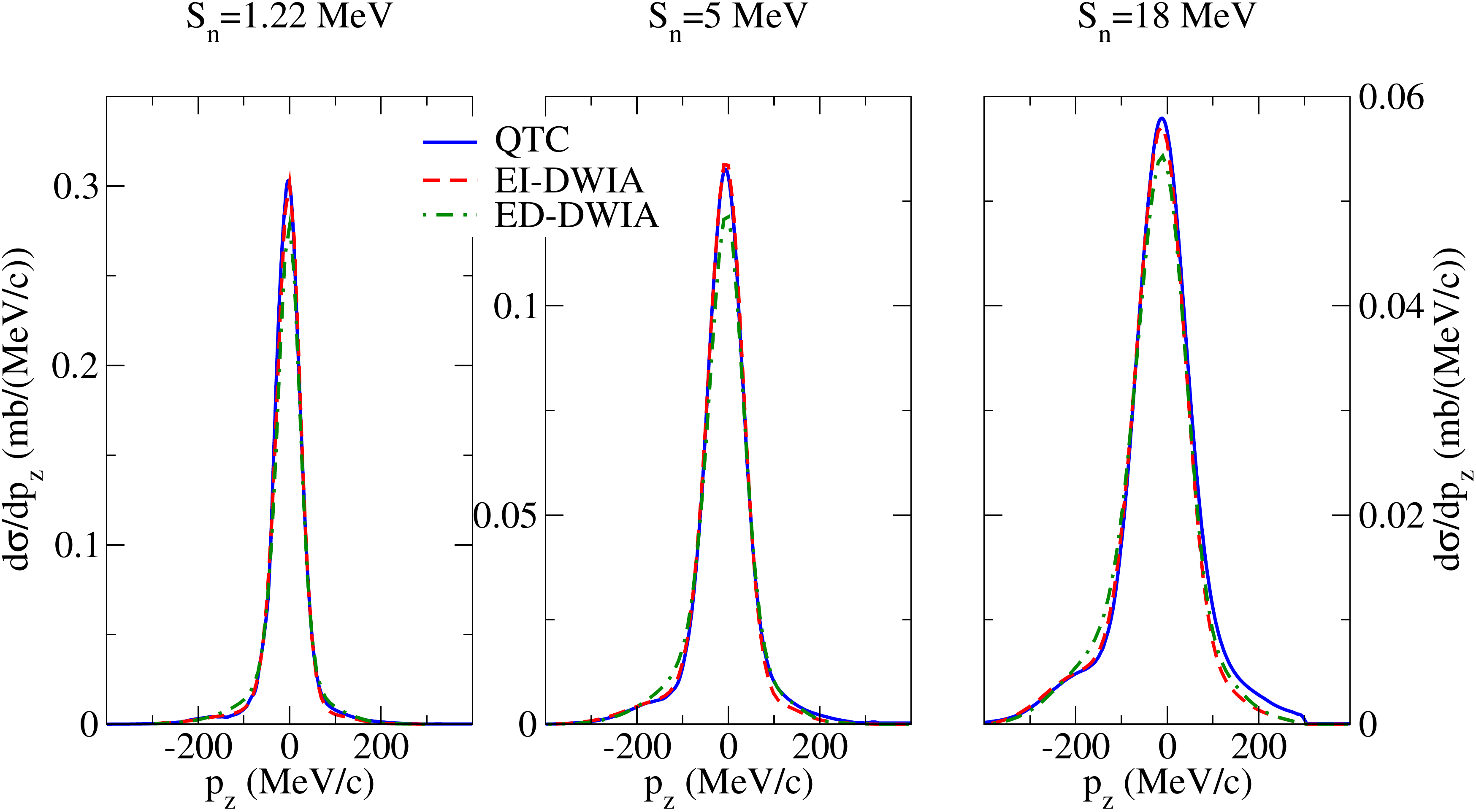}
\caption{\label{fig:bench_DWIA} (Color online) Longitudinal momentum distribution for the be $^{15}$C$(p,pn)$ at 420~MeV/nucleon. The removal of a neutron bound by 1.22, 5 and 18~MeV is shown in the figures from left to right. The blue solid line corresponds to the QTC calculation, while the dashed red line corresponds to a DWIA calculation using the same prescription of optical potentials as QTC (at half incident energy), while the green dot-dashed line corresponds to DWIA with energy-dependent potentials. Figure adapted with permission from \cite{Yos18}\textcopyright2018 by the American Physical Society.}
\end{figure}

Since the analysis  of $(p,pN)$ data has traditionally been performed with the DWIA method,  benchmark comparisons have been performed between it and the QTC method \cite{Yos18}. An example of such a comparison is shown in Fig.~\ref{fig:bench_DWIA}, corresponding to the $^{15}$C$(p,pn)$ reaction at 420~MeV. The removal of a neutron from the $2s_{1/2}$ orbital has been assumed, and the dependence on the binding energy has been assessed by comparing calculations with a neutron bound by 1.22, 5 and 18 MeV.

The comparison of the results of both methods has shown an excellent degree of agreement, with a difference of at most 5\% in the cross section, in all considered cases. In this benchmark calculation, the effect of the prescription for the choice of the energy for the evaluation of the potentials has also been explored, finding that the choice taken in QTC calculations (evaluate the potentials for the outgoing nucleons at half the incident energy per nucleon, labelled EI in Fig.~\ref{fig:bench_DWIA}) leads to changes of $\sim$3\% in the overall cross section, and at most of 8\% in the peak, when compared to the calculations with fully energy-dependent potentials (labelled ED). 

\begin{figure}[t]
\centering
\includegraphics[width=5.in]{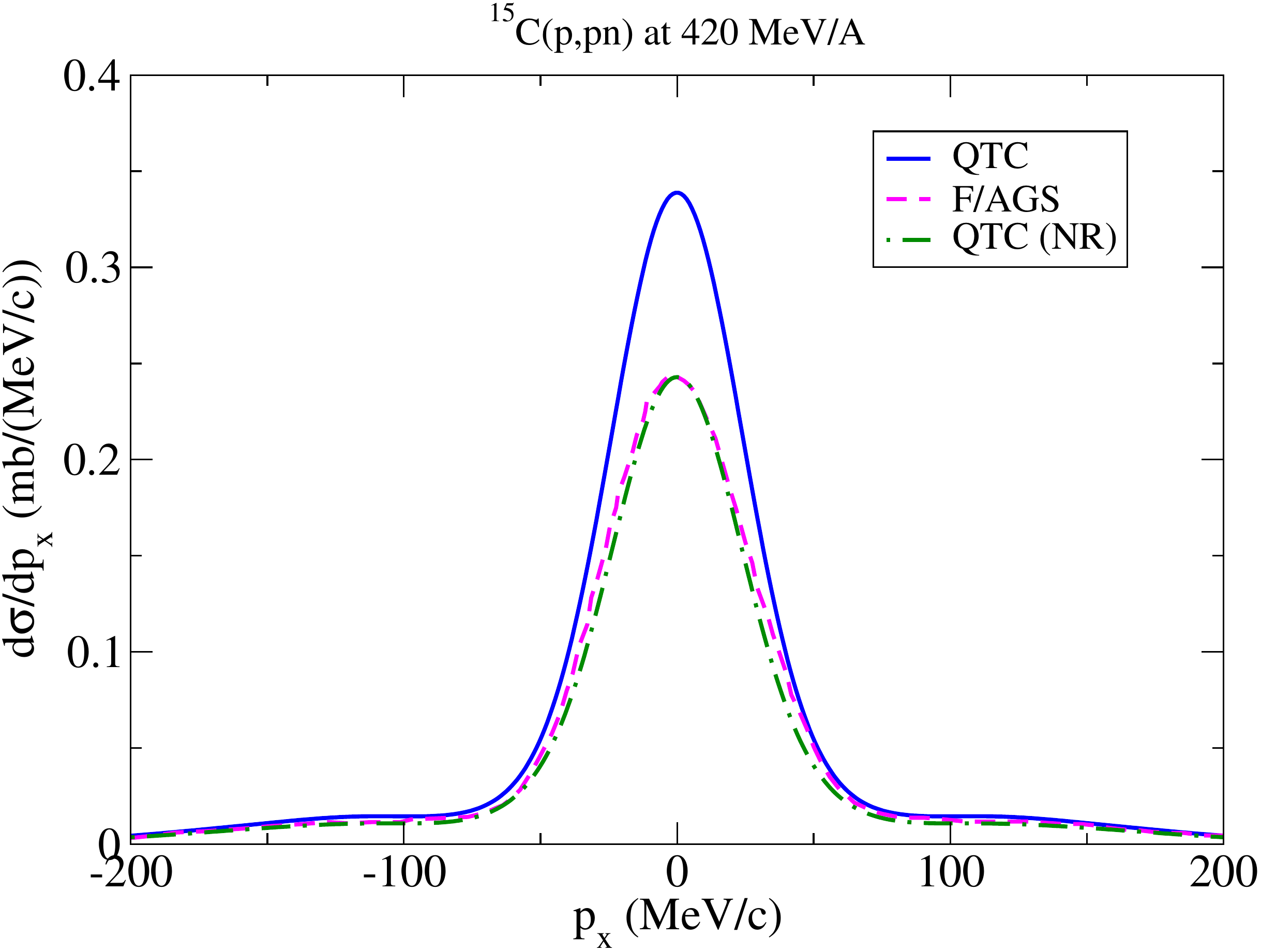}
\caption{\label{fig:c15ppn_tc_fags}  (Color online) Transverse momentum distribution of the $^{14}$C nucleus resulting from the  $^{15}$C($p,pn$)$^{14}$C reaction at 420~MeV/nucleon. The dashed magenta and dot-dashed green lines are the Faddeev/AGS calculation (quoted from Ref.~\cite{cravo2016}) and the QTC calculation, without relativistic corrections. The solid blue line is the QTC calculation including relativistic corrections in the kinematics. Figure adapted with permission from Ref. \cite{Yos18}\textcopyright2018 by the American Physical Society.}
\end{figure}

The reaction and energy for this benchmark calculation were chosen to be compatible with a calculation presented in \cite{cravo2016} with a binding energy of 1.22 MeV, where the non-relativistic Faddeev/AGS formalism was used to analyze the reaction. As such, we present in Fig.~\ref{fig:c15ppn_tc_fags} a comparison between QTC and Faddeev/AGS methods. Due to the non-relativistic nature of the Faddeev/AGS calculations in Ref. \cite{cravo2016}, for this comparison relativistic corrections were removed from the QTC calculations (the blue solid line corresponds to the full calculation, while the green dot-dashed line corresponds to the non-relativistic one). As can be seen in the figure an excellent agreement is obtained with the Faddeev/AGS results (magenta dashed line), showing the consistency of both methods when relativistic effects are removed. This comparison highlights also the importance of these relativistic effects at typical incident energies used at GSI.

\subsection{Application to lower energies: final-state interactions \label{sec:fsiTC}}

\begin{figure}[t]
 \centering
\includegraphics[width=4in]{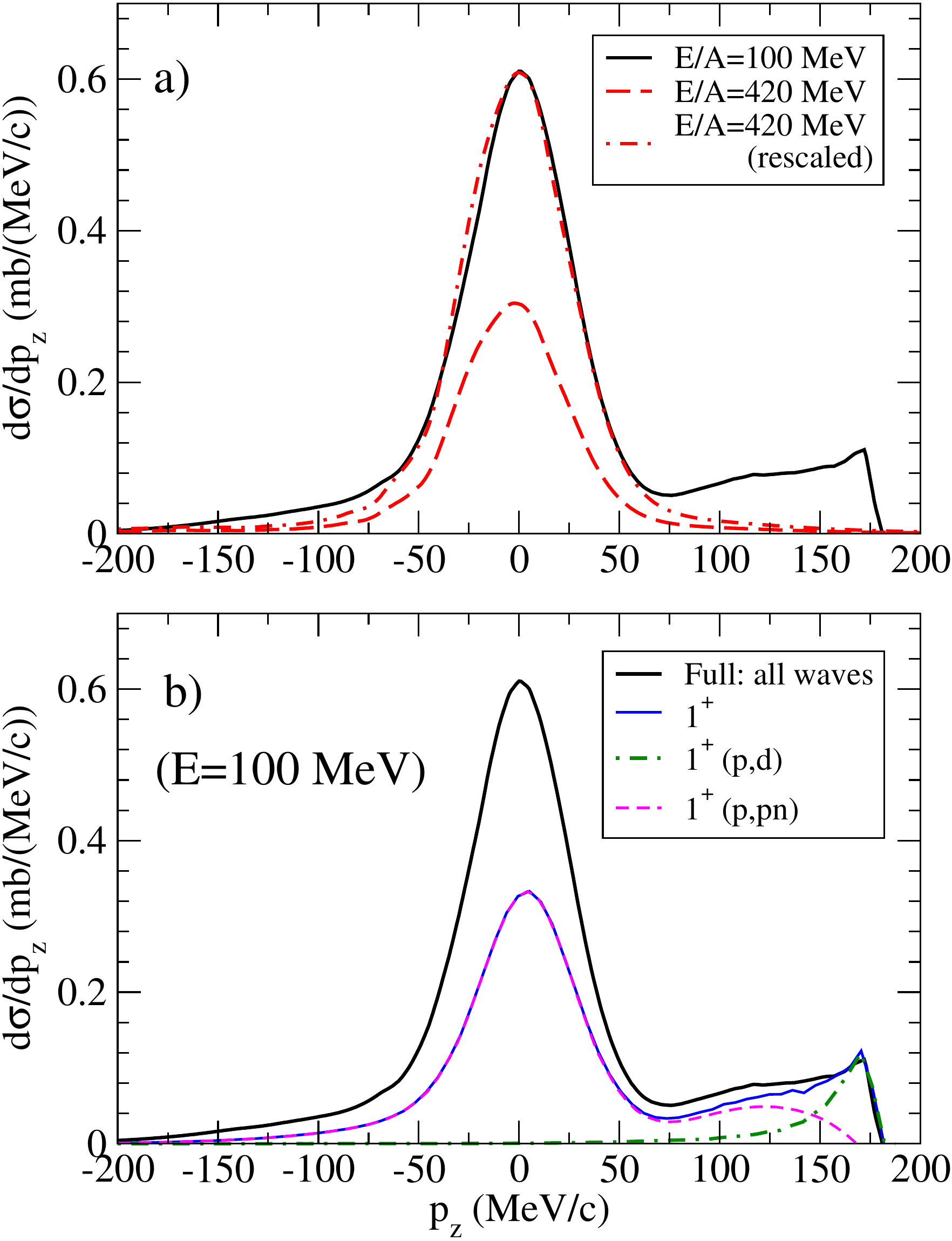}
\caption{\label{fig:fsi}  (Color online) a) Longitudinal $p_z$ momentum distribution in the projectile rest frame of reference for the $^{15}$C$(p,pn)$ reaction at an incident energy of 100~MeV/nucleon (black curve) and of 420~MeV/nucleon (red dashed curve). The distribution at 420~MeV/nucleon is also shown rescaled in the dot-dashed curve for better comparison. b) $p_z$ momentum distribution at 100~MeV/nucleon. The distribution leading to final $p-n$ states in the 1$^+$ configuration is shown in solid blue, indicating its components corresponding to $(p,pn)$ breakup (green dot-dashed) and to $(p,d)$ transfer (magenta dashed).}
\end{figure}

Due to its use of the prior-form transition amplitude, QTC is especially suitable to deal with the so-called final-state interaction between the outgoing proton and nucleon, which gains relevance as the incident energy of the projectile decreases. This is illustrated in the top panel of  Fig.~\ref{fig:fsi}, where the longitudinal momentum distribution is shown for the same reaction, $^{15}$C$(p,pn)$, at two different incident energies, 100 and 420~MeV/nucleon. It is apparent in the figure that the main peak in the distribution reflects the internal momentum distribution of the removed neutron, showing the same width for both incident energies. However, a remarkable increase in the cross section is seen at the high-momentum tail of the reaction at 100 MeV, which is further explored in the bottom panel, where it is shown that it is mainly due to the strong interaction of the proton-neutron pair in the $1^+$ configuration. This strong interaction leads to the increase of the $(p,pn)$ cross section at high momenta, where the relative energy of the proton-neutron pair is smaller. The importance of the $(p,d)$ transfer channel is especially remarkable at this low energy, which produces a sharp peak at the end of the high-momentum tail. In the bottom panel $(p,pn)$ neutron removal and $(p,d)$ transfer are shown separately to highlight their different contributions to the overall distribution.

\begin{figure}[t]
 \centering
\includegraphics[width=6in]{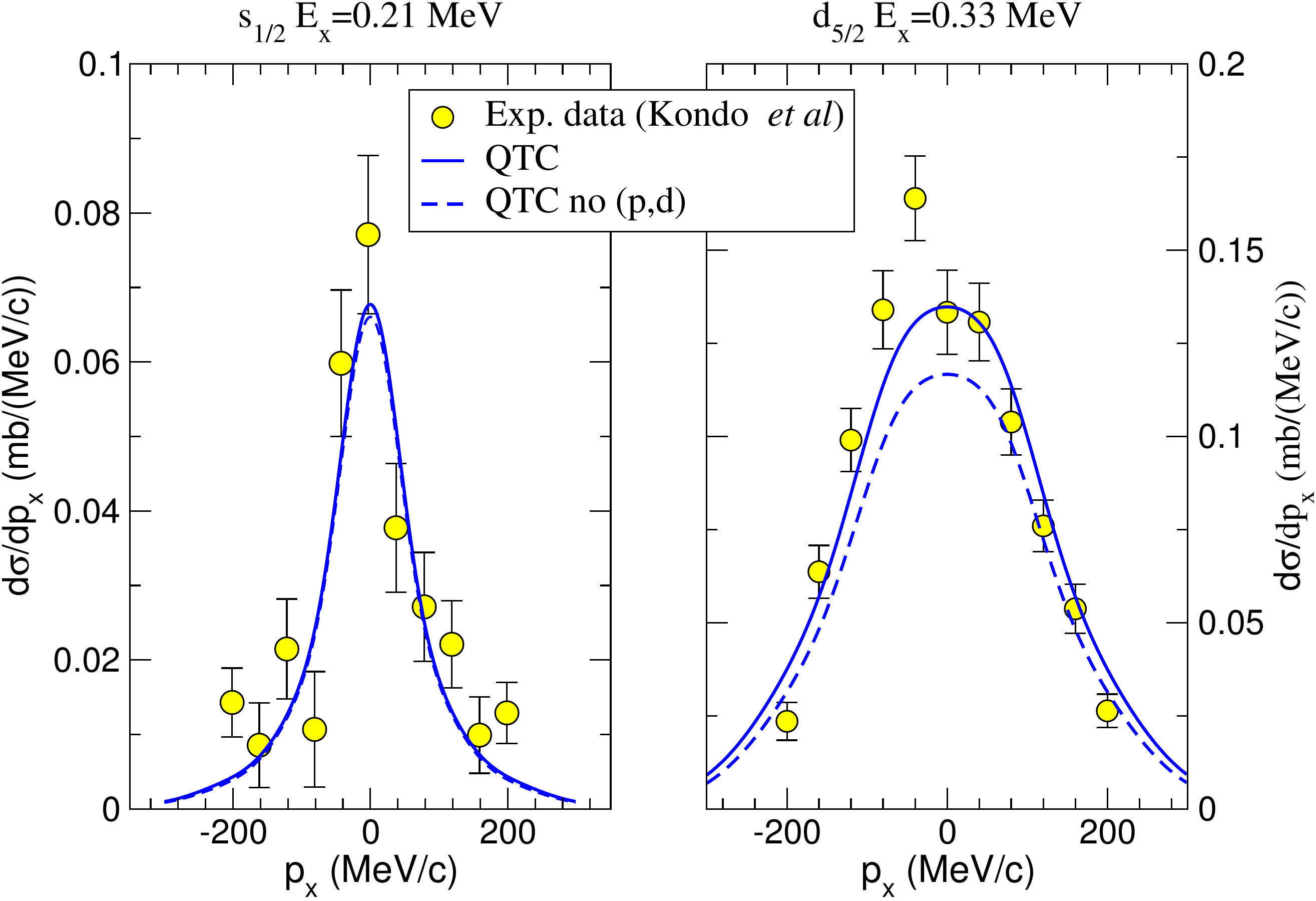}
\caption{\label{fig:c18}  (Color online) Transverse $p_x$ momentum distributions for the $^{18}$C$(p,pn)^{17}$C$^*$ reaction at 81~MeV/nucleon. Experimental data are from Ref. \cite{Kon09}. The left panel corresponds to $(p,pn)$ leading to the $E_x=0.21$ MeV $^{17}$C excited state, while the right panel corresponds to the $E_x=0.33$ MeV excited state. The blue solid line corresponds to the full calculation rescaled to give the experimental total cross section. The blue dashed line corresponds to the calculation removing the contribution from the $(p,d)$ transfer reaction, rescaled by the same factor as the full calculation. All theoretical calculations have been convoluted with the experimental resolution.}
\end{figure}

The capacity of the QTC formalism to treat $(p,d)$ transfer consistently with the $(p,pn)$ cross section is especially relevant in $A(p,pn)A-1$ reactions in which the outgoing proton and neutron are not detected, or where the proton can not be resolved from a deuteron, since in these reactions both the $(p,d)$ and $(p,pn)$ reactions contribute to the measured cross section. The relevance of the transfer channel is usually neglected, which may be justified at higher energies (100s of MeV), since the momentum matching conditions favouring transfer are not fulfilled. However, there are experiments at lower energies, $\lesssim$100 MeV, where the $(p,d)$ channel can contribute to the cross section and must be assessed. As an example, we present in Fig.~\ref{fig:c18} an analysis of the $^{18}$C$(p,pn)$ reaction, measured at 81~MeV/nucleon \cite{Kon09} leading to different excited states of $^{17}$C.

In this analysis, one finds a sizable contribution of the $(p,d)$ transfer channel leading to the population of the $^{17}$C excited state at $E_x=0.33$ MeV, contributing as much as 15\% of the overall cross section, while for the $E_x=0.21$ MeV state, the contribution is almost negligible, around 3\%. At these rather low energies, methods such as the Continuum-Discretized Coupled-Channel (CDCC) have been used \cite{Kon09,Oza11}, which do not include the transfer channel contribution, so they are only reliable if the transfer cross section can be assessed to be small enough. In contrast, QTC can be used confidently for $(p,pn)$ reactions at these low energies even when $(p,d)$ transfer gives a sizable contribution. The conspicuous peak at the tail of the longitudinal momentum distribution shown in Fig.~\ref{fig:fsi} could serve as a telltale sign of the significance of the transfer channel in these cases.

\subsection{Application to $R^3B$ inclusive data}
The QTC model has been employed to analyse recent experimental data from several $(p,2p)$ and $(p,pn)$ reactions measured at GSI by the $R^3B$ collaboration. One of the purposes of these experiments is to investigate the depletion of the $s.p.$ strength as a function of the separation energy of the extracted nucleon. As discussed in previous sections, in the case of nucleon-removal reactions in nucleus-nucleus collisions, analyses performed with the eikonal reaction theory suggest a significant departure of this depletion (quantified in terms of the spectroscopic factors) with respect to the small-scale shell-model calculations and, more notably, this deviation increases for increasing separation energies \cite{gade08b}. Although the effect has been attributed by some authors to the presence of additional, short-range correlations not present in standard shell-model calculations, some other authors have put into question the adequacy of the reaction model used in these analyses. Proton-induced nucleon-removal cross sections have been put forward as a simpler alternative to study this effect, and the R$^3$B experimental campaign has provided a systematic of $(p,pN)$ reactions on oxygen isotopes. These data have been analysed also with the DWIA \cite{atar2018} and  Faddeev/AGS \cite{Dia18} frameworks.

In the study performed with the QTC method, $s.p.$ one-nucleon removal cross section were multiplied by shell-model spectroscopic factors and the resultant cross sections compared with the data. Since most of these data are inclusive with respect to the excitation of the residual nucleus, when several bound states are present, the cross section is computed and added for  all these states. 

\begin{figure}[!t]
\centering
\includegraphics[width=5.3in]{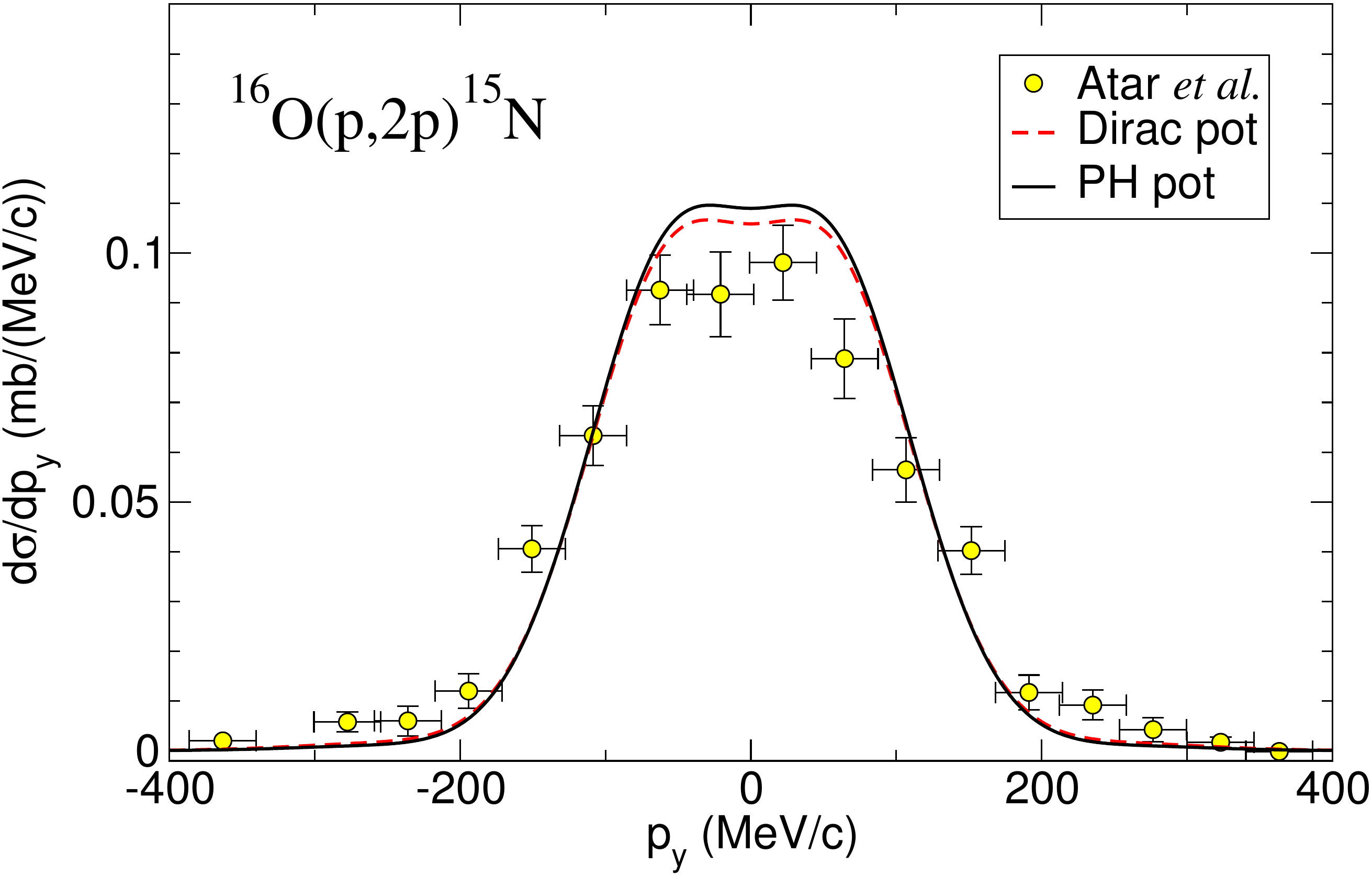}
\caption{\label{fig:o16p2p} (Color online) Transverse momentum distribution for $^{16}$O$(p,2p)$. Experimental data are taken from Ref.~\cite{atar2018}. The solid and dashed  lines correspond to the QTC calculations using Paris-Hamburg and Dirac potentials for nucleon-nucleus distorted waves, respectively. The calculations have been rescaled by the quenching factors required to reproduce the integrated $(p,2p)$ cross section, giving rise to $R_s=0.78$ and 0.74 for the Paris-Hamburg and Dirac potentials, respectively. Figure adapted with permission from Ref.~\cite{Gom18}\textcopyright2018 by Elsevier.}
\end{figure}

In addition to the integrated cross sections, momentum distributions have been also compared in selected cases.  An example is shown in Fig.~\ref{fig:o16p2p}, corresponding to the transverse momentum distribution of the $^{15}$N residual nucleus following the reaction  $^{16}$O$(p,2p)$. Two calculations are shown, corresponding to two different choices of the nucleon-nucleus optical potentials: the phenomenological Dirac parametrization \cite{Cooper:1993} and the microscopic g-matrix folding calculation, based on the effective Paris-Hamburg nucleon-nucleon g-matrix \cite{Ger83,Rik84}. The shape of the calculated momentum distribution is very similar for both potential sets, and turns out to be slightly narrower, but consistent, than the experimental one. However, both of the calculations overestimate the magnitude of the cross section, attesting to the well-known depletion of the $s.p.$ strength. In the plot, the calculations have been rescaled by a so-called reduction factor ($R_s$), defined as the ratio between the experimental and theoretical integrated cross sections. The extracted values ($R_s=0.78$ and $R_s=0.74$ for the Paris-Hamburg and Dirac potentials respectively) indicate a reduction of the spectroscopic factor (or the $s.p.$ occupancy) with respect to the assumed shell-model calculation.

\begin{figure}[t]
\centering
\includegraphics[width=5.3in]{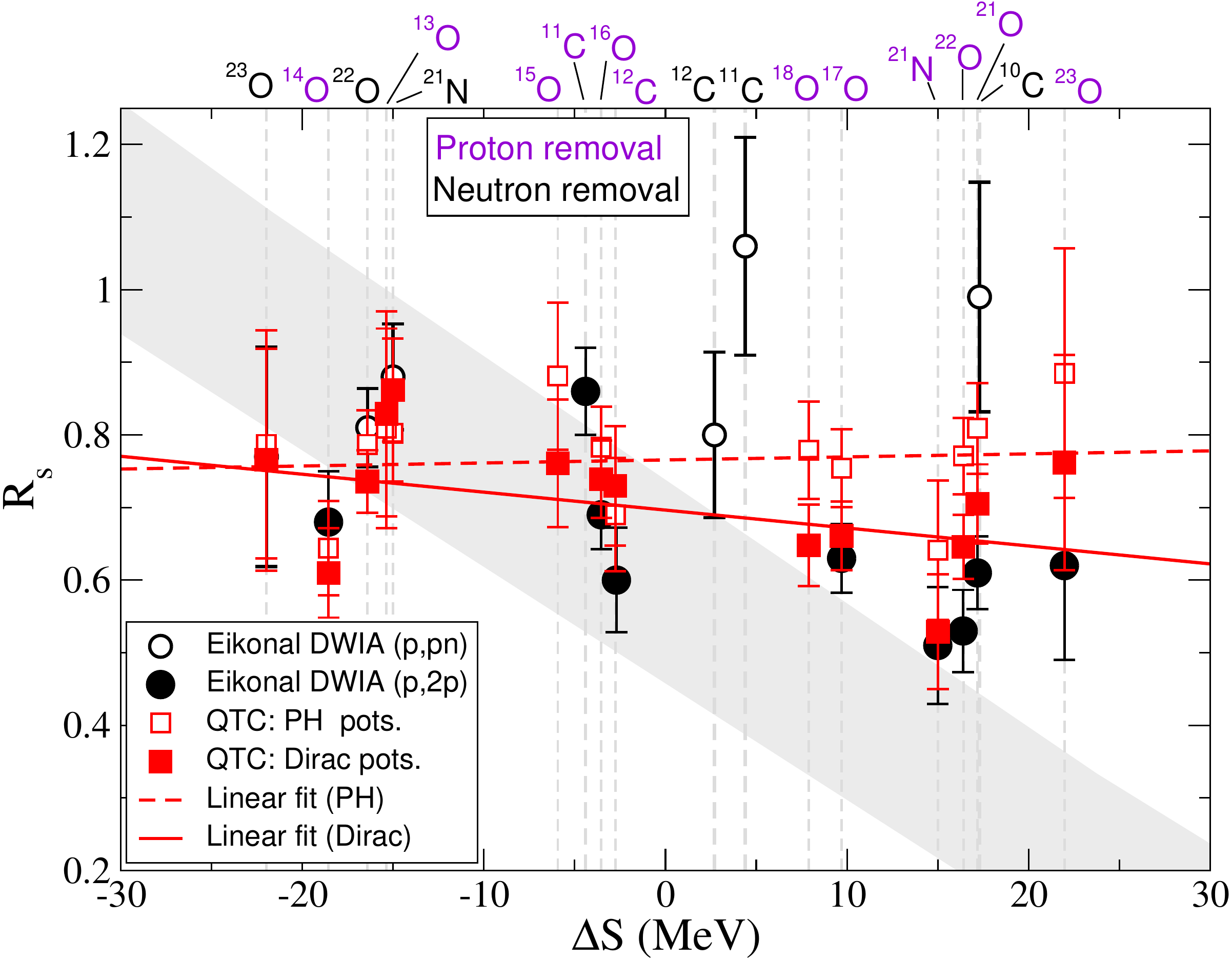}
\caption{\label{fig:Rs}  (Color online) Reduction factors obtained for different $(p,pn)$ and $(p,2p)$ reactions as a function of $\Delta S$ (see text). Red filled (open) squares correspond to calculations using Dirac (Paris-Hamburg) potential. Linear fits of each set are represented by the read solid and dashed dashed lines, respectively. Filled (open) black circles correspond to the eikonal DWIA analysis of $(p,2p)$ ($(p,pn)$) cross sections performed in \cite{atar2018} (\cite{HOLL2019682}). The grey band indicates the trend found for intermediate-energy nucleon-knockout reactions induced by $^9$Be and $^{12}$C target nuclei \cite{gade08b}. Figure extended and adapted with permission from Ref.~\cite{Gom18}\textcopyright2018 by Elsevier.}
\end{figure}

The dependence of the reduction factor with respect to the difference of the proton and neutron separation energies ($\Delta S$) is displayed in Fig.~\ref{fig:Rs}. Experimental cross sections are from Refs~\cite{atar2018,Dia18,HOLL2019682}.  As before, calculations using the phenomenological (Dirac) and microscopic (Paris-Hamburg) potentials are displayed. For comparison purposes, the results obtained from the analysis based on the eikonal DWIA reaction theory of Ref.~\cite{aumann2013}, and quoted in Refs.~\cite{atar2018,HOLL2019682}, are also shown. We see that the $R_s$ values extracted from the comparison of the experimental and theoretical (QTC) cross sections are systematically lower than unity and display a small separation energy dependence, at variance with the values from nucleus-nucleus knockout (shaded area), but  consistent with the findings in Ref. \cite{atar2018} and with systematic analyses of transfer reactions~\cite{Flavigny13} and other recent $(p,2p)$ measurements \cite{Kaw18}. For the smaller separation energies  ($\Delta S <0$) the extracted $R_s$ values are essentially independent of the optical model prescription, but this dependence becomes larger as the separation energy increases. Interestingly, the $R_s$ values extracted with the Dirac potential turn out to be rather close to those obtained with the eikonal DWIA results, despite the markedly different origins and assumptions of this formalism and QTC. Comparing the eikonal DWIA analysis for $(p,2p)$ (full black symbols) and $(p,pn)$ (open black symbols), we recognize that reduction factors for neutron-knockout are systematically larger compared those for proton knockout, with average values of $R_s=0.85(10)$ and $R_s=0.65(4)$, respectively \cite{HOLL2019682}. However, the $(p,pn)$ data are still scarce and not very precise in particular for large $\Delta S$, calling for new measurements to corroborate and better quantify this result.     

\subsection{Discussion on the DWIA description of nucleon knockout reactions by proton in normal and inverse kinematics}
\label{C12}
In this section, we briefly recapitulate the result of the distorted-wave impulse approximation (DWIA) analysis of the proton-induced proton knockout, $(p,2p)$, reactions shown in a recent review article~\cite{Wakasa17}. We focus on the aspect of $(p,2p)$ reactions as a spectroscopic tool to extract the proton $s.p.$ structure of nuclei, as an alternative to ($e,e'p$) reactions. Then, we apply the DWIA framework to the analysis of the $(p,2p)$ cross section on $^{12}$C measured in inverse kinematics~\cite{Pan16}. The detailed formalism of DWIA, which is essentially the same as in Ref.~\cite{ChantRooss83}, is given in Section~3.1 of Ref.~\cite{Wakasa17}.

\begin{figure}[ht]
\begin{center}
\includegraphics[width=5.3in,clip]{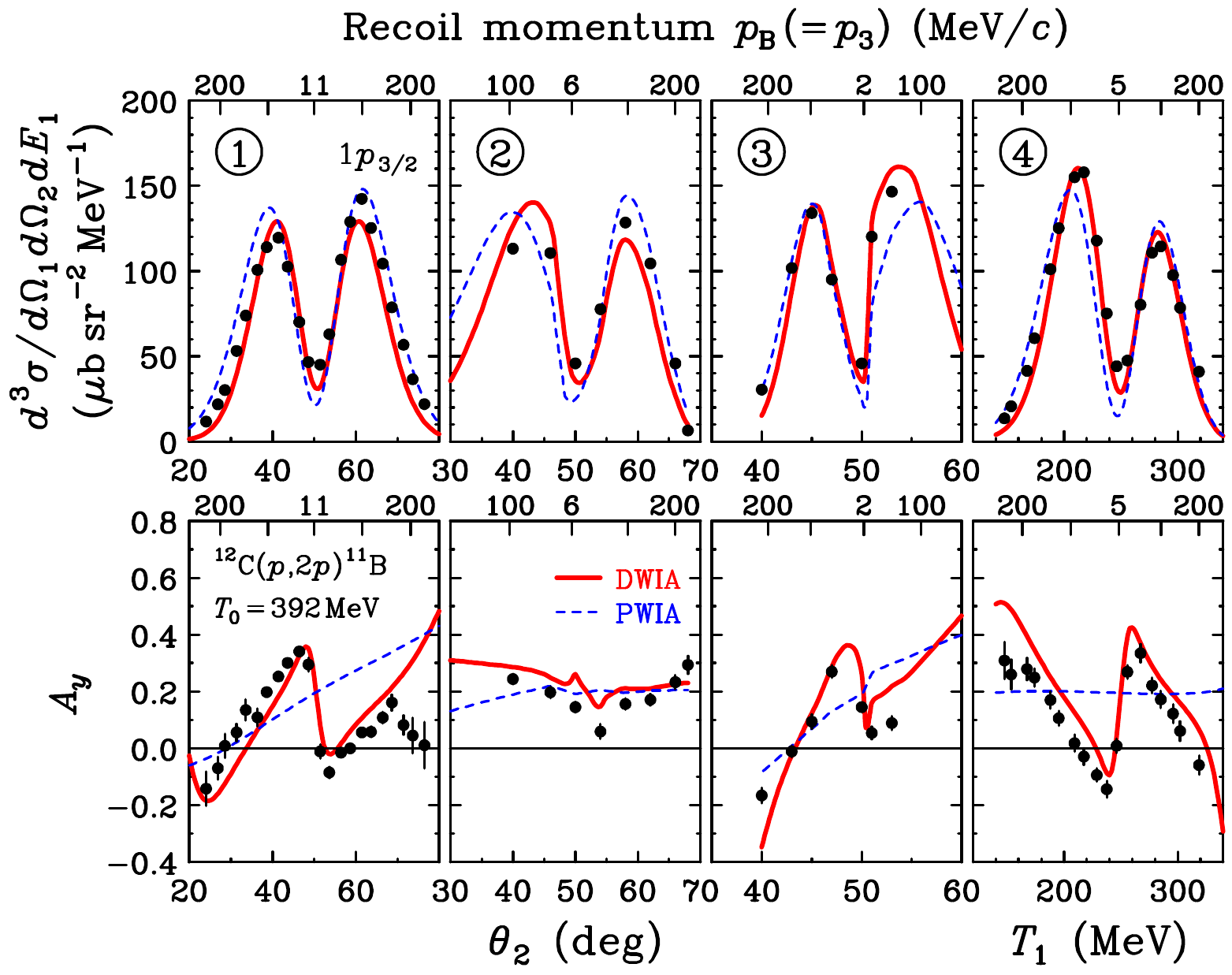}
\end{center}
\caption{TDX and $A_y$ for ${}^{12}{\rm C}(p,2p){}^{11}{\rm B}_{\rm g.s.}$ at 392~MeV calculated with DWIA (solid lines). The results calculated with the plane-wave impulse approximation (PWIA) are also shown by the dashed lines for comparison. Figure adapted with permission from \cite{Wakasa17}. Copyright (2020) by Elsevier.}
\label{12cp2pdwia}
\end{figure}
In Fig.~\ref{12cp2pdwia}, we show the result of the DWIA analysis on ${}^{12}{\rm C}(p,2p){}^{11}{\rm B}_{\rm g.s.}$ at 392~MeV. The upper and the lower panels correspond to the triple differential cross section (TDX) and the analyzing power ($A_y$), respectively. \textcircled{\scriptsize 1}--\textcircled{\scriptsize 4} represent four different choices of the kinematical setting. For instance, in \textcircled{\scriptsize 1}, the scattering angle and the kinetic energy of one of the two emitted protons are fixed and those for the other are varied. In \textcircled{\scriptsize 4}, the outgoing energies of the two protons are varied with fixing the scattering angles of them; the resulting observable in this setting is sometimes called the energy sharing distribution. For more details, see Fig.~16 of Ref.~\cite{Wakasa17}. The observables are plotted as a function of the scattering angle $\theta_2$ and the kinetic energy $T_1$ of one of the two protons for \textcircled{\scriptsize 1}--\textcircled{\scriptsize 3} and \textcircled{\scriptsize 4}, respectively. The upper abscissa in each panel represents the momentum of the residue ($^{11}$B in this case). In all the four kinematical conditions, the recoilless condition, in which the reaction residue is at rest, is achieved as much as possible. One sees clearly that the DWIA describes the experimental data very well, though a slight deviation is found for $A_y$. More importantly, the spectroscopic factor $S(p,2p)$ deduced from the comparison with the data is 1.82(3), which well agrees with that deduced from the $(e,e'p)$, that is, $S(e,e'p)=1.72(11)$~\cite{Kramer:1989,Kramer:2001}. This is the case also with the components for the $1/2^-$ and $3/2^-$ excited states of $^{11}$B; $S(p,2p)$ and $S(e,e'p)$ are shown in Table~\ref{sfacdwia} below. A systematic (re)analysis of the existing $(p,2p)$ data has been done in Ref.~\cite{Wakasa17} and it was concluded that the $(p,2p)$ reactions work as a spectroscopic tool; $S(p,2p)$ were found to be consistent with $S(e,e'p)$ within 15\% at energies above 200~MeV and within 30\% in the vicinity of 100~MeV.

Based on the results of \cite{atar2018}, a phenomenological approach to examine the role of short- and long-range nucleon-nucleon correlations in the quenching of {\it s.p.} strength and in particular how they evolve in asymmetric nuclei has been proposed \cite{paschalis2020}.  It has been shown that the $(p,2p)$ data analysed with the eikonal is in agreement with a isospin dependence predicted in this simple model assuming a 20\% contribution of SRC as extracted from the JLab data \cite{duer2019}. However, the systematics of quenching in the $(p,2p)$ depends slightly on the reaction-model used to analyse the data (see Fig. \ref{fig:Rs}).

As emphasized in Ref.~\cite{Wakasa17}, the success in reproducing not only the shape of the observables but also the absolute value of the TDX relies on the use of a reliable global optical potential that is applicable to various nuclei and in a wide range of energies; the global Dirac-type parameter, developed by Cooper {\it et al.} \cite{Cooper:1993} was employed: the parametrization EDAD1 for the exit channel and the parametrization EDAI for the incident one. Another important feature of the DWIA analysis is that the same proton $s.p.$ wavefunction as in the ($e,e'p$) analysis was adopted. Furthermore, the correction to the bound-state wavefunction and the distorted waves due to the nonlocality of the nucleon-nucleus potential is crucial to obtain a quantitatively reliable TDX. Details of the correction for nonlocality and its outcome are given in Ref.~\cite{Wakasa17}. It should be noted, however, that the prescription for the correction used so far is based on a phenomenological approach, which can be insufficient for fully accounting for the effect of nonlocality. Recently, the dispersive optical model (DOM)~\cite{Dickhoff:17}, which explicitly computes the wavefunctions with directly treating the nonlocality, was applied to the $^{40}$Ca($e,e'p$) reaction and a larger $S(e,e'p)$ than that obtained in the preceding analysis by about 10\% was obtained~\cite{Atkinson:2018}. It will be interesting and important to use the DOM wavefunctions in the DWIA analysis of the ($p,2p$) reactions. The last piece of the ingredients important for the quantitative calculation is the M{\o }ller factor \cite{Mol45} that transforms the nucleon-nucleon (NN) transition matrix from the NN c.m. frame to that in the c.m. frame of the ($p,2p$) reaction system. The importance of the M{\o }ller factor has been discussed in the literature, e.g. in the multiple-scattering-theory paper by Kerman, McManus and Thaler \cite{KERMAN1959551}, and included also in the preceding studies on the charge exchange ($p,n$) reactions~\cite{Ichimura:2006mq}. It is known to somewhat reduce the cross section at higher energies; we will return to this point below.

For further investigation on the nucleon knockout reactions from a theoretical point of view, we carried out a benchmark study on the ${}^{15}{\rm C}(p,pn){}^{14}{\rm C}_{\rm g.s.}$ at 420~MeV~\cite{Yos18}, with three different reaction framework, that is, the DWIA, the quantum transfer to the continuum model (QTC)~\cite{Mor15}, and the transition amplitude formulation of the Faddeev equations due to Alt, Grassberger, and Sandhas (FAGS)~\cite{AGS}. As shown in Fig.~\ref{fig:bench_DWIA}, the momentum distribution of $^{14}$C calculated with these three are in excellent agreement with each other, if the same ingredients are adopted. A remark on this benchmark study is that the QTC with relativistic kinematics includes the effect of the M{\o }ller factor, though it does not appear in the formalism. This is because the QTC does not use the NN transition matrix but solves a three-body equation with a Hamiltonian that contains a bare NN interaction. It will be very important that the DWIA with the M{\o }ller factor gives the same numerical result as that obtained with the QTC with a different prescription of the relativistic kinematics, when the same inputs are used and the non-locality corrections are ignored. Further benchmark studies on $(p,pN)$ reactions in a wide range of energies are ongoing.\\

Recently, the ($p,2p$) reactions in inverse kinematics have been utilized to extract $s.p.$ structure of unstable nuclei. To establish such measurements as a reliable spectroscopic tool as the ($p,2p$) in normal kinematics, a measurement of ${}^{12}{\rm C}(p,2p){}^{11}{\rm B}$ at around 400~MeV/nucleon has been performed at GSI~\cite{Pan16} in inverse kinematics. Because the same reaction in normal kinematics at almost the same energy has been successfully described by the DWIA as discussed above, we applied the same framework to the ${}^{12}{\rm C}(p,2p){}^{11}{\rm B}$ in inverse kinematics.
\begin{table*}[ht]
\begin{center}
\caption{$S(p,2p)$ obtained from the DWIA analysis of the $^{12}$C($p,2p$)$^{11}$B data for the specific
spin-parity $J^\pi$ and the excitation energy $E_{\rm ex}$ of $^{11}$B. For the analysis of the GSI data taken
in inverse kinematics, the results calculated with the DWIA are shown; w/o NL and w/o M{\o }ller mean the calculation
neglecting the effect of nonlocality and the M{\o }ller factor, respectively, and both are neglected in the w/o NL\&M{\o }ller
case. $S(p,2p)$ determined with the analysis of the RCNP data taken in normal kinematics \cite{Wakasa17} and $S(e,e'p)$ \cite{Ste88} are also displayed.}
\vspace*{3mm}
\begin{tabular}{lcccccc}
\hline
      & \multicolumn{4}{c}{GSI} & RCNP & $(e,e'p)$ \\
\hline
$J^\pi(E_{\rm ex} \text{ (MeV)})$ & DWIA & w/o NL & w/o M{\o }ller & w/o NL\&M{\o }ller & DWIA & DWIA \\
$3/2^-(0)$    & 3.04 & 2.66 & 2.41 & 2.11 & 1.82(3) & 1.72(11) \\
$1/2^-(2.13)$ & 0.32 & 0.29 & 0.26 & 0.23 & 0.30(2) & 0.26(2)  \\
$3/2^-(5.02)$ & 0.28 & 0.25 & 0.22 & 0.20 & 0.23(3) & 0.20(2)  \\
\hline
\end{tabular}
\label{sfacdwia}
\end{center}
\end{table*}
%

In Table~\ref{sfacdwia}, we show $S(p,2p)$ extracted by taking the ratio of the measured ($p,2p$) cross sections for the $3/2^-$ ground state and the $1/2^-$ and $3/2^-$ excited states of $^{11}$B~\cite{Pan16}, to the values obtained with the DWIA that adopts a normalized proton $s.p.$ wavefunction. The same ingredients as in the calculation shown in Fig.~\ref{12cp2pdwia} are used. The results of $S(p,2p)$ based on the analysis of the data taken at RCNP in normal kinematics \cite{Wakasa17} are also listed, as well as $S(e,e'p)$ \cite{Ste88}. In what follows, we denote the $S(p,2p)$ based on the DWIA analysis of the GSI and RCNP data by $S(p,2p)_{\rm GSI}$ and $S(p,2p)_{\rm RCNP}$, respectively. One sees that $S(p,2p)_{\rm GSI}$ are larger than $S(p,2p)_{\rm RCNP}$ for all the three states of $^{11}$B. In particular, $S(p,2p)_{\rm GSI}/S(p,2p)_{\rm RCNP}$ is even 1.67 for the ground state component of $^{11}$B. For each component, neglect of the effect of nonlocality and the M{\o }ller factor increases the cross section, hence reduces the $S(p,2p)$, by about 10\% and 20\%, respectively. If we disregard both, about 30\% reduction of the $S(p,2p)$ is obtained. Although the {\lq\lq}w/o NL\&M{\o }ller'' calculation seems to result in much better consistency with $S(p,2p)_{\rm RCNP}$ and $S(e,e'p)$, obviously, such a calculation alters also $S(p,2p)_{\rm RCNP}$. Therefore it will be difficult for the DWIA framework, which was successfully applied to various $(p,2p)$ processes in Ref.~\cite{Wakasa17}, to consistently explain the cross sections of the ${}^{12}{\rm C}(p,2p){}^{11}{\rm B}$ reaction measured in normal and inverse kinematics.

\begin{figure}[t]
\begin{center}
\includegraphics[width=6in,clip]{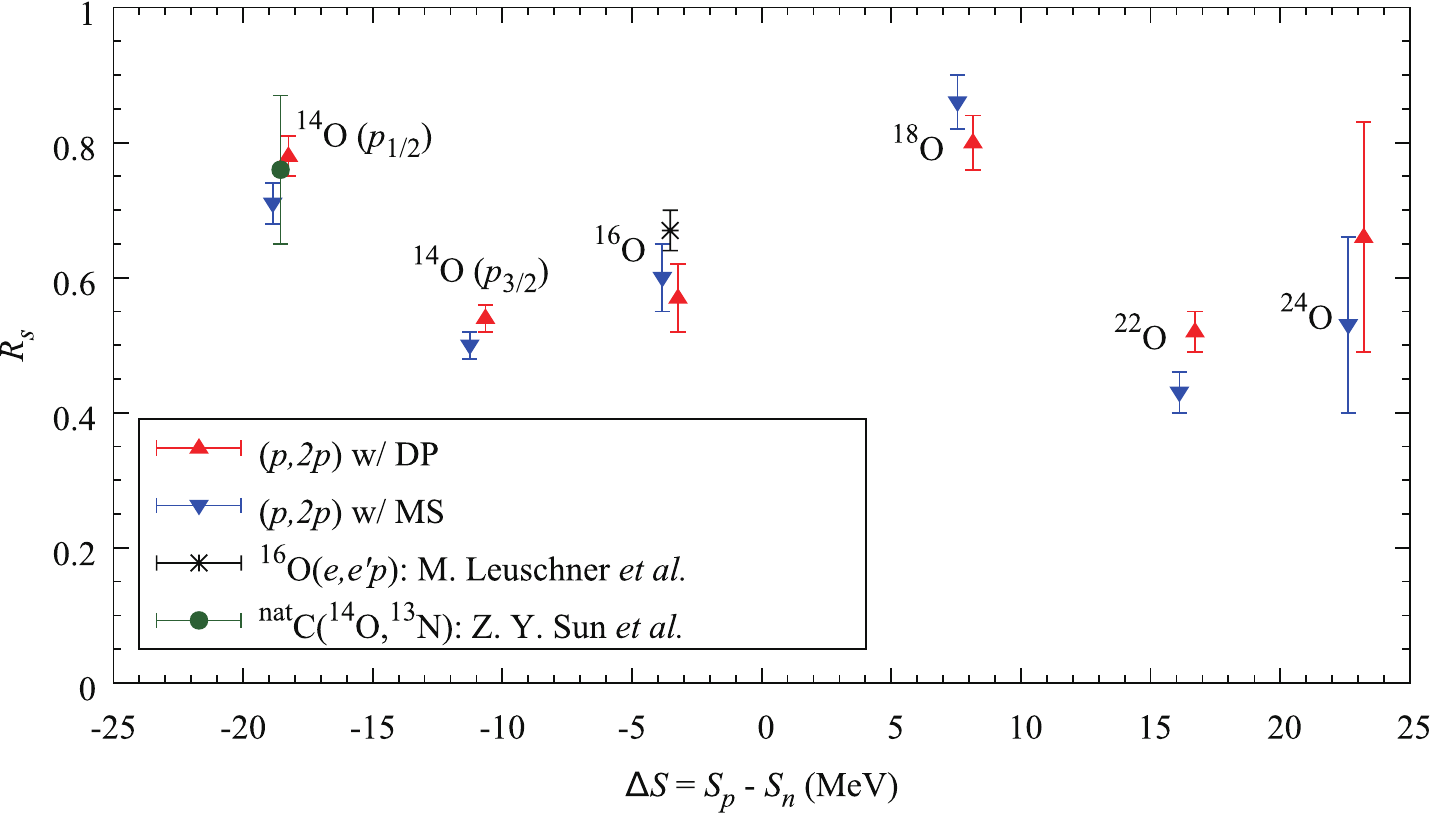}
\end{center}
\caption{ (Color online) Reduction factor $R_s$, the ratio of $S(p,2p)$ to the spectroscopic factor calculated with the shell model, for oxygen isotopes. The horizontal axis is the difference between the proton and neutron separation energies. The triangles (inverted triangles) correspond to the results of the DWIA with the optical potential of the Dirac phenomenology~\cite{Cooper:1993} (the $g$-matrix folding model~\cite{Toyokawa:2013}). For $^{16}$O, $R_s$ based on $S(e,e'p)$ is shown by the asterisk. The circle shows the result obtained by a proton removal cross section for $^{14}$O with a carbon target. Figure adapted with permission from Ref.~\cite{Kaw18}\textcopyright2018 by the Physical Society of Japan.}
\label{kawase}
\end{figure}
Recently, a systematic analysis of the $(p,2p)$ reactions on oxygen isotopes at around 250~MeV has been accomplished~\cite{Kaw18}. The results are summarized in Fig.~\ref{kawase}. An important finding is that the reduction factor $R_s$ has a weaker dependence on $\Delta S=S_p - S_n$, where $S_p$ ($S_n$) is the proton (neutron) separation energy, than in Ref.~\cite{tostevin14}. However, we here focus on another aspect of the result. In Fig.~\ref{kawase}, the results for $(p,2p)$ reactions in normal kinematics (for $^{16}$O and $^{18}$O) and those in inverse kinematics (for $^{14}$O, $^{22}$O, and $^{24}$O) are both included, and the same DWIA as in Ref.~\cite{Wakasa17} was employed. Even though the $R_s$ for the $p_{3/2}$ proton in $^{14}$O, $^{14}$O$_{p_{3/2}}$, will be still not conclusive, because $|\Delta S|$ is not so large, one can expect that its $R_s$ will not so deviate from that for $^{16}$O. Then, if the DWIA calculation for some reasons would give a significantly smaller cross section in case of the inverse kinematics measurement, $R_s$ for $^{14}$O$_{p_{3/2}}$ would be affected accordingly. In reality, as shown in Fig.~\ref{kawase}, $R_s$ deduced from the $(p,2p)$ data for $^{16}$O in normal kinematics is almost the same as that for $^{14}$O$_{p_{3/2}}$ in inverse kinematics. A key feature of the measurement in Ref.~\cite{Kaw18} is that the kinematics of the two outgoing protons are restricted rather severely, that is, the polar and azimuthal angles, $\theta$ and $\phi$, respectively, of each proton are limited as $20^\circ < \theta < 65^\circ$ and $|\phi|<15^\circ$, and its kinetic energy is required to be larger than 30~MeV. Such a selection can be a reason for the consistency between the results of the DWIA analysis on the $(p,2p)$ data in normal and inverse kinematics in Ref.~\cite{Kaw18}. To draw a conclusion, however, further investigations will be necessary.

The $(p,2p)$ cross sections on oxygen isotopes systematically measured at GSI~\cite{atar2018} will need to be analyzed with the same DWIA framework. Very recently, a systematic DWIA analysis has been carried out \cite{Phu19} on the $(p,pN)$ data for carbon, oxygen and nitrogen. The $\Delta S$ dependence of $R_s$ was found to be very similar to that evaluated with QTC. On the other hand, the absolute value of $R_s$ obtained by the DWIA analysis was somewhat larger than the values by QTC and eikonal DWIA. The difference between DWIA and QTC, which appears to contradicts the agreement in Fig. 43, may be due to the nonlocality correction and/or the difference in the treatment of the NN interaction. Further benchmark study is expected to reveal the source of the disagreement. There is currently a collaboration plan for pinning down the effect of the kinematical cut on the comparison between the DWIA calculation and measured value \cite{UT19}.

As a further evaluation of consistency of $(p,2p)$ calculations, we present in Table~\ref{table:54cap2p} the $s.p.$ cross sections for the $^{54}$Ca$(p,2p)^{53}$K reaction at 250~MeV/nucleon, as in \cite{Sun20}, for the ground and first two excited states of $^{53}$K. We must note that in \cite{Sun20}, $(p,2p)$, the DWIA cross sections were averaged over the beam energy along the thick target. The results of DWIA shown in Table~\ref{table:54cap2p} are therefore different from the $s.p.$ cross sections in Table 1 of \cite{Sun20}.

\begin{table*}[ht]
\begin{center}
\caption{Single-particle cross sections for the $^{54}$Ca($p,2p$)$^{53}$K reaction at 250~MeV/nucleon for the ground and excited states of $^{53}$K, computed using the DWIA, eikonal DWIA formailism and QTC formalisms.}
\vspace*{3mm}
\begin{tabular}{lcccccc}
$J^\pi(E_{\text{ex}} \text{ (MeV)})$ & $\sigma_\text{s.p.}$ DWIA (mb) & $\sigma_\text{s.p.}$ Eik. DWIA (mb) & $\sigma_\text{s.p.}$ QTC (mb) \\
\hline
$3/2^+(0)$    & 1.72 & 1.95 & 1.44   \\
$1/2^+(0.837)$ & 1.87 & 2.18 & 1.64 \\
$5/2^+(1.143)$ & 1.66 & 2.08 & 1.38  \\
\hline
\end{tabular}
\label{table:54cap2p}
\end{center}
\end{table*}

One sees from Table~\ref{table:54cap2p} that the DWIA result is in between the results of eikonal DWIA and QTC for each state and the standard deviation is about 15\%. This value is consistent with the uncertainty of the DWIA calculation estimated in \cite{Wakasa17} at energies higher than 200 MeV. Although there exist some differences in the inputs and prescriptions in the three calculations, for $^{54}$Ca$(p,2p)^{53}$K at 250~MeV/nucleon, the results are reasonably consistent with each other, though the difference in the agreement in this case and that in the $^{12}$C$(p,2p)$ case points to a dependence on the beam energy of this agreement.

In \cite{Sun20}, the $R_s$ was found to be 0.62--0.81. Because $\Delta S=16.3$ MeV for the proton knockout from $^{54}$Ca, one sees from Fig. 48 that this result of $R_s$ is consistent with the weaker $\Delta S$ dependence obtained for the $(p,pN)$ reactions than for the heavy-ion-induced-nucleon-removal reactions.
In Table~\ref{table:54cappn}, a similar comparison is presented for the $^{54}$Ca$(p,pn)$ reaction at 200~MeV/nucleon, as in \cite{Che19}. The agreement between the models deteriorates in this case with differences of $\sim 30\%$ between DWIA and eikonal DWIA and QTC and $\sim20\%$ between QTC and eikonal DWIA.

\begin{table*}[ht]
\begin{center}
\caption{Single-particle cross sections for the $^{54}$Ca($p,pn$)$^{53}$Ca reaction at 200~MeV/nucleon for the ground and excited states of $^{53}$Ca, computed using the DWIA, eikonal DWIA formailism and QTC formalisms.}
\vspace*{3mm}
\begin{tabular}{lcccccc}
$J^\pi(E_{\text{ex}} \text{ (MeV)})$ & $\sigma_\text{s.p.}$ DWIA (mb) & $\sigma_\text{s.p.}$ Eik. DWIA (mb) & $\sigma_\text{s.p.}$ QTC (mb) \\
\hline
$1/2^-(0)$    & 7.25 & 10.67 & 8.48   \\
$3/2^-(2.220)$ & 6.21 & 8.62 & 7.43 \\
$5/2^-(1.738)$ & 3.38 & 4.99 & 5.29  \\
\hline
\end{tabular}
\label{table:54cappn}
\end{center}
\end{table*}



\subsection{Conclusion}
The analysis of $(p,2p)$ and $(p,pn)$ data with stable nuclei and rare-isotope beams seem to indicate a weak or even no dependence on the proton-neutron asymmetry. This finding is compatible with the \emph{ab initio} Green’s function and coupled-cluster calculations but contradicts the trend derived from intermediate-energy one-nucleon removal cross section measurements.  In the future, quasifree knockout reactions in inverse kinematics will allow for a systematic investigation of proton and neutron knockout from exotic nuclei covering a wider range of neutron-to-proton asymmetry, which will be important to corroborate the observed trend and to improve our understanding on the evolution of the $s.p.$ structure as a function of neutron-to-proton asymmetry.

On the theory side, the challenges with quasifree reactions are still mounting. The most difficult one is to justify quasifree scattering in an environment with many nucleons. It is well known that the main effect of final-state interactions (FSI) in $(e,e'p)$ is a reduction of the corresponding cross section due to attenuation caused by multiple collisions of the nucleon in the nuclear matter. In $(p,2p)$ and $(p,pn)$ reactions this challenge is augmented by similar effects in the entrance channel. On the other hand, since the effect of attenuation is thus larger in $(p,2p)$ compared to $(e,e'p)$, where two protons have to leave the nucleus, the consistency of the reduction factors for both reactions as discussed in this section provides evidence, that the eikonal DWIA used to calculate the FSI of the outgoing protons in both cases, is a reasonably-good approximation at high energies. 

In DWIA, the multiple collisions are commonly treated by introducing optical potentials. A relativistic treatment of these potentials is necessary because covariance requires not only a scalar but also a vector potential. Dirac phenomenology can fix some of these issues, but  the scalar and vector part of the optical potentials  are often treated phenomenologically by adjusting free parameters.  Three-body models suffer from the same issues and they are also simplified by treating the daughter nucleus as a third particle, despite its complex compositeness. Glauber, or eikonal,  methods cure some of these deficiencies by resorting to nucleon-nucleon cross sections as inputs, bypassing nucleus-nucleus potentials and requirements set by covariance. Dealing with the medium modifications of the nucleon-nucleon cross sections is also a formidable theoretical problem, with different models being sought. It is worthwhile noticing  that some of these theoretical challenges are similarly encountered in heavy-ion knockout reactions. At present, nuclear reaction theory with high-energy probes is not amenable to an ``ab-initio" treatment, starting with a nucleon-nucleon interaction and the ensuing development of a numerical ``nearly exact" solution for a reaction channel in a fully covariant form. Therefore, the results presented here for quasifree and heavy-ion knockout reactions are clearly dependent on the adopted theoretical model and thus require further investigations and comparisons for consistency.

\section{Analysis of proton-removal reactions from {\boldmath$^{16}$}O}
\label{sec7}

\begin{figure}[t]
\begin{center}
\includegraphics[width=5.3in]{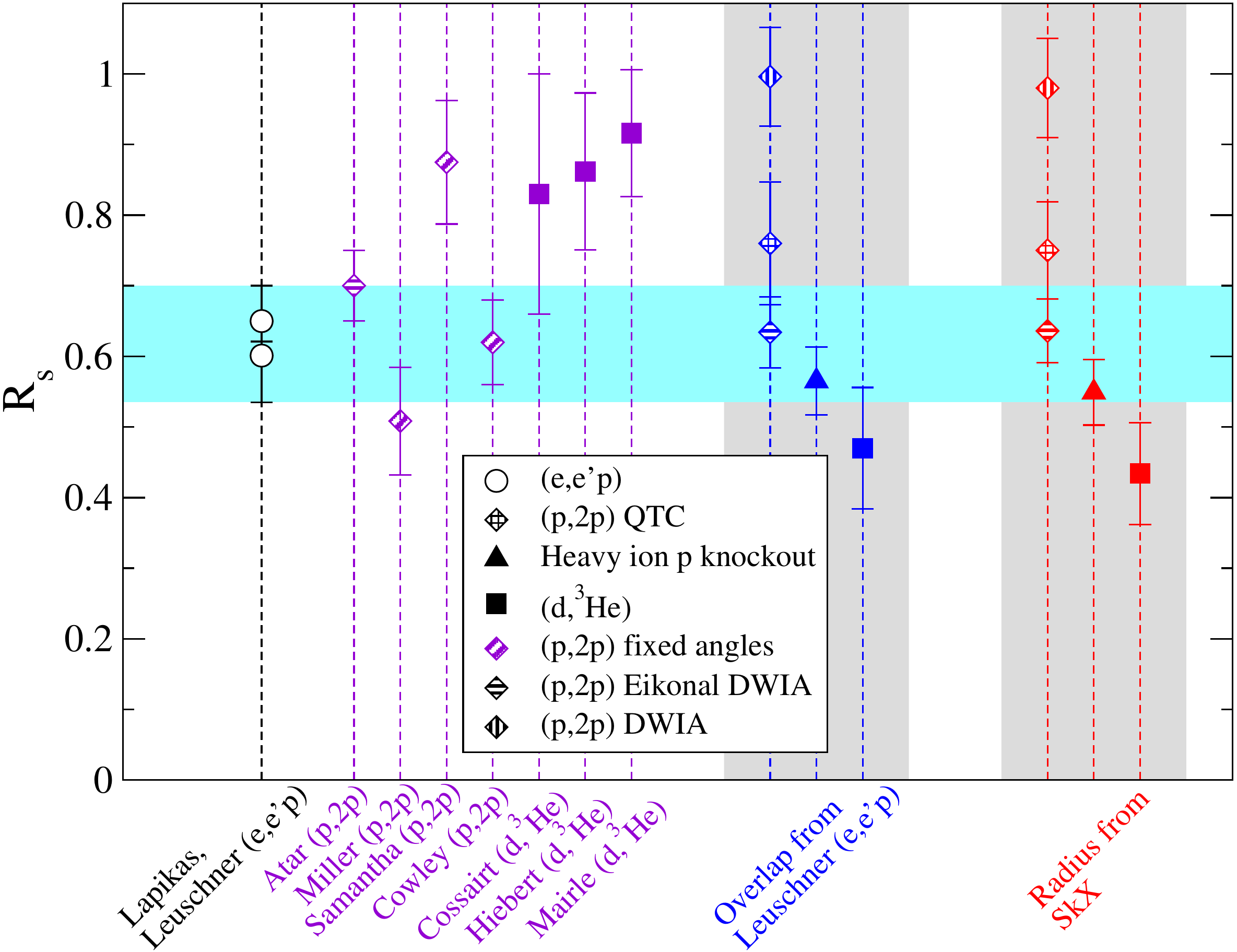}
\caption{\label{comp_o16}  (Color online) Comparison of the quenching factors $R_s$ for one-proton removal from $^{16}$O through different reactions: $(e,e'p)$ (circles), $(p,2p)$ (diamonds), heavy-ion knockout (triangles) and transfer (squares). Dashed diagonal diamonds are meant to represent $(p,2p)$ reactions in which the measurements were performed fixing the outgoing angles of the outgoing protons. $(e,e'p)$ $R_s$ have been taken from \cite{Lapikas:1993} and \cite{Leu94}. Experimental data are taken from \cite{atar2018} $(p,2p)$, \cite{Ols83} (heavy-ion knockout) and \cite{Bec77} $(d,\nuc{3}{He})$. The $R_s$ factor from \cite{atar2018,Mil98,Sam86,Cow91,Cos78,Hie67,Mai73} are presented in purple. Calculations using the same one-proton overlaps are presented within shaded bands and with the same color: blue correspond to overlaps with rms radius from \cite{Leu94} and red to rms radius computed from Hartree-Fock using SkX interaction. (More details in text).} 
\end{center}
\end{figure}

Transfer, heavy-ion induced knockout, and quasifree scattering are reviewed in the previous sections with emphasis on their comparison to data obtained with stable and radioactive beams. Here, we compare the ratio of predictions to data for these three reaction mechanisms for one specific case benchmarked with $(e,e'p)$. In Fig.~\ref{comp_o16} we show the quenching factors $(R_s)$ obtained from proton removal from \nuc{16}{O} through different reactions: $(e,e'p)$, $(p,2p)$, proton removal from a carbon target, and $(d,^3\text{He})$ transfer. The considered data are inclusive cross sections, \emph{i.e.}, they include the $1/2^-$ ground state of $^{15}$N and and $3/2^-$ bound excited states among which the first one at 6.32 MeV carries the main spectroscopic strength. Note that $(d,^3\text{He})$ data only consider the ground state and 6.32 MeV excited state. The quenching factors are computed with respect to the independent-particle model (IPM) occupation number (6 in this case). Note that large-scale shell model calculations would lead to an integrated spectroscopic strength for the $p$ orbitals very close to the IMP value of 6 when summed up to the proton separation energy of 10.2 MeV. Special care is taken to assure the consistency of the structure inputs (i.e. the one-proton overlap) in all performed calculations, so as to highlight the effect of the reaction description on the extracted quenching factors. The quenching factors from $(e,e'p)$ (circles) have not been computed, but instead have been taken from the corresponding references of Lapikas \cite{Lapikas:1993} and Leuschner \cite{Leu94} (at a beam energy of $\sim$450 MeV). We note that both references present different analyses of the same set of experimental data. The error bars for \cite{Lapikas:1993} correspond to the quoted values while for \cite{Leu94}, since multiple values are presented, the chosen error bar represents the spread of quoted values in the reference.

The calculations for heavy-ion knockout (triangles) and transfer (squares) have been performed using the sudden eikonal \cite{hansen03} and coupled-reaction channel (CRC) \cite{Satchler:1983} formalisms, respectively, and are compared to experimental data from \cite{Ols83} (at 2100~MeV/nucleon) and \cite{Bec77} (at 26~MeV/nucleon). For $(p,2p)$ the data from \cite{atar2018} (at 451 MeV/nucleon) is analyzed using three different formalisms: Quantum Transfer to the Continuum (QTC) \cite{Mor15} (diamonds with a grid), eikonal DWIA \cite{aumann2013} (dashed horizontal diamonds) and DWIA (dashed vertical diamonds) \cite{Cha77}. The $R_s$ presented in \cite{atar2018} is displayed as a purple diamond.  $R_s$ for the spectroscopic factors reported in \cite{Mil98,Sam86,Cow91,Cos78,Hie67,Mai73} are also presented.

The shaded areas correspond to calculations using the same one-proton overlap. From left to right, the blue symbols correspond to an overlap with rms radius of 2.943 and 2.719 fm for the $p_{1/2}$ and $p_{3/2}$ states respectively, taken from $(e,e'p)$ measurements \cite{Leu94}, while the red symbols represent calculations with rms radii computed from Hartree-Fock calculations using the SkX interaction \cite{brown98} (2.902 and 2.767 fm for $p_{1/2}$ and $p_{3/2}$).

The error bars have been computed through the quadratic sum of the experimental errors and theoretical errors. For $(p,2p)$ QTC calculations theoretical errors have been computed as the difference in the calculations using Dirac \cite{Cooper:1993} and PH \cite{Ger83,Rik84} parametrizations, while for eikonal DWIA and DWIA only experimental errors have been considered. Theoretical errors have been computed as the difference in results using Hartree-Fock and Gaussian \cite{brown02b} densities in heavy-ion knockout and K\"oning-Delaroche \cite{KD03} and CH89 \cite{Varner91} potentials in transfer. The $R_s$ factors in the figure correspond to the choice of potentials mentioned first for each reaction.

It can be seen that the $R_s$ factors obtained from $(p,2p)$ using QTC ($\sim$0.7) and DWIA ($\sim$1.0) are significantly larger than the ones from heavy-ion knockout ($\sim$0.5-0.6) and transfer ($\sim$0.4-0.5), while the result of the eikonal DWIA ($~\sim$0.65) is closer to them, and the $(p,2p)$ results from eikonal DWIA calculations and the results from heavy-ion knockout are more compatible with the $(e,e'p)$ result from \cite{Lapikas:1993,Leu94}, while the $R_s$ from transfer and $(p,2p)$ using QTC are in worse agreement, and those from DWIA are clearly incompatible.

The different proton overlaps increase or decrease consistently the $R_s$ factors for the three reactions, with no significant change in their differences. Given the only moderate sensitivity to the structure input, we may compare quenching factors (when referred to the IPM) from previous publications: 0.7 $(p,2p)$ \cite{atar2018}, 0.59 (heavy-ion knockout) \cite{brown02b} and 0.46 (transfer) \cite{Flavigny13}, which are consistent with the present results and trend: using the same structure inputs the quenching factor is larger for $(p,2p)$ and smaller for $(d,^3$He$)$ in the considered frameworks.

Although the scatter of published results of spectroscopic factors vary substantially as shown in the multiple references presented in Fig.~\ref{comp_o16}, most of the values are in agreement with the scatter of data from $(e,e'p)$ (indicated by the horizontal band), in particular if we compare only the results using the same overlap function (Leuschner), with the exception of the DWIA analysis, which results in a significantly larger $R_s$ value consistent with the discussion of the $^{12}$C case discussed in sec. \ref{sec6}.     


\section{Conclusions and Outlook}
\label{sec8}
It is well established and understood that the single-particle strength distribution is reduced by 30-40\% compared to shell-model predictions due correlations beyond that accounted for in the model. This picture is well established for stable nuclei and consistent with experimental nucleon-removal cross sections from $(e,e'p)$ and from other reactions like transfer, knockout, and quasifree scattering when compared to predictions of cross sections based on single-particle wave functions and shell-model correlations. For the nuclear probes, however, a scatter of up to $\sim$25\% is observed (see for example Sec. \ref{sec7} for $^{16}$O, and subsection \ref{C12} for $^{12}$C).

In this review, we discuss an extension of this picture which results from a large body of data using rare-isotope beams of very asymmetric nuclei, suggesting a strong dependence of this reduction as function of the proton-to-neutron binding-energy asymmetry. This asymmetry dependence has been deduced from measurements of Be- and C-induced nucleon-removal reactions at intermediate energies of $\sim$100 MeV/nucleon (see Fig. \ref{Rsplot}). The trend could originate from additional correlation effects in asymmetric nuclei, beyond our current understanding of nuclear structure. For example, a stronger spread of the single-particle strength for deeply-bound orbitals, or an increase of short-range correlations for the minority nucleon species would result in reduced spectroscopic factors. Under the assumption that the discussed direct reaction mechanisms probe the same nuclear structure information, It is a consensus that if they were understood properly, they should lead to the same conclusions on the asymmetry dependence of correlations.

\begin{figure*}[t]
\centering
\includegraphics[trim=0cm 0cm 0cm 0cm,clip,scale=0.58]{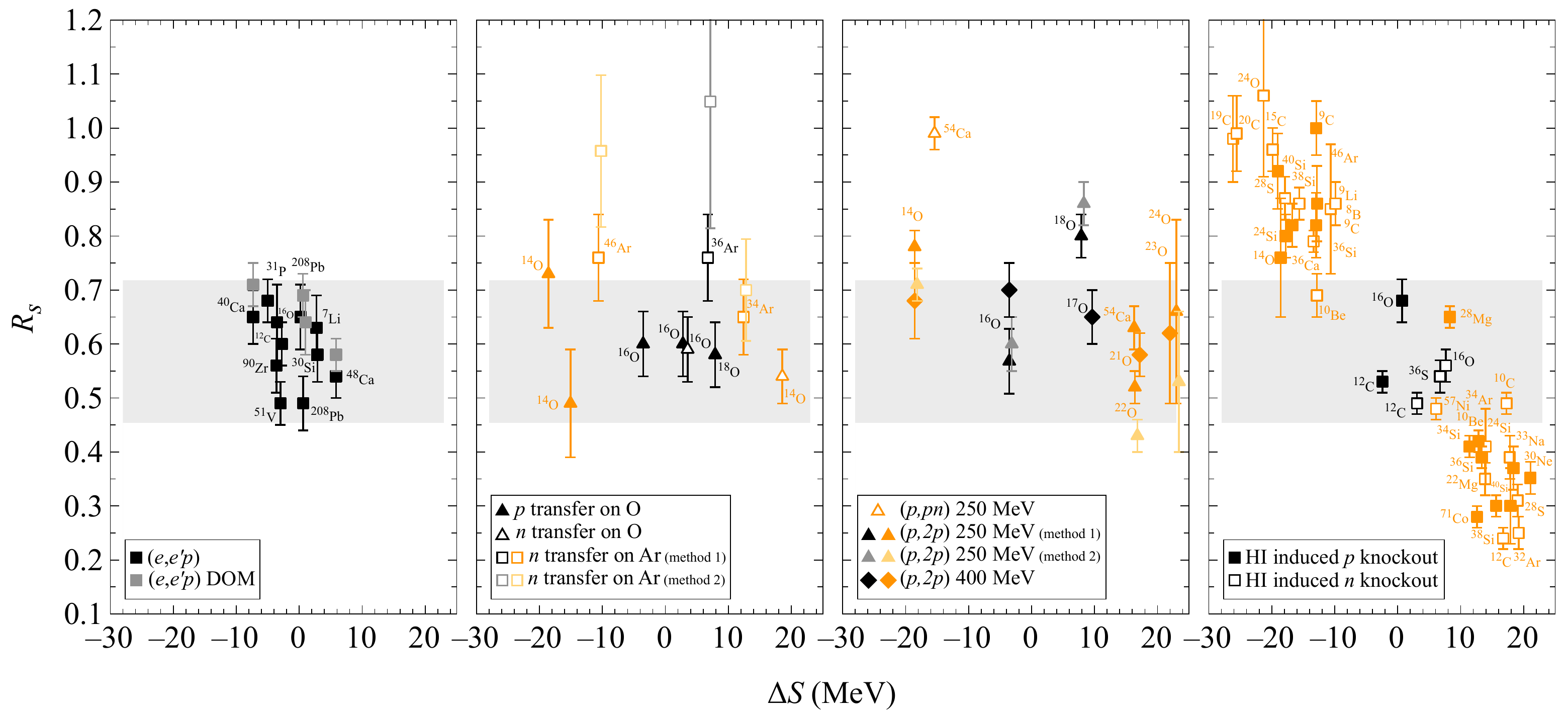}
\caption{\label{fig_all}  (Color online) The four panels of this plot show the quenching (reduction) factors for (a) electron-induced knockout reactions~\cite{Kramer01,deWitt3,Atkinson:2020,pruitt2020}, (b) transfer reactions with radioactive ion beams~\cite{Flavigny13,Lee10,nunes2011}, (c) quasifree $(p,2p)$ proton knockout on stable nuclei (from the compilation in~\cite{Wakasa17}) and radioactive nuclei~\cite{atar2018,Kaw18}, and (d) the inclusive intermediate-energy knockout data~\cite{tostevin14}. The measurements are compared to predictions based on effective-interaction shell-model SFs while, in the case of $(e,e'p)$, the integrated strength is compared to the independent-particle expectation.}
\end{figure*}

In recent years, different direct reactions, namely  transfer and quasifree scattering, have been employed to investigate this effect, although the data  are still scarce for very asymmetric nuclei. So far, these measurements could not confirm such a strong asymmetry dependence as observed in Be- or C-induced nucleon removal at intermediate energies. Figure~\ref{fig_all} shows a summary of various data from the different probes discussed in this review. For modest values of $-12\lesssim\Delta S\lesssim12$~MeV, all probes quantitatively agree that there is a quenching of single-particle strength, reduced to around 40-70\% of the total. This is arguably true for $-15\leq\Delta S\leq15$~MeV, within the moderately-large experimental and analytical uncertainties. 

The analyses of neutron-pickup transfer data with neutron-deficient $^{34}$Ar ($\Delta S$= MeV) and neutron-rich $^{46}$Ar ($\Delta S$= MeV) isotopes \cite{Lee10} using different methods and optical potentials lead to different conclusions regarding the presence or absence of a strong dependence of the data-to-prediction ratio with $\Delta S$ \cite{Lee10,nunes2011}.
Transfer studies based on three data sets of oxygen isotopes \cite{Flavigny18} and analysed within the coupled-channel formalism did not observe the strong trend from Be- and C-induced nucleon-removal reactions.
Recent $(p,2p)$ ~\cite{atar2018,Kaw18}, covering essentially the full range in $\Delta S$, do not confirm the strong trend from Be- and C-induced nucleon-removal reactions. These data have been meanwhile analyzed using different reaction models arriving at similar conclusions, although predictions of such reaction models can differ. Although Be- or C-induced nucleon-removal cross sections have been analysed with different models, the systematics of the quenching factor as a function of $\Delta S$ has been investigated with only one model so far.

A recent analysis of the $(e,e'p)$ reaction for both ${}^{40}$Ca ($\Delta S = -7.3$~MeV) and ${}^{48}$Ca ($\Delta S = 5.8$~MeV) employing the DOM predicts a reduction of the spectroscopic strength of 0.71 and 0.58, respectively (see Fig. \ref{fig_all}). The results are consistent with earlier analyses. Two recent DOM analysis for $^{208}$Pb give consistent results of 0.69 \cite{Atkinson:2020} and 0.64$\pm$0.06 \cite{pruitt2020}, in agreement with the value of Ref. \cite{deWitt3}, a re-analysis of the initial work of Ref. \cite{Kramer01} which led to a now-considered-too-low value for the quenching. All values are superimposed in the left panel of Fig. \ref{fig_all}. The DOM links both structure and reaction quantities and relies on experimental data to constrain removal probabilities as well as the optical potential for these isotopes. 
It therefore simultaneously allows for a change in the structure properties as a function of nucleon asymmetry but importantly also covers the change in the way continuum nucleons experience nuclei with different asymmetry.
This approach provides a distinct advantage over methods that rely on ingredients that are derived from free nucleon-nucleon scattering data or uncertain extrapolations of phenomenological optical potentials which are not constrained by experimental data.
A continued exploration of the DOM to generate results from data-driven extrapolations to the respective drip lines is therefore a promising approach to provide further clarification of the issues discussed in the review.
The DOM can also provide a liaison between \textit{ab initio} nuclear-structure calculations and experimental results by providing nonlocal optical potentials or conversely provide overlap functions to combine with \textit{ab initio} optical potentials that have become a focus of recent efforts some of which have been reviewed in Ref.~\cite{Dickhoff19}.

Independently of the origin of the observed trend, we can conclude that there are inconsistencies between the direct-reaction model conclusions. Reviewing the state-of-the-art reaction studies, we conclude that the problem cannot be resolved at this stage. It is of utmost importance to further understand the different reaction mechanisms by dedicated key experiments hand in hand with theory developments in the near future. 

\begin{figure}[t]
\centering
\includegraphics[scale=1.5]{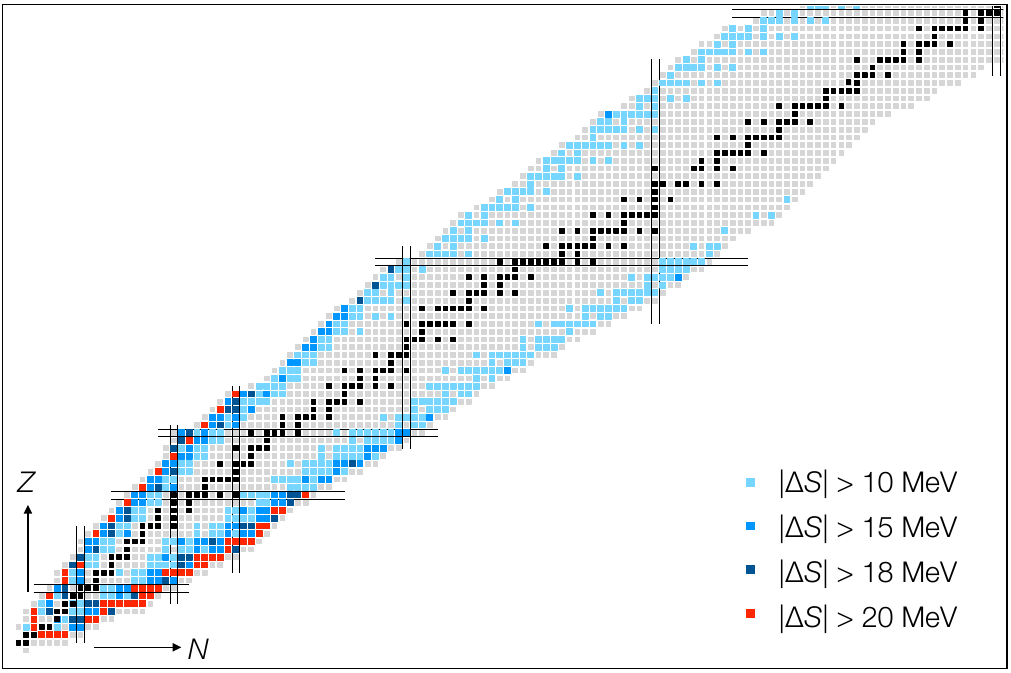}
\caption{\label{fig_delta_chart}  (Color online) A plot of $\Delta S$ values on the chart of nuclides for ground-state to ground-state transitions for all nuclei with known values of $S_n$ and $S_p$ in the 2016 Atomic Mass Evaluation~\cite{ame2016a}. About 20\% have $|\Delta S|\gtrsim12$~MeV.}
\end{figure}

So far, most of the investigations on reaction mechanisms with very asymmetric nuclei have been performed using Be-induced reactions. Some selected nuclei should be investigated with transfer and quasifree scattering at different energies as well. Figure~\ref{fig_delta_chart} shows the $\Delta S$ values for ground-state to ground-state transitions for nuclei across the chart of nuclides, for nuclei where both $S_n$ and $S_p$ are known. Those with $|\Delta S|\gtrsim15$~MeV are naturally confined to lighter (below $A\sim40$) neutron-rich or neutron-deficient systems. Many of these, and even those with $|\Delta S|\gtrsim18$~MeV, will be available at new or upcoming facilities and at suitable energies for different types of direct reactions. A systematic study of proton removal cross sections along the proton-closed shell of Ca and Ni isotopic chains would be of special interest.

Of particular relevance is the program started at the NSCL measuring mirror nuclei, which offers a promising prospect of disentangling the effects of the reaction-model approximations on the observed cross sections and resulting reduction factors. In particular, measurements on $p$-shell nuclei for which increasingly sophisticated \emph{ab initio} wave functions are now available would be of interest (see Sec.~\ref{sec:mirror}). This method should also be tested using other probes and reaction mechanisms used to deduce spectroscopic strength, such as the transfer reactions and quasifree proton scattering discussed in this review.

Measurements of the target excitation and breakup in Be- or C-induced nucleon-removal reactions are necessary to further understand the missing ingredients in the eikonal calculations.
Specific studies in quasifree scattering could be helpful to understand the re-interaction probability of scattered nucleons with the residual nucleus, as discussed in \cite{panin2019} and \cite{frotscher2020}. A direct measurement of the excitation-energy distribution of the target after knockout with identified $A-1$ fragments has not been performed so far.

A refined measurement of the energy dependence from $\sim$60 MeV/nucleon to $\sim$1 GeV/nucleon for both heavy-ion induced and quasifree $(p,pN)$ nucleon knockout on reference stable and asymmetric nuclei would provide a unique benchmark for reaction models. The $^{16}$O and $^{14}$O oxygen isotopes appear ideal cases. Different incident energies should result in a difference in the peripherality of the reaction. Since the strengths of the nuclear correlations depend on the density, the difference in peripherality can lead to differences in the obtained reduction factor. 

As discussed above, there are still significant discrepancies among quasifree scattering models. In particular, we note that they occur for integrated cross sections while for limited-acceptance measurements they disappear. A comparison of calculated angular distributions of the different models, and to measurements, should shed light on these differences. Along the same line, the detection of the knocked-out particle in heavy-ion-induced reactions, as benchmarked with $^{12}$C in \cite{panin2019}, over a broad range of neutron-to-proton asymmetries would provide further insight in the reaction mechanism and would allow a direct comparison to the $(p,pN)$ studies.

Another crucial type of measurement would be $(e,e'p)$ quasifree scattering off rare isotopes. Electron scattering experiments at the SCRIT facility of RIKEN will start soon. Although the present luminosity does not allow $(e,e'p)$ experiments with sufficient statistics, the opportunity will be opened by future upgrades, SCRIT-like experiments of new generation or by the realization of an electron-RI collider \cite{ELISE}. 

During the last two decades, direct reactions using rare-isotope beams have seen a fast development. In this review, we summarized the findings concerning the single-particle strength from direct reactions with rare isotopes. 
On the basis of these newly available data, together with the ongoing theoretical efforts and new experimental opportunities, we are confident that the remaining challenges will be addressed and will bring nuclear physics to a new level towards the understanding of nuclear structure from dripline to dripline.


\section*{Acknowledgements}
   We thank the ExtreMe Matter Institute EMMI at GSI, Darmstadt, for support in the framework of a dedicated EMMI Rapid Reaction Task Force meeting in Darmstadt, Germany, from July 30$^{\text{th}}$ to August 4$^{\text{th}}$, 2018, during which this work has been initiated. The authors thank H. Feldmeier, T. Neff, R. Roth and S. Typel for their participation to the meeting and discussions. C.~B. would like to thank A. Polls, V. Som{\`a} and T. Duguet for useful discussions.
   This work was also supported by the U.S. National Science Foundation under grant PHY-1912643, 1613362, 1415656 and the by the U.S. Department of Energy, Office of Science, Office of Nuclear Physics, grants DE-FG02-08ER41533 and DE-SC0020451 (MSU) and under Contract Number DE-AC02-06CH11357 (ANL). A.~M.~M. is partially supported by the Spanish Ministerio de Ciencia, Innovaci\'on y Universidades and FEDER funds under project FIS2017-88410-P and RTI2018-098117-B-C21 and by the European Union's Horizon 2020 research and innovation program under Grant Agreement No. 654002. K.~O. is supported in part by Grant-in-Aid of the Japan Society for the Promotion of Science (Grant No. JP16K05352). T.~N. acknowledges support from JSPS KAKENHI No. JP18H05404 and JP16H02179. C.~A.~B. acknowledges funding contributed through the LANL Collaborative Research Program by the Texas A\&M System National Laboratory Office and Los Alamos National Laboratory. T.~A. acknowledges support by the German Federal Ministry of Education and Research (BMBF project 05P2015RDFN1), and through the GSI-TU Darmstadt cooperation agreement. C.~B. acknowledges funding from the UK Science and Technology Facilities Council (STFC) through grants ST/P005314/1 and ST/L005516/1. First principle calculations discussed in Sec. \ref{sec:abinitio} used HPC resources at the DiRAC DiAL system at the University of Leicester, UK, (BIS National E-infrastructure Capital Grant No. ST/K000373/1 and STFC Grant No. ST/K0003259/1). T.~A. and A.~O. acknowledge partial support by the Deutsche Forschungsgemeinschaft (DFG, German Research Foundation) - Project-ID 279384907 - SFB 1245. A.~O. and M.~G.-R. acknowledge support by the Alexander von Humboldt foundation.
   
\appendix
\section{Appendix: Analysis of the nucleon self-energy and some pedagogical illustrations}
\label{sec:self}
\begin{figure}[tb]
\begin{center}
\includegraphics[width=7.85cm]{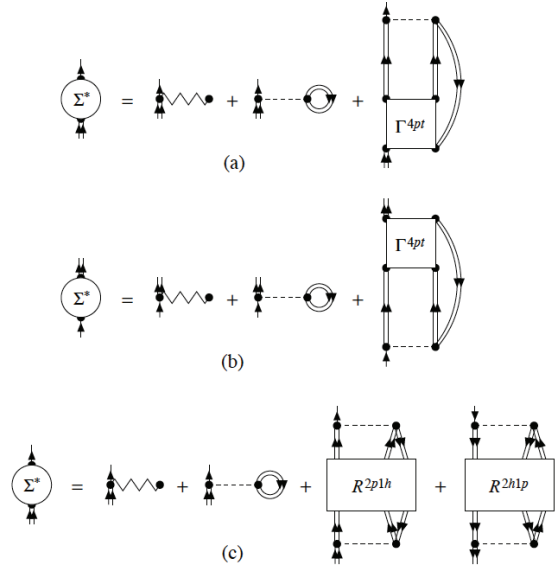}
\caption{Various possible expansions of the irreducible self-energy in terms  of higher-order Green's functions. In all cases the first two terms represent the mean-field  contributions $-\hat{U}$  and the HF self-energy $\Sigma^{HF}$. The dynamic part of the self-energy can be expressed in terms of the  four-points vertex function $\Gamma^{4pt}$. Part $a)$ represents the relation given in Eq.~(\ref{eq:selfen_Gamma}), while the part $b)$ corresponds to an alternative derivation involving the time derivative of $G$ w.r.t. $t'$. If $\Gamma^{4pt}$ is approximated in such a way that the corresponding two diagrams in $a)$ and $b)$ are  equivalent, the Dyson equation will satisfy appropriate sum rules. Part $c)$ gives the expansion in terms of the one-particle irreducible two-particle--one-hole (2p1h) / two-hole--one-particle (2h1p) propagator $R(\omega)$~\cite{Dickhoff04}.  Note that the full four-time dependence of $\Gamma^{4pt}$ is needed in principle, while in the $R(\omega)$ formulation one can specialize to a two-time quantity. Reprinted figure with permission from \cite{Dickhoff04} \textcopyright2004 by Elsevier.
\label{fig:SEDB}}
\end{center}
\end{figure}

\begin{figure}[tb]
\begin{center}
\includegraphics[origin=bl,angle=-90,width=0.45\textwidth]{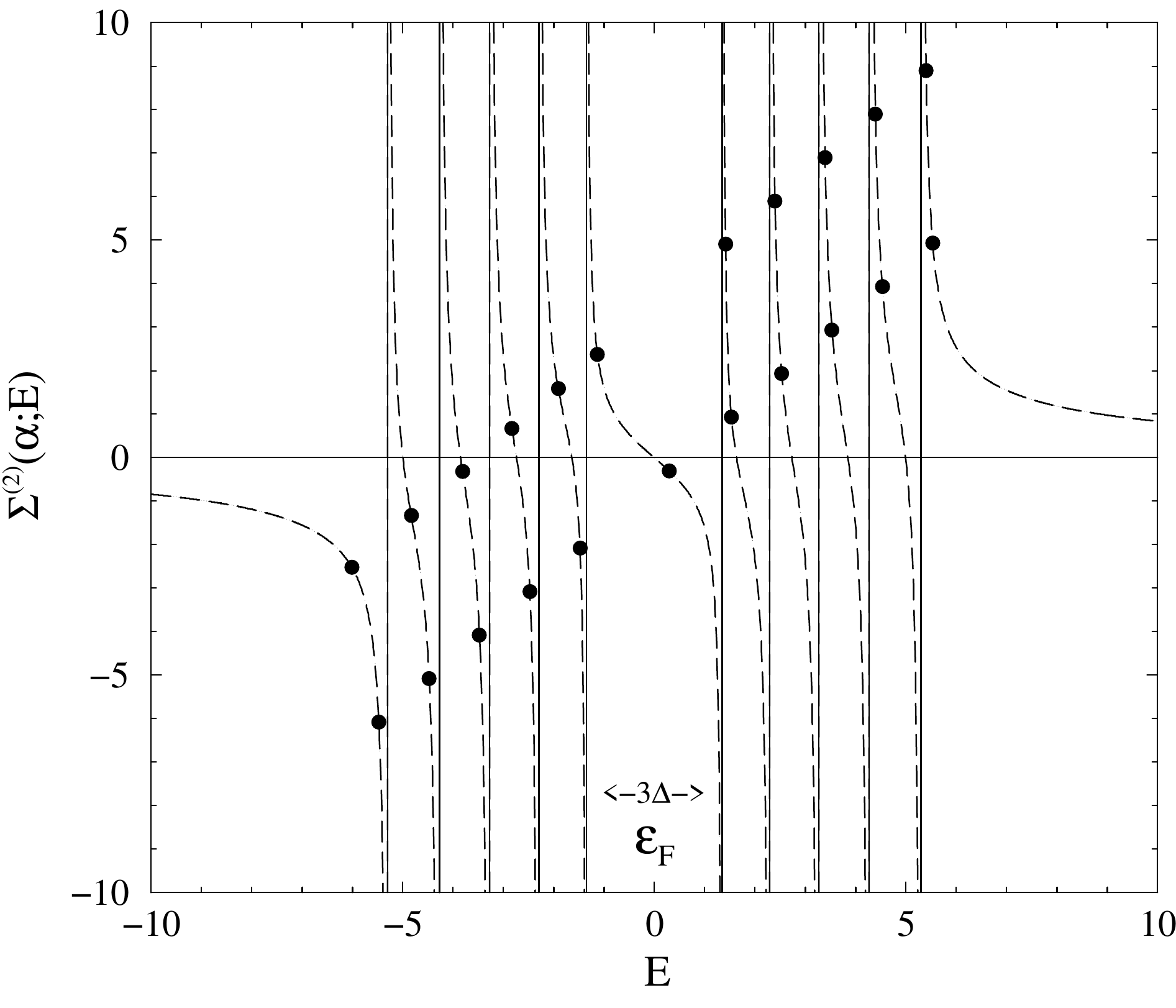}
\caption{\label{fig:13.2}
Graphical solution of Eq.~(\protect\ref{eq:13.21}).
The second-order self-energy $\Sigma^{(2)}(\alpha;E)$ 
of Eq.~(\protect\ref{eq:13.17}) is indicated by the 
dashed line. The roots of Eq.~(\protect\ref{eq:13.21}) are given by the
intersections points with the straight line $E-\varepsilon_\alpha$, drawn here
as dots, for two values of $\varepsilon_\alpha$.
Figure adapted from Ref.~\cite{Dickhoff:08}.
}
\end{center} 
\end{figure}

Calculations of the nucleon self-energy can proceed from a diagrammatic form for the self-energy.
Several possibilities are illustrated in Fig.~\ref{fig:SEDB} that are exact when only two-body interactions are present in the Hamiltonian.
Substantial modifications are involved when three-body interactions are included~\cite{Carbone13}.
Using the equation of motion for the sp propagator~\cite{Dickhoff:08} 
one obtains 
\begin{eqnarray}
 \Sigma^{\star}(\gamma, \delta;E)  &=& 
    - \bra{ \gamma} U \ket{ \delta}
  +  \bra{\gamma} \Sigma^{HF} \ket{\delta}
 \frac{1}{2} \int \frac{dE_1}{2 \pi}
                      \int \frac{dE_2}{2 \pi}
        \sum_{\mu , \nu , \lambda}    \sum_{\alpha , \beta , \varepsilon}
   \bra{\gamma \varepsilon} V \ket{ \alpha \beta} \; \nonumber \\
&\times&       G(\alpha , \mu; E_1)   G(\beta, \nu; E_2)
 \bra{\mu \nu} \Gamma^{4pt} 
          (E_1 , E_2 ; E , E_1 + E_2 - E ) \ket{ \delta \lambda} 
  G(\lambda , \varepsilon; E_1 + E_2 - E ) \nonumber \\  
\label{eq:selfen_Gamma}
\end{eqnarray}
where the term $-\langle \gamma \vert \hat{U} \vert \delta \rangle$
subtracts the auxiliary potential which was added
in the unperturbed hamiltonian $\hat{H}_0 =\hat{T} + \hat{U}$.
This result is shown in part $a)$ of Fig.~\ref{fig:SEDB}),
The second term in Eq.~(\ref{eq:selfen_Gamma}) is the HF
contribution to  the
self-energy,
\begin{equation}
  \bra{\alpha} \Sigma^{HF} \ket{\beta} =
 \sum_{\gamma \delta}
    \int \frac{dE}{2 \pi i} ~ e^{+i E \eta^+}
  ~ \bra{\alpha \gamma}V \ket{\beta \delta}  ~  G(\gamma, \delta; E)  \; \; .
\label{eq:selfen_HF}   
\end{equation}
which represent the (energy-independent) interaction of the nucleon
with the quasihole excitations inside the system and requires knowledge of the one-body density matrix given in Eq.~(\ref{eq:5.17}).
 When all the higher-order terms of the self-energy are neglected,
the solution of the Dyson equation $G$ takes the
 simple form as Eq.~(\ref{eq:g0}), which is equivalent to a 
Slater determinant description of the ground state filling the lowest-energy orbits.  
The corresponding unperturbed propagator associated with a Fermi-sea Slater determinant $\ket{\Phi^A_0}$ 
is given by
\begin{equation}
 G^{(0)}(\alpha, \beta; E) ~=~ \delta_{\alpha , \beta}   \left\{ 
 \frac{ \theta(\alpha - F) }{E - \varepsilon_\alpha + i \eta }  ~+~
 \frac{ \theta(F - \alpha) }{E - \varepsilon_\alpha - i \eta }
  \right\}   \; ,
\label{eq:g0}
\end{equation}
where $\theta(\alpha - F)$~(~$\theta(F - \alpha)$~) is equal to to 0~(1) if
$\alpha$ is an occupied state and it is 1~(0) otherwise.
In this case the energies $\varepsilon_\alpha$
contained in $G^{(0)}$ refer sp energies corresponding to completely filled or empty orbitals in $\ket{\Phi^A_0}$. 
The iterative solution of
Eqs.~(\ref{eq:10.2}) and~(\ref{eq:selfen_HF}) generates the standard
HF approximation.\\
All qualitative changes when higher-order contributions to the self-energy are incorporated, can already be illustrated with the introduction of the second-order term.
We illustrate this by forgoing a self-consistent formulation and employ HF propagators to evaluate the second-order diagram that can be obtained by replacing $\Gamma^{4pt}$ in Eq.~(\ref{eq:selfen_Gamma})
by the bare interaction $V$.
The result is given by 
\begin{eqnarray}
\Sigma^{(2)}(\gamma ,\delta ;E) =\frac{1}{2}
\left[ \sum_{p_1 p_2 h_3 }\frac{
\bra{\gamma h_3}V\ket{p_1 p_2 }\bra{p_1 p_2}V\ket{\delta h_3} } 
{E-(\varepsilon_{p_1}+\varepsilon_{p_2}-\varepsilon_{h_3})+i\eta } + \sum_{h_1 h_2 p_3 }\frac{
\bra{\gamma p_3}V\ket{h_1 h_2 }\bra{h_1 h_2}V\ket{\delta p_3} } 
{E+(\varepsilon_{p_3}-\varepsilon_{h_1}-\varepsilon_{h_2})-i\eta }\right] ,
\label{eq:13.16}
\end{eqnarray}
where labels identifying particle ($p$) and hole ($h$) states in the HF
approximation have been introduced.
We will proceed to examine the solution $G$ of the equation 
\begin{equation}
G(\alpha ,\beta;E) =G^{HF}(\alpha ,\beta;E) +
\sum_{\gamma\delta} G^{HF}(\alpha ,\gamma;E) \Sigma^{(2)} (\gamma ,\delta; E) 
G(\delta ,\beta;E).
\label{eq:13.16bis}
\end{equation}
It is clear from Eq.~(\ref{eq:13.16}) that the second-order self-energy  
in principle 
has non-diagonal contributions, even when evaluated with the diagonal HF sp 
propagator. However, in some cases it is a good approximation to neglect the 
off-diagonal terms. 
This happens \textit{e.g.}\ in closed-shell nuclei, where off-diagonal 
elements would require mixing between major shells having a large energy 
separation. 
Within this diagonal 
approximation, the self-energy (\ref{eq:13.16}) reads 
\begin{eqnarray}
\Sigma^{(2)}(\alpha ;E) 
= \frac{1}{2}
\left[\sum_{p_1 p_2 h_3 }\frac{
|\bra{\alpha h_3}V\ket{p_1 p_2 }|^2 }
{E-(\varepsilon_{p_1}+\varepsilon_{p_2}-\varepsilon_{h_3})+i\eta } + \sum_{h_1 h_2 p_3 }\frac{
|\bra{\alpha p_3}V\ket{h_1 h_2 }|^2 }
{E+(\varepsilon_{p_3}-\varepsilon_{h_1}-\varepsilon_{h_2})-i\eta }\right] ,
\label{eq:13.17}
\end{eqnarray}
and the Dyson equation (\ref{eq:13.16bis}) becomes 
\begin{equation}
G (\alpha ; E )=G^{HF}(\alpha ; E) +G (\alpha ; E )\Sigma^{(2)}(\alpha ;E) 
G^{HF}(\alpha ; E).
\label{eq:13.18}
\end{equation}
The latter has a simple algebraic solution, 
\begin{equation}
G (\alpha ; E)
=\frac{1}{\frac{1}{G^{HF}(\alpha ; E)} -\Sigma^{(2)}(\alpha ;E)}
=\frac{1}{E-\varepsilon_\alpha -\Sigma^{(2)}(\alpha ;E)} ,
\label{eq:13.19}
\end{equation}
using the inverse of the HF propagator  $[G^{HF}(\alpha ; E)]^{-1}= E-\varepsilon_{\alpha}$.\\
Extracting physical information from the sp propagator in general  
requires the knowledge of its poles and residues. 
We assume for simplicity that the   
self-energy $\Sigma^{(2)}$ has poles at a set of discrete energies 
(\textit{i.e.}, a set of isolated simple poles), while an extension to treat the case when branch-cuts are 
present can be included straightforwardly.    
Most realistic finite systems have branch-cuts, 
but since practical calculations are usually 
performed by introducing a finite and discrete sp basis, the self-energy is 
then automatically restricted to a discrete pole structure. \\
For the propagator $G(\alpha; E)$ given by the formal solution in 
Eq.~(\ref{eq:13.19}), the discrete poles $E_{n\alpha}$ obviously correspond to 
the roots of the nonlinear equation
\begin{equation}
E_{n\alpha}=\varepsilon_{\alpha} + \Sigma^{(2)}(\alpha; E_{n\alpha}),
\label{eq:13.21}
\end{equation}
with $\Sigma^{(2)}(\alpha ;E)$ defined in Eq.~(\ref{eq:13.17}).  
The residue $R_{n\alpha}$ at the pole $E_{n\alpha}$ of the propagator 
follows from 
\begin{equation}
R_{n\alpha} = \lim_{E\rightarrow E_{n\alpha}} (E-E_{n\alpha})G(\alpha ; E)
= 
= \left( 1- \left. \frac{d\Sigma^{(2)}(\alpha ;E)}{dE}
\right|_{E=E_{n\alpha}} \right)^{-1}.
\label{eq:13.22}
\end{equation}
To gain insight into the location of the roots of Eq.~(\ref{eq:13.21}), a 
graphical solution of the Dyson 
equation\index{graphical solution of the Dyson equation} is often helpful.
In Fig.~\ref{fig:13.2} the energy-dependence of the self-energy 
$\Sigma^{(2)}(\alpha ;E)$ of Eq.~(\ref{eq:13.17}) is shown. 
The case on display is for a typical confined finite 
system, having a discrete HF sp spectrum. The hole and particle HF energies 
are separated by the particle-hole gap, which has a width 
$\Delta = \varepsilon^{min}_{p} -\varepsilon^{max}_{h}$ and is centered on the 
HF Fermi energy 
\begin{equation}
\varepsilon_F = \frac{1}{2}(\varepsilon^{min}_{p} +\varepsilon^{max}_{h}). 
\label{eq:13.23}
\end{equation}
Since the poles in Eq.~(\ref{eq:13.17}) all have positive residues, 
$\Sigma^{(2)}(\alpha ;E)$ is 
monotonically decreasing where defined. There is a sequence of 
simple poles in the addition domain, located at the unperturbed HF 
2p1h energies, and another sequence in the removal domain, 
located at (minus) the unperturbed HF 
1p2h energies. The poles of the addition and removal sequence are separated 
by a gap of (at least) three times the HF particle--hole gap.   
The roots of Eq.~(\ref{eq:13.21}) are simply the intersection points of the self-energy $\Sigma^{(2)} (\alpha ;E)$ with the straight line $E-\varepsilon_{\alpha}$. 
It is obvious from the graph in Fig.~\ref{fig:13.2} that 
between any two successive poles of the self-energy a root is located. 
In addition, there is a root to the left and right of the sequence of self-energy poles. When a finite sp basis set is used, this implies that a self-energy having $D$ poles leads to a sp propagator with $D+1$ poles. 
The interpretation of these roots should by now be straightforward. The poles $E_{n\alpha}$ in the removal domain (below the Fermi energy) must be interpreted as approximate energies of the eigenstates in the $A-1$ 
system $E_{n\alpha}\approx E^A_0 -E^{A-1}_n$, 
that can be obtained by removing a particle in the sp state $\alpha$ from the $A$-particle ground state. The residue then corresponds to the (squared) removal amplitude,
\begin{equation}
R_{n\alpha}\approx |\bra{\Psi^{A-1}_n} a_{\alpha}\ket{\Psi_0^A}|^2 .
\label{eq:13.25}
\end{equation} 
Similarly, the poles $E_{n\alpha}$ in the addition domain (above the Fermi energy) correspond to eigenstates in the $A+1$ system $E_{n\alpha}\approx E^{A+1}_n -E^A_0$, 
having addition probabilities
\begin{equation}
R_{n\alpha}\approx |\bra{\Psi^{A+1}_n} a^{\dagger}_{\alpha}\ket{\Psi_0^A}|^2 .
\label{eq:13.27}
\end{equation} 
Note that since $d\Sigma^{(2)}(\alpha ;E)/dE < 0$, the residues $R_{n\alpha}$ 
that follow from Eq.~(\ref{eq:13.22}) obey 
\begin{equation}
0\leq R_{n\alpha} \leq 1 ,
\label{eq:13.28}
\end{equation}
in accordance with their relation to the physical addition or removal probabilities.
The sum of the removal probabilities then corresponds to the occupation of the state $\alpha$ as in Eq.~(\ref{eq:5.13}). The sum of the addition probabilities Eq.~(\ref{eq:5.14}) then complements the sum rule of Eq.~(\ref{eq:5.15}).\\ 
Adding the energy-dependent second-order self-energy
to the static HF self-energy therefore produces quite dramatic effects.
The removal from the ground state of a particle in an occupied HF sp state $\alpha$ no longer leads to a unique $A-1$ state, as in HF, but rather to a large number of $A-1$ states, each having a finite removal amplitude. Moreover, the removal from the ground state of a particle in an unoccupied HF sp state $\alpha$, clearly impossible in HF, 
is now allowed. Similar statements hold in the addition domain. Of course, any more sophisticated treatment of the self-energy will also include these fragmentation effects on the sp strength. Experimental information on physical spectral functions indicate that such features are indispensable for a meaningful comparison 
with data.\\
As a final remark on Fig.~\ref{fig:13.2}, we note that  
if the unperturbed sp energy $\varepsilon_{\alpha}$ is not too far removed from the Fermi energy, the root of Eq.~(\ref{eq:13.21}) lying in the interval which separates the removal and addition domain, has a special character. Since the self-energy $\Sigma^{(2)}(\alpha ;E)$ has no poles in this interval, the energy derivative is relatively small here, and as a consequence the residue corresponding to the solution will be quite close to (but still smaller than) unity. Such a solution represents a quasiparticle or quasihole excitation in a finite system, and corresponds to a $A\pm 1$ eigenstate which has a rather pure sp character. On the other hand, if the sp energy $\varepsilon_{\alpha}$ is far from the Fermi energy, it is in a region where the density of 2p1h or 1p2h states is high, and the strength of this sp state will be strongly fragmented over many $A-1$ states. The different fragmentation pattern observed for valence 
holes and deeply-bound hole states in finite nuclei is readily understood by these elementary considerations, and explains the qualitative behavior of the data discussed in Sec.~\ref{sec:eep}.

\section{Appendix: Momentum distributions in heavy-ion induced knockout}
\label{eikonal}
\label{intr} In this appendix we discuss in more details the usual eikonal formalism for determining the momentum distribution of teh core fragments.  The momentum $k_n$ of the struck nucleon in the projectile and that of the residue in the final state, $k_{A-1}$, are related by $k_n$
\begin{equation}
\mathbf{k}_{n}=\frac{A-1}{A}\mathbf{k}_{A}-\mathbf{k}_{A-1}. \label{kbal}%
\end{equation}
The  target state and that of the struck nucleon are usually not observed in the final state, but the four momenta of the residue and the gamma-rays following its de-excitation are often identified in coincidence measurements of in-flight decay. In the center-of-mass the transferred momentum will be denoted here by $\mathbf{k}_{c}$ and, in the sudden approximation defined bellow, it must be equal the momentum  of the struck nucleon before the collision. In analogy to angular distributions obtained in transfer reactions, the orbital angular momentum $l$ is revealed in the knockout reactions through the momentum distribution of the fragments. 

In collisions with $E_{\rm lab} \gtrsim 50$ MeV/nucleon, the longitudinal component of the momentum distributions  (along the beam direction $z$) yields the most accurate information and is rather insensitive to details of the collision, as first pointed out in Ref. \cite{bertulani92}. But the transverse momentum distributions of the core contain significant diffractive effects and Coulomb scattering leading to its broadening, as illustrated in Ref. \cite{ANNE1994125} by the measurement of the angular distribution of neutrons in the  $^{9}$Be($^{11}$Be,$^{10}$Be+n)X reaction. For light targets, in the absence of Coulomb scattering,  the width of the transverse distributions also reflects the size of the target.

\subsection{Differential stripping cross section}
\label{calc} 

\begin{figure}[!ht]
\begin{center}
\includegraphics[scale=0.3,angle=0]{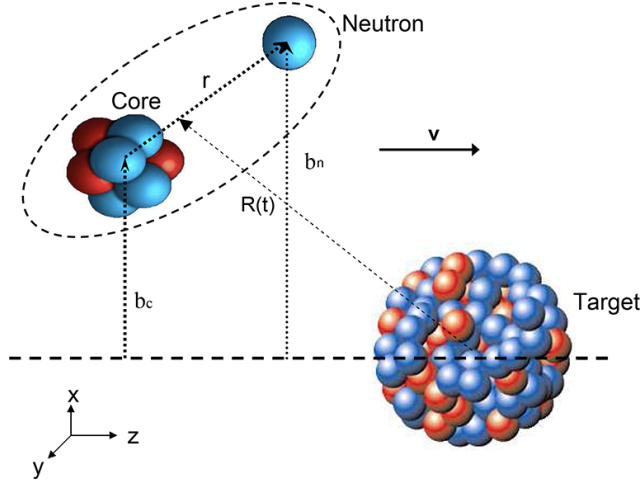}
\end{center}
\caption{ (Color online) Coordinates used in the text.} \label{coord2}
\end{figure}

Following Refs. \cite{bertulani04,hencken96}, the differential cross section for the stripping reaction $(c+n)+A\longrightarrow c+X$, where $c$ corresponds to the core in a specified single particle state, is 
\begin{eqnarray}
\frac{d\sigma_{\mathrm{str}}}{d^{3}k_{c}}=\frac{1}{\left(  2\pi\right)  ^{3}%
}\frac{1}{2l+1}\sum_{m}\int d^{2}b_{n}\left[  1-\left\vert S_{n}\left(
b_{n}\right)  \right\vert ^{2}\right]   \left\vert \int d^{3}r\ e^{-i\mathbf{k}%
_{c}\mathbf{.r}}S_{c}\left(  b_{c}\right)  \psi_{lm}\left(  \mathbf{r}\right)
\right\vert ^{2}, \label{sknock}%
\end{eqnarray}
with $\mathbf{r\equiv}\left(
\mathbf{\mbox{\boldmath$\rho$}},z,\phi\right)
=\mathbf{R}_{n}-\mathbf{R}_{c}$, and
\begin{eqnarray}
b_{c}    =\left\vert \mathbf{\mbox{\boldmath$\rho$}}-\mathbf{b}
_{n}\right\vert =\sqrt{\rho^{2}+b_{n}^{2}-2\rho\ b_{n}\cos\left(  \phi
-\phi_{n}\right)  }
  =\sqrt{r^{2}\sin^{2}\theta+b_{n}^{2}-2r\sin\theta\ b_{n}\cos\left(
\phi-\phi_{n}\right)  }.
\end{eqnarray}
with ${\bf b}_c$ and ${\bf b}_n$ denoting the two-dimensional vectors (see figure \ref{coord2}). Here, $S_{c}$
($S_{n}$) are the S-matrices for core+target and for the the removed nucleon+target scattering.

Integrating eq. (\ref{sknock}) over the transverse component of $\mathbf{k}$ yields the longitudinal momentum distributions \cite{bertulani04,hencken96}
\begin{eqnarray}
\frac{d\sigma_{\mathrm{str}}}{dk_z}  = \frac{1}{2\pi}\frac{1}%
{2l+1}\sum_{m}\int_{0}^{\infty}d^{2}b_{n}\ \left[  1-\left\vert S_{n}\left(
b_{n}\right)  \right\vert ^{2}\right]   \int_{0}^{\infty}d^{2}%
\rho\ \left\vert S_{c}\left(  b_{n}\right)  \right\vert ^{2}
 \left\vert \int_{-\infty}^{\infty}dz\ \exp\left[
-ik_{z}z\right]  \psi_{lm}\left(  \mathbf{r}\right) \right\vert
^{2}\ ,
\label{strL}%
\end{eqnarray}
where $k_z$ represents the longitudinal component of ${\bf k}_c$.

In the equation above, the single-particle bound state wave function for $(c+n)$, denoted by $\psi_{lm}\left(  \mathbf{r}\right)  $, are specified by\ $\psi_{lm}\left(
\mathbf{r}\right)  =R_{l}\left(  r\right)  Y_{lm}\left(
\widehat{\mathbf{r}}\right)  $, where $R_{l}\left(  r\right) $ is the single-particle radial wave function.The total single-particle angular momentum $j$ does not need to be specified if the interaction is spin-independent \cite{bertulani04}. 

Integrating eq. (\ref{sknock}) over the longitudinal component of $\mathbf{k}$ yields the transverse momentum distribution in cylindrical
coordinates (with $k_\bot=\sqrt{k_{x}^{2}+k_{y}^{2}}$) \cite{bertulani04}
\begin{eqnarray}
\frac{d\sigma_{\mathrm{str}}}{d^{2}k_\bot}  =
\frac{1}{2\pi}\frac {1}{2l+1}\ \int_{0}^{\infty}d^{2}b_{n}\ \left[ 1-\left\vert S_{n}\left( b_{n}\right)  \right\vert ^{2}\right]  \sum_{m,\ p}\ \int_{-\infty}^{\infty}dz\ \left\vert \int d^{2}\rho\ \exp\left(  -i\mathbf{k}_{c}^{\perp}\mathbf{.\mbox{\boldmath$\rho$}}%
\right)  S_{c}\left(  b_{n}\right)  \psi_{lm}\left(  \mathbf{r}\right) \right\vert ^{2}. \label{strT}%
\end{eqnarray}

In terms one of the Cartesian projected components of the transverse momentum, one has \cite{bertulani04}
\begin{equation}
\frac{d\sigma_{\mathrm{str}}}{dk_{y}}=\int dk_{x}\ \frac{d\sigma
_{\mathrm{str}}}{d^{2}k_\bot}\left(  k_{x},k_{y}\right)  \ . \label{sigtx}%
\end{equation}

If one integrates eq. \eqref{strL} over $k_z$ (or eq. \eqref{strT} over ${\bf k}_\bot$), using the properties of the Dirac delta function, one easily reproduces the eq. \eqref{strip1}.  Inclusion of the Coulomb scattering is obtained by modifying the eikonal scattering matrices appropriately, including nuclear size effects (for more details, see Refs. \cite{Book:Ber04,bertulani06}). 

\subsection{Differential diffraction dissociation cross section}
 The  theory of diffraction dissociation was developed for the first time by Akhiezer and Sitenko \cite{akhiezer1957} and by Glauber \cite{GlauberPRC99.1515} and Feinberg \cite{Feinberg1956}, independently, to study the deuteron breakup with just one bound state. Later it was also applied to halo nuclei, when only one bound state exists \cite{EVLANOV1986477,BERTULANI1988615,hencken96}.  Thus, it is worthwhile mentioning that the equation \eqref{dis1} has also been derived following these previous works, under the assumption that only one bound state, $\psi_{0}\left(  \mathbf{r}\right)$, exists \cite{YABANA1992295,hencken96}. However, it is commonly applied to several situations where more than one bound state exists.

Furthermore, it is commonly assumed that the momentum distributions due to diffraction dissociation have the same profile as those of stripping. Both assumptions, namely, one single bound state and equal shapes of momentum distributions for stripping and diffraction dissociation, have not been justified in the literature and other methods, such as the partial wave expansion method have also been used. Diffraction dissociation contributes to a sizable part of the nucleon knockout cross sections. The equations described in this section have been implemented numerically and made available in public codes using different techniques \cite{bertulani04,ABUIBRAHIM2003369,bertulani06}.

Ref. \cite{hencken96} generalized the stripping formula of Ref. \cite{YABANA1992295} in such a way to allow the calculation of parallel momentum distributions for stripping. The formalism to obtain P$_{\perp}$   distributions for stripping was presented in Ref. \cite{bertulani04} and reproduced above. At the moment there is no eikonal formalism to obtain P$_{\perp}$ nor P$_{//}$ distributions for diffraction. When necessary some authors use for diffraction the distributions obtained for stripping renormalizing them by the value obtained from the total closed form Eqs. (\ref{strip1}-\ref{dis1}) or relay on CDCC calculations for a cross check \cite{PhysRevC.66.024607}. The nuclear diffraction part should however be different because it cannot contain the so-called "recoil" term which is important for breakup of a halo neutron following the interaction with a heavy target \cite{Angela2018}. However there was in the past an attempt to calculate momentum distributions for both stripping and diffraction \cite{PhysRevC.70.054602} using an extended sudden method. Numerical results were in very good agreement with a large set of data presented in that work.

The recoil term arises from the difference  between the target- center-of-mass   coordinate of a halo projectile and the coordinate of the core-target system. This difference is negligible for all but neutron halo nuclei. The recoil effect gives rise to the so called Coulomb breakup which is due to the effective Coulomb force acting on the halo nucleon when the core is deflected by the Coulomb interaction with the target. The corresponding Coulomb potential is just the dipole potential \cite{MARGUERON2002105,PhysRevC.84.014613}. Because the Coulomb potential is long range while the nuclear potential is short range this effect, when present, must be treated   to all orders. Furthermore when the halo particle is a proton there is a direct Coulomb force between the proton and the target whose corresponding potential must be treated to all multipolarities. The Coulomb potential acts on a different time scale than the nuclear potential and thus a nucleon which breaks up in this way it is not expected to contribute to stripping. It will contribute to diffraction and actually the amplitudes for it and for nuclear diffraction  must be summed coherently but not in the way it is done by Eq. \eqref{dis1}. Therefore such equation, irrespective of the possible numerical values it might provide, is fundamentally inadequate because i) cannot be justified in presence of more than one bound excited state (see above, ii) it contains recoil but at the lowest order  and in the sudden approximation and thus it includes Coulomb breakup  but in a way which is inaccurate. Because of the possible interference effect with the nuclear part of the amplitude the total result can be unpredictably wrong.  Therefore such a formula  is not justifiable from the theoretical point of view (see subsection \ref{difdisrevisted}). 
The correct eikonal formalism which avoids Coulomb effects being treated in the wrong way has been presented in Ref. \cite{Angela2018}, and summarized later in this Appendix.
It is also worth mentioning that in Eq.~\eqref{dis1} both the real and imaginary parts of the core-target S-matrix play a role. This means that in principle the phase shift core-target should contain a Coulomb phase in its real part. For heavy targets Eq.~\eqref{dis1} would provide unreliably high cross sections and it is common practice to put such a phase equal to zero.

\begin{figure}[!th]
\centering
\includegraphics[scale=0.4]{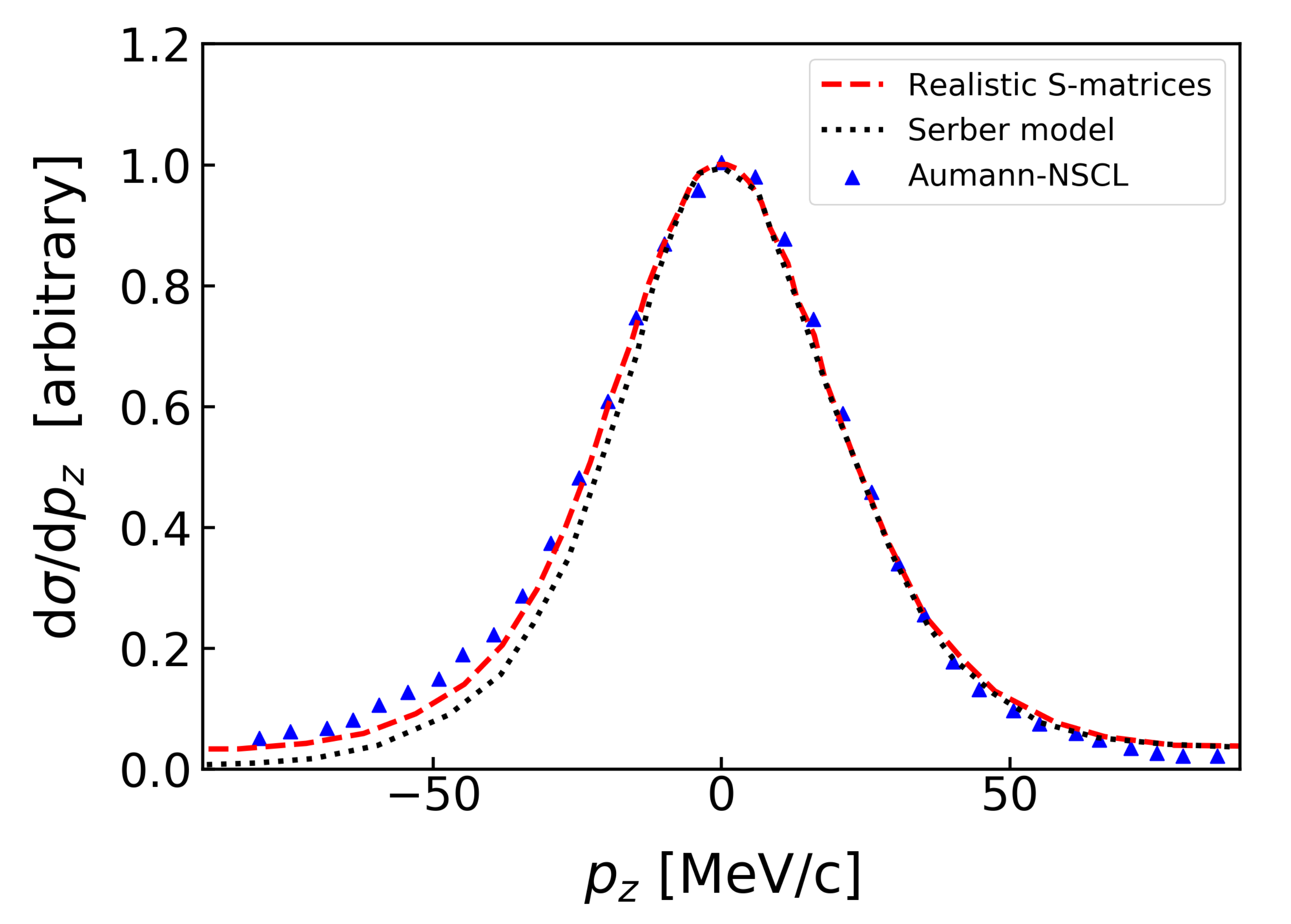}
\caption{\label{Serber} (Color online) Removal of $1s_{1/2}$ neutron, with separation energy $S_n = 0.503$ MeV in the reaction $^{9}$Be($^{11}$Be,$^{10}$Be) at 60 MeV/nucleon \cite{aumann00}. }
\end{figure}

\begin{figure}[!th]
\centering
\includegraphics[scale=0.45]{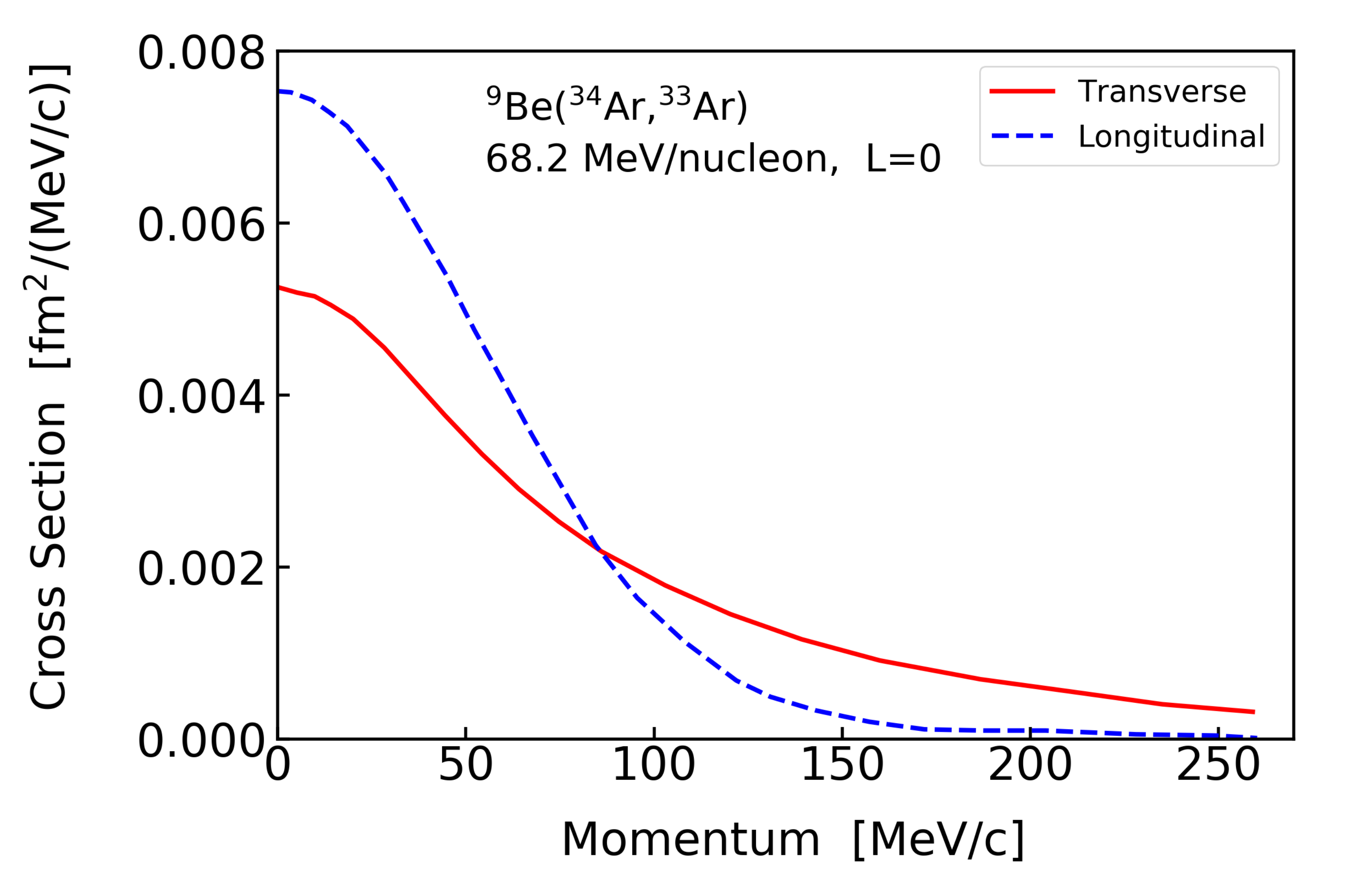}
\caption{(Color online) Transverse momentum distribution (full drawn) for the $l=0$ knockout reaction $^{9}$Be($^{34}$Ar,$^{33}$Ar(1/2$^{+}$)) at 68 MeV/nucleon and with a neutron separation energy of 17.06 MeV. The corresponding longitudinal momentum distribution (dashed line) is visibly narrower than the transverse one \cite{bertulani04}. }
\label{transvko}
\end{figure}

\subsection{Longitudinal vs. transverse momentum distributions}

The momentum distributions of core fragments of halo projectiles were often analyzed experimentally using the simple Serber formula \cite{Serber1947}
\begin{equation}
\frac{d \sigma_{c}}{d^{3} q} ={\cal C} \left| \psi_{{jl}}({\bf q})\right|^{2},
\end{equation}
where $\cal C$ is a kinematical constant and $\psi({\bf q})$ is the Fourier transform of the ground state wave function of the nucleus. In fact, this formalism works rather well for loosely-bound nuclei such as $^{11}$Li and $^{11}$Be \cite{orr92,aumann00,bertulani04}. 
It is easy to see that the Serber model arises by replacing the S-matrices by the unity in Eq. (\ref{sknock}). This means that absorption and the geometry of the nuclei  are neglected. For halo nuclei, for which most of the reaction occurs at large impact parameters,  this approximation is not so bad, as shown  in figure \ref{Serber} where the removal of $1s1/2$ neutron, with separation energy $S_n = 0.503$ MeV in the reaction $^{9}$Be($^{11}$Be,$^{10}$Be at 60 MeV/nucleon is shown \cite{aumann00}. For nucleon removal with larger separation energies, the Serber formalism is not appropriate and yields inaccurate results \cite{bertulani04,bertulani92,hencken96,MARGUERON2002105}. For example, the knock out of a deeply bound $l = 2$ neutron from Ar is shown in Fig. \ref{transvko}, which shows a transverse momentum distribution that is broader than the longitudinal distribution \cite{bertulani04}.

\begin{figure}[!ht]
\centering
\includegraphics[scale=0.45]{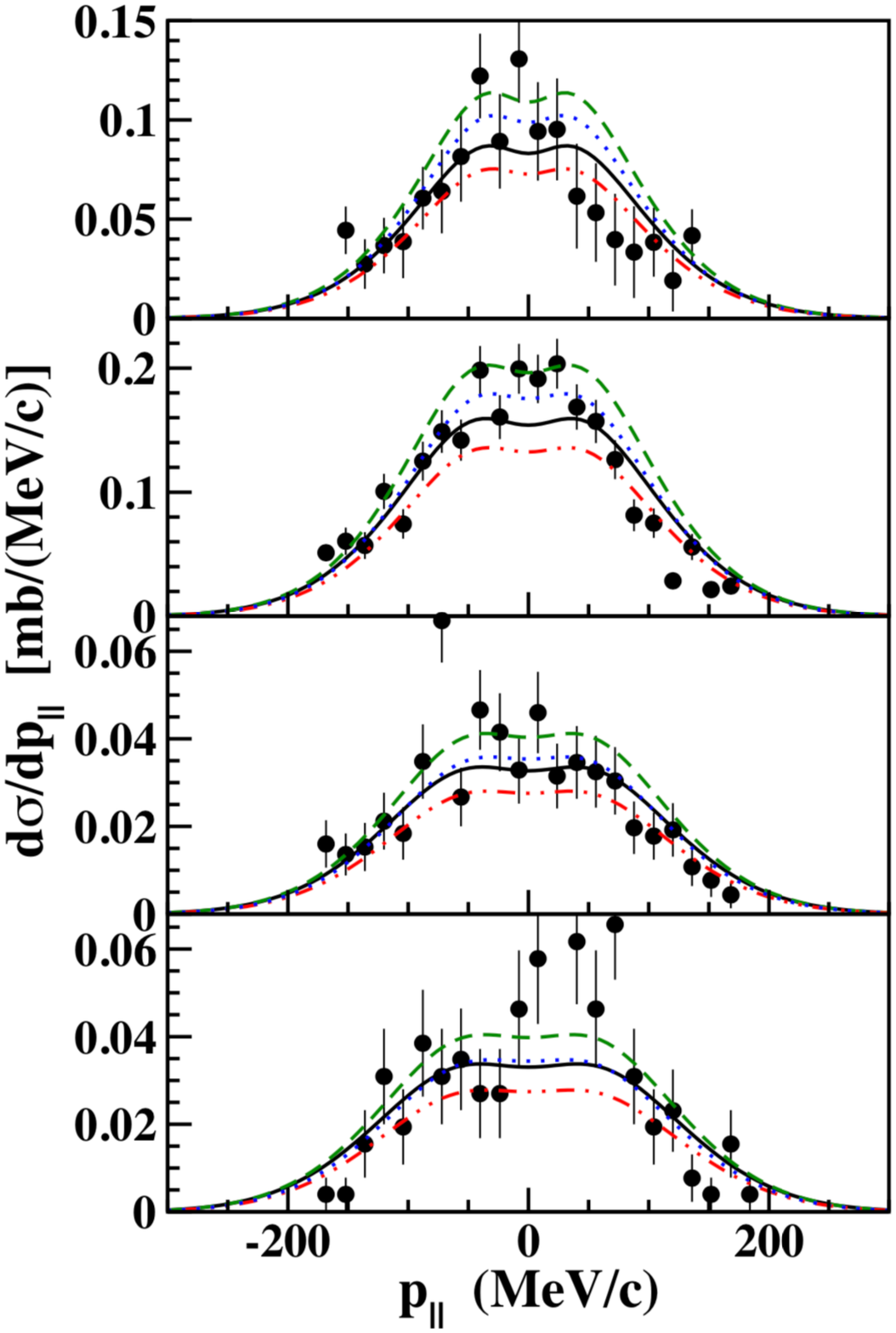}
\caption{\label{Banu} Comparison of experimental data of Ref. \cite{BanuPRC.84.015803} for proton removal from four different states
and calculations for exclusive longitudinal momentum distributions in the knockout reaction $^{12}$C($^{23}$Al,$^{22}$Mg)X   at 50~MeV/nucleon. The solid line has both Coulomb and medium corrections. The dashed-curve has no  medium corrections. The dashed-dotted line includes calculations without Coulomb corrections. The  dotted curve neither includes medium effects nor Coulomb corrections.}
\label{banu}
\end{figure}

The above discussion is of extreme relevance for knockout reactions as the spin assignment of the single particle orbitals is often accomplished through the comparison of experimental data to theoretically calculated momentum distributions. Large $l$ values yield wider distributions because the single-particle wave functions are more localized due to the centrifugal barrier. Even for the case of light targets, there exist non-negligible medium and Coulomb corrections which need to be accounted for. This is shown in Fig. \ref{Banu} with a comparison of experimental data of Ref. \cite{BanuPRC.84.015803} for proton removal from four different states and calculations for exclusive longitudinal momentum distributions in the knockout reaction $^{12}$C($^{23}$Al,$^{22}$Mg)X   at 50~MeV/nucleon. The solid line has both Coulomb and medium corrections. The dashed-curve has no  medium corrections. The dashed-dotted line includes calculations without Coulomb corrections. The  dotted curve neither includes medium effects nor Coulomb corrections \cite{Karakoc:PRC.87.024607}.

It has also been found that for some particular reactions  an accurate description of momentum distributions needs to include higher order effects that distort the tails of the momentum distributions and render them asymmetric \cite{tostevin14}. 

\subsection{Nuclear diffraction dissociation revisited}\label{difdisrevisted}

We give here a new derivation for the diffraction dissociation part in which nuclear and Coulomb effects are separated and also which allows calculations of momentum distributions. The final formula is different from Eq.(\ref{dis1})

Let us start with the eikonal breakup amplitude  defined by \cite{hencken96}
\begin{eqnarray}
A\left(  \mathbf{K,k}\right)  =\int d^{2}\mathbf{R}_{\perp}~e^{-i\mathbf{K_{\perp}\cdot R}%
_{\perp}}\int d^{3}\mathbf{r}~\psi_{\mathbf{k}}^{\ast}\left(  \mathbf{r}%
\right)  \left(  S_{c}\left(  \mathbf{b}_{c}\right)  S_{n}(\mathbf{b}%
_{n})-1\right)  \psi_{0}\left(\mathbf{r}\right).  \label{e4}%
\end{eqnarray}
The vectors
\begin{equation}
\mathbf{b}_{n}=\mathbf{R}_{\perp}+\beta_{2}\mathbf{r}_{\perp}\qquad
\mathrm{and}\qquad\mathbf{b}_{c}=\mathbf{R}_{\perp}-\beta_{1}\mathbf{r}%
_{\perp} \label{e2}%
\end{equation}
are the impact parameters of the neutron and the core with respect to the
target nucleus. Thus $\beta_{1}=m_{n}/m_{p}$, $\beta_{2}=m_{c}/m_{p}%
=1-\beta_{1}$, where $m_{n}$ is the neutron mass, $m_{c}$ is the mass of the
projectile core and m$_{p}=m_{n}+m_{c}$ is the projectile mass.  The quantities $\left(
\mathbf{K,k}\right)  $ are the momenta conjugate to the coordinates $\left(
\mathbf{R,r}\right).$ They are related to the final momenta of the core,
neutron and target by
\begin{equation}
\mathbf{k}_{c}=-\mathbf{k}+\beta_{2}\mathbf{K},\qquad\mathbf{k}_{n}%
=\mathbf{k}+\beta_{1}\mathbf{K,}\qquad\mathbf{k}_{T}=-\mathbf{K.} \label{e5}%
\end{equation}
The wave function $\psi_{\mathbf{k}}\left(  \mathbf{r}\right) $ is the final
continuum wave function of the neutron relative to the core. Eq. (\ref{e4}) can also be written as%

\begin{equation}
A\left(  \mathbf{K,k}\right)  =\int d^{2}\mathbf{R}_{\perp}~e^{-i\mathbf{K_{\perp} \cdot R}%
_{\perp}}\int d^{3}\mathbf{r}~\psi_{\mathbf{k}}^{\ast}\left(  \mathbf{r}%
\right)  S_{c}\left(  \mathbf{b}_{c}\right)  S_{n}\left(  \mathbf{b}_{n}\right)\psi
_{0}\left(  \mathbf{r}\right)  \label{e7}%
\end{equation}
because of the orthogonality of $\psi_{\mathbf{k}}\left(  \mathbf{r}\right) $
and $\psi_{0}\left(  \mathbf{r}\right)  $ (cf. Eq. (8) of Ref. \cite{hencken96}). 
Now we change the integration variable $\mathbf{R}_{\perp}$ in Eq. (\ref{e7})  to
$\mathbf{b}_{c}$ using Eq. (\ref{e2}) and then the amplitude (\ref{e7}) can also be
written as
\begin{eqnarray}
A\left(  \mathbf{K,k}\right)  =\int d^{2}\mathbf{b}_{c}~e^{-i\mathbf{K_{\perp} \cdot b}_{c}%
}S_{c}\left(  \mathbf{b}_{c}\right) \int d^{3}\mathbf{r}~\psi_{\mathbf{k}%
}^{\ast}\left(  \mathbf{r}\right)  e^{\left(  -i\beta_{1}\mathbf{K}_{\perp
}\cdot \mathbf{r}_{\perp}\right)  }S_{n}\left(  \mathbf{b}_{n}\right)  \psi
_{0}\left(  \mathbf{r}\right)  \label{e8}%
\end{eqnarray}
 
Write Eq. (\ref{e8}) as%
\begin{eqnarray}
A\left(  \mathbf{K,k}\right)  =\int d^{2}\mathbf{b}_{c}~e^{-i\mathbf{K_{\perp} \cdot b}_{c}%
}S_{c}\left(  \mathbf{b}_{c}\right)\int d^{3}\mathbf{r}~\psi_{\mathbf{k}%
}^{\ast}\left(  \mathbf{r}\right)  (e^{\left(  -i\beta_{1}\mathbf{K}_{\perp
}\cdot \mathbf{r}_{\perp}\right)  }S_{n}\left(  \mathbf{b}_{n}\right)  -1)\psi
_{0}\left(  \mathbf{r}\right) ~~~
\label{e9}%
\end{eqnarray}
where we have again used the orthogonality of $\psi_{\mathbf{k}}\left(
\mathbf{r}\right)  $ and $\psi_{0}\left(  \mathbf{r}\right)  $. 
Then we obtain $\bf {k}$ from the definition of  ${\bf {k}}_n$ 
and the transverse
component of the final neutron momentum is $\mathbf{k}_{n\perp}=\mathbf{k}%
_{\perp}+\beta_{1}\mathbf{K}_{\perp}.$\textbf{\ }

 Thus the second integral in   Eq. (\ref{e9}) can be defined as an amplitude

\begin{equation}
g(\mathbf{k}_{n}\mathbf{,b}_{c})\approx\int d^{3}\mathbf{r}~\psi_{\mathbf{k}%
}^{\ast}%
\left(  e^{\left(  i\beta_{1}\mathbf{K}_{\perp}\cdot \mathbf{r}_{\perp
}\right)  }-S_{n}\left(  \mathbf{b}_{n}\right)  \right)  \psi_{0}\left(
\mathbf{r}\right).  \label{eb2}%
\end{equation}
This is the total breakup amplitude which can be further written as a sum
\begin{equation}
g(\mathbf{k}_{n}\mathbf{,b}_{c})=g_{n}(\mathbf{k}_{n}\mathbf{,b}_{c}%
)+g_{c}(\mathbf{k}_{n}\mathbf{,b}_{c})\label{eb3}%
\end{equation}
where
\begin{eqnarray}
g_{n}(\mathbf{k}_{n}\mathbf{,b}_{c}) &  =&\int d^{3}\mathbf{r}
~\psi_{\mathbf{k}}^{\ast}\left(  1-S_{n}\left(  \mathbf{b}
_{n}\right)  \right) {\psi}_{0}\left(  \mathbf{r}\right)
\label{eb4}\\
g_{c}(\mathbf{k}_{n},\mathbf{K}_{\perp}\mathbf{b}_{c}) &  =&\int
d^{3}\mathbf{r}~\psi_{\mathbf{k}}^{\ast}
\left(e^{\left(  i\beta_{1}\mathbf{K}_{\perp} \cdot \mathbf{r}_{\perp}\right)  }-1\right)
{ \psi}_{0}(  \mathbf{r}).  \label{eb5}%
\end{eqnarray}

 The amplitude $g_{n}$ is a nuclear eikonal amplitude  Ref. \cite{MARGUERON2003337}. It depends on
the target-neutron interaction through the profile function $S_{n}\left(
\mathbf{b}_{n}\right)  $. The second integral $g_{c}$ is the recoil breakup
amplitude. It depends on the recoil momentum $\mathbf{K}_{\perp}$. 

In order to clarify the physical meaning of  Eq. (\ref{eb5}) and disentagle its effect from that of the nuclear potential, let us make a
semi-classical approximation. If the core-target profile function
$S_{c}\left(  \mathbf{b}_{c}\right)  $ is smooth and $K_{\perp}$ is large
enough $\left(  K_{\perp}b_{c}>>1\right)  $ then the integral over
$\mathbf{b}_{c}$ in (\ref{e9}) can be estimated by the method of stationary
phase. The dominant contribution comes from $\mathbf{K}_{\perp}$ parallel to
$\mathbf{b}_{c}$ and at the stationary point it will be approximated by the
classical momentum transfer
\begin{equation}
\mathbf{K}_{\perp}\approx\mathcal{K}_{\perp}\left(  \mathbf{b}_{c}\right)
=\frac{1}{\hbar}\int\mathbf{F}_{cT}\left(  \mathbf{b}_{c},vt\right)
dt\label{eb6}%
\end{equation}
where $\mathbf{F}_{cT}=-\nabla V_{cT}$ is the classical force on the
projectile core due to the core-target interaction and the integral is
calculated along the path with impact parameter $\mathbf{b}_{c}$. For a full
semi-classical evaluation of the $\mathbf{b}_{c}$ integral in Eq. (\ref{e9})
we have to assume that for each value of $\mathbf{K}_{\perp}$ there is a
unique \ core-target impact parameter which satisfies Eq. (\ref{eb6}). At this
stage we do not do this but instead approximate $\mathbf{K}_{\perp}$ in the
integral in Eq. (\ref{eb5}) by its semi-classsical value. For each value of
$\mathbf{b}_{c}$ there is a unique $\mathcal{K}_{\perp}\left(  \mathbf{b}%
_{c}\right)  $ given by Eq. (\ref{eb6}). This approximation results in a
decoupling of the two integrals in (\ref{e9}) where now the recoil amplitude
\begin{equation}
g_{c}(\mathbf{k}_{n}\mathbf{,b}_{c})=\int d^{2}\mathbf{r}_{\perp
}~e^{-i\mathbf{k}_{n}\cdot \mathbf{r}_{\perp}}\left(  e^{\left(  i\beta
_{1}\mathcal{K}_{\perp}\left(  \mathbf{b}_{c}\right)  \cdot \mathbf{r}_{\perp
}\right)  }-1\right)  \psi_{0}\left(  \mathbf{r}_{\perp},k_{z}\right)
\label{eb8}%
\end{equation}
is a function of $\mathbf{k}_{n}$ and $\mathbf{b}_{c}.$

Here we have approximated the final continuum state of the neutron with a plane wave which is a good approximation if there are no resonances in the low energy continuum.

With this approximation the breakup cross-section as a function of the neutron
momentum $\mathbf{k}_{n}$ when $\mathbf{K}_{\perp}$ is not observed is
\begin{equation}
\frac{d\sigma}{d^{3}\mathbf{k}_{n}}=\int\left|  A\left(  \mathbf{K,k}\right)
\right|  ^{2}d^{2}\mathbf{K}_{\perp}=\int d^{2}\mathbf{b}_{c}~P_{el}\left(
b_{c}\right)  \left|  g(\mathbf{k}_{n}\mathbf{,b}_{c})\right|  ^{2}.
\label{ek3}%
\end{equation}

From this equation the parallel and transverse distributions for nuclear diffraction can be obtained using the nuclear part of the amplitude that will be given in the following. Furtheremore
 $P_{el}\left(  b_{c}\right)  =\left|  S_{c}\left(  \mathbf{b}_{c}\right)
\right|  ^{2}$ is the probability that the core remains in its ground state
during the collision which is the same as in the stripping formula. 
Different ways of calculating the core-target S-matrix and their respective accuracies have been recently revised in Ref. \cite{Bonaccorso:2016,Bonaccorso:2016a}.

In general the Coulomb breakup of an odd-neutron nucleus like $^{11}$Be is due to the
core-target Coulomb interaction. Its contribution is included in the recoil amplitude
(\ref{eb5}) or (\ref{eb8}).When the recoil effect is small enough the
exponential factor in Eq. (\ref{eb8}) can be expanded to first order in
$\beta_{1}$ and the recoil amplitude reduces to the standard dipole form in
the eikonal limit%
\begin{equation}
g_{c}(\mathbf{k}_{n}\mathbf{,b}_{c})=i\beta_{1}\int d^{2}\mathbf{r}_{\perp
}~e^{-i\mathbf{k}_{n}\cdot \mathbf{r}_{\perp}}~\mathcal{K}_{\perp}\left(
\mathbf{b}_{c}\right)  \cdot \mathbf{r}_{\perp}~{\tilde \psi}_{0}\left(  \mathbf{r}_{\perp
},k_{z}\right).
\label{20}\end{equation}
The explicit expression for the momentum transfer (\ref{eb6}) is
\begin{equation}
\mathcal{K}_{\perp}\left(  \mathbf{b}_{c}\right)  =\frac{2Z_{P}Z_{T}e^{2}%
}{\hbar vb_{c}^{2}}\mathbf{b}_{c}.%
\end{equation}
Choosing the x-axis in the direction of $\mathbf{b}_{c}$, Eq. (\ref{20}) reduces to%
\begin{equation}
g_{c}(\mathbf{k}_{n}\mathbf{,b}_{c})=\beta_{1}\frac{2Z_{P}Z_{T}e^{2}}{\hbar
vb_{c}}\frac{\partial}{\partial k_{x}}\tilde{\psi}_{0}\left(
\mathbf{k}\right).\label{22}
\end{equation}
But the  Coulomb amplitude calculated in the standard dipole  approximation by time dependent perturbation theory \cite{MARGUERON2002105} reads%
\begin{equation}
g_{c}(\mathbf{k,b}_{c})=\beta_{1}\frac{2Z_{P}Z_{T}e^{2}}{\hbar vb_{c}
}\left(  \bar{\omega}K_{1}\left(  \bar{\omega}\right)  \frac{\partial
}{\partial k_{x}}+i\bar{\omega}K_{0}\left(  \bar{\omega}\right)
\frac{\partial}{\partial k_{z}}\right)  \tilde{\psi}_0\left(\mathbf{k}\right).
\end{equation}

Thus Eq. (\ref{22}) is just the sudden limit of the usual dipole Born approximation.
In fact when the adiabaticity parameter%
\begin{equation}
\bar{\omega}=\frac{\varepsilon_{k}-\varepsilon_{0}}{\hbar v}b_{c}%
\end{equation}
is small  the sudden limit  $\bar \omega\to 0$ applies and $ \bar{\omega}K_{1}\left(  \bar{\omega}\right) \to 1$ and  $\bar{\omega}K_{0}\left(  \bar{\omega}\right) \to 0$.

\smallskip

\begin{table}[]
\centering
\begin{tabular}{lccccc}
\hline\hline
\multicolumn{1}{c}{{Reaction \& s.p. orbital}}   & $E_{lab}$        & $S_n$  & {Eq. \eqref{dis1}}  & Eq. \eqref{ek33} & Eq. \eqref{dis1} no $\chi_C$ \\
\multicolumn{1}{c}{\textbf{}}    & (A.MeV)              & (MeV)                & (mb)                 & (mb)       &(mb)                \\
\hline\hline
\multicolumn{1}{c}{$^{11}$Be($2s_{1/2}$) + $^9$Be}           & 40                   & 0.54                 & 84.94                & 87.06      &78.41                \\
                                                & 70                   & 0.54                 & 65.44                & 65.49    & 61.21                  \\
                                                & 150                  & 0.54                 & 36.97                & 36.01     & 35.07                 \\
                                                & 1000                 & 0.54                 & 24.78                & 24.42    &23.9 \\ \hline
\multicolumn{1}{c}{$^{11}$Be($2s_{1/2}$) + $^{208}$Pb}            & 40                   & 0.54                 & 1311                 & 308.7  &
229.3
                  \\
                                                & 70                   & 0.54                 & 925.4                & 277.3     &208.2
                 \\
                                                & 150                  & 0.54                 & 603.4                & 269.9    & 196.0
              \\
                                                & 1000                 & 0.54                 & 258.4                & 178.1    &129.8 \\ \hline
\multicolumn{1}{c}{$^{14}$O(1p$_{3/2}$) + $^9$Be}          & 53                   & 23.18                & 3.431                & 3.321 &1.988                     \\
\hline
\multicolumn{1}{c}{$^{32}$Ar(1d$_{5/2}$) + $^9$Be}            & 70                   & 21.56                & 2.942                & 1.681   &
1.212 \\
\hline
\multicolumn{1}{c}{$^{32}$Ar(1d$_{5/2}$) + $^{208}$Pb}         & 70                   & 21.56                & 397.5                & 5.747 & 3.815                        \\
\hline
\multicolumn{1}{c}{$^{24}$Si(1d$_{5/2}$) + $^9$Be}        & 85.3                 & 21.09                & 2.962                & 2.33      &1.494                 \\
\hline\hline
\end{tabular}
\caption{Diffraction dissociation cross sections for the indicated initial states in a number of reactions, first column; incident energy, second column; initial state separation energy, and cross sections values as indicated.}\label{angtab}
\end{table}

If we consider only the nuclear part of the amplitude Eq. (\ref{eb4}), use again the approximation of a plane wave for the final wave function of the neutron and integrate over the neutron momentum $d^3{\bf k}_n$ the total  nuclear diffraction 
cross section becomes \cite{MARGUERON2002105,MARGUERON2003337}

\begin{equation}
\sigma_{el}=\int d^{2}\mathbf{b}_{c}~|S_{ct}\left(\mathbf 
{b}_{c}\right) |^2\int d^{2}\mathbf{r}_{\perp}~\left(| 1-S_{n}\left( \mathbf{b}
_{n}\right ) \right|^2 |{\tilde \psi}_{0}\left(  \mathbf{r}_{\perp}\right)|^2.
\label{ek33}%
\end{equation}
The general eikonal expressions Eq.
(\ref{e7}) and (\ref{eb2})  have been used instead in
\cite{YABANA1992295,hencken96,bertulani06} and by many other authors, without the steps discussed above to separate the nuclear and Coulomb parts thus  leading to Eq. (2.19) of \cite{YABANA1992295}, Eq. (10) of \cite{hencken96} and  Eq. (13) of \cite{bertulani06} and Eq. \ref{dis1} of this paper which are all equivalent and read
\begin{eqnarray}
\sigma_{el}=\int &d^{2}\mathbf{b}_{c}\int d^{2}\mathbf{r}~| S_{ct}\left(
\mathbf{b}_{c}) S_{n}\left(  \mathbf{b}%
_{n}\right ) \right|^2 |{\tilde \psi}_{0}\left(  \mathbf{r}\right)|^2-\int d^{2}\mathbf{b}_{c} ~\left| \int d^{2}\mathbf{r}~S_{ct}(
\mathbf{b}_{c}) S_{n}\left(  \mathbf{b}%
_{n}\right ) \ |{\tilde \psi}_{0}\left(  \mathbf{r}\right)|^2\right |^2.
\label{ek44}%
\end{eqnarray}

 On the basis of the previous equations  if one goes from Eq. (\ref{e7})  to Eq. (\ref{ek3}) and then  to Eq.  (\ref{ek44}) the effect of recoil is automatically included, which corresponds to what is also usually called Coulomb breakup calculated in the sudden limit. It is well known that the sudden approximation to the Coulomb breakup gives too large cross sections \cite{MARGUERON2002105,MARGUERON2003337,BAUR1986188,Baur_2007}. Therefore when applied to breakup on a heavy target results  from Eq. (\ref{ek44})  will be in general larger than what one would get from Eq. (\ref{ek33}) where only the nuclear part of the amplitude has been used. 

In Table \ref{angtab} we present a few results for Eq. (\ref{dis1}) calculated with and without the Coulomb phase and Eq. (\ref{ek33}), for the reactions indicated in the first column.  The results were obtained using a modified version of the code MOMDIS \cite{bertulani06}. It is clear that the Coulomb phase cannot be used in Eq. (\ref{dis1}) in particular for the heavy targets.
The results are different from those of Eq. (\ref{ek33}) because of the interference between the nuclear and recoil amplitude in Eq. (\ref{dis1}). Eq. (\ref{ek33}) can be used for any type of target because it is free from ambiguities on its implementation, furthermore it has the same structure as the stripping formula Eq. (\ref{strip1}) and as we said above, from its derivation one can also obtain momentum distribution

We have  therefore  clarified that Eq. (\ref{ek44}) which is the same as Eq. (\ref{dis1})  cannot be defined as the equation representing  only the nuclear elastic breakup part. From the point of view of the formalism it contains also the Coulomb breakup calculated in the sudden approximation. On light targets it  gives  close results to Eq. (\ref{ek33}), within other numerical uncertainties. On the other hand on heavy targets it  can  give unreliable cross sections,  because it contains an interference term and it would be particularly unreliable to make predictions on the nuclear part of elastic breakup while  Eq. (\ref{ek33}) provides a safer alternative.  Finally we propose to follow the method introduced in \cite{MARGUERON2002105} and \cite{MARGUERON2003337} to calculate consistently nuclear and Coulomb breakup for a neutron, while for proton breakup we propose to follow \cite{PhysRevC.76.014607,PhysRevC.84.014613}. In these references Coulomb breakup is treated to all orders and all multipolarities.


\bibliographystyle{unsrt}
\bibliography{EMMInoURL,alex,daniel,transfer_v3,16o,hi}

\end{document}